\documentclass[12pt]{article}
\usepackage{eqsection,subeqnarray,indent,amsfonts,amssymb,amsmath}
\usepackage{bm}    
\usepackage{cite}  
\usepackage{pstricks,pst-node,pst-text,pst-3d,pst-plot}  
\usepackage{epic,eepic}
\usepackage{rotating}  
\usepackage{graphicx}
\usepackage{url}       

\begin{document}

\bibliographystyle{plain}

\date{November 9, 2007 \\[1.5mm] 
revised February 6, 2009}

\title{\vspace*{-1cm}
       Transfer Matrices and Partition-Function Zeros \\
       for Antiferromagnetic Potts Models \\[5mm]
   \hspace*{-10mm}  
   \large\bf V.~Further Results for the Square-Lattice Chromatic Polynomial}

\author{
  \\[-0.5cm]
  {\small Jes\'us Salas}                                    \\[-0.2cm]
  {\small\it Instituto Gregorio Mill\'an}\\[-0.2cm]  
  {\small\it and}\\[-0.2cm]  
  {\small\it Grupo de Modelizaci\'on, Simulaci\'on Num\'erica y 
             Matem\'atica Industrial} \\[-0.2cm]
  {\small\it Universidad Carlos III de Madrid} \\[-0.2cm]
  {\small\it Avda.\  de la Universidad, 30}    \\[-0.2cm]
  {\small\it 28911 Legan\'es, SPAIN}           \\[-0.2cm]
  {\small\tt JSALAS@MATH.UC3M.ES}              \\[4mm]
  {\small Alan D.~Sokal\thanks{Also at Department of Mathematics,
           University College London, London WC1E 6BT, England.}} \\[-0.2cm]
  {\small\it Department of Physics}       \\[-0.2cm]
  {\small\it New York University}         \\[-0.2cm]
  {\small\it 4 Washington Place}          \\[-0.2cm]
  {\small\it New York, NY 10003 USA}      \\[-0.2cm]
  {\small\tt SOKAL@NYU.EDU}               \\[-0.2cm]
  {\protect\makebox[5in]{\quad}}  
  \\
}

\maketitle
\thispagestyle{empty}   


\vspace*{-0.5cm}

\begin{abstract}
We derive some new structural results for the transfer matrix
of square-lattice Potts models with free and cylindrical boundary conditions.
In particular, we obtain explicit closed-form expressions for the
dominant (at large $|q|$) diagonal entry in the transfer matrix,
for arbitrary widths $m$, as the solution of a special one-dimensional 
polymer model.
We also obtain the large-$q$ expansion of 
the bulk and surface (resp.\ corner) free energies
for the zero-temperature antiferromagnet (= chromatic polynomial)
through order $q^{-47}$ (resp.\ $q^{-46}$).
Finally, we compute chromatic roots for strips of widths $9\le m\le 12$
with free boundary conditions and locate roughly the limiting curves.
\end{abstract}

\bigskip
\noindent
{\bf Key Words:}  Chromatic polynomial; chromatic root;
antiferromagnetic Potts model; square lattice;
transfer matrix; Fortuin--Kasteleyn representation; 
Beraha--Kahane--Weiss theorem; large-$q$ expansion;
one-dimensional polymer model; finite-lattice method.

\clearpage

\newcommand{\be}{\begin{equation}}
\newcommand{\ee}{\end{equation}}
\newcommand{\<}{\langle}
\renewcommand{\>}{\rangle}
\newcommand{\widebar}{\overline}
\def\reff#1{(\protect\ref{#1})}
\def\spose#1{\hbox to 0pt{#1\hss}}
\def\ltapprox{\mathrel{\spose{\lower 3pt\hbox{$\mathchar"218$}}
 \raise 2.0pt\hbox{$\mathchar"13C$}}}
\def\gtapprox{\mathrel{\spose{\lower 3pt\hbox{$\mathchar"218$}}
 \raise 2.0pt\hbox{$\mathchar"13E$}}}
\def\textprime{${}^\prime$}
\def\proof{\par\medskip\noindent{\sc Proof.\ }}
\newcommand{\qed}{\quad $\Box$ \medskip \medskip}
\def\proofof#1{\bigskip\noindent{\sc Proof of #1.\ }}
\def\half{ {1 \over 2} }
\def\third{ {1 \over 3} }
\def\twothird{ {2 \over 3} }
\def\smfrac#1#2{\textstyle{#1\over #2}}
\def\smhalf{ \smfrac{1}{2} }
\newcommand{\real}{\mathop{\rm Re}\nolimits}
\renewcommand{\Re}{\mathop{\rm Re}\nolimits}
\newcommand{\imag}{\mathop{\rm Im}\nolimits}
\renewcommand{\Im}{\mathop{\rm Im}\nolimits}
\newcommand{\sgn}{\mathop{\rm sgn}\nolimits}
\newcommand{\tr}{\mathop{\rm tr}\nolimits}
\newcommand{\diag}{\mathop{\rm diag}\nolimits}
\newcommand{\Gal}{\mathop{\rm Gal}\nolimits}
\newcommand{\mycup}{\mathop{\cup}}
\def\hboxscript#1{ {\hbox{\scriptsize\em #1}} }
\def\zhat{ {\widehat{Z}} }
\def\phat{ {\widehat{P}} }
\def\qtilde{ {\widetilde{q}} }
\def\hboxrm#1{ {\hbox{\scriptsize\rm #1}} }
\renewcommand{\emptyset}{\varnothing}

\def\scra{\mathcal{A}}
\def\scrb{\mathcal{B}}
\def\scrc{\mathcal{C}}
\def\scrd{\mathcal{D}}
\def\scrf{\mathcal{F}}
\def\scrg{\mathcal{G}}
\def\scrl{\mathcal{L}}
\def\scro{\mathcal{O}}
\def\scrp{\mathcal{P}}
\def\scrq{\mathcal{Q}}
\def\scrr{\mathcal{R}}
\def\scrs{\mathcal{S}}
\def\scrt{\mathcal{T}}
\def\scrv{\mathcal{V}}
\def\scrz{\mathcal{Z}}

\def\q{{\sf q}}

\def\Z{{\mathbb Z}}
\def\R{{\mathbb R}}
\def\C{{\mathbb C}}
\def\Q{{\mathbb Q}}

\def\T{{\mathsf T}}
\def\H{{\mathsf H}}
\def\V{{\mathsf V}}
\def\D{{\mathsf D}}
\def\J{{\mathsf J}}
\def\P{{\mathsf P}}
\def\QQ{{\mathsf Q}}
\def\RR{{\mathsf R}}

\def\bone{{\mathbf 1}}
\def\bv{{\bf v}}
\def\basise{{\bf e}}   
\def\basisf{{\bf f}}   
\def\startv{{\boldsymbol{\alpha}}}   
\def\endv{{\boldsymbol{\omega}}}     
\def\bsigma{{\boldsymbol{\sigma}}}

\newtheorem{theorem}{Theorem}[section]
\newtheorem{definition}[theorem]{Definition}
\newtheorem{proposition}[theorem]{Proposition}
\newtheorem{lemma}[theorem]{Lemma}
\newtheorem{corollary}[theorem]{Corollary}
\newtheorem{conjecture}[theorem]{Conjecture}
\newtheorem{question}[theorem]{Question}

%
%
\newcommand{\ofo}{\, {{}_1 \! F_1} }
\newcommand{\tfo}{\, {{}_2 \!\!\: F_1} }
\newcommand{\tforeg}{\, {{}_2 \!\!\: \widetilde{F}_1} }
\newcommand{\pfq}{\, {{}_p \!\!\: F_q} }
\newcommand{\pfqreg}{\, {{}_p \!\!\: \widetilde{F}_q} }

\newenvironment{sarray}{
          \textfont0=\scriptfont0
          \scriptfont0=\scriptscriptfont0
          \textfont1=\scriptfont1
          \scriptfont1=\scriptscriptfont1
          \textfont2=\scriptfont2
          \scriptfont2=\scriptscriptfont2
          \textfont3=\scriptfont3
          \scriptfont3=\scriptscriptfont3
        \renewcommand{\arraystretch}{0.7}
        \begin{array}{l}}{\end{array}}

\newenvironment{scarray}{
          \textfont0=\scriptfont0
          \scriptfont0=\scriptscriptfont0
          \textfont1=\scriptfont1
          \scriptfont1=\scriptscriptfont1
          \textfont2=\scriptfont2
          \scriptfont2=\scriptscriptfont2
          \textfont3=\scriptfont3
          \scriptfont3=\scriptscriptfont3
        \renewcommand{\arraystretch}{0.7}
        \begin{array}{c}}{\end{array}}

%
%
\newcommand{\stirlingsubset}[2]{\genfrac{\{}{\}}{0pt}{}{#1}{#2}}
\newcommand{\stirlingcycle}[2]{\genfrac{[}{]}{0pt}{}{#1}{#2}}
\newcommand{\assocstirlingsubset}[3]{%
{\genfrac{\{}{\}}{0pt}{}{#1}{#2}}_{\! \ge #3}}
\newcommand{\assocstirlingcycle}[3]{{\genfrac{[}{]}{0pt}{}{#1}{#2}}_{\ge #3}}
\newcommand{\euler}[2]{\genfrac{\langle}{\rangle}{0pt}{}{#1}{#2}}
\newcommand{\eulergen}[3]{{\genfrac{\langle}{\rangle}{0pt}{}{#1}{#2}}_{\! #3}}
\newcommand{\eulersecond}[2]{\left\langle\!\! \euler{#1}{#2} \!\!\right\rangle}
\newcommand{\eulersecondgen}[3]{%
{\left\langle\!\! \euler{#1}{#2} \!\!\right\rangle}_{\! #3}}
\newcommand{\binomvert}[2]{\genfrac{\vert}{\vert}{0pt}{}{#1}{#2}}

%
%

\tableofcontents  
\clearpage        

%
%
\section{Introduction}   \label{sec.intro}

The Potts model \cite{Potts_52,Wu_82,Wu_82E,Wu_84}
on a regular lattice ${\cal L}$ is characterized by two parameters:
the number $q$ of Potts spin states,
and the nearest-neighbor coupling $v = e^{\beta J}-1$.\footnote{
   Here we are considering only the {\em isotropic}\/ model,
   in which each nearest-neighbor edge is assigned the same coupling $v$.
   In a more refined analysis, one could put (for example)
   different couplings $v_1,v_2$ on the horizontal and vertical edges
   of the square lattice, different couplings $v_1,v_2,v_3$ on the
   three orientations of edges of the triangular or hexagonal lattice, etc.
}
Initially $q$ is a positive integer
and $v$ is a real number in the interval $[-1,+\infty)$,
but the Fortuin--Kasteleyn representation
(reviewed in Section~\ref{sec.setup.FK} below)
shows that the partition function $Z_G(q,v)$ of the $q$-state Potts model
on any finite graph $G$ is in fact a {\em polynomial}\/ in $q$ and $v$.
This allows us to interpret $q$ and $v$ as taking arbitrary
real or even complex values,
and to study the phase diagram of the Potts model in
the real $(q,v)$-plane or in complex $(q,v)$-space.

According to the Yang--Lee picture of phase transitions \cite{Yang-Lee_52},
information about the possible loci of phase transitions can be obtained
by investigating the zeros of the partition function
for finite subsets of the lattice ${\cal L}$
when one or more physical parameters (e.g.\ temperature or magnetic field)
are allowed to take {\em complex}\/ values;
the accumulation points of these zeros in the infinite-volume limit
constitute the (possible) phase boundaries.
For the Potts model, therefore,
by studying the zeros of $Z_G(q,v)$ in complex $(q,v)$-space
for larger and larger pieces of the lattice ${\cal L}$,
we can learn about the phase diagram of the Potts model in
the real $(q,v)$-plane and more generally in complex $(q,v)$-space.

The partition function for $m \times n$ lattices
can be efficiently computed using {\em transfer matrices}\/.
Though the dimension of the transfer matrix
(and thus the computational complexity)
grows exponentially in the width $m$ ---
thereby restricting us in practice to widths $m \ltapprox 10$--30 ---
it is straightforward, by iterating the transfer matrix,
to handle quite large lengths $n$.
Indeed, by implementing the transfer-matrix method {\em symbolically}\/
(i.e., as polynomials in $q$ and/or $v$)
and using the Beraha--Kahane--Weiss theorem
(reviewed in Section~\ref{sec.setup.BKW}),
we can handle directly the limit $n \to\infty$
and compute the limiting curves $\scrb_m$ of partition-function zeros.
At a second stage we attempt to extrapolate these curves to $m=\infty$.

Since the problem of computing the phase diagram in
complex $(q,v)$-space is difficult, it has proven convenient
to study first certain ``slices'' through $(q,v)$-space,
in which one parameter is fixed (usually at a real value)
while the remaining parameter is allowed to vary in the complex plane.
One very interesting special case is the chromatic polynomial ($v=-1$),
which corresponds to the zero-temperature limit
of the Potts antiferromagnet ($\beta J=-\infty$).
In previous papers
\cite{transfer1,transfer2,transfer3,transfer4,Jacobsen-Salas_toroidal}
we have used symbolic transfer-matrix methods
to study the square-lattice and triangular-lattice chromatic polynomials
for free, cylindrical, cyclic and toroidal boundary conditions.\footnote{
   See also the bibliographies of
   \cite{transfer1,transfer2,transfer3,transfer4,Jacobsen-Salas_toroidal}
   for reference to the important related works of
   Shrock and collaborators.
}${}^,$\footnote{
   We adopt Shrock's \cite{Shrock_01_review} terminology
   for boundary conditions:
   free ($m_{\rm F} \times n_{\rm F}$),
   cylindrical ($m_{\rm P} \times n_{\rm F}$),
   cyclic ($m_{\rm F} \times n_{\rm P}$),
   toroidal ($m_{\rm P} \times n_{\rm P}$),
   M\"obius ($m_{\rm F} \times n_{\rm TP}$) and
   Klein bottle ($m_{\rm P} \times n_{\rm TP}$).
   Here the first dimension ($m$) corresponds to the transverse (``short'')
   direction, while the second dimension ($n$) corresponds to the
   longitudinal (``long'') direction.
   The subscripts F, P and TP denote free, periodic and twisted-periodic
   boundary conditions, respectively.
}
In this paper we provide some new structural results
for the transfer matrices with free and cylindrical boundary conditions.
(For simplicity we restrict attention to the square lattice,
 but the methods could easily be adapted to handle the triangular lattice.)
In particular, we shall obtain explicit closed-form expressions for the
dominant (at large $|q|$) diagonal entry in the transfer matrix,
for arbitrary widths $m$, by solving a special one-dimensional polymer gas.
We shall also obtain similar but weaker results for the dominant eigenvalue.
Finally, we shall obtain
the large-$q$ expansion of the bulk free energy
through order $q^{-47}$,
extending by 11 terms the previous computation of
Bakaev and Kabanovich \cite{Bakaev_94},
and the large-$q$ expansions of the surface and corner free energies
through orders $q^{-47}$ and $q^{-46}$, respectively.

This paper is organized as follows:
In Section~\ref{sec.setup} we 
review the Fortuin--Kasteleyn representation \cite{Kasteleyn_69,Fortuin_72}
of the $q$-state Potts model, the basic facts about transfer matrices
in this representation \cite{Blote_82,transfer1}, and 
the Beraha--Kahane--Weiss theorem
\cite{BKW_75,BKW_78,Beraha_79,Beraha_80,Sokal_chromatic_roots}. 
In Section~\ref{sec.struc} we prove some new structural properties of the
transfer matrix for a square-lattice strip of width $m$ and 
free or cylindrical boundary conditions;
in particular, we obtain closed-form expressions
for the dominant entry of the transfer matrix,
for arbitrary widths $m$.
In Section~\ref{sec.eigen} we study the large-$q$ expansion 
of the leading eigenvalue of the transfer matrix;
our results are similar to those obtained for the dominant entry,
but less explicit.
In Section~\ref{sec.thermo} we study the limit $m \to \infty$
of the strip free energy;
among other things, we compute the large-$q$ expansions
of the bulk, surface and corner free energies,
and we carry out a differential-approximant analysis
to locate the singularities of those free energies in the complex $q$-plane.
In Section~\ref{sec.numer} we provide some additional information
concerning the chromatic roots of strips of widths $9\le m\le 12$
with free boundary conditions.
Finally, in Section~\ref{sec.summary}
we list some open problems for future research.
In Appendix~\ref{appendix.combin} we prove some identities
arising in the study of the dominant transfer-matrix entry
for both boundary conditions.
In Appendix~\ref{appendix.upper}
we discuss a conjecture concerning the upper zero-free interval
for real chromatic roots of bipartite planar graphs.

%
%
\section{Preliminaries} \label{sec.setup}

In this section we review briefly some needed background
on chromatic and Tutte polynomials (Section~\ref{sec.setup.FK}),
transfer matrices (Section~\ref{sec.setup.transfer})
and the Beraha--Kahane--Weiss theorem (Section~\ref{sec.setup.BKW}).

%
%
\subsection{Chromatic polynomials, Potts models, and all that}
\label{sec.setup.FK}

Let $G=(V,E)$ be a finite undirected graph, and let $q$ be a positive integer.
Then the {\bf \protect\boldmath{$q$}-state Potts-model partition function}\/
for the graph $G$ is defined by the Hamiltonian
\begin{equation}
   H_{\rm Potts}(\bsigma)  \;=\;
   - \sum_{e=ij \in E} J_e \, \delta(\sigma_i, \sigma_j)
   \;,
 \label{def.HPotts}
\end{equation}
where the spins $\bsigma = \{\sigma_i\}_{i \in V}$
take values in $\{ 1,2,\ldots,q \}$,
the $J_e$ are coupling constants, and the $\delta$ is the Kronecker delta
\begin{equation}
   \delta(a,b)   \;=\;
   \begin{cases} 1  & \quad \mbox{\rm if $a=b$} \\ 
                 0  & \quad \mbox{\rm if $a \neq b$} 
   \end{cases} 
\end{equation}
The partition function can then be written as
\begin{equation}
   Z_G^{\rm Potts}(q, {\bf v})   \;=\;
   \sum_{ \sigma \colon\, V \to \{ 1,2,\ldots,q \} }
   \; \prod_{e=ij \in E}  \,
      \bigl[ 1 + v_e \delta(\sigma_i, \sigma_j) \bigr]
   \;,
 \label{def.ZPotts}
\end{equation}
where $v_e = e^{\beta J_e} - 1$.
Please note, in particular, that if we set $v_e = -1$ for all edges $e$,
then $Z_G^{\rm Potts}$ gives weight 1 to each proper coloring
and weight 0 to each improper coloring,
and so counts the proper colorings.
Proper $q$-colorings ($v_e=-1$) thus correspond to
the zero-temperature ($\beta \to +\infty$) limit
of the antiferromagnetic ($J_e < 0$) Potts model.

It is far from obvious that $Z_G^{\rm Potts}(q, {\bf v})$,
which is defined separately for each positive integer $q$,
is in fact the restriction to $q \in \Z_+$
of a {\em polynomial}\/ in $q$.
But this is in fact the case, and indeed we have:

\begin{theorem}[Fortuin--Kasteleyn \protect\cite{Kasteleyn_69,Fortuin_72}
                representation of the Potts model]
   \label{thm.FK}
\hfill\break
\vspace*{-4mm}
\par\noindent
For integer $q \ge 1$,
\begin{equation}
   Z_G^{\rm Potts}(q, {\bf v}) \;=\;  
   \sum_{ A \subseteq E }  q^{k(A)}  \prod_{e \in A}  v_e
   \;,
 \label{eq.FK.identity}
\end{equation}
where $k(A)$ denotes the number of connected components
in the subgraph $(V,A)$.
\end{theorem}

\proof
In \reff{def.ZPotts}, expand out the product over $e \in E$,
and let $A \subseteq E$ be the set of edges for which the term
$v_e \delta(\sigma_i, \sigma_j)$ is taken.
Now perform the sum over maps $\sigma \colon\, V \to \{ 1,2,\ldots,q \}$:
in each component of the subgraph $(V,A)$
the color $\sigma_i$ must be constant,
and there are no other constraints.
We immediately obtain \reff{eq.FK.identity}.
\qed

{\bf Historical Remark.}
The subgraph expansion \reff{eq.FK.identity} was discovered by
Birkhoff \cite{Birkhoff_12} and Whitney \cite{Whitney_32a}
for the special case $v_e = -1$ (see also Tutte \cite{Tutte_47,Tutte_54});
in its general form it is due to
Fortuin and Kasteleyn \cite{Kasteleyn_69,Fortuin_72}
(see also \cite{Edwards-Sokal}).

\bigskip

The foregoing considerations motivate defining the
{\bf multivariate Tutte polynomial}\/ of the graph $G$:
\begin{equation}
   Z_G(q, {\bf v})   \;=\;
   \sum_{A \subseteq E}  q^{k(A)}  \prod_{e \in A}  v_e
   \;,
 \label{def.ZG}
\end{equation}
where $q$ and ${\bf v} = \{v_e\}_{e \in E}$ are commuting indeterminates.
If we set all the edge weights $v_e$ equal to the same value $v$,
we obtain a two-variable polynomial that is equivalent to the
standard Tutte polynomial $T_G(x,y)$ after a simple change of variables
(see \cite{Sokal_bcc2005}).
If we set all the edge weights $v_e$ equal to $-1$,
we obtain the {\bf chromatic polynomial} $P_G(q) = Z_G(q,-1)$.

Further information on the multivariate Tutte polynomial
$Z_G(q,{\bf v})$ can be found in a recent survey article
\cite{Sokal_bcc2005}.

%
%
\subsection{Transfer matrices}   \label{sec.setup.transfer}

For any family of graphs $G_n = (V_n,E_n)$
consisting of $n$ identical ``layers''
with identical connections between adjacent layers,
the multivariate Tutte polynomials of the $G_n$
(with edge weights likewise repeated from layer to layer)
can be written in terms of a transfer matrix \cite{Blote_82,transfer1}.
Here we briefly summarize the needed formalism \cite{transfer1}
specialized to the case of an $m \times n$ square lattice
with free and cylindrical boundary conditions.
As usual, free (resp.\ cylindrical) boundary conditions means
free (resp.\ periodic) boundary conditions in the transverse (``short'')
direction and
free boundary conditions in the longitudinal (``long'') direction.
These lattices are denoted by $m_{\rm F}\times n_{\rm F}$ and 
$m_{\rm P}\times n_{\rm F}$, respectively.

Consider the $m \times n$ square grid
with edge weights $v_{i,i+1}$ on the horizontal edges ($1 \le i \le m$)
and $v_i$ on the vertical edges ($1 \le i \le m$). 
Site $m+1$ is always to be understood as a synonym for site 1.
If the weight $v_{m,m+1}\equiv v_{m,1}$  is zero (resp.\ nonzero)
we are considering free (resp.\ periodic) transverse boundary 
conditions. 

We fix the ``width'' $m$ and consider the family of graphs $G_n$
obtained by varying the ``length'' $n$;
our goal is to calculate the multivariate Tutte polynomials
$Z_{G_n}(q, {\bf v})$ for this family by building up the graph $G_n$
layer by layer.
What makes this a bit tricky is the nonlocality of the factor
$q^{k(A)}$ in \reff{def.ZG}.
At the end we will need to know the number of connected components
in the subgraph $(V_n,A)$;  in order to be able to compute this,
we shall keep track, as we go along, of which sites in the
current ``top'' layer are connected to which other sites in that layer
by a path of occupied edges (i.e.\ edges of $A$) in lower layers.
Thus, we shall work in the basis of connectivities of the top layer,
whose basis elements $\basise_{\scrp}$ are indexed by
partitions $\scrp$ of the single-layer vertex set $\{1,\ldots,m\}$.
The elementary operators we shall need are:
\begin{itemize}
   \item The {\em join operators}\/
\begin{equation}
   \J_{ij} \basise_{\scrp} \;=\; \basise_{\scrp \bullet ij}  \;,
\end{equation}
where $\scrp \bullet ij$ is the partition obtained from $\scrp$
by amalgamating the blocks containing $i$ and $j$
(if they were not already in the same block).
Note that all these operators commute.
   \item The {\em detach operators}\/
\begin{equation}
   \D_i \basise_{\scrp} \;=\;
        \begin{cases}
        \basise_{\scrp \setminus i}   & \mbox{\rm if $\{i\} \notin \scrp$} \\ 
                \noalign{\vskip 2mm}
        q \basise_{\scrp}             & \mbox{\rm if $\{i\} \in \scrp$}     
         \end{cases}     
\label{def_detach}
\end{equation}
where $\scrp \setminus i$ is the partition obtained from $\scrp$
by detaching $i$ from its block (and thus making it a singleton).
Note that these operators commute as well.
\end{itemize}
Note, finally, that $\D_k$ commutes with $\J_{ij}$ whenever
$k \notin \{i,j\}$.

The horizontal transfer matrix, which adds a row of horizontal edges, depends
on the boundary conditions, and is given by  
\begin{subeqnarray}
   \H_{\rm F}  &=&    \prod_{i=1}^{m-1} (1 + v_{i,i+1} \J_{i,i+1})
   \slabel{def_HF} \\
   \H_{\rm P}  &=&    (1 + v_{m,1} \J_{m,1}) \H_{\rm F} 
   \slabel{def_HP} 
   \label{def_H}
\end{subeqnarray}
(note that all the operators in both products commute).
The vertical transfer matrix, which adds a new row of sites along
with the corresponding vertical edges, is
\begin{equation}
   \V  \;=\;  \prod_{i=1}^{m} (v_i I + \D_i)
\label{def_V}
\end{equation}
(note once again that all the operators commute).
The multivariate Tutte polynomial for $G_n$ is then given \cite{transfer1}
by the formula
\begin{equation}
   Z_{G_n}(q, {\bf v})   \;=\;
     \endv^{\rm T} \H (\V \H)^{n-1} \basise_{\rm id}
   \;,
   \label{Z_free_FK}
\end{equation}
where ``id'' denotes the partition in which each site $i \in \{1,\ldots,m\}$
is a singleton, and the ``end vector'' $\endv^{\rm T}$ is defined by
\begin{equation}
   \endv^{\rm T} \basise_{\scrp}   \;=\;  q^{|\scrp|}
   \;.
 \label{def_uT}
\end{equation}
The transfer matrix is thus
\begin{equation}
   \T  \;=\;  \V \H   \;.
\label{def_T}
\end{equation}

In principle we are working here in the space spanned by the basis vectors
$\basise_{\scrp}$ for all partitions $\scrp$ of $\{1,\ldots,m\}$;
the dimension of this space is given by the Bell number $B_m$
\cite{Stanley_86,Stanley_99,deBruijn_61,Sloane_on-line}.
However, it is easy to see,
on topological grounds (thanks to the planarity of the $G_n$),
that only {\em non-crossing}\/ partitions can arise.
(A partition is said to be {\em non-crossing}\/ if
 $a < b < c < d$ with $a,c$ in the same block and $b,d$ in the same block
 imply that $a,b,c,d$ are all in the same block.)
The number of non-crossing partitions of $\{1,\ldots,m\}$
is given by the Catalan number \cite{Stanley_99,Sloane_on-line}
\begin{equation}
   C_m  \;=\; {(2m)! \over m! \, (m+1)!}  \;=\; {1 \over m+1} \, {2m
\choose m}
   \;.
\label{def_Cm}
\end{equation}

When the horizontal couplings $v_{i,i+1}$ are all equal to 0 or $-1$
(which is the case for the chromatic polynomial
 with either free or periodic transverse boundary conditions),
then the horizontal operator $\H$ is a {\em projection}\/
(i.e., $\H^2 = \H$),
so that our vector space $\scrv$ splits up as a direct sum
$\scrv = \scrv_0 \oplus \scrv_1$,
where $\H v=0$ for $v \in \scrv_0$ and $\H v=v$ for $v \in \scrv_1$.
Then every vector $v \in \scrv_0$ is an eigenvector of $\T = \V \H$
with eigenvalue 0;
and for each eigenpair $(\lambda,v)$ of $\T = \V \H$ with $v \notin \scrv_0$,
the pair $(\lambda, \H v)$ is an eigenpair of $\T' = \H \V \H$.
In this situation, therefore,
we can work in the smaller vector space $\scrv_1$
by using the modified transfer matrix
$\T' = \H \V \H$ in place of $\T = \V \H$,
and using the basis vectors
\begin{equation}
   \basisf_{\scrp}   \;=\;   \H \basise_{\scrp}
 \label{def_wP}
\end{equation}
in place of $\basise_{\scrp}$.
Indeed, if
$\T \basise_{\scrp} = \sum_{\scrp'} t_{\scrp \scrp'} \basise_{\scrp'}$, then
$\T' \basisf_{\scrp} = \sum_{\scrp'} t_{\scrp \scrp'} \basisf_{\scrp'}$,
as is easily seen by applying $\H$ to both sides and using $\H^2 = \H$.
Please note that $\basisf_{\scrp} = 0$ if $\scrp$ has
any pair of nearest neighbors in the same block.
We thus work in the space spanned by the basis vectors $\basisf_{\scrp}$
where $\scrp$ is a non-crossing non-nearest-neighbor partition of
$\{1,\ldots,m\}$.
The dimensionality of this space depends on the transverse 
boundary conditions:

\begin{itemize}
\item {\em Free transverse boundary conditions}\/:
The number of non-crossing non-nearest-neighbor partitions of
$\{1,\ldots,m\}$ is given
\cite[Proposition 3.6]{Simion_91} \cite{Klazar_98}
by the Motzkin number $M_{m-1}$,
where \cite{Riordan_75,Donaghey_77,Gouyou_88,Bernhart_99,%
Stanley_99,Sloane_on-line}\footnote{
   {\em Warning:}\/  Several references use the notation $m_n$
   to denote what we call $M_n$;
   and one reference \cite{Donaghey_77} writes $M_n$
   to denote a {\em different}\/ sequence.
}
\begin{equation}
   M_n  \;=\;  \sum\limits_{j=0}^{\lfloor n/2 \rfloor}
                  {n \choose 2j} \, C_j
   \;.
 \label{def_Motzkin}
\end{equation}

\item {\em Periodic transverse boundary conditions}\/: 
The number of non-crossing non-nearest-neighbor partitions
of $\{ 1,\ldots,m \}$ when it is considered periodically
(i.e.\ when 1 and $m$ also are considered to be nearest neighbors)
is given by \cite[Section 3.2, family R2]{Bernhart_99}
\begin{equation}
d_m   \;=\; \begin{cases} 1     & \mbox{\rm for $m=1$} \\
                          R_m   & \mbox{\rm for $m \ge 2$} 
            \end{cases} 
\label{def_d} 
\end{equation}
where the {\em Riordan numbers}\/ (or Motzkin alternating sums) $R_m$
\cite{Riordan_75,Donaghey_77,Bernhart_99}\footnote{
   In most of the literature (e.g.\ \cite{Riordan_75,Donaghey_77})
   these numbers are called $\gamma_m$.
   We have adopted the recent proposal of Bernhart \cite{Bernhart_99}
   to name them after Riordan \cite{Riordan_75} and denote them $R_m$.
}
are defined by $R_0 = 1$, $R_1 = 0$ and
\begin{equation}
   R_m  \;=\;  \sum_{k=0}^{m-1}  (-1)^{m-k-1} M_k
    \qquad \hbox{for $m \ge 2$}
  \label{Motzkin_sum}
\end{equation}
\end{itemize}

Finally, spatial symmetries further restrict the subspace
whenever the couplings $v_{i,i+1}$ and $v_i$
are invariant under the symmetry. Again the symmetries depend on the
transverse boundary conditions:

\begin{itemize}
\item {\em Free transverse boundary conditions}\/: 
Here the relevant symmetry is reflection in the center line of the strip.
For reflection-invariant couplings,
we can work in the space of equivalence classes of
non-crossing non-nearest-neighbor partitions modulo reflection.
The dimension $\mbox{\rm SqFree}(m)$ of the transfer matrix
is then given by the number of these equivalence classes.
The exact expression for $\mbox{\rm SqFree}(m)$ was obtained in 
\cite[Theorem 2]{Tutte_sq_02}:
\begin{equation}
\mbox{\rm SqFree}(m) \;=\; {1\over 2} M_{m-1} + {(m'-1)! \over 2} 
 \sum\limits_{j=0}^{\lfloor m'/2 \rfloor} {m'-j \over (j!)^2 (m'-2j)!}
\label{def_SqFree}
\end{equation}
where 
\begin{equation}
m' \;=\; \left\lfloor {m+1 \over 2} \right\rfloor
\label{def_mprime}
\end{equation}
and $\lfloor p \rfloor$ stands for the largest integer $\le p$.\footnote{
 We have recently rederived this formula using a different method 
 \protect\cite{square_with_extra_sites}.
} 
The asymptotic behavior of
$\mbox{\rm SqFree}(m)$ is given by \cite[Corollary 1]{Tutte_sq_02}
\begin{equation}
\mbox{\rm SqFree}(m)\;\sim\; {\sqrt{3} \over 4\sqrt{\pi} } \, 3^m \,
m^{-3/2} 
\left[ 1 + O\left( {1\over m} \right) \right] \qquad \mbox{\rm as
$m\to\infty$}
\label{def_SqFree_asymptotics}
\end{equation}

\item {\em Periodic transverse boundary conditions}\/:
For the square lattice with periodic transverse boundary conditions,
both reflections and translations are symmetries.
We therefore define equivalence classes of non-crossing non-nearest-neighbor
partitions modulo reflections and translations
and the corresponding number $\mbox{\rm SqCyl}(m)$
of equivalence classes. To our knowledge there is no known closed
form for these numbers; but there is a conjectured formula
\cite[Conjecture 2.2]{Tutte_tri_04}
\begin{equation}
{\rm SqCyl}(m) \;=\;   \begin{cases} 
         \frac{1}{2} \left[  {\rm TriCyl}(m) +
                             \frac{1}{2} N_{\rm FP}\left(\frac{m}{2}\right) 
                     \right] & \mbox{\rm for even $m$} \\[4mm] 
        \frac{1}{2} \left[  {\rm TriCyl}(m) +
                             \frac{1}{4} N_{\rm FP}\left(\frac{m+1}{2}\right)
                            -\frac{1}{2} R_{\frac{m-1}{2}} 
                      \right] & \mbox{\rm for odd $m\geq 3$} 
                     \end{cases}
\label{def_SqCyl} 
\end{equation}
where 
$N_{\rm FP}(m)$ is the number of different eigenvalues for a 
strip of either square or triangular lattice with cyclic boundary 
conditions (i.e., free transverse and periodic longitudinal boundary
conditions),
and
${\rm TriCyl}(m)$ is the number of equivalence classes of
non-crossing non-nearest-neighbor partitions modulo translations.
It is known \cite{Shrock_01b} that
\begin{equation}
N_{\rm FP}(m) \;=\; 2(m-1)! \, \sum_{j=0}^{\lfloor m/2\rfloor}
\frac{(m-j)}{(j!)^2(m-2j)!}    \;.
\label{def_SqTriCyclic}
\end{equation}
It is conjectured \cite[Conjecture 2.1]{Tutte_tri_04} that
\begin{equation}
{\rm TriCyl}(m) \;=\; \frac{1}{m}\left[ d_m + \sum_{d|m; \ 1 \le d < m}
                    \phi(m/d)\, t_d \right] 
\label{def_TriCyl}
\end{equation}
where $t_d$ is the coefficient of $z^d$ in the expansion of
$(1+z+z^2)^d$, i.e., the central trinomial coefficient (given as sequence
A002426 in \cite{Sloane_on-line}), and $\phi(x)$ is Euler's totient function.  
\end{itemize}

The values of all these dimensions for $m\leq 16$ are displayed in 
Table~2 of Ref.~\cite{transfer1}.

%
%
\subsection{Beraha--Kahane--Weiss theorem}   \label{sec.setup.BKW}

A central role in our work is played by a theorem
on analytic functions due to
Beraha, Kahane and Weiss \cite{BKW_75,BKW_78,Beraha_79,Beraha_80}
and generalized slightly by one of us \cite{Sokal_chromatic_roots}.
The situation is as follows:
Let $D$ be a domain (connected open set) in the complex plane,
and let $\alpha_1,\ldots,\alpha_M,\lambda_1,\ldots,\lambda_M$ ($M \ge 2$)
be analytic functions on $D$, none of which is identically zero.
For each integer $n \ge 0$, define
\begin{equation}
   f_n(z)   \;=\;   \sum\limits_{k=1}^M \alpha_k(z) \, \lambda_k(z)^n
   \;.
   \label{def_fn}
\end{equation}
We are interested in the zero sets
\begin{equation}
   \scrz(f_n)   \;=\;   \{z \in D \colon\;  f_n(z) = 0 \}
\end{equation}
and in particular in their limit sets as $n\to\infty$:
\begin{eqnarray}
   \liminf \scrz(f_n)   & = &  \{z \in D \colon\;
   \hbox{every neighborhood $U \ni z$ has a nonempty intersection}
\nonumber\\
      & & \qquad \hbox{with all but finitely many of the sets }
\scrz(f_n) \}
   \\[4mm]
   \limsup \scrz(f_n)   & = &  \{z \in D \colon\;
   \hbox{every neighborhood $U \ni z$ has a nonempty intersection}
\nonumber\\
      & & \qquad \hbox{with infinitely many of the sets } \scrz(f_n) \}
\end{eqnarray}
Let us call an index $k$ {\em dominant at $z$}\/ if
$|\lambda_k(z)| \ge |\lambda_l(z)|$ for all $l$ ($1 \le l \le M$);
and let us write
\begin{equation}
   D_k  \;=\;  \{ z \in D \colon\;  k \hbox{ is dominant at } z  \}
   \;.
\end{equation}
Then the limiting zero sets can be completely characterized as follows:

\begin{theorem}
  {\bf
\protect\cite{BKW_75,BKW_78,Beraha_79,Beraha_80,Sokal_chromatic_roots}}
   \label{BKW_thm}
Let $D$ be a domain in $\C$,
and let $\alpha_1,\ldots,\alpha_M$, $\lambda_1,\ldots,\lambda_M$ ($M \ge 2$)
be analytic functions on $D$, none of which is identically zero.
Let us further assume a ``no-degenerate-dominance'' condition:
there do not exist indices $k \neq k'$
such that $\lambda_k \equiv \omega \lambda_{k'}$ for some constant $\omega$
with $|\omega| = 1$ and such that $D_k$ ($= D_{k'}$)
has nonempty interior.
For each integer $n \ge 0$, define $f_n$ by
$$
   f_n(z)   \;=\;   \sum\limits_{k=1}^M \alpha_k(z) \, \lambda_k(z)^n
   \;.
$$
Then $\liminf \scrz(f_n) = \limsup \scrz(f_n)$,
and a point $z$ lies in this set if and only if either
\begin{itemize}
   \item[(a)]  There is a unique dominant index $k$ at $z$,
       and $\alpha_k(z) =0$;  or
   \item[(b)]  There are two or more dominant indices at $z$.
\end{itemize}
\end{theorem}
Note that case (a) consists of isolated points in $D$,
while case (b) consists of curves
(plus possibly isolated points where
all the $\lambda_k$ vanish simultaneously).
Henceforth we shall denote by $\scrb$ the locus of points
satisfying condition (b).

We shall often refer to the functions $\lambda_k$ as ``eigenvalues'',
because that is how they arise in the transfer-matrix formalism.

%
%
\section{Structural properties of the square-lattice transfer matrix}
\label{sec.struc}

In this section we prove some structural results concerning
the transfer matrices of square-lattice Potts models
(and in particular chromatic polynomials)
with free or cylindrical boundary conditions.
We begin by proving some general results (Section~\ref{sec.general})
concerning the polynomial dependence in $q$ of the transfer-matrix entries
and the large-$q$ behavior of the eigenvalues.
Then we provide explicit closed-form expressions for the
dominant diagonal entry of the transfer matrix
with free or cylindrical boundary conditions
(Sections~\ref{sec.struc.F} and \ref{sec.struc.P}).

%
%
\subsection{General results}
\label{sec.general}

Let us begin by considering the case of the full Potts-model partition function.
Indeed, we can be quite a bit more general,
and consider any transfer matrix built out of operators of the form
\begin{subeqnarray}
   \H  & = &   \sum\limits_{A \subseteq \{1,\ldots,m\}}  c_A \,
                    \prod\limits_{i \in A} \J_{i,i+1}      \\[2mm]
   \V  & = &   \sum\limits_{B \subseteq \{1,\ldots,m\}}  d_B \,
                    \prod\limits_{i \in B} \D_i
 \label{eq.HV.general}
 \slabel{eq.HV.general.b}
\end{subeqnarray}
with arbitrary coefficients $\{c_A\}$ and $\{d_B\}$.
(Recall that site $m+1$ is to be understood as a synonym for site $1$.)
This general form includes as particular cases
the square-lattice transfer matrix with
free or cylindrical boundary conditions
and arbitrary couplings $\{ v_{i,i+1} \}$ and $\{ v_i \}$.

Let us now recall that ``id'' denotes the partition of
$\{1,\ldots,m\}$ in which each element is a singleton (i.e., 
$\{\{1\},\{2\},\ldots,\{m\}\}$).
Let us call a partition $\scrp$ of $\{1,\ldots,m\}$ {\em non-trivial}\/
if it is not ``id''. 

\begin{proposition}
\label{prop.TM.1}
For any operators $\H$ and $\V$ of the form \reff{eq.HV.general}, 
where the coefficients $\{c_A\}$ and $\{d_B\}$ are \emph{numbers}
(i.e., independent of $q$),
the diagonal entry $t_{\rm id}(m)$ of the transfer matrix $\T(m) = \V \H$
associated to the basis element $\basise_{\rm id}$
is a polynomial in $q$ of degree at most $m$, of the form
\begin{equation}
 t_{\rm id}  \;=\;  c_\emptyset\, d_{\{1,\ldots,m\}}\, q^m \,+\, 
             \hbox{terms of order at most $q^{m-1}$}
  \;.
 \label{eq.prop.TM}
\end{equation}
All other entries of the transfer matrix $\T(m)$ are polynomials
in $q$ of degree at most $m-1$.
\end{proposition}

{\bf Remark.}
If $\H$ is a projection, then the diagonal entry $t'_{\rm id}(m)$
of the modified transfer matrix $\T'(m) = \H \V \H$
associated to the basis element $\basisf_{\rm id} = \H \basise_{\rm id}$
is given by $t'_{\rm id} = t_{\rm id}$;
indeed, {\em all}\/ the entries of $\T'(m)$
are equal to the corresponding entries of $\T(m)$.
This follows immediately from the fact that
$\T \basise_{\scrp} = \sum_{\scrp'} t_{\scrp \scrp'} \basise_{\scrp'}$
implies
$\T' \basisf_{\scrp} = \sum_{\scrp'} t_{\scrp \scrp'} \basisf_{\scrp'}$.

\proofof{Proposition~\ref{prop.TM.1}}
First of all, it is obvious that each entry in the transfer matrix 
$\T(m)$ is a polynomial in $q$. Indeed, from 
\reff{def_detach}/\reff{def_V} it is clear that we get a factor of $q$
every time we apply the operator $\D_i$ to a partition in which $i$ is
a singleton. We can {\em maximize}\/ the number of factors
of $q$ by applying the vertical transfer matrix $\V$ to the vector  
$\basise_{\rm id}$ that corresponds to the partition in which every site is
a singleton. In particular, from \reff{eq.HV.general.b} we have
\begin{subeqnarray}
   \V \basise_{\rm id} &=& 
      \sum_{B\subseteq \{1,\ldots,m\}} d_B q^{|B|} \, \basise_{\rm id} \\
   &=&  \bigl( d_{\{1,\ldots,m\}} q^m + 
           \hbox{terms of order at most $q^{m-1}$} \bigr) \, \basise_{\rm id}
        \;. \qquad
  \label{eq.tdom.1}
\end{subeqnarray}
If we apply the vertical transfer matrix to any other partition,
we get a polynomial in $q$ of degree at most $m-1$. 

Let us now consider the quantity $\H \basise_{\rm id}$:
\begin{subeqnarray}
   \H \basise_{\rm id} &=&  \sum\limits_{A \subseteq \{1,\ldots,m\}}  c_A
       \left( \prod\limits_{i \in A} \J_{i,i+1} \right)
       \basise_{\rm id} \\ 
   &=& c_{\emptyset}\, \basise_{\rm id} +
       \sum\limits_{\scrp\, \hboxrm{non-trivial}}
       a_{\scrp} \, \basise_{\scrp} 
   \;,
 \label{eq.He}
\end{subeqnarray}  
for some quantities $a_{\scrp}$ that are polynomials in $\{c_A\}$
(and of course independent of $q$).
Using \reff{eq.tdom.1}/\reff{eq.He} it is obvious that
\begin{equation}
    \V \H \basise_{\rm id}    \;=\;
  c_\emptyset \sum_{B\subseteq \{1,\ldots,m\}} d_B \, q^{|B|} \,  
     \basise_{\rm id} + 
     \sum\limits_{\scrp} a'_{\scrp}(q)\,  \basise_{\scrp} 
   \;,
\end{equation}
where the coefficients $a'_{\scrp}(q)$ are polynomials in $q$ of
degree at most $m-1$.
\qed

In view of Proposition~\ref{prop.TM.1},
we shall henceforth refer to $t_{\rm id}$ (or $t'_{\rm id}$)
as the ``dominant diagonal entry'' in the transfer matrix,
as it is indeed dominant at large $|q|$.
Furthermore, we can deduce from Proposition~\ref{prop.TM.1} 
the leading large-$|q|$ behavior of the eigenvalues.
We begin with a simple perturbation lemma:

\begin{lemma}
   \label{lemma.perturbation}
Consider an $N \times N$ matrix $M(\xi) = (M_{ij}(\xi))_{i,j=1}^N$
whose entries are analytic functions of $\xi$ in some disc $|\xi| < R$.
Suppose that $M_{11} = 1$ and that $M_{ij} = O(\xi)$ for $(i,j) \neq (1,1)$.
Then, in some disc $|\xi| < R'$, $M(\xi)$ has a simple leading eigenvalue
$\mu_\star(\xi)$ that is given by a convergent expansion
\begin{equation}
   \mu_\star(\xi)  \;=\;  1 \,+\, \sum_{k=2}^\infty \alpha_k \xi^k
 \label{eq.lemma.perturbation.1}
\end{equation}
[note that $\alpha_1 = 0$] with associated eigenvector
\begin{equation}
   \bv_\star(\xi)  \;=\;  \basise_1 \,+\, \sum_{k=1}^\infty \bv_k \xi^k
   \;,
 \label{eq.lemma.perturbation.2}
\end{equation}
while all other eigenvalues are $O(\xi)$.
\end{lemma}

The key fact here is that the eigenvalue shift
in \reff{eq.lemma.perturbation.1} begins at order $\xi^2$, not order $\xi$.

\proof
We have
\begin{equation}
   \det[\mu - M(\xi)]  \;=\;
   (\mu-1) \prod_{i=2}^N [\mu - M_{ii}(\xi)]  \,+\, \xi^2 F(\mu,\xi)
   \;,
\end{equation}
where $F(\mu,\xi)$ is a polynomial in $\mu$ whose coefficients
are analytic functions of $\xi$ for $|\xi| < R$.
The polynomial $P(\mu) = (\mu-1) \prod_{i=2}^N [\mu - M_{ii}(\xi)]$
has, for all sufficiently small $|\xi|$, a simple root at $\mu=1$
and roots (not necessarily simple) at $\mu = M_{ii}(\xi) = O(\xi)$.
The simple root at $\mu=1$ moves analytically \cite{Kato_80}
under the perturbation $\xi^2 F(\mu,\xi)$
--- let us call this root $\mu_\star(\xi)$ ---
and so is given by the convergent expansion \reff{eq.lemma.perturbation.1}.
The corresponding eigenvector also moves analytically under the
perturbation $M(\xi) = \diag(1,0,\ldots,0) + O(\xi)$,
which proves \reff{eq.lemma.perturbation.2}.
To see that all other eigenvalues are $O(\xi)$,
it suffices to consider the reduced matrix
\begin{equation}
   M(\xi)  \,-\,  \mu_\star(\xi) \, {\bv_\star(\xi) \, \bv_\star(\xi)^{\rm T}
                                     \over
                                     \bv_\star(\xi)^{\rm T} \, \bv_\star(\xi)}
\end{equation}
and observe that all its entries are $O(\xi)$.
\qed

\noindent
{\bf Remark.}  The ``small'' eigenvalues need not be analytic in $\xi$.
For instance,
\begin{equation}
   M(\xi)  \;=\;   \left( \! \begin{array}{ccc}
                                  1 & 0 & 0  \\
                                  0 & 0 & \xi \\
                                  0 & \xi^2 & 0
                             \end{array}
                   \!\right)
\end{equation}
has eigenvalues $\mu=1$ and $\mu = \pm \xi^{3/2}$.
\qed

To apply Lemma~\ref{lemma.perturbation} to our transfer matrix $\T$,
we set $\xi = q^{-1}$ and $M = \T/t_{\rm id}$.  We then have:

\begin{corollary}
   \label{cor.TMblockdiag.1and2}
Consider operators $\H$ and $\V$ of the form \reff{eq.HV.general}
where the coefficients $\{c_A\}$ and $\{d_B\}$ are \emph{numbers}
(i.e., independent of $q$) and $c_\emptyset d_{\{1,\ldots,m\}} \neq 0$.
Then $\T$ has a simple eigenvalue $\lambda_\star$
that is analytic for large $|q|$ and behaves there like
$c_\emptyset d_{\{1,\ldots,m\}} q^m$:
more precisely, it has a convergent expansion
\begin{equation}
   {\lambda_\star \over t_{\rm id}}  \;=\;
       1 \,+\, \sum_{k=2}^\infty \alpha_k q^{-k}
 \label{eq.cor.perturbation.1}
\end{equation}
[so that, in particular, $\lambda_{\star} - t_{\rm id} = O(q^{m-2})$].
All other eigenvalues are $O(q^{m-1})$.
\end{corollary}

Let us now return to the case of main interest,
in which $\H$ and $\V$ are the transfer matrices
\reff{def_H}/\reff{def_V} for the chromatic polynomial $v_i=v_{i,j}=-1$.
In this case we can sharpen \reff{eq.prop.TM}
by providing explicit expressions for the lower-order terms.
We must now distinguish between free and cylindrical boundary conditions,
and we shall treat each case in a separate subsection. 

%
%
\subsection{Free boundary conditions}
\label{sec.struc.F}

Let us consider a square-lattice grid of fixed width $m \ge 1$ and free boundary
conditions. Let us also assume that all horizontal edges have weight $v$
and all vertical edges have weight $v'$; they need not be $-1$.
The horizontal transfer matrix \reff{def_HF} is thus
\begin{equation}
   \H   \;=\; \prod_{i=1}^{m-1} (1 + v \J_{i,i+1})
   \label{def_HF_bis}
\end{equation}
and the vertical transfer matrix is
\begin{equation}
   \V   \;=\; \prod_{i=1}^{m} (v' I + \D_i)
   \label{def_V_bis}
   \;.
\end{equation}

Consider first the action of $\H$ on the start vector $\basise_{\rm id}$.
It generates $2^{m-1}$ terms, each of which corresponds to a partition
$\scrp$ in which all the blocks are sequential sets of vertices
in $\{1,\ldots,m\}$ (we shall call these sets ``polymers'').
Furthermore, each polymer of size $\ell$ picks up a factor $v^{\ell-1}$.

Consider next the action of $\V$ on a basis vector $\basise_{\scrp}$
corresponding to an arbitrary partition $\scrp = \{P_1,\ldots,P_k\}$.
If we are to end up with the partition $\basise_{\rm id}$,
then for each block $P_j$ we must either choose the
delete operator $\D_i$ for all $i \in P_j$
(the last deletion gives a factor $q$)
or else choose the delete operator for all but one $i \in P_j$
and choose $v' I$ for the last site (this can be done in $|P_j|$ ways).
We therefore have
\begin{equation}
   \V \basise_{\scrp}
   \;=\;
   \left[ \prod_{j=1}^k (q + v' |P_j|) \right] \basise_{\rm id}
   \,+\,
   \hbox{other terms}
\end{equation}
where ``other terms'' means terms involving $\basise_{\scrp'}$
with $\scrp' \neq \hbox{id}$.
We thus obtain a factor $q + \ell v'$ for each block of size 
$\ell = |P_j| \ge 1$.

Putting these facts together, we conclude that
\begin{equation}
   \V \H \basise_{\rm id}
   \;=\;
   t_{\rm F}(m) \, \basise_{\rm id}  \,+\,  \hbox{other terms}
   \;,
\end{equation}
where $t_{\rm F}(m)$ is the partition function for a
one-dimensional $m$-site polymer gas (with free boundary conditions)
in which each site must be occupied by exactly one polymer,
and each polymer of length $\ell \ge 1$ gets a fugacity 
$\mu_\ell = v^{\ell-1} (q + \ell v')$, i.e.
\begin{equation}
   t_{\rm F}(m)
   \;=\;
   \sum_{k=1}^\infty
   \sum_{\begin{scarray}
            \ell_1,\ldots,\ell_k \ge 1  \\
            \ell_1+\ldots+\ell_k = m
         \end{scarray}}
   \prod_{j=1}^k  v^{\ell_k-1} (q + \ell_k v')
   \;.
\label{def_t_F}
\end{equation}
To solve this polymer model, let us introduce the generating function
(``grand partition function'')
\begin{subeqnarray}
   \slabel{def_Phi_F}
   \Phi_{\rm F}(z)  & \equiv &  \sum_{m=1}^\infty z^m t_{\rm F}(m)   \\[2mm]
   & = &
   \sum_{k=1}^\infty
   \sum_{ \ell_1,\ldots,\ell_k \ge 1 }
   \prod_{j=1}^k  z^{\ell_j} v^{\ell_j-1} (q + \ell_j v')
      \\[2mm]
   & = &
     {\Psi(z) \over 1-\Psi(z)}
\end{subeqnarray}
where
\begin{equation}
   \Psi(z)  \;\equiv\;  \sum_{\ell=1}^\infty z^\ell v^{\ell-1} (q+\ell v')
      \;=\;
   z \left[ {q \over 1-zv} \,+\, {v' \over (1-zv)^2} \right]
   \label{def_Psi}
\end{equation}
is the total weight for a single polymer of arbitrary size.
We therefore have
\begin{equation}
   \Phi_{\rm F}(z)   \;=\;
   {(q+v')z - qvz^2  \over 1 - (q+2v+v')z + v(q+v) z^2}
   \;.
 \label{eq.Phi_F.general}
\end{equation}
When $v=v'=-1$ this reduces to
\begin{equation}
   \Phi_{\rm F}(z)   \;=\;
   {(q-1)z + qz^2  \over 1 - (q-3)z - (q-1) z^2}
   \;.
 \label{eq.Phi_F.chromatic}
\end{equation}
As a check we have expanded \reff{eq.Phi_F.chromatic} in powers of $z$,
and verified that it agrees with available dominant diagonal elements
$t_{\rm F}(m)$ for $1\leq m \leq 12$ (see \cite{transfer1} for $m\leq 8$). 

The next step is to get an explicit expression for $t_{\rm F}(m)$.
Using the notation $[z^m] P(z)$ for the $m$-th coefficient in a
polynomial or power series, we have the alternate representations
\begin{subeqnarray}
   [z^m] {1 \over 1-az-bz^2}
   & = &
   \sum_{j=0}^{\lfloor m/2 \rfloor} {m-j \choose j} a^{m-2j} b^j
       \slabel{eq.trinomial.a} \\[2mm]
   & = &
   2^{-m}
   \sum_{j=0}^{\lfloor m/2 \rfloor} {m+1 \choose 2j+1} a^{m-2j} (a^2+4b)^j
   \;.
       \slabel{eq.trinomial.b}
\end{subeqnarray}
The first of these comes directly from $[z^m] \sum_{k=0}^\infty (az+bz^2)^k$,
while the second is obtained by factoring the quadratic and using partial 
fractions. Using \reff{eq.Phi_F.chromatic} we have
\begin{equation}
   t_{\rm F}(m)  \;=\; [z^m] \Phi_{\rm F}(z)  \;=\;
   (q-1) [z^{m-1}] {1 \over 1-az-bz^2}  \,+\, q [z^{m-2}] {1 \over 1-az-bz^2}
\end{equation}
where $a=q-3$ and $b=q-1$.
Inserting this into \reff{eq.trinomial.a} we have
\begin{eqnarray}
   t_{\rm F}(m)
   & = &
   \sum_{j=0}^{\lfloor (m-1)/2 \rfloor}
         {m-1-j \choose j} (q-3)^{m-1-2j} (q-1)^{j+1}
       \nonumber \\
   & &  +\;
   q \sum_{j=0}^{\lfloor (m-2)/2 \rfloor}
         {m-2-j \choose j} (q-3)^{m-2-2j} (q-1)^{j}
   \;,
 \label{eq.tFm.chromatic}
\end{eqnarray}
which is manifestly a polynomial in $q$ of degree at most $m$.
Furthermore, the term of order $q^m$ comes only from $j=0$
in the first sum and has coefficient 1
(thereby confirming explicitly what we already knew from
Proposition~\ref{prop.TM.1}), so the degree is exactly $m$.

We may therefore write $t_F(m)$ explicitly as a polynomial in $q$
with certain coefficients $a^{\rm F}_k(m)$:
\begin{equation}
   t_{\rm F}(m) \;=\;  \sum\limits_{k=0}^m  (-1)^k a^{\rm F}_{k}(m) \, q^{m-k}
   \;.
 \label{eq.T_F}
\end{equation}
The next step is to obtain a closed formula for the coefficients  
$a^{\rm F}_k(m)$ with $m\geq 1$ and  $0 \le k \le m$.
We shall prove that, for each fixed $k \ge 0$,
the coefficient $a^{\rm F}_k(m)$ is in fact a {\em polynomial}\/ in $m$
of degree exactly $k$.\footnote{
   More precisely, $a^{\rm F}_k(m)$ is the restriction to integers
   $m \ge \max(k,1)$ of such a polynomial.
}
We begin by expanding the binomials in \reff{eq.tFm.chromatic}:
\begin{subeqnarray}
   a^{\rm F}_k(m)  & \equiv &  
           (-1)^k [q^{m-k}] t_{\rm F}(m) \\[4mm]
   & = &
   (-1)^k \sum_{j=0}^{\lfloor (m-1)/2 \rfloor}  \sum_{\ell=0}^\infty
       {m-1-j \choose j} {m-1-2j \choose m-k-\ell} {j+1 \choose \ell}
       (-3)^{k+\ell-1-2j} (-1)^{j+1-\ell}
   \nonumber \\
   &   & \quad +\;
   (-1)^k \sum_{j=0}^{\lfloor (m-2)/2 \rfloor}  \sum_{\ell=0}^\infty
       {m-2-j \choose j} {m-2-2j \choose m-k-\ell-1} {j \choose \ell}
       (-3)^{k+\ell-1-2j} (-1)^{j-\ell}
   \nonumber \\ 
   &   &  \\
   & = &
   \sum_{j=0}^{\lfloor (m-1)/2 \rfloor}  \sum_{\ell=0}^\infty
       {m-1-j \choose j} {m-1-2j \choose k+\ell-1-2j} {j+1 \choose \ell}
       (-1)^j 3^{k+\ell-1-2j}
   \nonumber \\
   &   & \quad -\;
   \sum_{j=0}^{\lfloor (m-2)/2 \rfloor}  \sum_{\ell=0}^\infty
       {m-2-j \choose j} {m-2-2j \choose k+\ell-1-2j} {j \choose \ell}
       (-1)^j 3^{k+\ell-1-2j}    \nonumber \\ 
   &   &  
 \slabel{eq.akm_sum_F}
\end{subeqnarray}

Let us consider the first sum in \reff{eq.akm_sum_F}: 
\begin{eqnarray}
S^{(1)}(m,k) &=& \sum_{j=0}^{\lfloor (m-1)/2 \rfloor}  \sum_{\ell=0}^\infty
       {m-1-j \choose j} {m-1-2j \choose k+\ell-1-2j} {j+1 \choose \ell}
       (-1)^j 3^{k+\ell-1-2j} \nonumber \\ 
        &\equiv& \sum_{j=0}^{\lfloor (m-1)/2 \rfloor}  
            \sum_{\ell=0}^\infty S^{(1)}_{j,\ell}(m,k)
   \;.
\label{eq_S1}
\end{eqnarray} 
The goal is to substitute $\sum_{j=0}^{\lfloor (m-1)/2 \rfloor}$ 
by something independent of $m$, e.g., $\sum_{j=0}^{k}$.
Indeed, if $k=\lfloor (m-1)/2 \rfloor$, the identity is trivial. Let us 
consider next the case $k<\lfloor (m-1)/2 \rfloor$. 
Then, if we prove that the sum 
\begin{equation}
\sum_{j=k+1}^{\lfloor (m-1)/2 \rfloor} S^{(1)}_{j,\ell}(m,k) \;=\; 0\,, 
\end{equation}
then, we have not modified the result of \reff{eq_S1} by changing the upper 
index in the sum over the variable $j$. The second binomial appearing in 
\reff{eq_S1} vanishes whenever $k+\ell-1-2j<0$. On the other hand, the third
binomial is {\em non-vanishing} only if $j\geq \ell-1$. Therefore, if 
$j>k$ and $j\geq \ell-1$, we have $k+\ell-1-2j<k+\ell-1-k-(\ell-1)=0$.
So, $S^{(1)}(m,k)=0$ whenever $j>k$.\footnote{
  This is true even if $m$ is treated as an algebraic indeterminate.
} 
Finally, let us consider the third case $k>\lfloor (m-1)/2 \rfloor$. Then, by 
making this change in the upper index in the sum over $j$, we are adding 
some extra terms 
\begin{equation}
\sum_{j=\lfloor (m-1)/2 \rfloor+1}^{k} S^{(1)}_{j,\ell}(m,k) 
   \;.
\label{eq_S1_case3} 
\end{equation}
In this case we have to focus on the first binomial of \reff{eq_S1}. This 
binomial is nonzero only when $0\leq j \leq \lfloor (m-1)/2 \rfloor$ or 
when $j\geq m$. The first of these do not appear in \reff{eq_S1_case3};
and since $m\geq k$, the second appears only when $j=k=m$. 
The contribution of this extra term is $1$. Thus, \reff{eq_S1} reduces to
\begin{equation}
S^{(1)}(m,k) = \sum_{j=0}^k  \sum_{\ell=0}^\infty
       {m-1-j \choose j} {m-1-2j \choose k+\ell-1-2j} {j+1 \choose \ell}
       (-1)^j 3^{k+\ell-1-2j} - \delta_{km}
   \;.
\label{eq_S1_ok}
\end{equation}

Let us now consider the second sum in \reff{eq.akm_sum_F}: 
\begin{eqnarray}
S^{(2)}(m,k) &=& \sum_{j=0}^{\lfloor (m-2)/2 \rfloor}  \sum_{\ell=0}^\infty
       {m-2-j \choose j} {m-2-2j \choose k+\ell-1-2j} {j \choose \ell}
       (-1)^j 3^{k+\ell-1-2j}
        \nonumber \\ 
        &\equiv& \sum_{j=0}^{\lfloor (m-2)/2 \rfloor}  
\sum_{\ell=0}^\infty S^{(2)}_{j,\ell}(m,k)
   \;.
\label{eq_S2}
\end{eqnarray} 
The goal is now to substitute $\sum_{j=0}^{\lfloor (m-2)/2 \rfloor}$ by 
$\sum_{j=0}^{k-1}$.
As before, the case $k-1=\lfloor (m-2)/2 \rfloor$ is trivial. Let us now 
suppose that $k-1<\lfloor (m-2)/2 \rfloor$. Then, the second binomial 
vanishes whenever $k+\ell-1-2j<0$, the third binomial is {\em non-vanishing}
only if $j\geq \ell$. Therefore for $j>k-1$ and $j\geq \ell$ we have
$k+\ell-1-2j<k+\ell-1-(k-1)-\ell=0$. So $S^{(2)}_{j,\ell}(m,k)=0$ whenever 
$j> k-1$. Let us finally consider the third case 
$k-1>\lfloor (m-2)/2 \rfloor$. Then, we should  consider the extra terms 
\begin{equation}
\sum_{j=\lfloor (m-2)/2 \rfloor+1}^{k-1} S^{(2)}_{j,\ell}(m,k) 
   \;.
\label{eq_S2_case3} 
\end{equation}
In this case we have to focus on the first binomial of \reff{eq_S2}. This 
binomial is nonzero only when $0\leq j \leq \lfloor (m-2)/2 \rfloor$ or 
when $j\geq m-1$. The first of these do not appear in 
\reff{eq_S2_case3}; and since $m\geq k$, the second appears only when 
$j=k-1=m-1$. The contribution of this extra term is again $1$. 
Thus, \reff{eq_S2} reduces to 
\begin{equation}
S^{(2)}(m,k) = \sum_{j=0}^{k-1}  \sum_{\ell=0}^\infty   
       {m-2-j \choose j} {m-2-2j \choose k+\ell-1-2j} {j \choose \ell}
       (-1)^j 3^{k+\ell-1-2j} - \delta_{km}
   \;.
\label{eq_S2_ok}
\end{equation}

Putting together \reff{eq_S1_ok}/\reff{eq_S2_ok}, we find that the 
two contributions $\delta_{km}$ cancel exactly,
and that \reff{eq.akm_sum_F} can be written as
\begin{eqnarray}
   a^{\rm F}_k(m)  & = &
   \sum_{j=0}^{k}  \sum_{\ell=0}^{j+1}
       {m-1-j \choose j} {m-1-2j \choose k+\ell-1-2j} {j+1 \choose \ell}
       (-1)^j 3^{k+\ell-1-2j}
   \nonumber \\
   &   & \quad -\;
   \sum_{j=0}^{k-1} \sum_{\ell=0}^{j}
       {m-2-j \choose j} {m-2-2j \choose k+\ell-1-2j} {j \choose \ell}
       (-1)^j 3^{k+\ell-1-2j}
   \;,
   \nonumber \\
 \label{eq.akm_sum_bis_F}
\end{eqnarray}
where the independent variable $m$ does not appear in the summation limits. 
After some straightforward but lengthy algebra we can rewrite the above 
formula in the more compact form 
\begin{eqnarray}
   a^{\rm F}_k(m)  &  = &
   \sum_{p=0}^{k}  (-1)^p {m-1-p\choose p} \sum_{r=0}^{k-p} 3^r
       {m-1-2p \choose r} {p+1 \choose k-p-r}
   \nonumber \\
& & \quad +
   \sum_{p=1}^{k}  (-1)^p {m-1-p\choose p-1} \sum_{r=0}^{k-p} 3^r
       {m-2p \choose r} {p-1 \choose k-p-r}
   \;. \qquad
\label{eq.akm_final_F}
\end{eqnarray}

{}From \reff{eq.akm_sum_bis_F} or \reff{eq.akm_final_F} we see
that $a^{\rm F}_k(m)$ is (the restriction of)
a polynomial in $m$ of degree at most $k$,
as $m$ appears only in the upper index of the binomials and
\begin{equation}
{m \choose j} \;=\; \frac{ m^{\underline{j}} }{j! } \;=\; 
\frac{m(m-1)(m-2)\cdots(m-j+1)}{j!}
\end{equation}
is a polynomial in $m$ of degree $j$.
[Here and in what follows, we use Knuth's \cite{Graham_94}
notation for falling powers:
$x^{\underline{j}} = x(x-1)(x-2)\cdots(x-j+1)$.]

The degree of the polynomial $a^{\rm F}_k(m)$ is in fact exactly $k$.
To see this, let us extract the term of order $m^k$ from \reff{eq.akm_final_F}.
The second sum in \reff{eq.akm_final_F} does not contribute,
as it is a polynomial in $m$ of order at most $k-1$;
the only contribution comes from the first sum:
\begin{equation}
 [m^k] a^{\rm F}_k(m) \;=\; 
       \sum_{p=0}^k \frac{ (-1)^p 3^{k-p} }{p! (k-p)!} 
              \;=\; \frac{2^k}{k!} \;\neq \; 0 
   \;.
\label{def_mk_ak_F} 
\end{equation}

We can summarize all this into the following proposition:

\begin{proposition}
   \label{prop.TM_F}
Let $\H$ and $\V$ be the transfer matrices
\reff{def_HF}/\reff{def_V} for the chromatic polynomial $v_i=v_{i,i+1}=-1$
with free boundary conditions.
Then the dominant diagonal entry in the transfer matrix can be written as 
\begin{equation}
t_{\rm F}(m) \;=\;  
    \sum\limits_{k=0}^m  (-1)^k a^{\rm F}_{k}(m) \, q^{m-k}
\label{eq.T_F_bis}
\end{equation} 
where each $a^{\rm F}_k(m)$ is a polynomial in $m$ of degree 
$k$ given by \reff{eq.akm_final_F}.
\end{proposition}

For the first few values of $k$, we obtain  
\begin{subeqnarray}
a^{\rm F}_0(m)  & = &   1       \\[2mm]
a^{\rm F}_1(m)  & = &   2m-1    \\[2mm]
a^{\rm F}_2(m)  & = &   2m^2 - 3m + 1       \\[2mm]
a^{\rm F}_3(m)  & = &   \smfrac{4}{3} m^3 - 4 m^2 + \smfrac{8}{3} m
                \\[2mm]
a^{\rm F}_4(m)  & = &   \smfrac{2}{3} m^4 - \smfrac{10}{3} m^3 +
                 \smfrac{23}{6} m^2 + \smfrac{5}{6} m - 1    \\[2mm]
a^{\rm F}_5(m)  & = &   \smfrac{4}{15} m^5 - 2 m^4 + \smfrac{11}{3} m^3
              + \smfrac{3}{2} m^2 - \smfrac{133}{30} m - 2   \\[2mm]
a^{\rm F}_6(m)  & = &   \smfrac{4}{45} m^6 - \smfrac{14}{15} m^5 
              + \smfrac{23}{9} m^4  + \smfrac{7}{6} m^3
              - \smfrac{733}{90} m^2 - \smfrac{71}{15} m + 12
\end{subeqnarray}

We can also prove the following result
concerning the polynomials $a^{\rm F}_k(m)$:\footnote{
   This Proposition refers, however, to the regime $m < k$ that does not
   contribute to the sum \reff{eq.T_F_bis}.
}

\begin{proposition} \label{prop.eq.empirical.1}
For each integer $s\ge 1$, the quantity
\begin{eqnarray}
   a^{\rm F}_k(k-s)  & \!=\! &
   \sum_{p=0}^{k}  (-1)^p {k-s-1-p\choose p} \sum_{r=0}^{k-p} 3^r
       {k-s-1-2p \choose r} {p+1 \choose k-p-r}
   \nonumber \\
& & \, +
   \sum_{p=1}^{k}  (-1)^p {k-s-1-p\choose p-1} \sum_{r=0}^{k-p} 3^r
       {k-s-2p \choose r} {p-1 \choose k-p-r}
   \qquad\quad
\label{akm_F}
\end{eqnarray}
is, when restricted to $k \ge s+1$,
given by a polynomial in $k$ of degree $\max(0,s-3)$,
with leading coefficient
\begin{eqnarray}
[k^{s-3}] a^{\rm F}_k(k-s) &=& \frac{(-1)^{s+1}}{(s-3)!}  \quad 
  \text{for $s\ge 3$}
 \label{akm_F.i=3}
\end{eqnarray}
and first subleading coefficient
\begin{eqnarray}
[k^{s-4}] a^{\rm F}_k(k-s) &=& \frac{(-1)^s s}{2(s-4)!}  \quad 
  \text{for $s\ge 4$ \;.} 
 \label{akm_F.i=4}
\end{eqnarray}
\end{proposition}

The proof of this Proposition can be found in Appendix~\ref{appendix.A2}.
For the first values of $s$, we have 
\begin{subeqnarray}
   a^{\rm F}_k(k-1)  & = &  0  \qquad\hbox{for $k \ge 2$}    \\[2mm]
   a^{\rm F}_k(k-2)  & = &  0  \qquad\hbox{for $k \ge 3$}    \\[2mm]
   a^{\rm F}_k(k-3)  & = &  1  \qquad\hbox{for $k \ge 4$}    \\[2mm]
   a^{\rm F}_k(k-4)  & = &  -(k-2)  \quad\hbox{for $k \ge 5$}    \\[2mm]
   a^{\rm F}_k(k-5)  & = &  \smfrac{1}{2} (k^2 - 5k - 2)
           \qquad\hbox{for $k \ge 6$}  \\[2mm]
   a^{\rm F}_k(k-6)  & = &  -\smfrac{1}{6} (k^3 - 9k^2 - 4k + 66) \nonumber \\
                     & = & -\smfrac{1}{6} (k-3) (k^2 -6k -22)
           \qquad\hbox{for $k \ge 7$}  \\[2mm]
   a^{\rm F}_k(k-7)  & = &  \smfrac{1}{24} (k^4 - 14k^3 - k^2 + 350k - 384)
           \qquad\hbox{for $k \ge 8$}  \\[2mm]
   a^{\rm F}_k(k-8)  & = &  \smfrac{1}{120} (k^5 - 20k^4 + 15k^3 + 1100k^2
                              - 3016k - 2400)  \nonumber \\
                     & = &  \smfrac{1}{120} (k-4) (k^4 - 16k^3 - 49k^2 +
                              904k + 600)
           \quad\hbox{for $k \ge 9$} \quad 
 \label{eq.empirical.1}
\end{subeqnarray}
In particular, we see from (\ref{eq.empirical.1}a,b) that
$(m-k+1)(m-k+2)$ is a factor of $a^{\rm F}_k(m)$ for $k \ge 3$.

We find empirically that, for each integer $s \ge 3$, the polynomial
$p_s^{\rm F}(k)$ that matches $(-1)^{s-1} (s-3)! \, a^{\rm F}_k(k-s)$
for integer $k \ge s+1$ has all {\em integer}\/ coefficients;
and we further find empirically that for \emph{even}\/ integers $s \ge 4$,
we have $p_s^{\rm F}(s/2) = 0$, 
so that the polynomial $p_s^{\rm F}(k)$ has $k-s/2$ as a factor.

We also find
\begin{subeqnarray}
   a^{\rm F}_k(k)    & = &  F_{2k}  \qquad\hbox{for $k \ge 1$}
       \slabel{eq.fibonacci.F.s=0}    \\[2mm]
   a^{\rm F}_k(k+1)  & = &  {(2k+1) F_{2k+2} - (k-4) F_{2k+1}  \over 5}
       \slabel{eq.fibonacci.F.s=1}
       \label{eq.fibonacci.F.s=0and1}
\end{subeqnarray}
where
\begin{equation}
   F_n  \;=\;  \frac{1}{\sqrt{5}} \left[
                \biggl( \frac{1+\sqrt{5}}{2} \biggr)^{n}  \,-\,
                \biggl( \frac{1-\sqrt{5}}{2} \biggr)^{n}
                \right]
 \label{def.fibonacci}
\end{equation}
are the Fibonacci numbers:
see \cite[sequences A001906/A088305 and A038731]{Sloane_on-line}.
We prove \reff{eq.fibonacci.F.s=0} in Appendix~\ref{appendix.A4}.
We have checked \reff{eq.fibonacci.F.s=1} up to $k=100$,
but do not have any proof.\footnote{
   With some more work it might be possible to find
   a proof of \reff{eq.fibonacci.F.s=1} using the same strategy as
   was used for the proof of \reff{eq.fibonacci.F.s=0}
   in Appendix~\ref{appendix.A4}.
} 
Please note that (\ref{eq.fibonacci.F.s=0and1}a,b)
give the low-order coefficients in the polynomials $t_{\rm F}(m)$:
\begin{subeqnarray}
   a^{\rm F}_k(k)    & = &  (-1)^k [q^0] t_{\rm F}(k)  \\[2mm]
   a^{\rm F}_k(k+1)  & = &  (-1)^k [q^1] t_{\rm F}(k+1)
\end{subeqnarray}

Since we have proven that $a^{\rm F}_k(m)$ is a polynomial in $m$ 
of degree $k$, it is also of interest to obtain explicit expressions for the
coefficients of this polynomial, which we write as
\begin{equation}
a^{\rm F}_k(m) \;=\; \sum_{\ell=0}^k
             \frac{ (-1)^\ell 2^{k-2\ell+1}}{(k-\ell)! (\ell+2)!} \;
             a^{\rm F}_{k,\ell} \; m^{k-\ell}
   \;;
\label{def_akl}
\end{equation}
here the prefactors have been chosen to make many (though not all)
of the coefficients $a^{\rm F}_{k,\ell}$ integers
(in fact, all of them are integers for $\ell\leq 5$, see below).
Now we use the well-known expansion of the falling powers in terms of
Stirling cycle numbers \cite{Graham_94},
\begin{equation}
x^{\underline{r}} \;=\; \sum_{c\geq 0} \stirlingcycle{r}{c}  (-1)^{r-c} x^c
\label{def_xfalling}
   \;,
\end{equation}
and expand all the binomials in \reff{eq.akm_final_F} involving $m$.
We arrive after some algebra at the following expression:
\begin{subeqnarray}
a^{\rm F}_{k,\ell} & \equiv &
   \frac{ (k-\ell)!(\ell+2)!}{(-1)^\ell 2^{k-2\ell+1}} \,
               [m^{k-\ell}] a^{\rm F}_k(m) \\[2mm]
 &=& 
\frac{(k-\ell)!(\ell+2)! (-1)^k}{2^{k-2\ell+1}} \left\{  
           \sum_{p=0}^k \sum_{r=0}^{k-p}
           {p+1 \choose k-p-r} \frac{(-3)^r }{p! r!} \right. \nonumber \\
       & & \qquad\quad \times \sum_{a=0}^p \sum_{c=0}^r
            \stirlingcycle{p}{a}  \stirlingcycle{r}{c}
            \sum_{d=0}^{k-\ell} {a\choose k-\ell-d} 
             {c \choose d} (1+2p)^{c-d} (1+p)^{a+d-k+\ell} \nonumber \\
 & & \qquad -\;  
           \sum_{p=1}^k \sum_{r=0}^{k-p}
           {p-1 \choose k-p-r} \frac{(-3)^r }{(p-1)! r!}  
            \sum_{a=0}^{p-1} \sum_{c=0}^r
            \stirlingcycle{p-1}{a}  \stirlingcycle{r}{c}
           \nonumber \\
       & & \qquad\quad \times \left. 
            \sum_{d=0}^{k-\ell} {a\choose k-\ell-d}
             {c \choose d} (2p)^{c-d} (1+p)^{a+d-k+\ell}
            \right\}  \;.
\label{def_akl_bis_F}
\slabel{def_akl_bis_F.b}
\end{subeqnarray}
By computing \reff{def_akl_bis_F.b} for integers $k \ge \ell \ge 0$,
we find {\em empirically}\/ that $a^{\rm F}_{k,\ell}$ is in fact,
for each fixed $\ell$, (the restriction of)
a {\em polynomial}\/ in $k$ of degree $\ell$.
The first few of these polynomials are:
\begin{subeqnarray}
a^{\rm F}_{k,0} &=& 1    \slabel{eq.ak0.F} \\[2mm]
a^{\rm F}_{k,1} &=& 3k+3 \\[2mm]
a^{\rm F}_{k,2} &=& 6k^2-14k+52 \\[2mm]
a^{\rm F}_{k,3} &=& 10k^3-100k^2+130k +240 \\[2mm]
a^{\rm F}_{k,4} &=& 15k^4 - 330k^3 +845 k^2 - 18 k - 1928 \\[2mm]
a^{\rm F}_{k,5} &=& 21k^5 -805k^4+ 5005k^3 +749k^2 + 8358k -87360\\[2mm]
a^{\rm F}_{k,6} &=& 28k^6 - 1652k^5 + 20020 k^4 - \smfrac{128156}{9}k^3 
  + \smfrac{278096}{3}k^2 \nonumber \\[2mm]
                & & \qquad + \smfrac{3141872}{9}k - \smfrac{3838336}{3}
\end{subeqnarray}
The fact that $a^{\rm F}_{k,0} = 1$ for all $k \ge 0$
is just a restatement of \reff{def_mk_ak_F}
[compare \reff{def_akl}].

%
%
\subsection{Cylindrical boundary conditions}
\label{sec.struc.P}

Let us now consider a square-lattice grid of fixed width $m \ge 1$
and cylindrical boundary conditions.
(Please note that for $m=1$ the horizontal edges are loops,
 and that for $m=2$ there are {\em two}\/ horizontal edges
 connecting the pair of sites in each row.)
Let us also assume that all horizontal edges have weights $v$
and all vertical edges have weights $v'$; they need not be $-1$.
We proceed analogously to the preceding subsection,
making the changes necessary to handle cylindrical rather than
free boundary conditions.

Consider first the action of $\H$ on the start vector $\basise_{\rm id}$.
It generates $2^m$ terms, each of which corresponds to a partition
$\scrp$ in which all the blocks are sequential sets of vertices
on the $m$-cycle (we shall call these sets ``polymers'').
Furthermore, each polymer of size $\ell < m$ picks up a factor $v^{\ell-1}$,
while a polymer of size $m$ picks up a factor $v^m + m v^{m-1}$
(the $v^m$ comes from the case in which all edges are occupied,
 while the $m v^{m-1}$ comes from the $m$ cases in which
 all edges but one are occupied).
The action of $\V$ is identical to that for free boundary conditions.

The upshot is that we have
\begin{equation}
   \V \H \basise_{\rm id}
   \;=\;
   t_{\rm P}(m) \, \basise_{\rm id}  \,+\,  \hbox{other terms}
   \;,
\end{equation}
where $t_{\rm P}(m)$ is the partition function for a
polymer gas on the $m$-cycle in which each polymer
of length $\ell \ge 1$ gets a fugacity 
\begin{equation}
   \widehat{\mu}_\ell  \;=\;
   \begin{cases}
       v^{\ell-1} (q + \ell v')  & \quad \mbox{\rm for $1 \le \ell \le m-1$} \\
       v^{m-1} (v+m) (q+m v')    & \quad \mbox{\rm for $\ell = m$}
   \end{cases}
\end{equation}
Please note that
\begin{equation}
   \widehat{\mu}_\ell  \;=\;
   \begin{cases}
        \mu_\ell     & \quad \mbox{\rm for $1 \le \ell \le m-1$} \\
        (v+m) \mu_m  & \quad \mbox{\rm for $\ell = m$}
   \end{cases}
\end{equation}
where $\mu_\ell$ are the fugacities for free boundary conditions
considered in the preceding subsection.

We can obtain the $t_{\rm P}(m)$ by using a simple recursion
relating the periodic and free cases:
\begin{equation}
t_{\rm P}(m) \;=\; \sum\limits_{k=1}^{m-1} k \mu_k t_{\rm F}(m-k) 
                   + \widehat{\mu}_m 
   \;.
\label{def_t_P}
\end{equation}
To see this, single out a site (e.g.\ 1) and let $k \ge 1$ be
the size of the polymer placed on it.
If $k \le m-1$, we have $k$ ways of placing this polymer
such that the selected site belongs to it,
with fugacity $\mu_k$ for each such placement;
and for the rest of the ring, the total weight of
all admissible polymer configurations is simply $t_{\rm F}(m-k)$.
Finally, if $k=m$, there is only one way of placing the polymer,
and it receives fugacity $\widehat{\mu}_m$.
This proves \reff{def_t_P}.

In order to compute explicitly the $t_{\rm P}(m)$,
it is convenient to introduce the generating function
\begin{equation}
\Phi_{\rm P}(z) \;=\; \sum\limits_{m=1}^\infty z^m t_{\rm P}(m) 
   \;.
\label{def_Phi_P}
\end{equation}
Note next that the upper limit on the sum in \reff{def_t_P}
can be changed to $\infty$, provided that we define $t_{\rm F}(\ell)=0$
for $\ell \le 0$
[which is anyway implicit in the definition \reff{def_Phi_F}
 of the generating function $\Phi_{\rm F}(z)$].
Multiplying both sides of \reff{def_t_P} by $z^m$
and summing over $m$, we arrive easily at the equation
\begin{subeqnarray}
\Phi_{\rm P}(z) &=& z \frac{d\Psi(z)}{dz} 
                      \left[ 1 + \Phi_{\rm F}(z)\right] + v \Psi(z) \\[1mm]
                &=&   \frac{z}{1-\Psi(z)} \, \frac{d\Psi(z)}{dz} + v \Psi(z) 
\slabel{eq.Phi_P.general}
\label{eq.Phi_P.general.2}
\end{subeqnarray}
where $\Psi(z) = \sum_{\ell=1}^\infty z^\ell \mu_\ell$
is defined in \reff{def_Psi}.\footnote{
   Note that the term of order $z^1$ in \reff{eq.Phi_P.general.2}
   vanishes whenever $v=-1$ (irrespective of the values of $q$ and $v'$).
   This reflects the fact that $t_{\rm P}(1) = 0$ whenever $v=-1$
   because of the loops at each vertex.
}
When $v=v'=-1$, we obtain the final formula
\begin{subeqnarray}
\Phi_{\rm P}(z) &=& \left(\frac{z}{1+z}\right)^2 
   \frac{q^2 -3q+3 +2(q-1)^2 z + q(q-1)z^2}{1 - (q-3)z - (q-1)z^2} \\[2mm]
    &=& - \, \frac{qz^2 +(q+1)z +2}{(1+z)^2}  \,+\,
          \frac{2-(q-3)z}{1 - (q-3)z - (q-1)z^2}
      \;.
\slabel{eq.Phi_P.chromatic}
\end{subeqnarray}
By expanding this function in powers of $z$, we have checked that it
agrees with the known dominant diagonal elements $t_{\rm P}(m)$
for $m \le 13$ \cite{transfer1,transfer2}.

It is now easy to extract the partition function $t_{\rm P}(m)$:
using \reff{eq.trinomial.a} we get
\begin{eqnarray}
   t_{\rm P}(m)   \;=\;   [z^m] \Phi_{\rm P}(z)
   &=& (-1)^m (q-m-2) 
       \nonumber \\
   &  &\quad +\; 2 \sum_{j=0}^{\lfloor m/2 \rfloor}
         {m-j \choose j} (q-3)^{m-2j} (q-1)^j \nonumber \\
   & &  \quad -\;  \sum_{j=0}^{\lfloor (m-1)/2 \rfloor}
         {m-1-j \choose j} (q-3)^{m-2j} (q-1)^{j}
   \;, \qquad
 \label{eq.tPm.chromatic}
\end{eqnarray}
which is manifestly a polynomial in $q$ of degree $m$. 

We can now define the coefficients $a^{\rm P}_k(m)$ in the same way as
for free boundary conditions:  
\begin{equation}
   t_{\rm P}(m) \;=\;  \sum\limits_{k=0}^m  (-1)^k a^{\rm P}_{k}(m) \, q^{m-k}
   \;,
 \label{eq.T_P}
\end{equation}
where $k$ and $m$ are integers satisfying $m \ge 1$ and $0 \le k \le m$.
However, it is slightly more convenient to extract explicitly
{\em part of}\/ the term $(-1)^m (q -m-2)$ from 
\reff{eq.tPm.chromatic},
and define $\tilde{a}^{\rm P}_k(m)$ to be the coefficients 
in what remains:
\begin{equation}
   t_{\rm P}(m) \;=\;   (-1)^m (q -m-1)  \,+\,
      \sum\limits_{k=0}^m  (-1)^k  \tilde{a}^{\rm P}_k(m) \, q^{m-k}
   \;.
 \label{eq.T_P_atilde}
\end{equation}
Notice that the relation between $\tilde{a}^{\rm P}_{k}(m)$ and
$a^{\rm P}_{k}(m)$ is rather simple: for fixed $m\geq 1$ we have that 
\begin{equation}
a^{\rm P}_{k}(m) \;=\; \begin{cases}
            \tilde{a}^{\rm P}_{k}(m)   & \quad \text{for $0\leq k \leq m-2$} \\ 
            \tilde{a}^{\rm P}_{k}(m)-1 & \quad \text{for $k =m-1$} \\ 
            \tilde{a}^{\rm P}_{k}(m)-(m+1) & \quad \text{for $k =m$} 
                        \end{cases} \;. 
\label{def_atilde_vs_a_P}
\end{equation}
Expanding the binomials in \reff{eq.tPm.chromatic}, we have
\begin{eqnarray}
   \tilde{a}^{\rm P}_k(m)
   & = &
   2 \sum_{j=0}^{\lfloor m/2 \rfloor}  \sum_{\ell=0}^\infty
       {m-j \choose j} {m-2j \choose k+\ell-2j} {j \choose \ell}
       (-1)^j 3^{k+\ell-2j}
   \nonumber \\[1mm]
   &   & \quad -\;
   \sum_{j=0}^{\lfloor (m-1)/2 \rfloor}  \sum_{\ell=0}^\infty
       {m-1-j \choose j} {m-2j \choose k+\ell-2j} {j \choose \ell}
       (-1)^j 3^{k+\ell-2j}  \;. \nonumber \\[1mm]
   &  & \quad -\ \delta_{km}
 \label{eq.akm_sum_P}
\end{eqnarray}

Again we want to substitute the $m$-dependent upper index in the sum over 
$j$ by something independent of $m$: e.g., by $k$. 

In the first sum there are two non-trivial cases: 
a) If $k<\lfloor m/2\rfloor$, then the second binomial vanishes whenever
$k+\ell-2j<0$, and the third binomial is non-vanishing only if $j\geq \ell$.
Therefore for $j>k$ and $j\geq \ell$ we have that
$k+\ell-2j < k +\ell -k-\ell=0$. So all these terms vanish.  
b) If $k>\lfloor m/2\rfloor$, then the first binomial does not vanish only when
$0\leq j \leq \lfloor m/2\rfloor$ or when $j\geq m+1$. As we are adding
terms with $\lfloor m/2 \rfloor + 1 \leq j\leq k-1\leq m-1$, none of them
give rise to a non-vanishing contribution. 

In the second sum we play a similar game:
a) If $k<\lfloor (m-1)/2\rfloor$, the binomials involved are the same
as for the first sum, so the same result applies here too. 
b) If $k>\lfloor (m-1)/2\rfloor$, then we are adding terms with  
$\lfloor (m-1)/2 \rfloor + 1 \leq j\leq k-1$. 
The first binomial does not vanish only when 
$0\leq j \leq \lfloor (m-1)/2\rfloor$ or when $j\geq m$. 
The first of these do not appear in the extra terms; and since $m\geq k$,
the second appears only when $j=k=m$, giving rise to an extra contribution 
equal to $1$. This contribution cancels out exactly the term 
$-\delta_{km}$ in \reff{eq.akm_sum_P}.  

Putting all the pieces together, we end with the following expression for
$\tilde{a}^{\rm P}_k(m)$:
\begin{equation}
   \tilde{a}^{\rm P}_k(m)
   \;=\;
   \sum_{j=0}^{k}  (-1)^j \left[ 2{m-j \choose j} - {m-1-j \choose j} \right]  
   \sum_{\ell=0}^{\infty}
       {m-2j \choose k+\ell-2j} {j \choose \ell}
       3^{k+\ell-2j}
   \;,
 \label{eq.akm_sum_bis_P}
\end{equation}
where the independent variable $m$ does not appear in the summation limits.
After some straightforward but lengthy algebra we can rewrite the above
formula in a more compact form:
\begin{equation}
   \tilde{a}^{\rm P}_k(m)  \;=\;
   3^k {m\choose k}   \,+\,
   \sum_{p=1}^k (-1)^p \, \frac{m}{p} \, {m-p-1 \choose p-1}
         \sum_{r=0}^{k-p}  3^r {m-2p \choose r} {p \choose k-p-r} \,.\quad
 \label{eq.atildekm_final_P}
\end{equation}
It is clear from \reff{eq.atildekm_final_P} that $\tilde{a}^{\rm P}_k(m)$
is (the restriction of) a polynomial in $m$ of degree at most $k$.
To see that its degree is exactly $k$,
let us extract the term of order $m^k$:
\begin{equation}
[m^k] \tilde{a}^{\rm P}_k(m) \;=\; \frac{3^k}{k!} + 
             \sum_{p=1}^k \frac{ (-1)^p 3^{k-p} }{p! (k-p)! } 
         \;=\; \frac{3^k}{k!} + \frac{2^k - 3^k}{k!}  
         \;=\; \frac{2^k}{k!} \;\neq\; 0 
   \;.
\label{def_mk_ak_P}
\end{equation}
Let us also remark that the constant term in $\tilde{a}^{\rm P}_k(m)$
vanishes whenever $k \ge 1$:
\begin{equation}
   [m^0] \tilde{a}^{\rm P}_k(m) \;=\; \delta_{k0}
   \;=\;  \begin{cases}
              1   & \quad \mbox{\rm if $k=0$} \\
              0   & \quad \mbox{\rm if $k \ge 1$}
          \end{cases}
 \label{eq.APconst}
\end{equation}

We can summarize the foregoing results in the following proposition:

\begin{proposition}
   \label{prop.TM_P}
Let $\H$ and $\V$ be the transfer matrices
\reff{def_HP}/\reff{def_V} for the chromatic polynomial $v_i=v_{i,i+1}=-1$
with cylindrical boundary conditions.
Then the dominant diagonal entry in the transfer matrix can be written as 
\begin{equation}
t_{\rm P}(m) \;=\;  (-1)^m (q -m-1) \,+\,
  \sum\limits_{k=0}^m  (-1)^k \tilde{a}^{\rm P}_{k}(m) \, q^{m-k}
\label{eq.T_P_bis}
\end{equation} 
where each $\tilde{a}^{\rm P}_k(m)$ is a polynomial in $m$ of degree $k$
given by \reff{eq.atildekm_final_P}.
\end{proposition}

The first polynomials $\tilde{a}^{\rm P}_k(m)$ are given by
\begin{subeqnarray}
   \tilde{a}^{\rm P}_0(m) & = &   1       \\[2mm]
   \tilde{a}^{\rm P}_1(m) & = &   2m         \\[2mm]
   \tilde{a}^{\rm P}_2(m) & = &   2m^2 - m   \\[2mm]
   \tilde{a}^{\rm P}_3(m) & = &   \smfrac{4}{3} m^3 - 2 m^2 
                                 -\smfrac{1}{3} m \\[2mm]
   \tilde{a}^{\rm P}_4(m) & = &   \smfrac{2}{3} m^4 - 2 m^3 -\smfrac{1}{6} m^2
                                + \smfrac{3}{2} m   \\[2mm]
   \tilde{a}^{\rm P}_5(m) & = &   \smfrac{4}{15} m^5 - \smfrac{4}{3} m^4  
                                + \smfrac{1}{3} m^3
                                + \smfrac{10}{3} m^2 - \smfrac{3}{5} m \\[2mm]
   \tilde{a}^{\rm P}_6(m) & = &   \smfrac{4}{45} m^6 - \smfrac{2}{3} m^5
                                + \smfrac{5}{9} m^4  + \smfrac{7}{2} m^3
                                - \smfrac{119}{45} m^2 - \smfrac{23}{6} m 
\end{subeqnarray}

We can also prove the following result
concerning the polynomials $\tilde{a}^{\rm P}_k(m)$:\footnote{
   This Proposition refers, however, to the regime $m < k$ that does not
   contribute to the sum \reff{eq.T_P_bis}.
}

\begin{proposition} \label{prop.eq.empirical_P.1}
For each integer $s\ge 1$, the quantity
\begin{equation}
   \!\!
   \tilde{a}^{\rm P}_k(k-s) \;=\; 
   3^k \binom{k-s}{k} \,+\,
   \sum_{p=1}^k (-1)^p \, \frac{k-s}{p} \, {k-s-p-1 \choose p-1}
   \sum_{r=0}^{k-p}  3^r {k-s-2p \choose r} {p \choose k-p-r}
\label{akm_P}
\end{equation}
is, when restricted to $k \ge s$,
given by a polynomial in $k$ of degree $s$,
with leading coefficient
\begin{eqnarray}
[k^s] \tilde{a}^{\rm P}_k(k-s) &=& \frac{(-1)^{s+1}}{s!}
\end{eqnarray}
and first subleading coefficient
\begin{eqnarray}
[k^{s-1}] \tilde{a}^{\rm P}_k(k-s)  &=& \frac{(-1)^s (s+1)}{2(s-1)!} \;. 
\end{eqnarray}
Furthermore, $\tilde{a}^{\rm P}_k(0)=0$,
so that the polynomial representing $\tilde{a}^{\rm P}_k(k-s)$
for $k \ge s$ has a factor $k-s$.
\end{proposition}

The proof of this Proposition can be found in Appendix~\ref{appendix.A3}.
For the first values of $s$, we have
\begin{subeqnarray}
 \tilde{a}^{\rm P}_k(k-1)  & = &  k-1  \qquad\hbox{for $k \ge 1$}    \\[2mm]
 \tilde{a}^{\rm P}_k(k-2)  & = &  -\smfrac{1}{2} (k^2 - 3k + 2)  \nonumber \\
                   & = &  -\smfrac{1}{2} (k-1)(k-2)
                             \qquad\hbox{for $k \ge 2$}    \\[2mm]
 \tilde{a}^{\rm P}_k(k-3)  & = &  \smfrac{1}{6} (k^3 - 6k^2 + 5k + 12) 
 \nonumber \\
                   & = &  \smfrac{1}{6} (k+1)(k-3)(k-4)
                             \qquad\hbox{for $k \ge 3$}    \\[2mm]
 \tilde{a}^{\rm P}_k(k-4)  & = &  -\smfrac{1}{24} 
         (k^4 - 10k^3 + 11k^2 + 94k -168)  \nonumber \\
                   & = &  -\smfrac{1}{24} (k+3)(k-2)(k-4)(k-7)
                             \quad\hbox{for $k \ge 4$}    \\[2mm]
 \tilde{a}^{\rm P}_k(k-5)  & = &  \smfrac{1}{120} 
         (k^5 - 15k^4 + 25k^3 + 375k^2 - 1346k + 480)  \nonumber \\
                   & = &  \smfrac{1}{120} (k-5)(k^4-10k^3-25k^2+250k-96)
                             \qquad\hbox{for $k \ge 5$}  \\[2mm]
 \tilde{a}^{\rm P}_k(k-6)  & = &  -\smfrac{1}{720} 
         (k^6 - 21k^5 + 55k^4 + 1065k^3- 6176k^2 + 3636k + 15840) \nonumber \\
                   & = &  -\smfrac{1}{720} (k-3)(k-6)(k^4 - 12k^3 - 71k^2
                                 + 642k + 880)
                             \qquad\hbox{for $k \ge 6$} \nonumber \\ 
 \label{eq.empirical_P.1}
\end{subeqnarray}

We find empirically that, for each integer $s \ge 1$, the polynomial 
$p_s^{\rm P}(k)$ that matches $(-1)^{s-1} s! \, \tilde{a}^{\rm P}_k(k-s)$
for $k \ge s$ has all {\em  integer}\/ coefficients;
and we further find empirically
that for \emph{even}\/ integers $s \ge 2$, we have 
$p_s^{\rm P}(s/2) = 0$, so that the polynomial $p_s^{\rm P}(k)$ has 
$k-s/2$ as a factor.

We also find empirically
\begin{subeqnarray}
   \tilde{a}^{\rm P}_k(k)    & = &  F_{2k+1} + F_{2k-1} - 1   \\[2mm]
   \tilde{a}^{\rm P}_k(k+1)  & = &  \sum\limits_{j=0}^{k+1} \binom{k+1}{j} j F_j
 \label{eq.empirical_P.2}
\end{subeqnarray}
where $F_n$ are again the Fibonacci numbers \reff{def.fibonacci}:
see \cite[sequences A005592 and A117202]{Sloane_on-line}.
We have checked these relationships up to $k=100$,
but do not have any proof.\footnote{
   With some more work it might be possible to find
   a proof of (\ref{eq.empirical_P.2}a,b) using the same strategy as
   was used for the proof of \reff{eq.fibonacci.F.s=0}
   in Appendix~\ref{appendix.A4}.
}
Please note that (\ref{eq.empirical_P.2}a,b) give the low-order coefficients
in the polynomials $t_{\rm P}(m)$:
\begin{subeqnarray}
   \tilde{a}^{\rm P}_k(k)    & = &  (-1)^k [q^0] t_{\rm P}(k) 
         \,+\, (k+1) \\[2mm]
   \tilde{a}^{\rm P}_k(k+1)  & = &  (-1)^k [q^1] t_{\rm P}(k+1)  \,+\, 1
\end{subeqnarray}

Since $\tilde{a}^{\rm P}_k(m)$ is a polynomial in $m$ of degree $k$,
we are again interested in obtaining the coefficients 
$\tilde{a}^{\rm P}_{k,\ell}$ defined by
\begin{equation}
\tilde{a}^{\rm P}_k(m) \;=\; \sum_{\ell=0}^k
             \frac{ (-1)^\ell 2^{k-2\ell+1}}{(k-\ell)! (\ell+2)!} \,
             \tilde{a}^{\rm P}_{k,\ell} \; m^{k-\ell}
   \;.
\label{def_akl_P}
\end{equation}
With the help of \reff{def_xfalling},
we obtain after some algebra the following result:
\begin{eqnarray}
\tilde{a}^{\rm P}_{k,\ell} &=& 
   \frac{(k-\ell)!(\ell+2)!}{2^{k-2\ell+1}} \left\{ 
   \frac{3^k}{k!} \stirlingcycle{k}{k-\ell} \right. \nonumber \\
& & \quad +\; (-1)^k \sum_{p=1}^k \sum_{r=0}^{k-p} \frac{(-3)^r}{p!r!}
   {p\choose k-p-r} \sum_{a=0}^{p-1}\sum_{c=0}^r \stirlingcycle{p-1}{a}  
    \stirlingcycle{r}{c}
    \nonumber \\
& & \left. 
    \qquad \times\;   \sum_{d=0}^{k-\ell-1} 
    {c\choose d} {a\choose k-\ell-d-1} (2p)^{c-d} (1+p)^{a-k+\ell+d+1}
    \right\}  \;. \qquad
 \label{eq.atilde_kl}
\end{eqnarray}
By computing \reff{eq.atilde_kl} for integers $k \ge \ell \ge 0$,
we find {\em empirically}\/ that $\tilde{a}^{\rm P}_{k,\ell}$ is in fact,
for each fixed $\ell$, (the restriction of)
a {\em polynomial}\/ in $k$ of degree $\ell$.
The first few are:
\begin{subeqnarray}
   \tilde{a}^{\rm P}_{k,0}  & = & 1       \slabel{eq.ak0.P} \\[2mm]
   \tilde{a}^{\rm P}_{k,1}  & = & 3k-3         \\[2mm]
   \tilde{a}^{\rm P}_{k,2}  & = & 6k^2 -38k+52   \\[2mm]
   \tilde{a}^{\rm P}_{k,3}  & = & 10k^3 -160k^2 + 390k + 0 \\[2mm]  
   \tilde{a}^{\rm P}_{k,4}  & = & 15k^4 -450k^3 +2405k^2 -1450k -7688  \\[2mm]
   \tilde{a}^{\rm P}_{k,5}  & = & 21k^5 -1015k^4 + 10465 k^3 
                                - 16121 k^2 -37030 k -151200 \\[2mm]
   \tilde{a}^{\rm P}_{k,6}  & = & 28k^6 - 1988 k^5 + 34580k^4 
                     - \smfrac{1100876}{9} k^3 
                     - \smfrac{327376}{3} k^2 \nonumber \\
        &   & \quad  + \smfrac{480752}{9} k   - \smfrac{1902976}{3} 
\end{subeqnarray}
The fact that $\tilde{a}^{\rm P}_{k,0} = 1$
for all $k \ge 0$ is just a restatement of \reff{def_mk_ak_P}.

%
%
\section{Large-$\bm q$ expansion of the leading eigenvalue}
\label{sec.eigen}

In this section we compute the large-$q$ expansion of
the leading eigenvalue $\lambda_\star(m)$
for both free and cylindrical boundary conditions,
and determine empirically some of its remarkable properties.
In Section~\ref{sec.thermo} we shall provide theoretical explanations
of some (but not all!)\ of these empirical observations.

%
%
\subsection{Overview of method and results}

In the preceding section we computed in closed form
the dominant diagonal entry in the transfer matrix, $t_{\rm id}$,
for a strip of width $m\geq 1$ with either free or cylindrical
boundary conditions (denoted $t_{\rm F}$ and $t_{\rm P}$, respectively).
We found that this entry is in each case a polynomial in $q$ of degree $m$:
\begin{subeqnarray}
   t_{\rm F}(m)  & = &  \sum\limits_{k=0}^m  (-1)^k a^{\rm F}_{k}(m) \, q^{m-k}
  \slabel{tF_atilde_c}
        \\[4mm]
   t_{\rm P}(m)  & = &  \sum\limits_{k=0}^m  (-1)^k a^{\rm P}_{k}(m) \, q^{m-k}
                            \\
                 & = &  (-1)^m (q -m-1) \,+\,
         \sum\limits_{k=0}^m  (-1)^k \tilde{a}^{\rm P}_{k}(m) \, q^{m-k}
  \slabel{tP_atilde_c}
\end{subeqnarray}
We furthermore computed in closed form the coefficients
$a^{\rm F}_{k}(m)$ and $\tilde{a}^{\rm P}_{k}(m)$,
which are in fact polynomials in $m$ of degree $k$
[cf.\ \reff{eq.akm_final_F} and \reff{eq.atildekm_final_P}].
For instance, the leading few terms for large $|q|$ are
\begin{subeqnarray}
   t_{\rm F}(m)  & = &
        q^m \,-\, (2m-1) q^{m-1}  \,+\, (2m^2 - 3m + 1) q^{m-2}
            \nonumber \\
                 &   & \qquad
            \,-\, (\smfrac{4}{3} m^3 - 4 m^2 + \smfrac{8}{3} m) q^{m-3} 
            \,+\, \ldots \quad\hbox{[for $m \ge 1$]}
        \\[2mm]
   t_{\rm P}(m)  & = &
        q^m \,-\, 2m q^{m-1}  \,+\, (2m^2 - m) q^{m-2}
            \nonumber \\
                 &   & \qquad
            \,-\, (\smfrac{4}{3} m^3 - 2 m^2 - \smfrac{1}{3} m) q^{m-3}
            \,+\, \ldots \quad\hbox{[for $m \ge 5$]}
\end{subeqnarray}

In this section we want to carry out an analogous computation
for the dominant {\em eigenvalue}\/ of the transfer matrix,
which we call $\lambda_\star^{\rm F/P}$.
Already from Corollary~\ref{cor.TMblockdiag.1and2}
we can conclude that $\lambda_\star(m)$ has, for large $|q|$,
a convergent expansion in powers of $q^{-1}$,
\begin{subeqnarray}
   \lambda_{\star}^{\rm F}(m)  & = &
       \sum\limits_{k=0}^\infty  (-1)^k b^{\rm F}_{k}(m) \, q^{m-k}
      \slabel{def_bkm_F}   \\[2mm]
   \lambda_{\star}^{\rm P}(m)  & = &
       \sum\limits_{k=0}^\infty  (-1)^k b^{\rm P}_{k}(m) \, q^{m-k}
      \slabel{def_bkm_P}
      \label{def_bkm_FP}
\end{subeqnarray}
and that the first two terms in this expansion coincide with those
in the dominant diagonal entry:
\begin{equation}
   \lambda_{\star}^{\rm F/P}(m)  \,-\, t_{\rm F/P}(m)   \;=\;  O(q^{m-2})
\end{equation}
and hence
\begin{equation}
   b^{\rm F/P}_k(m)  \;=\;  a^{\rm F/P}_k(m)
   \qquad\hbox{for $k=0,1$} \;.
\end{equation}
Here we shall go further and compute the coefficients $b^{\rm F/P}_k(m)$
for $1 \le m \le 12_{\rm F}, 13_{\rm P}$ and $0 \le k \le 40$.\footnote{
   It would not be difficult to extend this computation to
   much larger values of $k$, if we really cared.
   Extension to larger values of $m$ is, however, an extremely demanding
   computational task.
   The dimension of the transfer matrix for $m=12_{\rm F}$ is $2947$;
   for $m=13_{\rm F}$ it is $7889$, which is beyond the capabilities
   of our current computer facilities.
}
Somewhat surprisingly, we shall find
(for the $m$ values we were able to study) that
\begin{subeqnarray}
   b^{\rm F}_k(m)  & = &  a^{\rm F}_k(m)   \qquad\hbox{for $k=2,3$}  \\[1mm]
   b^{\rm P}_k(m)  & = &  a^{\rm P}_k(m)
                           \qquad\hbox{for $k=2,3$ and $m \ge k+2$}
 \label{eq.bkm.surprise}
\end{subeqnarray}
so that
\begin{subeqnarray}
   \lambda_{\star}^{\rm F}(m)  \,-\, t_{\rm F}(m)   & = &   O(q^{m-4})  \\[1mm]
   \lambda_{\star}^{\rm P}(m)  \,-\, t_{\rm P}(m)   & = &   O(q^{m-4})
         \qquad\hbox{for $m \ge 5$}
 \label{eq.ckm.surprise}
\end{subeqnarray}
rather than merely $O(q^{m-2})$ as Corollary~\ref{cor.TMblockdiag.1and2} shows.
We {\em conjecture}\/ that this behavior holds for larger $m$ as well.

For some purposes it is slightly more convenient to use,
in place of the coefficients $b^{\rm P}_k(m)$,
the modified coefficients $\tilde{b}^{\rm P}_k(m)$ defined by
\begin{equation}
   \lambda_{\star}^{\rm P}(m)   \;=\;
      (-1)^m (q-m-1) \,+\,
      \sum\limits_{k=0}^\infty  (-1)^k \tilde{b}^{\rm P}_{k}(m) \, q^{m-k}
  \label{def_btilde}
\end{equation}
[analogously to \reff{tP_atilde_c} for $t_{\rm P}(m)$].
Note that the relation between the coefficients 
$b^{\rm P}_k(m)$ and $\tilde{b}^{\rm P}_k(m)$ is the same as for the 
coefficients  $a_k^{\rm P}(m)$ and $\tilde{a}_k^{\rm P}(m)$ 
[cf.\ \reff{def_atilde_vs_a_P}].

Most importantly, however, it is enlightening to pass from
the eigenvalue $\lambda_\star^{\rm F/P}(m)$ to its logarithm,
which is a free energy, and define
\begin{equation}
   \log {\lambda_{\star}^{\rm F/P}(m)  \over q^m}
   \;=\;
   \sum\limits_{k=1}^\infty  c^{\rm F/P}_k(m) \, q^{-k}
   \;.
 \label{def_ckm}
\end{equation}
For cylindrical boundary conditions it is slightly more efficient to
define the modified coefficients $\tilde{c}_k^{\rm P}(m)$ by
\begin{equation}
\log {\lambda_{\star}^{\rm P}(m) - (-1)^m (q-m-1)  \over q^m}
   \;=\;
\sum\limits_{k=1}^\infty  \tilde{c}_k^{\rm P}(m) \, q^{-k} \;. 
\label{def_ctildekm}
\end{equation}

In this section we shall see {\em empirically}\/ that the
coefficients $c_k(m)$ behave in a much simpler way than the $b_k(m)$:
namely, while $b_k(m)$ is, for large enough $m$,
(the restriction of) a polynomial {\em of degree $k$}\/ in $m$,
we shall find that $c_k(m)$ is, for large enough $m$,
(the restriction of) a polynomial {\em of degree 1}\/ in $m$.
In Section~\ref{sec.thermo} we shall discuss the theoretical interpretation
of this empirical observation.

We shall proceed as follows:
Using the methods of \cite{transfer1,transfer2}
we shall compute the transfer matrices for strips of width
$m \le 12$ for free boundary conditions and
$m \le 13$ for cylindrical boundary conditions.\footnote{
   In fact, this was already done in ref.~\cite{transfer1}
   for $m \le 8$ with both boundary conditions,
   and in ref.~\cite{transfer2} for $9 \le m \le 13$ with
   cylindrical boundary conditions.
   Therefore, the only new transfer matrices we need to compute here
   are $m=9,10,11,12$ with free boundary conditions.
   See also Sections~\ref{sec.free_energy_series} and \ref{sec.numer} below
   for further results from this computation.
}
{}From these we can extract the dominant eigenvalue as a power series
in $q^{-1}$, i.e.\ for each available $m$ we can easily compute
as many coefficients $b_k(m)$ and $c_k(m)$ as we please.\footnote{
   To compute the dominant eigenvalue as a power series in $q^{-1}$,
   we have applied the power method \cite[Section~7.3.1]{Golub_96}
   {\em in symbolic form}\/ to the transfer matrix.
   Each iteration gives one additional term
   in the expansion of the dominant eigenvalue in powers of $q^{-1}$.
   We can therefore compute the {\em exact}\/ expansion up to
   any desired order in a {\em finite}\/ number of steps.
   \protect\label{footnote_symbolic_power}
}
We then observe {\em empirically}\/ that, for each $k \ge 0$,
the coefficient $b_k(m)$ [resp.\ $c_k(m)$]
is a polynomial $B_k$ [resp.\ $C_k$] in $m$ of degree $k$ [resp.\ degree 1]
{\em provided that we restrict to integers}\/
$m \ge$ some $m_{\rm min}(k)$.\footnote{
   By contrast, for the dominant diagonal entry we have {\em proven}\/
   that $a^{\rm F}_k(m)$ and $\tilde{a}^{\rm P}_k(m)$ are polynomials
   in $m$ of degree $k$;
   and in this case the polynomial form holds for {\em all}\/ allowable
   integers $m$, i.e.\ $m \ge \max(k,1)$.
}
{\em Assuming}\/ that this empirical observation is accurate
(i.e., that the polynomial behavior persists to all larger $m$),
we can infer the expressions for the polynomials $B_k$ and $C_k$
for $k \le 31$ (resp.\ $k\leq 16$) for free (resp.\  cylindrical)
boundary conditions.

%
%
\subsection{Free boundary conditions}
\label{sec.eigen.F} 

Using the methods just described, we have obtained
the leading eigenvalue $\lambda_{\star}^{\rm F}(m)$ 
for $0\leq m \leq 12$ as a power series in $q^{-1}$
[cf.\ \reff{def_bkm_F}] through order $k=40$.
The resulting coefficients $b^{\rm F}_k(m)$ are displayed
in Table~\ref{table_coef_b_F},
and the corresponding coefficients $c^{\rm F}_k(m)$ [cf.\ \reff{def_ckm}]
are displayed in Table~\ref{table_coef_c_F}.
It is interesting to note that for all $(k,m)$ that we have computed (i.e., 
$1\leq m \leq 12$ and $0\leq k \leq 40$), the coefficients   
$b^{\rm F}_k(m)$ and $k c^{\rm F}_k(m)$ are integers. 
We observe {\em empirically}\/ that, for each fixed $k$,
the coefficients $b^{\rm F}_k(m)$ are the restriction to integers $m$
of a polynomial $B_k^{\rm F}$ in $m$ of degree $k$,
and that the coefficients $c^{\rm F}_k(m)$ are the restriction to integers $m$
of a polynomial $C_k^{\rm F}$ in $m$ of degree 1,
{\em provided that we restrict attention to 
$m\geq m_{\rm min}^{\rm F}(k)$}\/  with  
\begin{equation}
m_{\rm min}^{\rm F}(k) \;=\; \begin{cases} 
    1                          & \quad \text{if $0\le k \le 6$}\\[2mm]
    \left\lceil \frac{k}{2} \right\rceil -2 & \quad \text{if $k\ge 7$} 
    \end{cases} \,. 
\label{def_m_min_F}
\end{equation}
Below this threshold $m^{\rm F}_{\rm min}(k)$,
the coefficients deviate from polynomial behavior.
With our available data together with a few tricks described below,
we are able to determine these polynomials for $0\leq k \leq 33$.

First we start by trying to fit the coefficients $b^{\rm F}_k(m)$ with
$m \ge m^{\rm F}_{\rm min}(k)$ to a polynomial $B_k^{\rm F}$ 
in $m$ of degree $k$. As we need $k+1$
coefficients for such a polynomial, we are able to obtain these 
polynomials only up to $k=8$.
Please note that in all cases we have at least one data point 
more than the number of unknowns, so every fit can be tested at least on 
one extra data point. Our results are: 
\begin{subeqnarray}
   B^{\rm F}_0(m)  & = &   1       \\[2mm]
   B^{\rm F}_1(m)  & = &   2m-1       \\[2mm]
   B^{\rm F}_2(m)  & = &   2m^2 - 3m + 1       \\[2mm]
   B^{\rm F}_3(m)  & = &   \smfrac{4}{3} m^3 - 4 m^2 + \smfrac{8}{3} m   \\[2mm]
   B^{\rm F}_4(m)  & = &   \smfrac{2}{3} m^4 - \smfrac{10}{3} m^3
                        + \smfrac{23}{6} m^2 + \smfrac{11}{6} m - 3    \\[2mm]
   B^{\rm F}_5(m)  & = &   \smfrac{4}{15} m^5 - 2 m^4 + \smfrac{11}{3} m^3
                         + \smfrac{7}{2} m^2 - \smfrac{433}{30} m + 9  \\[2mm] 
   B^{\rm F}_6(m)  & = &  \smfrac{4}{45} m^6 - \smfrac{14}{15} m^5 
                + \smfrac{23}{9} m^4
                + \smfrac{19}{6} m^3 - \smfrac{2263}{90} m^2
                + \smfrac{574}{15} m - 18\\[2mm]
   B^{\rm F}_7(m)  & = &  \smfrac{8}{315} m^7 - \smfrac{16}{45} m^6 +
                  \smfrac{62}{45} m^5 + \smfrac{16}{9} m^4 -
                  \smfrac{1144}{45} m^3 + \smfrac{5947}{90} m^2 \nonumber\\
           & & \quad  -\; \smfrac{15011}{210} m + 29 \\[2mm]
   B^{\rm F}_8(m)  & = &  \smfrac{2}{315} m^8 - \smfrac{4}{35}m^7 +
                  \smfrac{3}{5} m^6 + \smfrac{2}{3}m^5
                 -\smfrac{2131}{120}m^4 + \smfrac{4129}{60}m^3 \nonumber \\ 
           & & \quad -\; \smfrac{302017}{2520}m^2 
                     +  \smfrac{9041}{84}m  -49
\label{def_bk_poly_F}
\end{subeqnarray}
Notice that the three highest-order coefficients agree with those of the
corresponding polynomial $a_k^{\rm F}(m)$, i.e.
\begin{equation}
B^{\rm F}_k(m) \;=\;
     \begin{cases}
        a^{\rm F}_k(m)                   & \text{for $0 \le k \le 3$} \\[2mm]
        a^{\rm F}_k(m) + O(m^{k-3})      & \text{for $k\geq 4$} 
     \end{cases}
\end{equation}

However, there is a better way of extracting the desired information from
our numerical data: instead of using the coefficients $b^{\rm F}_k(m)$ as
our basic quantities, we can use the related coefficients $c^{\rm F}_k(m)$
[cf.\ \reff{def_ckm}].
The latter coefficients are {\em empirically}\/ found to be,
for each fixed $k$, 
the restriction to integer $m$ of a polynomial $C_k^{\rm F}$ 
in $m$ {\em of degree 1}\/, i.e.\ 
\begin{equation}
   C_k^{\rm F}(m)  \;=\;  \alpha_k^{\rm F} m + \beta_k^{\rm F}  \;,
 \label{def.CkF}
\end{equation}
provided that $m\ge$ the same $m^{\rm F}_{\rm min}(k)$
defined in \reff{def_m_min_F}. 
As we need only {\em two}\/ coefficients for such a polynomial
(i.e., $\alpha_k^{\rm F}$ and $\beta_k^{\rm F}$),
we are able to obtain these polynomials
up to $k=24$ (if we want at least one extra data point to test the fit)
or $k=26$ (if we don't).
Our results for $k\le 8$ are:
\begin{subeqnarray}
   C^{\rm F}_1(m)  & = &              -2m + 1              \\[2mm]
   C^{\rm F}_2(m)  & = &              - m + \smfrac{1}{2}  \\[2mm]
   C^{\rm F}_3(m)  & = & \smfrac{1}{3}  m - \smfrac{2}{3}  \\[2mm]
   C^{\rm F}_4(m)  & = & \smfrac{5}{2}  m - \smfrac{11}{4} \\[2mm]  
   C^{\rm F}_5(m)  & = & \smfrac{28}{5} m - \smfrac{29}{5} \\[2mm]  
   C^{\rm F}_6(m)  & = & \smfrac{55}{6} m - \smfrac{28}{3} \\[2mm]  
   C^{\rm F}_7(m)  & = & \smfrac{89}{7} m - \smfrac{97}{7} \\[2mm]  
   C^{\rm F}_8(m)  & = & \smfrac{81}{4} m - \smfrac{243}{8}
\label{def_ck_poly_F}
\end{subeqnarray}
The polynomials $C^{\rm F}_k$ for $9\leq k \leq 24$
are reported in the {\sc Mathematica} file {\tt data\_FREE.m}
that is included with the preprint version of this article at arXiv.org;
they can also be read off from the results of Section~\ref{sec.thermo} below
[cf.\ \reff{series_fbulk}/\reff{series_fsurf}]. 
Finally, the polynomials $B^{\rm F}_k$ for $9\le k\le 24$ can be
determined from the $C^{\rm F}_k$ using \reff{def_ckm}.

Actually, we can do better than this. We believe that the coefficients
$c_k^{\rm F}(m)$ are, for each fixed $k \ge 0$,
the restriction to integers $m \ge m^{\rm F}_{\rm min}(k)$ 
of a polynomial $C_k^{\rm F}$ in $m$ of degree 1.
If we compute the difference 
\begin{equation}
\Delta^{\rm F}_k(m) \;=\; c^{\rm F}_k(m) - C_k^{\rm F}(m)
\label{def_Delta}  
\end{equation}
between the numerical coefficients $c^{\rm F}_k(m)$ 
and the corresponding polynomials $C_k^{\rm F}$,
we find, not surprisingly, that they are nonzero whenever 
$m< m_{\rm min}(k)$:  see Table~\ref{table_diff_coef_c_F}. 
If we could somehow guess an analytic form for at least some of
these coefficients $\Delta^{\rm F}_k(m)$,
we could then define improved coefficients $\hat{c}^{\rm F}_k(m)$ by
\begin{equation}
\hat{c}^{\rm F}_k(m) \;=\; c^{\rm F}_k(m) - \Delta^{\rm F}_k(m) \;,
\label{def_chat}
\end{equation} 
so that these coefficients $\hat{c}_k^{\rm F}(m)$ would be,
for each fixed $k$, the restriction to integers
$m \ge \widehat{m}^{\rm F}_{\rm min}(k)$
of the same polynomial $C_k^{\rm F}$,
with a {\em smaller}\/ threshold
$\widehat{m}^{\rm F}_{\rm min}(k) < m^{\rm F}_{\rm min}(k)$. 
The important point here is that a smaller threshold 
$\widehat{m}^{\rm F}_{\rm min}(k)$ implies that we can obtain more 
polynomials $C_k^{\rm F}$ with the same raw data.

By inspecting Table~\ref{table_diff_coef_c_F}, it is not difficult 
to realize that there are some patterns in $\Delta^{\rm F}_k(m)$
immediately below the threshold $m^{\rm F}_{\rm min}(k)$:
for instance, for odd $k=2p+1$ with $3\le p\le 10$, we have
$\Delta^{\rm F}_{2p+1}(p-2)=1$;
and for even $k=2p$ with $4\le p\le 10$, we have
$\Delta^{\rm F}_{2p}(p-3) = 3p-2$.
We then {\em assume}\/ that this behavior holds true for all larger $p$.
For other subsets of the nonzero values of $\Delta^{\rm F}_k(m)$
slightly farther below the boundary $m^{\rm F}_{\rm min}(k)$,
we likewise find simple polynomial Ans\"atze.
Our {\em empirical}\/ results are: 
\begin{subeqnarray}
\Delta^{\rm F}_{2p+1}(p-2) &=& 1 \,, \quad \hbox{\rm for $p\geq 3$}\\[2mm]
\Delta^{\rm F}_{2p+1}(p-3) &=& \smfrac{13}{2}p^2 - \smfrac{37}{2}p + 27 
  \,, \quad \hbox{\rm for $p\geq 4$} \\[2mm]
\Delta^{\rm F}_{2p+1}(p-4) &=& \smfrac{95}{8}p^4 - \smfrac{1265}{12}p^3 
                             + \smfrac{2961}{8}p^2 - \smfrac{8515}{12}p + 917 
               \,, \quad \hbox{\rm for $p\geq 5$} \\[5mm]
\Delta^{\rm F}_{2p}(p-3) &=& 3p-2  \,, \quad \hbox{\rm for $p\geq 4$} \\[2mm]
\Delta^{\rm F}_{2p}(p-4) &=& \smfrac{59}{6}p^3 - 66p^2 + \smfrac{817}{6}p-17 
               \,, \quad \hbox{\rm for $p\geq 5$} \\[2mm]
\Delta^{\rm F}_{2p}(p-5) &=& \smfrac{473}{40} p^5 - \smfrac{2207}{12}p^4 
                           + \smfrac{9023}{8}p^3  - \smfrac{38413}{12}p^2 
                           + \smfrac{29763}{10}p + 2941 \,, \nonumber \\
                         & & \qquad \qquad \quad \hbox{\rm for $p\geq 7$} 
\slabel{eq.1} 
\end{subeqnarray} 
We are able to test each fit on at least one additional data point. 
Notice that in \reff{eq.1}, the condition $p\ge 7$ does {\em not}\/ follow the 
expected behavior from the previous correction terms (i.e., one would have 
expected $p\ge 6$).  
The new threshold $\widehat{m}^{\rm F}_{\rm min}(k)$ is given by  
\begin{equation}
\widehat{m}^{\rm F}_{\rm min}(k) \;=\; \begin{cases} 
    1                          & \quad \text{if $0\le k \le 11$}\\[2mm]
    2                          & \quad \text{if $12\le k \le 14$}\\[2mm]
    \left\lceil \frac{k}{2} \right\rceil -5 & \quad \text{if $k\ge 15$} 
    \end{cases}
\label{def_mhat_min_F}
\end{equation}
By this method, we can obtain the polynomials $C^{\rm F}_k$
(and therefore the polynomials $B^{\rm F}_k$) up to $k=30$.
Indeed, the fits to obtain the polynomials $C^{\rm F}_k$ with $k\le 30$
were tested on at least one additional data point. 

If we do not demand to have at least one extra data point to test the fits,
we can extend this computation up to $k=32$.
We can then guess one further correction term $\Delta^{\rm F}_k(m)$
(this time with an additional data point to test the fit):
\begin{eqnarray}
\Delta^{\rm F}_{2p+1}(p-5) &=& \smfrac{161}{16}p^6 - \smfrac{43693}{240}p^5
   + \smfrac{21757}{16}p^4 - \smfrac{86909}{16}p^3 + \smfrac{110297}{8}p^2 
   - \smfrac{433772}{15}p \nonumber \\
                           & & \qquad + 42719 \,, 
   \quad \hbox{\rm for $p\geq 8$} \,.  
\end{eqnarray}         
With this additional correction, the new threshold
$\widehat{m}^{\rm F}_{\rm min}$ is 
\begin{equation}
\widehat{m}^{\rm F}_{\rm min}(k) \;=\; \begin{cases} 
    1                          & \quad \text{if  $0\le k \le 11$}\\[2mm]
    2                          & \quad \text{if $12\le k \le 14$}\\[2mm]
    3                          & \quad \text{if $15\le k \le 17$}\\[2mm]
    \left\lfloor \frac{k}{2} \right\rfloor -5 & \quad \text{if $k\ge 18$} 
    \end{cases}
\label{def_mhat_min_F_bis}
\end{equation}
so the computation of the polynomials $C^{\rm F}_k$ can be extended 
up to $k=33$ (with no extra data points to test the fit).
The polynomials $C^{\rm F}_k$ with $1\le k\le 33$ are included in the 
{\tt mathematica} file {\tt data\_FREE.m}. 

%
%
\subsection{Cylindrical boundary conditions}
\label{sec.eigen.P} 

We have likewise obtained
the leading eigenvalue $\lambda_{\star}^{\rm P}(m)$ 
for $0\leq m \leq 13$ as a power series in $q^{-1}$
[cf.\ \reff{def_bkm_P}] through order $k=40$.
The resulting coefficients $b^{\rm P}_k(m)$ are displayed
in Table~\ref{table_coef_b_P},
and the corresponding coefficients $c^{\rm P}_k(m)$ are displayed
in Table~\ref{table_coef_c_P}.
As for free boundary conditions, we note that for all $(k,m)$ that we 
have computed (i.e., $1\leq m \leq 13$ and $0\leq k \leq 40$), the coefficients
$b^{\rm P}_k(m)$ and $k c^{\rm P}_k(m)$ are integers.

We observe {\em empirically}\/ that, for each fixed $k$,
the coefficients $b^{\rm P}_k(m)$ are
the restriction to integers $m$ of a polynomial $B_k^{\rm P}$ 
in $m$ of degree $k$,
and that the coefficients $c^{\rm P}_k(m)$ are the restriction to integers $m$
of a polynomial $C_k^{\rm P}$ in $m$ of degree 1, 
{\em provided that we restrict attention 
to $m \ge m^{\rm P}_{\rm min}(k)$}\/ with
\begin{equation}
   m^{\rm P}_{\rm min}(k)   \;=\;  k+2  \;.
\label{def_m_min_P}
\end{equation}
Below this threshold, the coefficients deviate from polynomial behavior.
The polynomial behavior can be extended downwards by two steps,
i.e.\ to $m = k$, if we use the coefficients $\tilde{b}^{\rm P}_k(m)$
[cf.\ \reff{def_btilde}] in place of $b^{\rm P}_k(m)$.
With our available data together with the tricks described in
the preceding subsection, we are able to determine these polynomials
for $0\leq k \leq 15$.
The coefficients $\tilde{b}^{\rm P}_k(m)$ are displayed
in Table~\ref{table_coef_btilde_P}.

We begin, as before, by fitting $\tilde{b}^{\rm P}_k(m)$
for $m \ge k$ to a polynomial $B_k^{\rm P}$ of degree $k$.
With our data we can do this for $0 \le k \le 6$;
we also have, in each case,
at least one extra data point to test the fit. Our results are: 
\begin{subeqnarray}
 B^{\rm P}_0(m)  & = & 1       \\[2mm]
 B^{\rm P}_1(m)  & = & 2m       \\[2mm]
 B^{\rm P}_2(m)  & = & 2m^2 - m \\[2mm]
 B^{\rm P}_3(m)  & = & \smfrac{4}{3} m^3 - 2 m^2 
                     - \smfrac{1}{3} m \\[2mm]
 B^{\rm P}_4(m)  & = & \smfrac{2}{3} m^4 - 2 m^3 - \smfrac{1}{6} m^2
                     + \smfrac{5}{2} m  \\[2mm]
 B^{\rm P}_5(m)  & = & \smfrac{4}{15} m^5 - \smfrac{4}{3} m^4 
                     + \smfrac{1}{3} m^3
                     + \smfrac{16}{3} m^2 - \smfrac{28}{5} m \\[2mm]
 B^{\rm P}_6(m)  & = & \smfrac{4}{45} m^6 - \smfrac{2}{3} m^5 
                     + \smfrac{5}{9} m^4  + \smfrac{11}{2} m^3
                     - \smfrac{614}{45} m^2  + \smfrac{55}{6} m 
\label{def_bk_poly_P}
\end{subeqnarray}
Notice that the three highest-order coefficients agree with those of the
corresponding polynomial $\tilde{a}^{\rm P}_k(m)$, i.e.
\begin{equation}
B^{\rm P}_k(m) \;=\;
   \begin{cases}
      \tilde{a}^{\rm P}_k(m)               & \text{for $0 \le k \le 3$} \\[2mm]
      \tilde{a}^{\rm P}_k(m) + O(m^{k-3})  & \text{for $k\geq 4$} 
   \end{cases}
\end{equation}
Note also that the constant term vanishes in all these polynomials
except $B^{\rm P}_0$.

As in the previous subsection, we can extract the desired information
more efficiently by analyzing the coefficients $c^{\rm P}_k(m)$,
which are found empirically to be, for each fixed $k$,
a polynomial $C_k^{\rm P}$ in $m$ {\em of degree 1}\/
for $m\ge m_{\rm min}^{\rm P}(k) = k+2$, i.e.\
\begin{equation}
   C_k^{\rm P}(m)  \;=\;  \alpha_k^{\rm P} m + \beta_k^{\rm P}  \;,
 \label{def.CkP}
\end{equation}
As we need only two coefficients for such a polynomial,
we can obtain these polynomials
up to $k=9$ (if we want at least one extra data point to test the fit)
or $k=10$ (if we don't).

However, we can do slightly better if we consider instead of the coefficients
$c^{\rm P}_k(m)$ the modified coefficients
$\tilde{c}^{\rm P}_k(m)$ defined by \reff{def_ctildekm}.
We find empirically that $k \tilde{c}^{\rm P}_k(m)$ is an integer
for all the computed values of $(k,m)$:
see Table~\ref{table_coef_ctilde_P}.
We also find empirically that $\tilde{c}^{\rm P}_k(m)$ is,
for each fixed $k$, the restriction to integers 
$m\ge \widetilde{m}_{\rm min}^{\rm P}(k)$ of the same polynomial $C_k^{\rm P}$ 
with a {\em smaller}\/ threshold
$\widetilde{m}_{\rm min}^{\rm P}(k) = \max(k,2) < m_{\rm min}^{\rm P}(k)$. 
In this way we can obtain the polynomials $C_k^{\rm P}$ up to $k=11$ or $k=12$, 
depending on whether or not we insist on having an extra data point to test 
the fit. Our results for $k\le 8$ are: 
\begin{subeqnarray}
   C^{\rm P}_1(m)  & = &              -2m \\[2mm]
   C^{\rm P}_2(m)  & = &              - m \\[2mm]
   C^{\rm P}_3(m)  & = & \smfrac{1}{3}  m \\[2mm]
   C^{\rm P}_4(m)  & = & \smfrac{5}{2}  m \\[2mm]
   C^{\rm P}_5(m)  & = & \smfrac{28}{5} m \\[2mm]
   C^{\rm P}_6(m)  & = & \smfrac{55}{6} m \\[2mm]
   C^{\rm P}_7(m)  & = & \smfrac{89}{7} m \\[2mm]
   C^{\rm P}_8(m)  & = & \smfrac{81}{4} m
\label{def_ck_poly_P}
\end{subeqnarray}
The polynomials $C_k^{\rm P}$ for $9 \le k \le 12$ are reported in the 
{\sc Mathematica} file {\tt data\_CYL.m} that is included with the preprint
version of this article at arXiv.org;
they can also be read off from the results of Section~\ref{sec.thermo} below
[cf.\ \reff{series_fbulk}]. 
Please note that the constant term vanishes in all these polynomials,
while the term linear in $m$ is the same as
for free boundary conditions [cf.\ \reff{def_ck_poly_F}]:
\begin{subeqnarray}
   \alpha_k^{\rm P}  & = &  \alpha_k^{\rm F}   \\[2mm]
   \beta_k^{\rm P}   & = &  0
\slabel{eq_equal_bulk_term.b}
\label{eq_equal_bulk_term}
\end{subeqnarray}
in all cases that we are able to test (namely, $1 \le k \le 12$).
Finally, the polynomials $B^{\rm P}_k$ for $7\le k\le 12$ can be
determined from the $C_k^{\rm P}$ using \reff{def_ctildekm}.

These results can be improved in the same way as we did in the previous
subsection. First we compute the difference 
$\Delta^{\rm P}_k(m) = \tilde{c}^{\rm P}_k(m) - C_k^{\rm P}(m)$:
see Table~\ref{table_diff_coef_ctilde_P}. 
We then try to guess an analytic form for some of the coefficients
$\Delta^{\rm P}_k(m)$, and we define improved coefficients
$\hat{c}^{\rm P}_k(m)$ [as in \reff{def_chat}]
so that the $\hat{c}_k^{\rm P}(m)$ will be, for each fixed $k$,
the restriction to integers $m \ge \widehat{m}^{\rm P}_{\rm min}(k)$
of the polynomial $C_k^{\rm P}$,
with a {\em smaller}\/ threshold
$\widehat{m}^{\rm P}_{\rm min}(k)< \widetilde{m}^{\rm P}_{\rm min}(k)$. 
As in the case of free boundary conditions, we find empirically
that the coefficients $\Delta^{\rm P}_k(m)$ closest to the boundary
$\widetilde{m}^{\rm P}_{\rm min}(k)$ are the restriction to integers $m$ 
of certain polynomials:
\begin{subeqnarray}
\Delta^{\rm P}_{k}(k-1) &=& (-1)^{k-1}\left( \smfrac{3}{2}k^2 
         - \smfrac{11}{2}k + 4 \right) \,, \quad \hbox{\rm for $k\ge 4$}\\[2mm]
\Delta^{\rm P}_{k}(k-2) &=& (-1)^{k-2}\left( 
           \smfrac{11}{6}k^3 -10k^2 + \smfrac{73}{6}k +1 \right)
           \,, \quad \hbox{\rm for $k\ge 6$} \\[2mm] 
\Delta^{\rm P}_{k}(k-3) &=& (-1)^{k-3}\left( \smfrac{35}{24}k^4 - 
           \smfrac{199}{12}k^3 -\smfrac{1837}{24}k^2 - 
           \smfrac{2201}{12}k +191 \right) 
           \,, \quad \hbox{\rm for $k\ge 8$} \nonumber \\
  & &  
 \label{def_Delta_P}
\end{subeqnarray} 
Again, each fit can be tested on at least an extra data point. 
The new threshold $\widehat{m}^{\rm P}_{\rm min}$ is 
\begin{equation}
\widehat{m}^{\rm P}_{\rm min}(k) \;=\; \begin{cases} 
    2    & \quad \text{if $k\leq 2$}\\[2mm]
 \left\lfloor \smfrac{k}{2} \right\rfloor +1 & 
                             \quad \text{if $3\le k \le 8$}\\[2mm]
    k-3  & \quad \text{if $k\ge 9$} \,,  
    \end{cases}
\label{def_mhat_min_P}
\end{equation}
By this method, we can obtain the polynomials $C^{\rm P}_k$
(and therefore the polynomials $B^{\rm P}_k$) up to $k=14$,
with at least one extra data point to test the fit.

If we do not insist on having an extra data point to test the fit,
we can extend this computation of $C^{\rm P}_k$ up to $k=15$.
We can then guess one further correction term
$\Delta^{\rm P}_k(m)$ (again with no additional test for the fits): 
\begin{equation}
\Delta^{\rm P}_{k}(k-4) \;=\; (-1)^{k-4}\left( \smfrac{33}{40}k^5 - 
  \smfrac{115}{8}k^4 + \smfrac{741}{8}k^3 - \smfrac{1821}{8}k^2 
  -\smfrac{729}{20}k + 695\right)\,,  
  \quad \hbox{\rm for $k\ge 10$} \,.  
\end{equation} 
With this correction the new threshold is 
\begin{equation}
\widehat{m}^{\rm P}_{\rm min}(k) \;=\; \begin{cases} 
    2    & \quad \text{if $k\le 2$}\\[2mm]
\left\lfloor \smfrac{k}{2} \right\rfloor +1 & 
                             \quad \text{if $3\le k \le 10$}\\[2mm]
    k-4  & \quad \text{if $k\ge 11$} 
    \end{cases}
\label{def_mhat_min_P_bis}
\end{equation}
so the computation of the polynomials  $C^{\rm P}_k$ can be extended 
up to $k=16$ (with no extra data points to test the fit).

The polynomials $C^{\rm P}_{k}$ for $10\le k\le 16$ also have a zero
constant term, and the relation \reff{eq_equal_bulk_term}
between periodic and free boundary conditions continues to hold.

%
%
\section{Thermodynamic limit ($\bm {m \to \infty}$)}
\label{sec.thermo}

In previous sections we have dealt with semi-infinite square-lattice 
strips of fixed width $m$. In this section we will study
the thermodynamic limit $m\to\infty$ of the free energy of our model. 

In Section~\ref{sec.thermo.gen} we introduce some preliminary definitions 
and discuss the expected behavior of the strip free energies per
unit length, $f_m^{\rm F}(q)$ and $f_m^{\rm P}(q)$,
as a function of the strip width $m$.
In Section~\ref{sec.free_energy_series}
we discuss the large-$|q|$ expansion of the bulk free energy.  
Bakaev and Kabanovich \cite{Bakaev_94}
have calculated the first 36 terms of this expansion;
we confirm their computation and extend it by providing 11 more terms,
i.e.\ through order $q^{-47}$.
We also compute the large-$|q|$ expansion of the surface (resp.\ corner)
free energy through order $q^{-47}$ (resp.\ $q^{-46}$).
These computations are based on the finite-lattice method
\cite{Neef-Enting_77,Enting_78,Kim-Enting_79,Enting_96,Enting_05,Enting_06}.

In Section~\ref{sec.thermo.F} we obtain (using the polynomials $C_k^{\rm F}$)
a large-$|q|$ expansion for the limiting free energy $f_m^{\rm F}(q)$
of a semi-infinite strip of width $m$ and free boundary conditions. We find
that this expansion contains two terms: a bulk term (independent of $m$)
and a surface term (linear in $1/m$).
This computation gives an independent check of the first 33 terms
of the series expansions for the bulk and surface free energies for
the square lattice.  
In Section~\ref{sec.thermo.P} we repeat the computation with 
cylindrical boundary conditions.  We find that the bulk 
contribution is the same as for free boundary conditions and that there 
is {\em no}\/ surface contribution.
These results provide a theoretical interpretation
for some aspects of the behavior found in the preceding section for
$C_k^{\rm F}$ and $C_k^{\rm P}$
[cf.\ \reff{def_ck_poly_F}/\reff{def_ck_poly_P}
 and \reff{eq_equal_bulk_term}].

Finally, in Section~\ref{sec.series} we perform a series-extrapolation
analysis of the large-$|q|$ series for the bulk, surface and corner
free energies, in an effort to locate their singular points in the 
complex $q$-plane.

%
%
\subsection{Generalities and finite-size-scaling theory} 
\label{sec.thermo.gen}

Corollary~\ref{cor.TMblockdiag.1and2} shows that,
for each width $m$ and each boundary condition (free or cylindrical),
the transfer matrix has, for sufficiently large $|q|$,
a {\em single}\/ dominant eigenvalue $\lambda_\star(q)$
that moreover is an analytic function of $q$
(in fact, it is $q^m$ times an analytic function of $q^{-1}$).
However, Corollary~\ref{cor.TMblockdiag.1and2} gives no information
about {\em how large}\/ $|q|$ has to be for this behavior to occur;
in particular, there is no guarantee that this large-$q$ domain
is uniform in $m$.
However, the uniformity in $m$ can be proven by invoking
the following theorem:

\begin{theorem}
{\bf \protect \cite[Corollary~5.3 and Proposition~5.4]{Sokal_chromatic_bounds}}
   \label{thm.sokal2}
Let $G=(V,E)$ be a loopless\footnote{
   {\bf Warning:}  We are here using the graph theorists' terminology,
   in which a {\em loop}\/ is an edge connecting a vertex to itself.
   Obviously, if $G$ has a loop, then $P_G(q)$ is identically zero.
   What physicists often call ``loops'' ---
   particularly when referring to Feynman diagrams ---
   are called ``cycles'' or ``circuits'' by graph theorists.
   Thus, a ``3-loop Feynman diagram''
   is a graph with with cyclomatic number 3;
   it may or may not have loops.
} 
finite undirected graph of maximum degree $\Delta$.
Then all the zeros of the chromatic polynomial $P_G(q)$
lie in the disc $|q| < 7.963907 \Delta$.
\end{theorem}

\noindent
Let us remark that this theorem has been recently improved
by Fern\'andez and Procacci \cite[Corollary 2]{Fernandez_07}:
they showed that the constant $7.963907$ in Theorem~\ref{thm.sokal2}
can be replaced by $6.907652$.\footnote{
   Jackson, Procacci and Sokal \cite{jps_08} have recently observed
   that the Fern\'andez--Procacci constant is in fact
   $$
      K^*  \;=\;  W(e/2)/[1-W(e/2)]^2
    \;\approx\; 6.907\:651\:697\:774\:449\:218\:\ldots\;\,
   $$
   where $W$ is the Lambert $W$ function \cite{Corless_96},
   i.e.\ the inverse function to $x \mapsto x e^x$.
}

For the square lattice with any of the standard boundary conditions
(free, cylindrical, cyclic or toroidal)
we have $\Delta = 4$, so we can conclude that
all chromatic roots lie inside the disc
$|q| <  7.963907 \times 4 = 31.855628$.
Actually, by \cite[Corollary~5.3 and Table~1]{Sokal_chromatic_bounds},
we have for $\Delta = 4$ the slightly stronger bound
$|q| < C(4) \le 29.081607$.
With the improved result of Fern\'andez and Procacci,
we get $|q| < 6.907652\times 4 = 27.630607$;
and for the particular case $\Delta=4$ these authors 
\cite[Corollary~1]{Fernandez_07} obtained the slightly better bound
$|q| < C^\star(4) \le 24.443218$.

It follows that the limiting curves $\scrb_m$
must also lie inside the disc $|q| \le 24.443218$ for all widths $m$:
for if part of the limiting curve were to lie outside the (closed) disc, 
then the Beraha--Kahane--Weiss theorem 
(Theorem~\ref{BKW_thm})
would imply that chromatic roots would also lie outside the disc
for $m \times n$ strips of all sufficiently large lengths $n$.
Furthermore, using again the Beraha--Kahane--Weiss theorem
we can conclude that, outside this disc,
the transfer matrix for each width $m$
must have one and only one eigenvalue of largest modulus.
Since this dominant eigenvalue cannot collide with any other eigenvalue,
it must be an analytic function of $q$ outside the given disc.

In summary, the transfer matrix for a square-lattice strip of width $m$ 
and with free or cylindrical boundary conditions has a single dominant 
eigenvalue $\lambda_{\star,m}^{\rm F/P}(q)$
that is an analytic function of $q$
(in fact, $q^m$ times an analytic function of $q^{-1}$)
whenever $|q| > 24.443218$.

Let us now introduce the free energy per site for a finite strip
with free or cylindrical boundary conditions,
\begin{equation}
 f_{m,n}^{\rm F/P}(q)  \;=\;
     {1 \over m n} \log P_{G_{m_{\rm F/P}\times n_{\rm F}}}(q)   \;,
\label{def_Fmn}
\end{equation}
and its limiting value for a semi-infinite strip,
\begin{equation}
 f_{m}^{\rm F/P}(q)  \;=\;
   \lim_{n\to\infty} {1 \over m n} \log P_{G_{m_{\rm F/P}\times n_{\rm F}}}(q)
 \;.
\label{def_Fm}
\end{equation}
Finally, let us introduce the free energy per site for the infinite lattice,
\begin{equation}
 f^{\rm F/P}(q)  \;=\;
  \lim_{m,n\to\infty} {1 \over m n} \log P_{G_{m_{\rm F/P}\times n_{\rm F}}}(q)
  \;.
\label{def_F}
\end{equation}
Here we are assuming that the indicated limits exist and that in
\reff{def_F} the limit is independent of the way that
$m$ and $n$ tend to infinity.
Furthermore, it is natural to expect that in \reff{def_F}
the limiting free energy is independent of boundary conditions,
in which case we can omit the superscripts F or P and write simply $f(q)$.

In fact, some of these assumptions can be proven.
Indeed, the above discussion guarantees that at least for $|q| > 24.443218$,
the limiting strip free energy $f_m(q)$ exists for all $m$ and is given by
\begin{equation}
f_m^{\rm F/P}(q) \;=\; \frac{1}{m} \log\lambda_{\star,m}^{\rm F/P}(q)
   \;,
\label{def_Fm_bis}
\end{equation}
which in particular is an analytic function of $q$
in the indicated domain.
Moreover, Procacci {\em et al.}\/ \cite[Theorem 2]{Procacci_03}
have proven that, when $|q|$ is large enough 
(namely, $|q|> 8e^3 \approx 160.684295$),
the infinite-volume limiting free energy $f(q)$ exists and is analytic in $1/q$
and is the same for all reasonable sequences of graphs $G_{m\times n}$
(in particular, it is the same for free, cylindrical, cyclic and toroidal
 boundary conditions and is independent of the way that $m$ and $n$
 tend to infinity).
In this paper we will take $n \to\infty$ first
and then take $m\to\infty$, so that
\begin{equation}
 f(q)  \;=\;  \lim_{m\to\infty} f_m^{\rm F/P}(q)  \;.
\label{def_F_bis}
\end{equation}

Finite-size-scaling theory \cite[Section~2.5]{Privman_90} 
gives a rather precise prediction for the form
of the free energy \reff{def_Fmn}/\reff{def_Fm} 
for a finite or semi-infinite system
away from a critical point (and in the absence of soft modes). 
In particular, for an $m\times n$ strip with free or cylindrical
boundary conditions and bulk correlation length $\xi_{\rm bulk} \ll m,n$,
the predicted behavior is
\begin{subeqnarray}
   f^{\rm F}_{m,n}  & = &
        f_{\rm bulk} \,+\, \frac{m+n}{mn} \, f_{\rm surf}
                     \,+\, \frac{1}{mn} \, f_{\rm corner}
                     \,+\, O( e^{-\min(m,n)/\xi_{\rm bulk}} )
      \slabel{def_FSS_Ansatz.free} \\[2mm]
   f^{\rm P}_{m,n}  & = &
        f_{\rm bulk} \,+\, \frac{1}{n} \, f_{\rm surf}
                     \,+\, O( e^{-\min(m,n)/\xi_{\rm bulk}} )
      \slabel{def_FSS_Ansatz.cyl}
  \label{def_FSS_Ansatz}
\end{subeqnarray}
where $f_{\rm bulk} = f$, $f_{\rm surf}$ and $f_{\rm corner}$ are, respectively,
the bulk, surface and corner free energies.
(More precisely, $f_{\rm surf}$ is the free energy for {\em two}\/ units
 of surface, and $f_{\rm corner}$ is the free energy for {\em four}\/ corners.)
For a semi-infinite strip $m\times\infty$ with free or cylindrical
boundary conditions, we have
\begin{subeqnarray}
   f^{\rm F}_{m}  & = &
        f_{\rm bulk} \,+\, \frac{1}{m} \, f_{\rm surf}
                     \,+\, O( e^{-m/\xi_{\rm bulk}} )
      \slabel{def_FSS_Ansatz_Bis.free} \\[2mm]
   f^{\rm P}_{m}  & = &
        f_{\rm bulk} \,+\, O( e^{-m/\xi_{\rm bulk}} )
      \slabel{def_FSS_Ansatz_Bis.cyl}
  \label{def_FSS_Ansatz_Bis}
  \label{def_FSS_Ansatz2}
\end{subeqnarray}

The relations \reff{def_FSS_Ansatz2} of course hold for the
chromatic polynomials at {\em fixed}\/ large $q$.
But we can also argue heuristically what they should imply
for the series expansion in powers of $1/q$.
It is not difficult to see that,
for large $q$, we have
\begin{equation}
   e^{-1/\xi_{\rm bulk}(q)}  \;=\;
   {1 \over q} \,+\, O\Bigl( {1 \over q^2} \Bigr)
 \label{eq.xibulk}
\end{equation}
\nopagebreak
(just as for a {\em one-dimensional}\/
 Potts antiferromagnet at zero temperature).\footnote{
   Let $G=(V,E)$ be a finite graph (let us suppose for simplicity
   that it has no loops or multiple edges) with $n$ vertices and
   $m$ edges;  then one sees immediately from the Fortuin--Kasteleyn
   representation \protect\reff{eq.FK.identity} that its Potts-model partition
   function has the large-$q$ expansion
   $$ 
      Z_G(q,v)  \;=\;  q^n \,+\, m q^{n-1} \,+\, \frac{m(m-1)}{2} q^{n-2}
                       \,+\, O(q^{n-3}) \;.
   $$ 
   When $G$ is a finite piece of a regular lattice,
   the corresponding expansion for $|V|^{-1} \log Z_G(q,v)$
   gives in the infinite-volume limit the large-$q$ expansion
   for the bulk free energy $f(q,v)$.
   
   Now let $i,j$ be vertices of $G$;  then the unnormalized 2-point
   correlation function
   $Z_G \< \bsigma_i \cdot \bsigma_j \> = 
    Z_G \< (q \delta_{\sigma_i,\sigma_j} - 1)/(q-1) \>$
   is given by a representation like \reff{eq.FK.identity}
   but with the constraint that $i$ and $j$ must belong to
   the same connected component. The dominant terms of the
   large-$q$ expansion are the ones in which this component has
   the minimal number of vertices and all other components
   are isolated vertices;  one therefore gets
   $$ 
      Z_G \< \bsigma_i \cdot \bsigma_j \>  \;=\;
      Q_G(v;i,j) \, \Bigl( \frac{v}{q} \Bigr)^{d_G(i,j)} \, q^n
         \left[ 1 + O(q^{-1}) \right]
   $$ 
   where $d_G(i,j)$ is the length of the shortest path in $G$
   from $i$ to $j$, and $Q_G(v;i,j)$ is a polynomial in $v$
   that enumerates the connected subgraphs of $G$ that contain $i$ and $j$
   and have exactly $d_G(i,j) + 1$ vertices
   [with a weight $v$ for each edge beyond the minimum number $d_G(i,j)$].
   In particular, if $G$ is triangle-free, these subgraphs are simply
   shortest paths from $i$ to $j$.
   In general $Q_G(v;i,j)$ can grow exponentially in $d_G(i,j)$
   [when e.g.\ $G$ is an infinite regular lattice];
   but if $G$ is a piece of the square lattice and
   $i-j$ lies {\em along an axis direction}\/, then $Q_G(v;i,j) = 1$.
   It follows that the exponential decay rate of correlations along an axis is
   $$ 
      e^{-1/\xi_{\rm bulk}(q,v)}  \;=\;
      \Bigl| {v \over q} \Bigr| \,+\, O\Bigl( {1 \over |q|^2} \Bigr)
      \;.
   $$ 
   For $v=-1$ this gives \reff{eq.xibulk}.

   It is instructive to ask how these results would be seen in the
   transfer-matrix formalism.
   Ordinarily one has $e^{-1/\xi_{\rm bulk}} = |\lambda_2/\lambda_\star|$,
   where $\lambda_\star$ is the dominant eigenvalue and
   $\lambda_2$ is the first subleading eigenvalue.
   But when one performs the computation using our transfer matrices,
   one finds $\lambda_2/\lambda_\star = \alpha_m^{\rm F/P}\!(v)/q^2 +O(1/q^3)$
   for suitable polynomials $\alpha_m^{\rm F/P}$
   --- {\em not}\/ the predicted $v/q$.
   (Indeed, for width $m=1$ --- i.e., a one-dimensional Potts model ---
   the transfer matrix is of size $1 \times 1$, i.e.\ there is {\em no}\/
   subleading eigenvalue at all.)
   What is going on here?

   The point is that the exponential decay rate in the correlation function
   $\< \bsigma_0 \cdot \bsigma_x \>$ is controlled by a ``colored''
   intermediate state, i.e.\ the state obtained by applying the field
   $\bsigma_0$ to the vacuum.  The corresponding eigenvalue $\lambda_2$
   would be seen in a transfer matrix in the {\em spin representation}\/
   \cite[Section~3.1]{transfer1};
   but it is not seen in our transfer matrix in the
   {\em Fortuin--Kasteleyn representation}\/,
   which represents only ``colorless'' states
   (i.e., states invariant under the Potts global symmetry group $S_q$).
   Rather, the first subleading eigenvalue of the latter transfer matrix
   corresponds to a ``colorless'' two-particle state,
   hence has $\lambda_2/\lambda_\star$ of order $1/q^2$.
   More precisely, for square-lattice strips of widths
   $m \ge 4$ (resp.\ $m \ge 3$) with cylindrical (resp.\ free)
   boundary conditions, we find
   (at least up to $m = 9_{\rm P}, 7_{\rm F}$)
   that there are {\em at least two}\/ subleading eigenvalues of order $1/q^2$.
   Some of these satisfy $\lambda_2/\lambda_\star = (v/q)^2 + O(q^{-3})$,
   and the others satisfy $\lambda_2/\lambda_\star = (1+v)(v/q)^2 + O(q^{-3})$;
   note that the latter ones vanish at order $1/q^2$
   for the chromatic polynomial $v=-1$.
   All other eigenvalues are $O(q^{-3})$.
}
We can therefore interpret $O( e^{-m/\xi_{\rm bulk}(q)} )$
as meaning $O(q^{-m})$.  Therefore, we expect that
\begin{subeqnarray}
   f^{\rm F}_{m}  & = &
        f_{\rm bulk} \,+\, \frac{1}{m} \, f_{\rm surf}
                     \,+\, O(q^{-m})
      \slabel{eq.fm_prediction.free} \\[2mm]
   f^{\rm P}_{m}  & = &
        f_{\rm bulk} \,+\, O(q^{-m})
      \slabel{eq.fm_prediction.cyl}
 \label{eq.fm_prediction}
\end{subeqnarray}
This explains why the coefficients $c_k^{\rm F/P}(m)$
in the expansion of $\log \lambda_{\star,m}^{\rm F/P} = m f_m^{\rm F/P}$
[cf.\ \reff{def_ckm}] are polynomials {\em of degree 1}\/ in $m$
for large enough $m$, i.e.\ $m \ge m_{\rm min}^{\rm F/P}(k)$,
and consequently why the coefficients $b_k^{\rm F/P}(m)$
in the expansion of $\lambda_{\star,m}^{\rm F/P}$ [cf.\ \reff{def_bkm_FP}]
are polynomials of degree $k$ in $m$.
It also explains why, for cylindrical boundary conditions,
$c_k^{\rm P}(m)$ is strictly proportional to $m$,
i.e.\ the constant term vanishes
[cf.\ \reff{def_ck_poly_P}/\reff{eq_equal_bulk_term.b}].
The error term $O(q^{-m})$ in \reff{eq.fm_prediction}
furthermore suggests that $m_{\rm min}^{\rm F/P}(k) = k+1$.
Of course, we should not take too seriously the ``$+1$'' here,
since the {\em amplitude}\/ of the correction term in \reff{def_FSS_Ansatz2}
could be proportional to a positive or negative power of $q$.
But we do predict that $m_{\rm min}^{\rm F/P}(k) = k + O(1)$ as $k \to\infty$.
This is indeed what we found for cylindrical boundary conditions,
where we have $m_{\rm min}^{\rm P}(k) = k+2$ [cf.\ \reff{def_m_min_P}].
For free boundary conditions, however,
we found the {\em faster}\/ convergence $m_{\rm min}^{\rm F}(k) \approx k/2$
[cf.\ \reff{def_m_min_F}],
for which we lack at present any theoretical explanation.

In the next subsection we will use our transfer matrices
to compute the large-$q$ expansion for the bulk, surface 
and corner free energies, using the finite-lattice method
\cite{Neef-Enting_77,Enting_78,Kim-Enting_79,Enting_96,Enting_05,Enting_06}.
In the subsequent two subsections
we will compute the large-$q$ expansion of the strip free energy $f_m(q)$
for free and cylindrical boundary conditions, respectively,
using our polynomials $C_k^{\rm F}$ and $C_k^{\rm P}$.

%
%
\subsection[Large-$q$ expansion for the bulk, surface and corner free energies]
    {Large-$\bm{q}$ expansion for the bulk, surface and corner free energies}
\label{sec.free_energy_series}

First of all, as we are interested in the large-$q$ limit, it is
convenient to explicitly remove the leading term $\log q$ in the
free energy by considering the modified chromatic polynomial
$\widetilde{P}_{G}$ for a loopless graph $G=(V,E)$:
\begin{equation}
\widetilde{P}_{G}(q) \;=\; q^{-|V|} P_{G}(q) \;.
\label{def.Ptilde}
\end{equation}
Using the Fortuin--Kasteleyn representation \reff{eq.FK.identity} we get
\begin{subeqnarray}
\widetilde{P}_G(q) &=& \sum\limits_{A\subseteq E} (-1)^{|E|} \, q^{k(A)-|V|}
   \\[1mm]
         &=& \sum\limits_{A\subseteq E} (-1)^{|E|} \, (1/q)^{|A|-c(A)}  \;,
\slabel{eq.FK.identity.Bis.b}
\label{eq.FK.identity.Bis}
\end{subeqnarray}
where $c(A) = |A| - |V| + k(A)$ is the cyclomatic number
of the subgraph $(V,A)$.  

It is instructive to begin by computing ``by hand''
the first few terms of the large-$q$ expansion for
the bulk, surface and corner free energies.
To do this, let us first consider an $m\times n$ square lattice
with {\em free}\/ boundary conditions:
it has $|V|=mn$ sites, $|E|=2mn-m-n$ edges, and $|F|=(m-1)(n-1)$ square faces
(``plaquettes'').
We can compute the first few first terms in the large-$q$ expansion
for the modified chromatic polynomial
$\widetilde{P}_{m_{\rm F}\times n_{\rm F}}(q)$
by using \reff{eq.FK.identity.Bis.b} and explicitly identifying the
subsets $A$ having a given small value of $|A| - c(A)$:
\begin{quote}
\begin{itemize}
   \item[] $|A| - c(A) = 0$: Only $A = \emptyset$.
   \item[] $|A| - c(A) = 1$: $A = $ any single edge.
   \item[] $|A| - c(A) = 2$: $A = $ two distinct edges.
   \item[] $|A| - c(A) = 3$: $A = $ three distinct edges {\em or}\/
              four edges forming a plaquette.
   \item[] $|A| - c(A) = 4$: $A = $ four distinct edges not forming
              a plaquette {\em or}\/ four edges forming a plaquette
              together with one additional edge.
\end{itemize}
\end{quote}
We therefore have
\begin{eqnarray}
\widetilde{P}_{m_{\rm F}\times n_{\rm F}}(q) &=&  
  1 - \frac{|E|}{q} + \frac{|E|(|E|-1)}{2q^2} 
  - \left[ \frac{|E|(|E|-1)(|E|-2)}{6} - |F| \right] \frac{1}{q^3}   
  \nonumber \\
               & & \qquad + \left[ 
     \frac{|E|(|E|-1)(|E|-2)(|E|-3)}{24} - |F|(|E|-3) \right] \frac{1}{q^4} 
  \nonumber \\
               & & \qquad + O(q^{-5}) \,.  
 \label{eq.exp_Ptilde_mn}
\end{eqnarray}
Taking the logarithm, dividing by $|V|$, and putting back the leading term  
$\log q$, one finds the large-$q$ expansion for the free energy
\begin{eqnarray}
f_{m,n}^{\rm F}(q) &=& \log q -\frac{2}{q} - \frac{1}{q^2} + \frac{1}{3q^3} 
     + \frac{5}{2q^4} 
     + \left[ \frac{1}{q} + \frac{1}{2q^2} - \frac{2}{3q^3} 
             -\frac{11}{4q^4} \right] \left(\frac{1}{m}+\frac{1}{n}\right) 
  \nonumber \\
     & & \quad + \left[  \frac{1}{q^3}+\frac{3}{q^4} \right] \frac{1}{mn} 
   + O(q^{-5})\,.
\label{series.sq.free}
\end{eqnarray}
Comparing to the finite-size-scaling Ansatz \reff{def_FSS_Ansatz}, we obtain
\begin{subeqnarray}
f_{\rm bulk}(q) &=& \log q -\frac{2}{q} - \frac{1}{q^2} + \frac{1}{3q^3} 
     + \frac{5}{2q^4} + O(q^{-5})   \\[1mm]
f_{\rm surf}(q) &=&  \frac{1}{q} + \frac{1}{2q^2} - \frac{2}{3q^3} 
             -\frac{11}{4q^4} + O(q^{-5})   \\[1mm]
f_{\rm corner}(q) &=& \frac{1}{q^3}+\frac{3}{q^4} + O(q^{-5})
\label{series.sq.free.components}
\end{subeqnarray} 

Now consider an $m\times n$ square lattice with 
{\em cylindrical}\/ boundary conditions:
we have $|V|=mn$, $|E|=2mn-m$, and $|F|=m(n-1)$.
For small $m$ we have to worry about terms $A$
that wind horizontally around the lattice using the periodic
boundary conditions;  such terms start at order $q^{-(m-1)}$.
But we can avoid such terms simply by assuming that $m$
is large enough, i.e.\ $m \ge k+2$ if we want an expansion
valid through order $q^{-k}$.
Therefore, we can obtain the expansion through order $q^{-4}$
by assuming that $m \ge 6$;
then the contributing terms are exactly the same as those
for free boundary conditions, and we obtain the same expansion
\reff{eq.exp_Ptilde_mn} but with the modified values of $|E|$ and $|F|$.
A simple computation shows that
\begin{equation}
f_{m,n}^{P}(q) \;=\; f_{\rm bulk}(q)  
     \,+\, f_{\rm surf}(q) \, \frac{1}{n} \,+\, O(q^{-5})
\label{series.sq.cyl}
\end{equation}
where $f_{\rm bulk}$ and $f_{\rm surf}$ are the {\em same}\/ as those given in
\reff{series.sq.free.components},
in agreement with the finite-size-scaling prediction \reff{def_FSS_Ansatz}.

Unfortunately, this elementary graphical method for generating
the large-$q$ expansion is not very efficient if one wants to go to high order.
To obtain long expansions we will use a more sophisticated procedure:
the {\em finite-lattice method}\/ pioneered by Enting and collaborators
\cite{Neef-Enting_77,Enting_78,Kim-Enting_79,Enting_96,Enting_05,Enting_06}.
Indeed, Bakaev and Kabanovich \cite{Bakaev_94} used this method
to obtain, already in 1994, the
large-$q$ expansion of the bulk free energy through order $q^{-36}$.\footnote{
   For relevant earlier work on obtaining large-$q$ series for the
   infinite-volume limit of the chromatic polynomial,
   see Kim and Enting \cite{Kim-Enting_79}
   and the references cited therein.
}
Their results can be summarized as follows:
as $|q|\to\infty$, the exponential of the bulk free energy per site for
the zero-temperature Potts antiferromagnet (i.e., chromatic polynomial)
on the square lattice is given by the series expansion
\begin{eqnarray}
e^{f(q)} &=& \frac{(q-1)^2}{q} \left[ 1 + z^3 + z^7 + 3z^8 + 4z^9 +
 3z^{10} + 3z^{11} + 11z^{12} + 24z^{13} + 8z^{14} \right. 
\nonumber \\
         & & \quad -91z^{15} 
-261z^{16} -290z^{17} + \ldots -3068121066z^{36} + 
\left.   O(z^{37}) \right]
\label{series_1}
\end{eqnarray}
%
(the full series is given in Table~\ref{table_series}), where $z$ is defined as
\begin{equation}
z \;=\; \frac{1}{q-1}  \;.
\label{def_z}
\end{equation}
Before presenting our extensions of this result,
let us first briefly review the finite-lattice method.

In the finite-lattice method
\cite{Neef-Enting_77,Enting_78,Kim-Enting_79,Enting_96,Enting_05,Enting_06},
the large-$q$ expansion of the infinite-volume free energy
through a given order in $z=1/(q-1)$ is written as a
linear combination of free energies for rectangles of
various sizes $r \times s$:
\begin{subeqnarray}
f(q) - \log q &\equiv& \lim_{m,n\to\infty} \frac{1}{mn} 
       \log \widetilde{P}_{m_{\rm F} \times n_{\rm F}}(z)   \\[1mm]
              &=& \sum\limits_{(r,s)\in B(k)} \alpha_k(r,s) 
\log \widetilde{P}_{r_{\rm F}\times s_{\rm F}}(z)
              + O(z^{2k-3}) \,, 
\label{def_FLM}
\end{subeqnarray}
where we have used the variable $z = 1/(q-1)$ instead of $q=1+1/z$,
and the modified partition function $\widetilde{P}_{m_{\rm F}\times n_{\rm F}}$
[see \reff{def.Ptilde}].
The sum in \reff{def_FLM} is taken over all rectangles $r \times s$
belonging to the set
\begin{equation}
B(k) \;=\; \left\{ (r,s) \colon \, \text{$r\le s$ and $r+s\le k$}\right\}\,.
  \label{def_Bk}
\end{equation}
The weights $\alpha_k(r,s)$ are defined as
\begin{equation}
\alpha_k(r,s) \;=\; \begin{cases}
  2W_k(r,s) & \quad \text{for $r<s$}\\
   W_k(r,r) & \quad \text{for $r=s$}
\end{cases}
\end{equation}
where
\begin{equation}
 W_k(r,s) \;=\; \begin{cases}
   1 & \quad \text{for $r+s=k$}\\
  -3 & \quad \text{for $r+s=k-1$}\\
   3 & \quad \text{for $r+s=k-2$}\\
  -1 & \quad \text{for $r+s=k-3$}\\
   0 & \quad \text{otherwise}
\end{cases}
\end{equation}
The error term in \reff{def_FLM} is given by a
particular subclass of connected graphs 
that do not fit into any of the rectangles in $B(k)$
\cite{Neef-Enting_77} \cite[Chapter~12]{Biggs_93}.
In our case, these graphs are \cite{Bakaev_94} the convex polygons\footnote{
   We recall \cite{Guttmann_88} that a {\em convex polygon}\/ 
   (in the square lattice) is a self-avoiding polygon whose length
   equals the perimeter of its minimal bounding rectangle.
}
of perimeter $2k-2$ having the property that any pair of nearest-neighbor sites
belonging to the polygon must be connected by an edge of the polygon.\footnote{
   This condition excludes, for instance, polygons containing a
   ``bottleneck'' of width 1 or a ``protuberance'' of width 1.
}
Since any polygon of perimeter $2k-2$ has $|A| - c(A) = (2k-2) - 1 = 2k-3$,
we deduce that the error term in \reff{def_FLM} is of order  
$q^{-(2k-3)} \sim z^{2k-3}$.

Now suppose that we are somehow able to compute the transfer matrices
for widths $L \le L_{\rm max}$ with free boundary conditions.
We can then use these transfer matrices to compute
the partition functions $\widetilde{P}_{r_{\rm F}\times s_{\rm F}}$
for $r \le L_{\rm max}$ and $s$ arbitrary.
This means that if we set the cut-off $k$ equal to $2L_{\rm max}+1$, 
we will be able to compute the partition function
for all pairs $(r,s) \in B(k)$.
It follows that formula \reff{def_FLM} gives the bulk-free-energy series
correct through order $z^{4L_{\rm max}-2}$.  
We have empirically checked (by doing computations for different values of 
$L_{\rm max}$ and checking to what order they agree) that this formula
is correct. In our case $L_{\rm max}=12$, so we expect to obtain the 
free-energy series correct up to order $z^{46}$. 

If we compare the results coming from different values of $L_{\rm max}$,
we find empirically that the first incorrect term (of order 
$z^{4L_{\rm max}-1}$) is given by the generating function of the
aforementioned subclass of convex polygons, which is \cite{Bakaev_94}
\begin{equation}
\sum\limits_{k=4}^\infty \mu_k x^{2k} \;=\; 
x^8 \frac{2-2x^2-x^2\sqrt{1-4x^2}}{(1-4x^2)(2+x^2)} + 
x^{12} \frac{3-4x^2 - 4\sqrt{1-4x^2}}{(1-4x^2)^2} \;.
\label{def_gen_function}
\end{equation} 
If we add the term $\mu_{2L_{\rm max}} z^{4L_{\rm max}-1}$ to the
series obtained with a given value of $L_{\rm max}$, we get a series
correct up to (and including) $z^{4L_{\rm max}-1}$. In our case
$L_{\rm max}=12$, so we obtain a series expansion correct up to
order $z^{47}$.
The idea of adding this correction term
is due to Bakaev and Kabanovich \cite{Bakaev_94}
and was used by them to improve their own series by one term.\footnote{
   They actually went farther and improved their series by a second term,
   by enumerating a more complicated class of contributing graphs
   that we refrain from considering here.
}

In this way, we confirm the expansion \reff{series_1}
found by Bakaev and Kabanovich \cite{Bakaev_94} and extend it by 11 terms: 
\begin{eqnarray}
e^{f(q)} &=& \frac{(q-1)^2}{q} \left[ 1 + z^3 + z^7 + 3z^8 + 4z^9 +
 3z^{10} + 3z^{11} + 11z^{12} + 24z^{13} + 8z^{14} \right. 
\nonumber \\
         & & \quad -91z^{15} 
 -261z^{16} -290z^{17} + \ldots 
\left.  -598931311074z^{47} + O(z^{48}) \right] \,,
\label{series_2}
\end{eqnarray}
where the full series is given in Table~\ref{table_series}.
In terms of the variable $1/q$, we obtain
\begin{eqnarray}
e^{f(q)} &=&  q \left[ 1 - 2q^{-1} + q^{-2} + q^{-3} + q^{-4} + q^{-5}
  + q^{-6} + 2q^{-7} + 9q^{-8} + 38q^{-9} + 130 q^{-10}  \right.
\nonumber \\
         & & \quad + 378q^{-11} + 987q^{-12} + \ldots 
+1311159363081366872 q^{-47} + \left. O(q^{-48})\right] \,. \quad
\label{series_2bis}
\end{eqnarray}
%
Finally, for the bulk free energy $f(q)$ itself
(rather than its exponential)
in terms of the variable $1/q$, we obtain
\begin{eqnarray}
f(q) &=& \log q - \smfrac{2}{q} - \smfrac{1}{q^2} +
   \smfrac{1}{3q^3} + \smfrac{5}{2q^4} + \smfrac{28}{5q^5}+ \smfrac{55}{6q^6}+
   \smfrac{89}{7q^7} + \smfrac{81}{4q^8} + \smfrac{505}{9q^9} 
   +\smfrac{1029}{5q^{10}} + \smfrac{7742}{11q^{11}} \nonumber \\
    & & \qquad
   + \smfrac{25291}{12q^{12}}
   + \smfrac{73552}{13q^{13}} 
   + \smfrac{197755}{14q^{14}} + \ldots 
   + \smfrac{190018276619486037135}{47q^{47}}
+ O(q^{-48})
   \;. \quad
\label{series_fbulk}
\end{eqnarray}
Clearly, $e^{f(q)}$ has a much simpler expansion than $f(q)$;
in particular, its coefficients are integers
(at least through the order calculated thus far).\footnote{
   The coefficients of $f(q)$ are not integers,
   but $k \, [q^{-k}] f(q)$ is an integer
   (at least through the order calculated thus far).
   Indeed, it is not hard to show that if $F(z)$ is a power series
   with integer coefficients and constant term 1,
   then $k \, [z^{k}] \log F(z)$ is always an integer.
}
A further simplification is obtained by using the variable $z = 1/(q-1)$
in place of $1/q$:  the integer coefficients become much smaller.
Finally, a slight extra simplification arises from
extracting the prefactor $(q-1)^2/q$ in \reff{series_2}.

The finite-lattice method can be used for computing series expansions,
not only of the bulk free energy, but also of the surface and corner
free energies [cf.\ \reff{def_FSS_Ansatz}]. 
The main ideas were introduced by Enting \cite{Enting_78}
three decades ago and have been further developed by him
\cite{Enting_05,Enting_06}.
To our knowledge, however, no one has yet applied this technique
to the large-$q$ expansions of chromatic polynomials,
so the all the results to be presented below are new.   

As for the bulk case, the large-$q$ expansions of the surface and corner
free energies through a given order in $z=1/(q-1)$ are written as  
certain linear combinations of free energies for rectangles of
various sizes $r \times s$ [cf.\ \reff{def_FLM}]:
\begin{subeqnarray}
f_{\rm surf}(q) &=& \sum\limits_{(r,s)\in B(k)} \beta_k(r,s) 
\log \widetilde{P}_{r_{\rm F}\times s_{\rm F}}(z)
              + O(z^{2k-3})  \\[2mm]
f_{\rm corner}(q) &=& \sum\limits_{(r,s)\in B(k)} \gamma_k(r,s) 
\log \widetilde{P}_{r_{\rm F}\times s_{\rm F}}(z)
              + O(z^{2k-3}) \;, 
\label{def_FLMBis}
\end{subeqnarray}
where the set $B(k)$ is the same as for the bulk case [cf.\ \reff{def_Bk}].
The weights for the surface free energy are given by \cite{Enting_78}
\begin{equation}
\beta_k(r,s) \;=\; \begin{cases}
  S_k(r,s)+S_k(s,r) & \quad \text{for $r<s$}\\
  S_k(r,r)          & \quad \text{for $r=s$}
\end{cases} 
\end{equation}
where
\begin{equation}
 S_k(r,s) \;=\; \begin{cases}
   1-r & \quad \text{for $r+s=k$}\\
   3r-1 & \quad \text{for $r+s=k-1$}\\
  -3r-1 & \quad \text{for $r+s=k-2$}\\
   r+1 & \quad \text{for $r+s=k-3$}\\
   0 & \quad \text{otherwise}
\end{cases}
\end{equation}
The weights for the corner free energy are given by \cite{Enting_78}
\begin{equation}
\gamma_k(r,s) \;=\; \begin{cases}
  2V_k(r,s) & \quad \text{for $r<s$}\\
  V_k(r,r)  & \quad \text{for $r=s$}
\end{cases} 
\end{equation}
where 
\begin{equation}
 V_k(r,s) \;=\; \begin{cases}
   (r-1)(s-1) & \quad \text{for $r+s=k$}\\
   1+r+s-3rs & \quad \text{for $r+s=k-1$}\\
   3rs+r+s-1 & \quad \text{for $r+s=k-2$}\\
   -(r+1)(s+1) & \quad \text{for $r+s=k-3$}\\
   0 & \quad \text{otherwise}
\end{cases}
\end{equation}
The error term in (\ref{def_FLMBis}a,b) is given by the same reasoning as for
the bulk case, i.e.\ it is $O(z^{2k-3})$.
Therefore, by considering strips up to a maximum width $L_{\rm max}$,
we can use (\ref{def_FLMBis}a,b) to obtain the series for the
surface and corner free energies correct through order $z^{4L_{\rm max}-2}$.
We have empirically checked (by doing computations for different values of 
$L_{\rm max}$ and checking to what order they agree) that this formula
is correct. 
In our case we have $L_{\rm max} = 12$, hence series valid through 
order $z^{46}$.

For the surface free energy we can conjecturally extend the series
by one term by comparing the series for different values of $L_{\rm max}$
and noticing empirically that the needed correction for
the term of order $z^{4L_{\rm max}-1}$
is simply a multiple of the correction term \reff{def_gen_function}
for the bulk-free-energy series.
More precisely, if we add the term
$-L_{\rm max} \mu_{2L_{\rm max}} z^{4L_{\rm max}-1}$ 
to the series obtained with a given value of $L_{\rm max}$, we get a series
correct through order $z^{4L_{\rm max}-1}$.
In our case $L_{\rm max}=12$, so we obtain a series expansion for
$f_{\rm surf}$ correct (conjecturally) through order $z^{47}$.
Unfortunately, we have not yet succeeded in figuring out an analogous
correction for the corner free energy.

The results for the series expansions are given by
\begin{eqnarray}
e^{f_{\rm surf}} &=& 1 +z -z^3 -z^4 +z^6 -z^7 -8z^8 -16z^9 -16z^{10} -12z^{11}
     -41z^{12}  \nonumber \\
\label{series_exp_fsurf}
                 & & \qquad  -138z^{13} + \ldots -130312353695974 z^{47} 
     + O(z^{48}) \\[3mm]
e^{f_{\rm corner}} &=& 1 + z^3 + 4z^7 + 12z^8 + 20z^9 + 28z^{10} + 67z^{11}
    +208z^{12} + 484z^{13}  \nonumber \\
                   & & \qquad + 753z^{14} + \ldots  
   + 448320847685638 z^{46} + O(z^{47})  \;,
\label{series_exp_fcorner}
\end{eqnarray}
where the full list of coefficients is displayed in Table~\ref{table_series}. 
In terms of the variable $1/q$ we get 
\begin{eqnarray}
f_{\rm surf}   &=& \smfrac{1}{q} +\smfrac{1}{2q^2} -\smfrac{2}{3q^3} 
     -\smfrac{11}{4q^4} -\smfrac{29}{5q^5} -\smfrac{28}{3q^6} 
     -\smfrac{97}{7q^7} -\smfrac{243}{8q^8} -\smfrac{1019}{9q^9} 
     -\smfrac{4489}{10q^{10}} - \smfrac{17280}{11q^{11}} \nonumber \\
               & & \qquad -\smfrac{14654}{3q^{12}} -\smfrac{183143}{13q^{13}}
             + \ldots  -\smfrac{1103009229135728011786}{47q^{47}}
+ O(q^{-48})
     \label{series_fsurf}  \\[3mm]
f_{\rm corner} &=& \smfrac{1}{q^3} + \smfrac{3}{q^4} + \smfrac{6}{q^5}
+ \smfrac{19}{2q^6} + \smfrac{16}{q^7} + \smfrac{101}{2q^8} 
+ \smfrac{685}{3q^9} + \smfrac{948}{q^{10}} + \smfrac{3409}{q^{11}}\nonumber \\
              & & \qquad + \smfrac{44399}{4q^{12}} + \smfrac{34558}{q^{13}} 
             + \ldots +\smfrac{280315319437238591517}{2q^{46}}  
+ O(q^{-47})
     \label{series_fcorner}
\end{eqnarray}
%

%
%
\subsection[Large-$q$ expansion for the strip free energy:
               Free boundary conditions]
           {Large-$\bm{q}$ expansion for the strip free energy: \hfill\break
               Free boundary conditions} \label{sec.thermo.F}

We can check the results \reff{series_fbulk}/\reff{series_fsurf}
for the bulk and surface free energies
by computing the large-$|q|$ expansion of the limiting free energy
for a semi-infinite strip with free boundary conditions:
\begin{eqnarray}
f_{m}^{\rm F}(q)&=& \frac{1}{m} \log \lambda_{\star,m}^{\rm F}(q)  \nonumber \\
         &=& \log q + \frac{1}{m}\log\left[ \sum\limits_{k=0}^\infty
             (-1)^k b^{\rm F}_k(m) q^{-k} \right] \nonumber \\
&=& \log q + \frac{1}{m}  \sum\limits_{k=1}^\infty
             (-1)^k c^{\rm F}_k(m) q^{-k}
\label{def_fm_F_largeq}
\end{eqnarray}
If $m \ge m_{\rm min}^{\rm F}(k)$,
we can replace the coefficients $c_k^{\rm F}(m)$
by the corresponding polynomials $C_k^{\rm F}(m)$ [cf.\ \reff{def_ck_poly_F}].
These polynomials $C^{\rm F}_k$ are of degree 1 in $m$ for $1\leq k \leq 33$,
and we have conjectured that this behavior holds for all values of $k\geq 1$. 
Thus, $f_m^{\rm F}(q)$ contains
[as predicted in \reff{eq.fm_prediction.free}]
only two terms:
a bulk term $f_{\rm bulk}(q) = f(q)$ that is independent of $m$,
and a surface free energy energy $f^{\rm F}_{\rm surf}(q)$
that is of order $1/m$:
\begin{equation}
f_{m}^{\rm F}(q) \;\cong\;
  f_{\rm bulk}(q) \,+\, \frac{1}{m} f_{\rm surf}(q)  \;.
\label{def_fbulk_fsurf_F}
\end{equation}
Here $\cong$ denotes that the two sides agree at each order $q^{-k}$
of the expansion in powers of $q^{-1}$, but we require this only for
$m \ge m_{\rm min}^{\rm F}(k)$.
This computation using the polynomials $C_k^{\rm F}$ thus provides an 
independent check of the first $33$ terms of the series \reff{series_fbulk} 
and \reff{series_fsurf}.
Conversely, the finite-lattice computation of the preceding subsection
provides an independent confirmation of the empirically observed
regularities in the behavior of the polynomials $C_k^{\rm F}$
[cf.\ \reff{def_Delta}--\reff{def_mhat_min_F_bis}].

We can slightly extend the check for the surface-free-energy series 
by using the series \reff{series_fbulk} for $f_{\rm bulk}$ as an {\em input}\/:
in this case, each polynomial $C_k^{\rm F}$ contains a single unknown
coefficient to be determined (rather than two unknown coefficients).
We then obtain the coefficient of the term
$z^{34}$ in $f_{\rm surf}$, which agrees with the result from
the finite-lattice method displayed in Table~\ref{table_series}.

%
%
\subsection[Large-$q$ expansion for the strip free energy:
               Cylindrical boundary conditions]
           {Large-$\bm{q}$ expansion for the strip free energy: \hfill\break
               Cylindrical boundary conditions} \label{sec.thermo.P}

The large-$|q|$ expansion of the free energy for a semi-infinite strip with
cylindrical boundary conditions is
\begin{eqnarray}
f_{m}^{\rm P}(q)&=& \frac{1}{m} \log \lambda_{\star,m}^{\rm P}(q)  \nonumber \\
         &=& \log q + \frac{1}{m}\log\left[ \sum\limits_{k=0}^\infty
             (-1)^k b^{\rm P}_k(m) q^{-k} \right] \nonumber \\
&=& \log q + \frac{1}{m}   \sum\limits_{k=1}^\infty
             (-1)^k c^{\rm P}_k(m) q^{-k}
\label{def_fm_P_largeq}
\end{eqnarray}
Once again, if $m$ is large enough (depending on $k$),
we can replace the coefficients $c^{\rm P}_k(m)$
by the corresponding polynomials $C^{\rm P}_k(m)$ [cf.\ \reff{def_ck_poly_P}].
These polynomials $C_k^{\rm P}$ are of degree 1 in $m$
with a zero constant term for $1\leq k \leq 16$, and we conjecture
that this behavior holds for every $k\geq 1$.  
Thus, $f_m^{\rm P}(q)$ contains
[as predicted in \reff{eq.fm_prediction.cyl}]
only a single $m$-independent term $f_{\rm bulk}(q) = f(q)$;
there is no surface contribution.
Of course this result is to be expected,
as an infinitely long cylinder has no boundary.
This computation using the polynomials $C^{\rm P}_k$
yields a series for $f_{\rm bulk}$ that is the {\em same}\/,
up to the order we are able to compute (namely, $q^{-16}$),
as for free boundary conditions;
in particular, it provides an independent check of the first 16 terms
of the series \reff{series_fbulk}.
Conversely, the finite-lattice computation of the preceding subsection
provides an independent confirmation of the empirically observed
regularities in the behavior of the polynomials $C_k^{\rm P}$
[cf.\ \reff{def_Delta_P}--\reff{def_mhat_min_P_bis}].
 
If we use the bulk free-energy series \reff{series_fbulk} as input,
we can compute the term of order $q^{-17}$ in the surface free energy:
as expected, we find that it vanishes.

%
%
\subsection[Analysis of the large-$q$ series]
           {Analysis of the large-$\bm{q}$ series} \label{sec.series}

In this subsection we shall perform a series-extrapolation analysis
of the large-$q$ series \reff{series_2} ff.\ for the bulk free energy,
\reff{series_exp_fsurf} ff.\ for the surface free energy, 
and \reff{series_exp_fcorner} ff.\ for the corner free energy. 
As a warm-up, we shall first perform an analogous analysis
for the large-$q$ series of the strip free energy $f_m(q)$
for two selected cases, $m=3_{\rm F}$ and $m=4_{\rm P}$.
In these cases we can easily generate long large-$q$ series
(up to 100 terms or more) using the symbolic power method
(see footnote~\ref{footnote_symbolic_power} above)
and can compare the predicted location and nature of singularities
with the exactly known answers \cite{transfer1}.
In this way, we can learn how many terms in the large-$q$ series
are likely to be needed in order to extract specific features
of the singularity structure.

Before beginning this analysis, it is useful to know what types
of singularities we are expecting.
For statistical-mechanical models
with only one dimension tending to infinity
(i.e., finite-width strips),
the answer is clear:
the free energy $f_m(q)$ [cf.\ \reff{def_Fm_bis}]
is an analytic function of $q$
except at branch points arising from the collision of two eigenvalues.
Therefore, the limiting curve $\scrb_m$ is {\em not}\/ a curve of singularities
of the free energy;  only its endpoints are singularities.

For statistical-mechanical models with two or more dimensions
tending to infinity, as in our infinite-width limit $m \to\infty$,
the situation is much less clear.
It is well established in a variety of cases
\cite{Isakov_84,Friedli_04a,Friedli_04b,Pfister_05}
that the infinite-volume free energy has a ``soft'' essential singularity
(i.e., one in which the free energy is infinitely differentiable
 but not analytic)
at a first-order phase transition in the physical region.
What is less clear, however, is how the free energy behaves on
{\em complex}\/ phase boundaries (such as our limiting curve $\scrb_\infty$).
For instance, the point $h=0$ is known rigorously to be
an essential singularity for the free energy of a
low-temperature Ising ferromagnet in dimension $d \ge 2$
\cite{Isakov_84,Friedli_04a,Pfister_05}:
it is not possible to analytically continue {\em through}\/ this point,
because the derivatives at $h=0$ grow too fast
(the Taylor series has zero radius of convergence).
But might it be possible to analytically continue {\em around}\/ this point
in the complex $h$-plane?
If so, then the imaginary axis $\real h = 0$, $\imag h \neq 0$
would be merely the curve where two analytic functions
{\em defined in a neighborhood of that curve}\/
--- namely, the free energies that are initially defined
for $\real h > 0$ and for $\real h < 0$ but are analytically continuable
to at least part of the opposite half-plane, the point $h=0$ excluded ---
have equal real part.
By contrast, the alternative scenario is that these two analytic functions
{\em cannot}\/ be continued into any part of the opposite half-plane:
that is, they would have the full imaginary axis $\real h = 0$ 
as a {\em natural boundary}\/.
It does not seem to be known, even heuristically,
which of these two scenarios is correct.\footnote{
   See also \cite{Biskup_04a,Biskup_04b} for related work
   on complex phase boundaries and partition-function zeros.
}

%
%
\subsubsection{Differential approximants}

To perform the series-extrapolation analysis of our large-$q$ series,
we shall use the {\em differential-approximant (DA) method}\/:
see the review \cite{Guttmann_89} and the references cited therein. 
The idea is simple: The $K$th-order approximant 
$[N_0,N_1,\ldots,N_K;M]$ to a power series $F(z) = \sum_{k=0}^\infty f_k z^k$
is built by choosing polynomials $Q_0,Q_1,\ldots,Q_K$ and $P$ of degrees 
$N_0,N_1,\ldots,N_K$ and $M$, respectively, so that the solution 
$\widetilde{F}$ of the inhomogenenous linear differential equation
\begin{equation}
\sum\limits_{j=0}^K Q_j(z) \left( z \frac{d}{dz}\right)^j \, 
\widetilde{F}(z) \;=\; P(z)
\label{def_DA}
\end{equation}
agrees with the initial coefficients of the series $F(z)$
through order $z^N$, where $N = \sum_{i=0}^K (N_i + 1) + M$
(notice that we fix the overall normalization with $Q_K(0)=1$).
The singularities of $\widetilde{F}(z)$ are located 
at the zeros $\{z_\ell\}$ of the polynomial $Q_K$
(and possibly also at $z=0$ and $z=\infty$),
and the critical exponent associated to a {\em simple}\/ zero 
$z_\ell$ of $Q_K$ is given by 
\begin{equation}
\lambda_\ell \;=\; K-1 - \frac{ Q_{K-1}(z_\ell)} {z_\ell \, Q'_K(z_\ell)} \;,
\end{equation}
in the sense that
$\widetilde{F}(z) \sim (z-z_\ell)^{\lambda_\ell}$ as $z \to z_\ell$.

In practice, we have used a {\sc Mathematica} code to obtain the polynomials 
$Q_j$ {\em exactly}\/ (i.e., with exact rational arithmetic)
from the coefficients $\{ f_k \}$,
and then used the program {\sc MPSolve} \cite{MPSolve,Bini_00} to compute the 
$N_K$ zeros of the polynomial $Q_K$ to very high precision (i.e., 100 digits).
Our code and the numerical accuracy of its results were checked 
in a previous work \cite{forest} with the help of an independent 
{\sc C++} program written by Y.~Chan, A.J.~Guttmann and A.~Rechnitzer. 

In many cases, one is principally interested in the singularities 
that are located 
on the real axis. This assumption simplifies a bit the practical procedure,
as one needs to search only on a one-dimensional space.
However, in our case we want to locate all the singularities (real or complex)
of the approximants. In fact, we expect from the sign pattern
of the series coefficients \reff{series_2}--\reff{series_fbulk}
[cf.\ Table~\ref{table_series}]
that the leading singularity of the bulk free energy is {\em not}\/
on the real $z$-axis.  
Therefore, we need to slightly change the usual definition of a non-defective 
approximant (i.e., those that are taken into account in the computation)
\cite{Guttmann_89,forest}.  
Roughly speaking, we want to consider an approximant $[N_0,N_1,\ldots,N_K;M]$  
to be non-defective if there is a zero of $Q_K$ sufficiently near to the 
expected (complex) singularity, and this zero is sufficiently well separated 
from all other zeros of $Q_K$. 

Of course, this definition needs to be made more precise;
and things are made more difficult by the fact that we do not know
even roughly the position of the singularities in the complex $z$-plane.
In order to find a rough estimate of these positions,
we first make a histogram of all the ``well-separated'' complex zeros
$\{z_\ell\}$ coming from all computed approximants $[N_0,N_1,\ldots,N_K;M]$. 
A pair of complex zeros $z_1, z_2$ of $Q_K$ is defined to be well-separated if  
$|z_1 - z_2| > R$ for some (small) free parameter $R$;
a zero $z_1$ is then defined to be well-separated
if it is well-separated from all other zeros
(in particular, it is required to be simple).
Given a fixed value of $R$, we select from each polynomial $Q_K$
the well-separated zeros and make the corresponding histogram,
using square cells of side $R' = 0.04$.
This histogram is expected to display peaks at the singularities
of the function $F$. 
We further expect that the final result should not depend strongly
on the chosen value of $R$.
In practice, we started with $R=0.2$ and then compared the histogram to
those obtained with smaller values of $R=0.1,0.05,0.02$, etc. We considered 
that the procedure had converged when the number and positions of the peaks
did not vary when ``halving'' the value of $R$. Usually, the choice $R=0.05$ 
was optimal.  

Each peak displayed in the histogram corresponds to a cell in the
complex $z$-plane: $\Re z \in [A_1,A_2]$ and $\Im z \in [B_1,B_2]$
with $A_2 - A_1 = B_2 - B_1 = R' = 0.04$.
Therefore, for a given singularity $z_\ell$, 
we consider an approximant $[N_0,N_1,\ldots,N_K;M]$  
to be non-defective if there is a (simple) zero of $Q_K$ inside the above cell, 
and there is no other zero of $Q_K$ inside the larger region 
$\Re z \in [A_1-0.05,A_2+0.05]$ and $\Im z \in [B_1-0.05,B_2+0.05]$.\footnote{
  For real singularities, we used a slightly different definition:
  namely, for zeros near $z=0.5$ (resp.\ $z=-1$)
  we asked that there should be no other zero in the region
  $|\Im z| \le 0.1$ and $\Re z \in [0,1.2]$
  (resp.\  $\Re z \in [-1.8,0]$).
}
Then we perform the analysis explained
by Guttmann in his review article \cite{Guttmann_89}.
Taking all the non-defective approximants of a given order $N$
corresponding to a given singularity $z_\ell$,
we form the mean and standard deviation
for each of the three quantities $\Re z$, $\Im z$ and $\lambda$.
Of course, this standard deviation is only pseudo-statistical;
for brevity we term it the ``error bar'',
but it is not necessarily indicative of the accuracy of the estimate.

Our initial goals are twofold:
to test that the above-described procedure gives the 
right answers in exactly soluble cases, and to determine how the estimates
for the positions $\{z_\ell\}$ and critical exponents $\{\lambda_\ell\}$ of the
singularities vary with the available order $N$ of the series.\footnote{
   That is, we assume that we know the coefficients of the series $F(z)$
   through order $z^N$.
   Sometimes we shall refer loosely to $N$ as the
   ``number of available coefficients''
   even though, strictly speaking, the number of available coefficients
   (including the constant term) is $N+1$.
}
In order to achieve these goals, we have chosen two simple examples for 
which we know the position and type of all the singularities, and for which
we can easily obtain many terms of the corresponding free energies
(say, 100 terms):
namely, the strip free energies $m = 3_{\rm F}$ and $m = 4_{\rm P}$.
As in Ref.~\cite{forest}, we have computed the approximants of 
first and second order ($K=1,2$) satisfying $|N_i-N_j|\le 1$. There are 
four free parameters in this computation:
the maximum degree $N_{0,\rm{max}}$ of the polynomial $Q_0$, 
the maximum degree $M_{\rm max}$ of the inhomogeneous term $P$, 
the minimum number of series coefficients used in the analysis 
$N_{\rm min}$, and the number of available coefficients $N$. 
We have chosen the first three parameters in terms of $N$ in the following way: 
$N_{0,\rm{max}}=0.8N$, $M_{\rm max}=0.5N$, and $N_{\rm min}=0.3N$. Thus, 
the only free parameter in the computation is $N$. In our test cases, we have
chosen $N$ to range between $30$ and $100$ in steps of $10$. The two test
cases are described in the next subsection. 

%
%
\subsubsection[Test cases: ${m=3_{\rm F}}$ and ${m=4_{\rm P}}$]
              {Test cases: $\bm{m=3_{\rm F}}$ and $\bm{m=4_{\rm P}}$}
 
The eigenvalues for the case $m=3_{\rm F}$ are the solutions
of the quadratic equation \cite{transfer1}
\begin{equation}
\lambda^2 - (q^3-5q^2+11q-10)\lambda + (q^4-7q^3+19q^2-24q+11) \;=\; 0
  \;.
\end{equation}
Therefore, the only singularities are those corresponding to the six endpoints
of the limiting curve $\mathcal{B}_3$,
where the discriminant of the quadratic vanishes.
In the complex plane of $z=1/(q-1)$, they are located at
\begin{subeqnarray}
z_1 &\approx& -0.2811172691\pm 0.7752009092\,i \\
z_2 &\approx&  0.4477839366\pm 0.5382490441\,i \\
z_3 &=& \frac{3+\sqrt{3}\,i}{6} \;\approx\; 0.5 \pm 0.2886751346 \,i  
\end{subeqnarray}

In Figures~\ref{figure_histo_3F}(a,b) we plot the histograms obtained with
$N=50$ and $N=100$, respectively, using $R=0.05$.
The number of zeros contained in each square cell of linear size $0.04$
is indicated with a gray scale of $16$ tones: those cells with the 
maximum number of zeros are depicted in black, while those with the smallest
number of zeros (i.e., less than 7\% of that maximum number) are depicted 
in white.  
We have labelled the peaks with $\Im z \ge 0$ as $z_1, \ldots, z_7$.
We observe several empirical properties:
as $N$ is increased, the number of peaks increases
(e.g.\ $z_6$ and $z_7$ appear for $N=100$ but not for $N=50$)
and some of the peaks become sharper (e.g.\ $z_4$);
furthermore, we find more peaks than the actual number of singularities
(we expect three pairs of complex-conjugate peaks $z_1,z_2,z_3$, 
but we also find two further pairs $z_4,z_7$ of complex-conjugate peaks
and two peaks $z_5, z_6$ on the real axis).\footnote{
 In some cases, we find a ``broad'' peak that has large counts in two 
 (or more) neighboring cells. These peaks are counted as one peak; 
 but we have always looked carefully to see whether they have a finer
 structure or not (i.e., if they correspond to two nearby singularities).
}
For each of the seven peaks,
we have performed the analysis described in the preceding subsection,
in order to estimate
the position of the singularity and the corresponding exponent.
In Tables~\ref{table_sq_3F_series1} and \ref{table_sq_3F_series2}
we display the results.  

We indeed find the three ``correct'' singularities associated to each of 
three endpoints of the limiting curve in the upper half-plane (see 
Table~\ref{table_sq_3F_series1}). As the
number of coefficients is increased, the accuracy of the results increase:
with $N=100$ we attain a
maximum accuracy of order $10^{-7}$ (resp.\  $10^{-5}$) 
for the position of the singularity (resp.\ the critical exponent), and in 
all cases the estimate agrees with the known exact value
within 3 times the claimed error bar (and usually much less).

In Table~\ref{table_sq_3F_series2}, we show the ``singularities'' found 
in the series analysis that do {\em not}\/ correspond
to any of the endpoints of the limiting curve.
Once again, as we increase the number of coefficients $N$ in 
the analysis, the claimed error bars decrease.
Let us stress that the exponents of the first three singularities
($z_4,z_5,z_6$) are consistent with $\lambda=0$;
therefore, the DA analysis is asserting (correctly) that
there is no true singularity at these points.
The same conclusion holds for the point $z_7$, for which
estimated exponent is very close to the positive-integer value $\lambda=2$.
Finally, it seems that $z_5=-1$ (i.e., $q=0$).  
For this example, at least, the DA method has itself told us (correctly)
that the only true singularities are those shown in
Table~\ref{table_sq_3F_series1}.

In Figure~\ref{figure_singularities_test}(a) we plot the limiting curve 
for $m=3_{\rm F}$ in the complex $z$-plane together with the
singularities found by the series analysis.
For comparison, we have also estimated the 
radius of convergence of the series from the formula
$r_{\rm conv} = \liminf_{n \to\infty} |a_n|^{-1/n}$:
despite some oscillations, we are able to obtain the estimate
$r_{\rm conv}\approx 0.5982$, which is close to (and slightly larger than)
the modulus $|z_3| \approx 0.5774$ of the singularity closest to the origin
(see Table~\ref{table_sq_3F_series1}).

The case $m=4_{\rm P}$ is more involved from a numerical point of view
because, even though the free-energy singularities are of the same type
as for $m=3_{\rm F}$ (namely, square-root branch points with $\lambda=1/2$),
two of them lie extremely close to one another.
The eigenvalues for this strip are given by the solutions of the
quadratic equation \cite{transfer1}
\begin{equation}
\lambda^2 - (q^4-8q^3+29q^2-55q+46)\lambda +
   (q^6-12q^5+61q^4 -169q^3+269q^2-231q+85)
\;=\; 0 \;.
\end{equation} 
The eight endpoints of the corresponding limiting curve are located 
in the $z$-plane at
\begin{subeqnarray}
z_1 &\approx& 0.3316354418 \pm 0.2371152471\,i  \\
z_2 &\approx&-0.0681712693 \pm 0.4798609413\,i  \\
z_3 &\approx& 0.2814289723 \pm 0.4521062477\,i \\
z_4 &\approx& 0.7398155434 \\ 
z_5 &\approx& 0.7978491474  
\end{subeqnarray}

In Tables~\ref{table_sq_4P_series1} and \ref{table_sq_4P_series2}
we display our results as a function of $N$.
(For brevity we refrain from showing the histograms.)
The three complex singularities ($z_1,z_2,z_3$)
and their corresponding exponents are well determined with
small error bars. However, the determination of the two closely-separated
real singularities ($z_4,z_5$) is (not surprisingly) rather poor:
these singularities can be seen only if we have at least 80 coefficients,
and even with 90 coefficients we obtain only a single singularity located
in-between the two endpoints.
Only for $N=100$ do we start seeing a (noisy) signal of the second endpoint. 

In this case the analysis also gives two singularities that do not
correspond to any of the endpoints of the limiting curve
(see Table~\ref{table_sq_4P_series2}). One is located at $z=-1$
(i.e.\ $q=0$) and the other at $z\approx 0.285\pm 0.518\,i$. In
both cases, the critical-exponent estimate is consistent with
$\lambda=0$. Therefore, the DA analysis is itself telling us
that they are not true singularities.

In Figure~\ref{figure_singularities_test}(b) we plot the limiting curve 
for $m=4_{\rm P}$ in the complex $z$-plane together with the
singularities found by the series analysis. We have also estimated the 
radius of convergence of the series
using $r_{\rm conv} = \liminf_{n \to\infty} |a_n|^{-1/n}$:
the result
is $r_{\rm conv}\approx 0.4216$, which is close to (and slightly larger than)
the modulus $|z_1| \approx 0.4077$ of the singularity closest to the origin
(see Table~\ref{table_sq_4P_series1}).

%
%
\subsubsection{Analysis of the series for the bulk, surface and corner 
free energies}   \label{subsec.series_analysis}

We are now ready to analyze the large-$q$ series
\reff{series_2} ff.\ for the bulk free energy,
\reff{series_exp_fsurf} ff.\ for the surface free energy,
and \reff{series_exp_fcorner} ff.\ for the corner free energy. 
We began by analyzing the series expansions for the 
exponential of the free energy $e^{f_i(z)}$
[$i={\rm bulk},{\rm surface},{\rm corner}$]
using the variable $z=1/(q-1)$.  
Later we tried analogous analyses for the free energy $f_i(z)$ itself,
and analyses using $1/q$ as the variable instead of $z$.

\begin{figure}
\centering
\begin{tabular}{cc}
  \includegraphics[width=200pt]{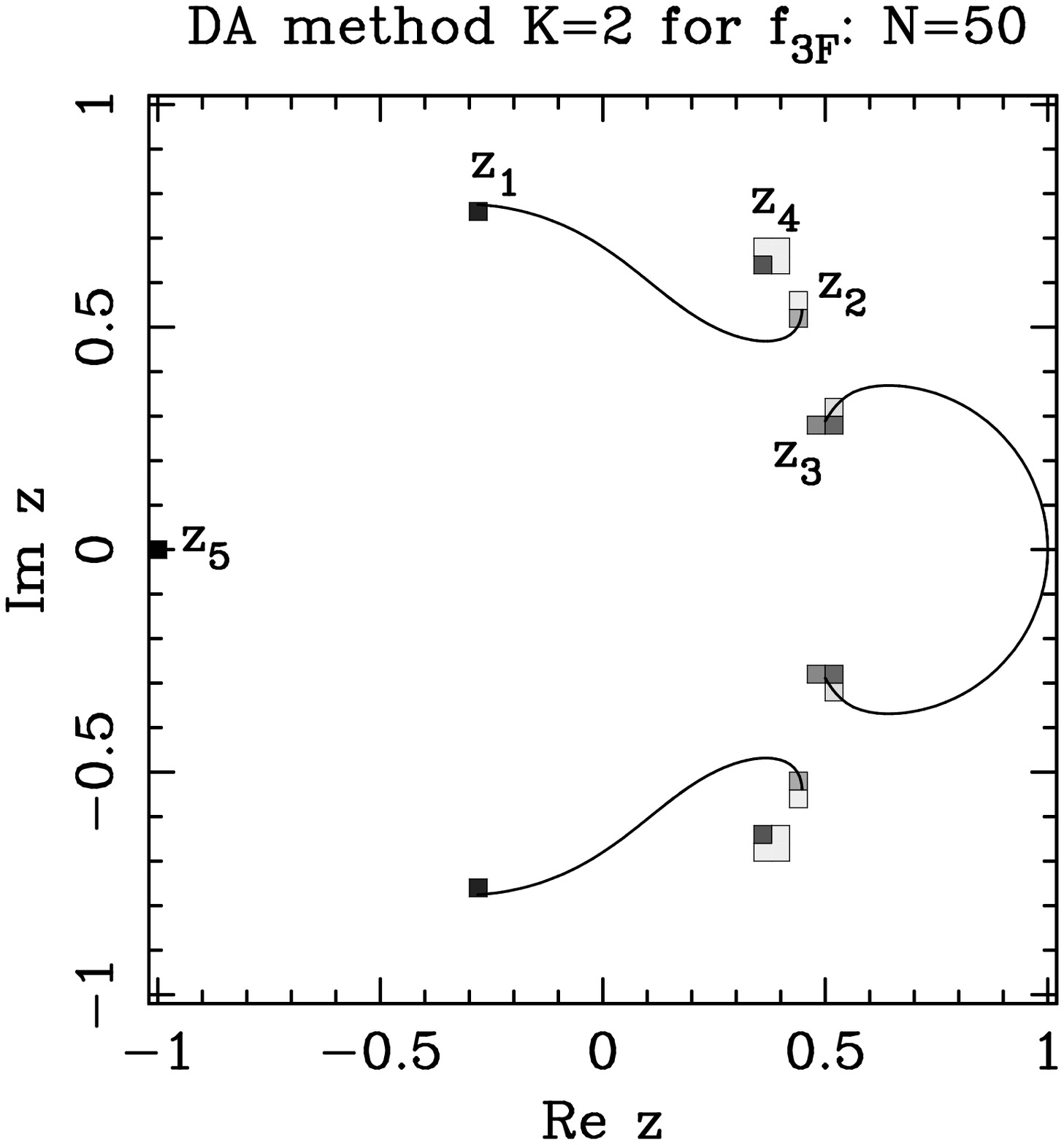} & 
  \includegraphics[width=200pt]{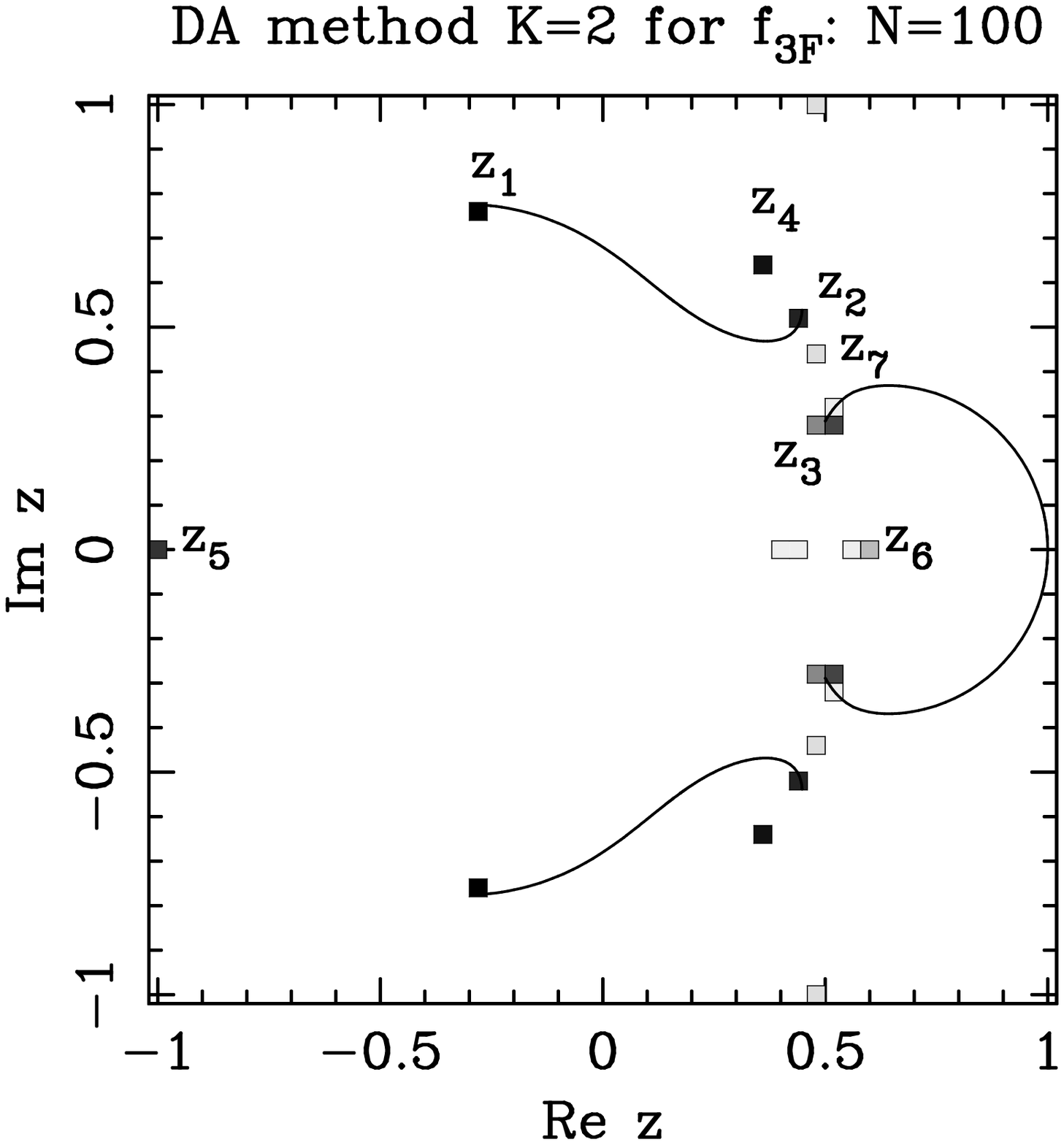}\\
  \phantom{(((a)}(a) & \phantom{(((a)}(b)\\[1mm]
\end{tabular}
\caption{\label{figure_histo_3F}
  Histograms of the non-defective zeros for the square lattice
  strip of width $L=3_{\rm F}$ using $K=2$ differential
  approximants with $N=50$ (a) and $N=100$ (b) coefficients
  of the free-energy series expansion. 
  Each cell is a square of size $0.04$, and the bin count 
  is indicated with a gray-scale code: there are 16 gray tones ranging from 
  black (largest counts) to white (smallest counts).
  We also show the limiting curve $\mathcal{B}_{3_{\rm F}}$.  
}
\end{figure}

\begin{figure}
\centering
\begin{tabular}{c@{\hspace{1cm}}c}
  \includegraphics[width=200pt]{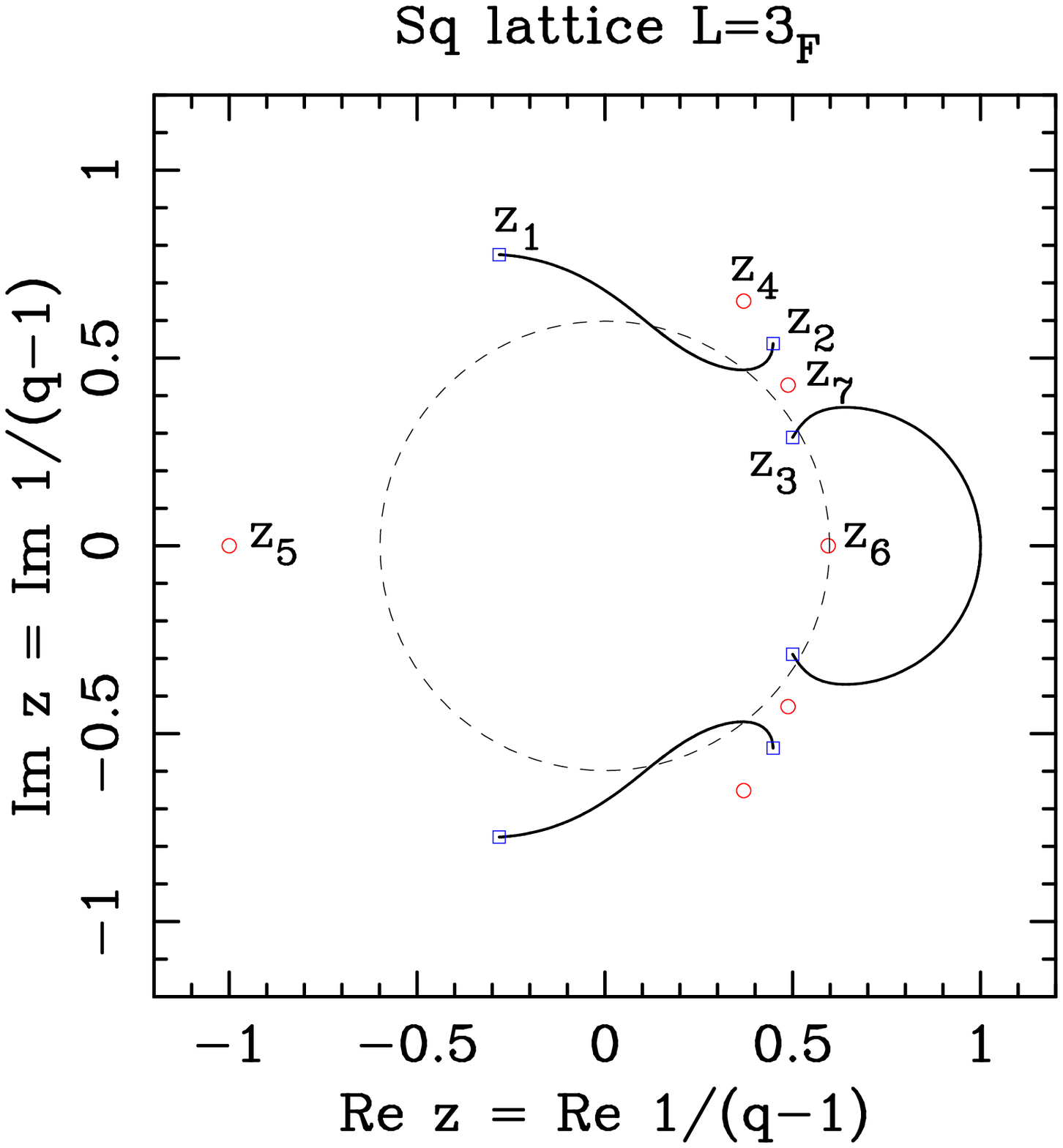} &
  \includegraphics[width=200pt]{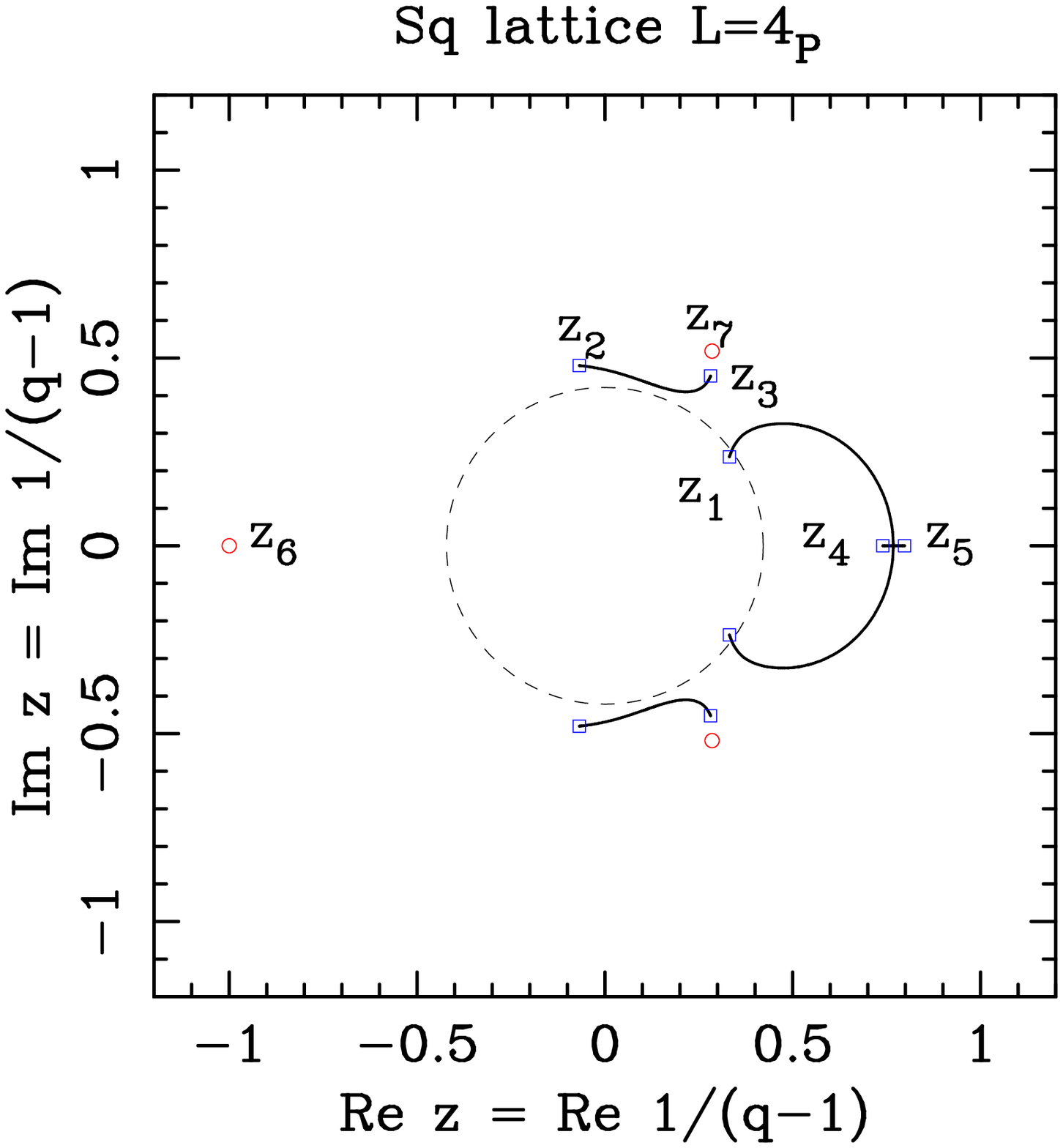} \\
  \phantom{(((a)}(a) & \phantom{(((a)}(b)\\[1mm]
\end{tabular}
\caption{\label{figure_singularities_test}
  Plot of the limiting curves for the square-lattice strips of
  widths $3_{\rm F}$ (a) and $4_{\rm P}$ (b) in the complex
  $z=1/(q-1)$ plane. We also show the
  location of the physical singularities (i.e., the endpoints of the
  limiting curve, depicted in blue $\square$) and
  of the spurious ones (red $\circ$).
  The dashed line shows the circle with the radius of convergence
  obtained by a raw fit using
  $r_{\rm conv} = \liminf_{n \to\infty} |a_n|^{-1/n}$.
  The point labels correspond to those of the tables.
}
\end{figure}

\begin{figure}
\centering
\begin{tabular}{cc}
  \includegraphics[width=200pt]{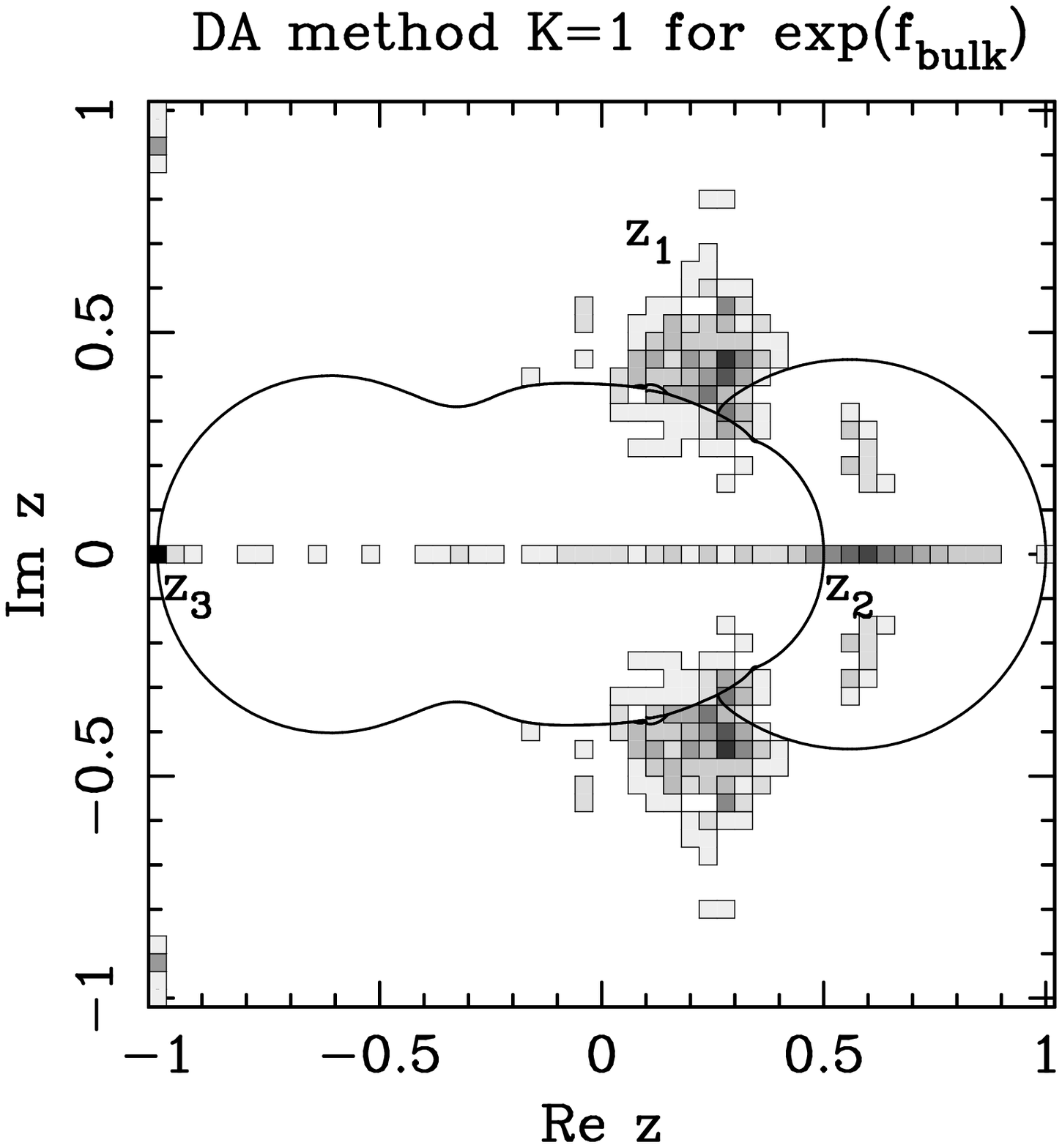} &
  \includegraphics[width=200pt]{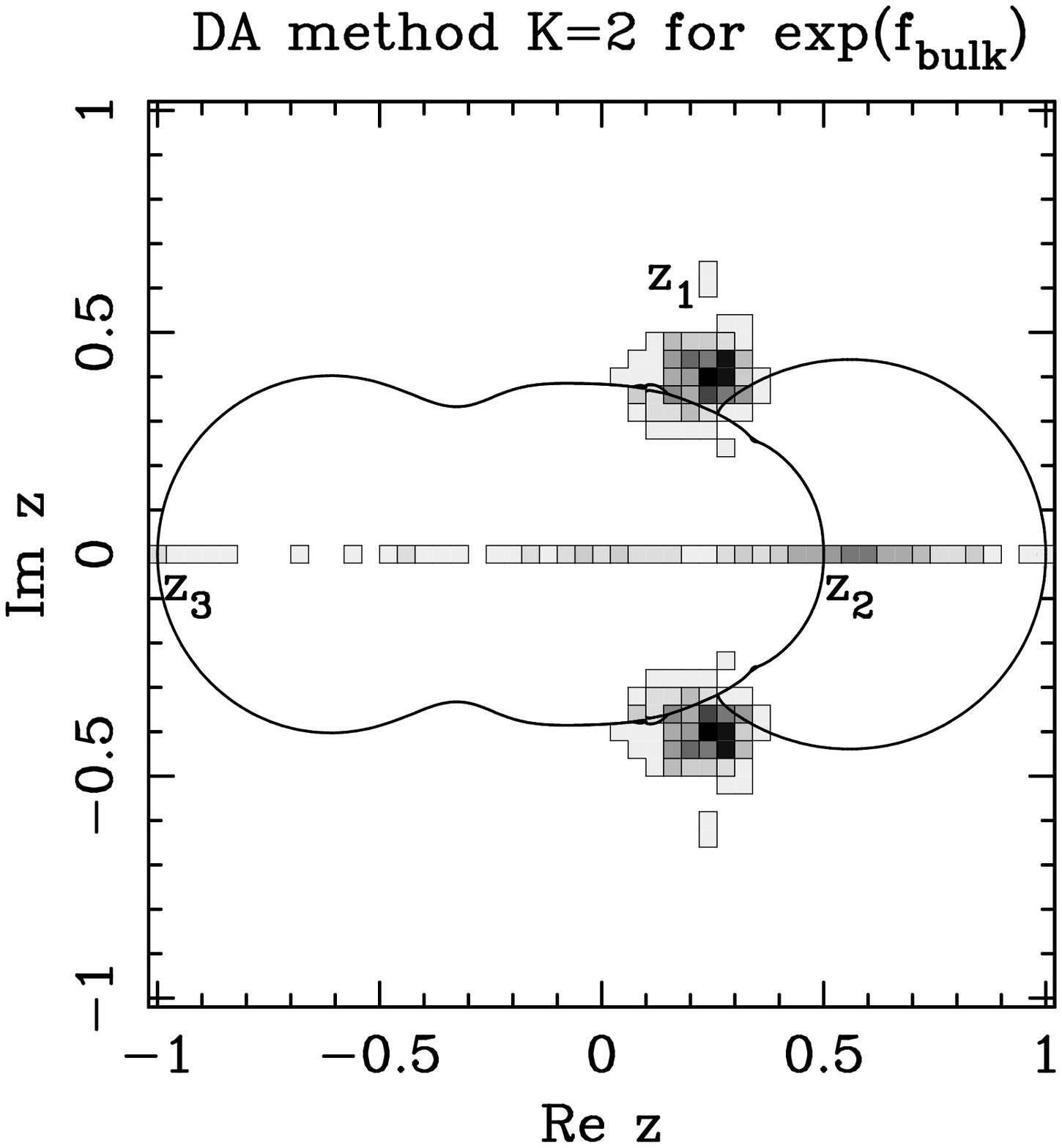} \\
  \phantom{(((a)}(a) & \phantom{(((a)}(b)\\[1mm]
\end{tabular}
\caption{\label{figure_histo_bulk}
  Histograms of the non-defective zeros for the bulk-free-energy  
  series \protect\reff{series_2} using $K=1$ (a) and $K=2$ (b) differential
  approximants with $N=47$ coefficients.
  Gray-scale codes are as in Figure~\protect\ref{figure_histo_3F}. 
  For comparison, we show the limiting curve $\scrb_m$
  for the square-lattice strip of width $m=7$ with
  toroidal boundary conditions \protect\cite{Jacobsen-Salas_toroidal}.
}
\end{figure}

For the bulk free energy, we have $N=47$ coefficients.
The histograms for the $K=1$ and $K=2$ approximants are displayed in
Figure~\ref{figure_histo_bulk}(a,b),
using the gray-scale coding described earlier.
If we compare these histograms to that of the test case of 
Figure~\ref{figure_histo_3F}(a) with a similar number of coefficients ($N=50$), 
we see that the zeros do not accumulate on small regions on the complex 
$z$-plane as they did for the test case;
rather, we find a complex-conjugate pair of broad peaks $z_1$,
a broad peak $z_2$ on the real axis,
and for $K=1$ a rather sharp peak $z_3 \approx -1$.
It is unclear (at least to us) what this behavior of the
differential approximants is telling us about the singularity structure
of the function $e^{f_{\rm bulk}(z)}$;
suffice it to say that we do not expect to obtain terribly accurate estimates
for the locations and exponents of the singularities.

For the point $z_1$ using the second-order ($K=2$) approximants, our protocol 
gives the estimates
\begin{subeqnarray}
z_1 &=& 0.25(4) \pm 0.40(4) \,i \\
\lambda_1 &=& -2.4(6)
\end{subeqnarray}
For the first-order ($K=1$) approximants, the estimates are
\begin{subeqnarray}
z_1 &=& 0.28(7) \pm 0.44(7) \,i \\
\lambda_1 &=& -1.6(6)
\end{subeqnarray}

For the broad peak $z_2$ on the real axis, all the non-defective zeros 
are precisely real (if they weren't, they would come in complex-conjugate pairs
so close to the real axis that they would fail to be well-separated),
and for $K=2$ our protocol gives the estimates  
\begin{subeqnarray}
z_2 &=& 0.60(8)  \\
\lambda_2 &=& -1.5(14)
\end{subeqnarray}
For $K=1$, the estimates are
\begin{subeqnarray}
z_2 &=& 0.64(7)  \\
\lambda_2 &=& -1.7(6)
\end{subeqnarray}

The peak $z_3$ is seen only in the $K=1$ approximants, and we estimate
\begin{subeqnarray}
z_3 &=& -1.00(2)
 \slabel{z3.K=1.expfbulk} \\
\lambda_3 &=& 1.0(1)
 \slabel{lambda3.K=1.expfbulk}
\end{subeqnarray}
The exponent $\lambda_3\approx 1$ suggests that $z_3$ may not be a singularity
of $e^{f_{\rm bulk}(z)}$, but rather a simple zero; if so, it would be 
a logarithmic singularity of the free energy $f_{\rm bulk}(z)$. 

\begin{figure}
\centering
\includegraphics[width=400pt]{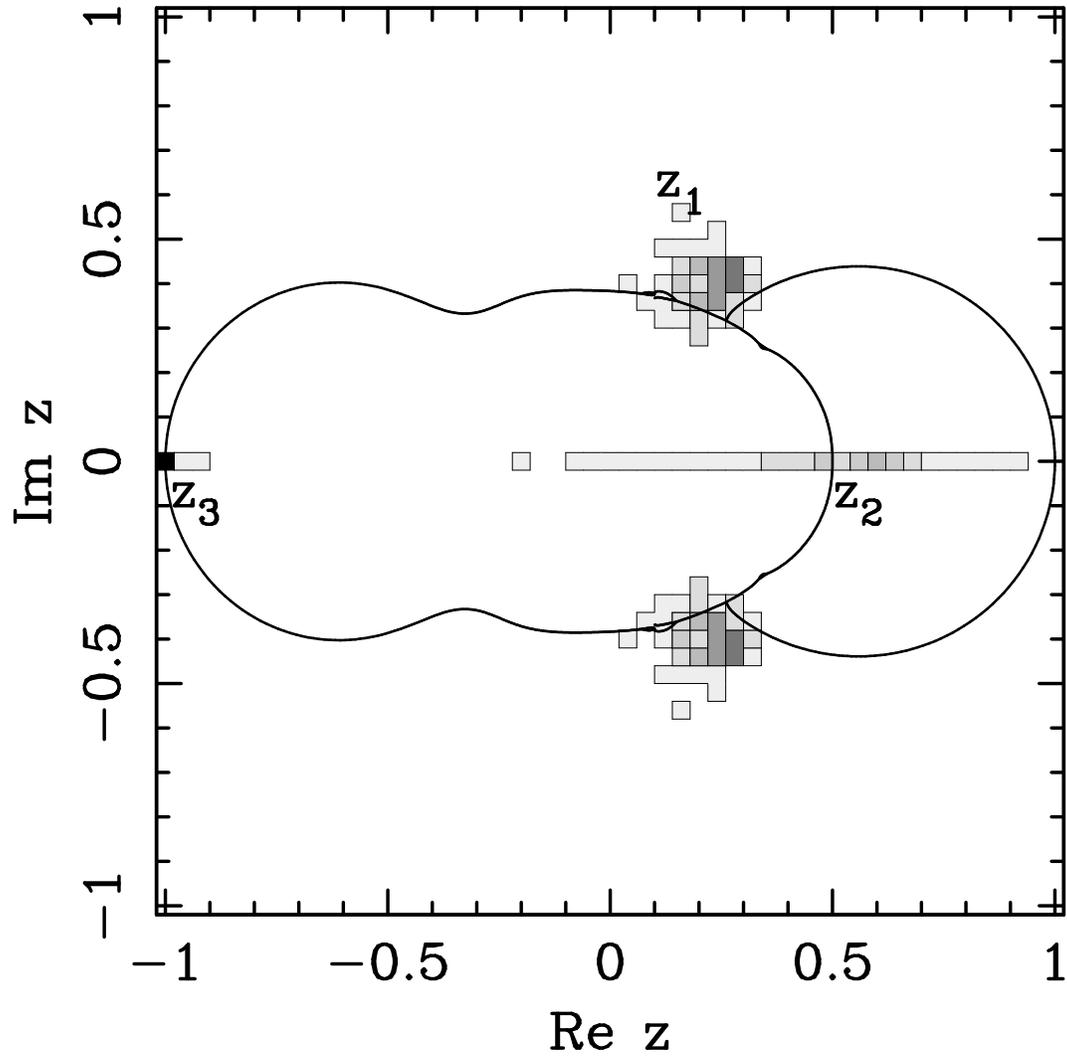}
\caption{\label{figure_estimates_bulk2}
  Histogram of the non-defective zeros for the unexponentiated 
  bulk-free-energy series \protect\reff{series_fbulk}
  using $K=2$ differential approximants with $N=47$ coefficients.
}
\end{figure}

We also analyzed the series expansion \reff{series_fbulk}
of the bulk free energy $f_{\rm bulk}(z)$
[as opposed to its exponential \reff{series_2},
which we have discussed until now].
In Figure~\ref{figure_estimates_bulk2} we show the
histogram for the $K=2$ approximants of $f_{\rm bulk}(z)$.
For $z_1$ and $z_2$ the picture is roughly similar to what was seen
in Figure~\ref{figure_histo_bulk}(b);
but for $z_3$ we see a much sharper peak. Our protocol gives the estimates
\begin{subeqnarray}
z_3 &=& -1.02(2)
 \slabel{z3.K=1.fbulk} \\
\lambda_3 &=& -0.6(5)
 \slabel{lambda3.K=2.fbulk}
\end{subeqnarray}
The estimate $\lambda=-0.6(5)$ in \reff{lambda3.K=2.fbulk}
is barely compatible with the value $\lambda=0$
corresponding to a logarithmic singularity in $f_{\rm bulk}$,
which is what we would expect if $e^{f_{\rm bulk}}$
has a singularity or zero of finite order
[e.g.\ the estimate \reff{lambda3.K=1.expfbulk}
 that suggests a simple zero].

\begin{figure}
\centering
\begin{tabular}{cc}
  \includegraphics[width=200pt]{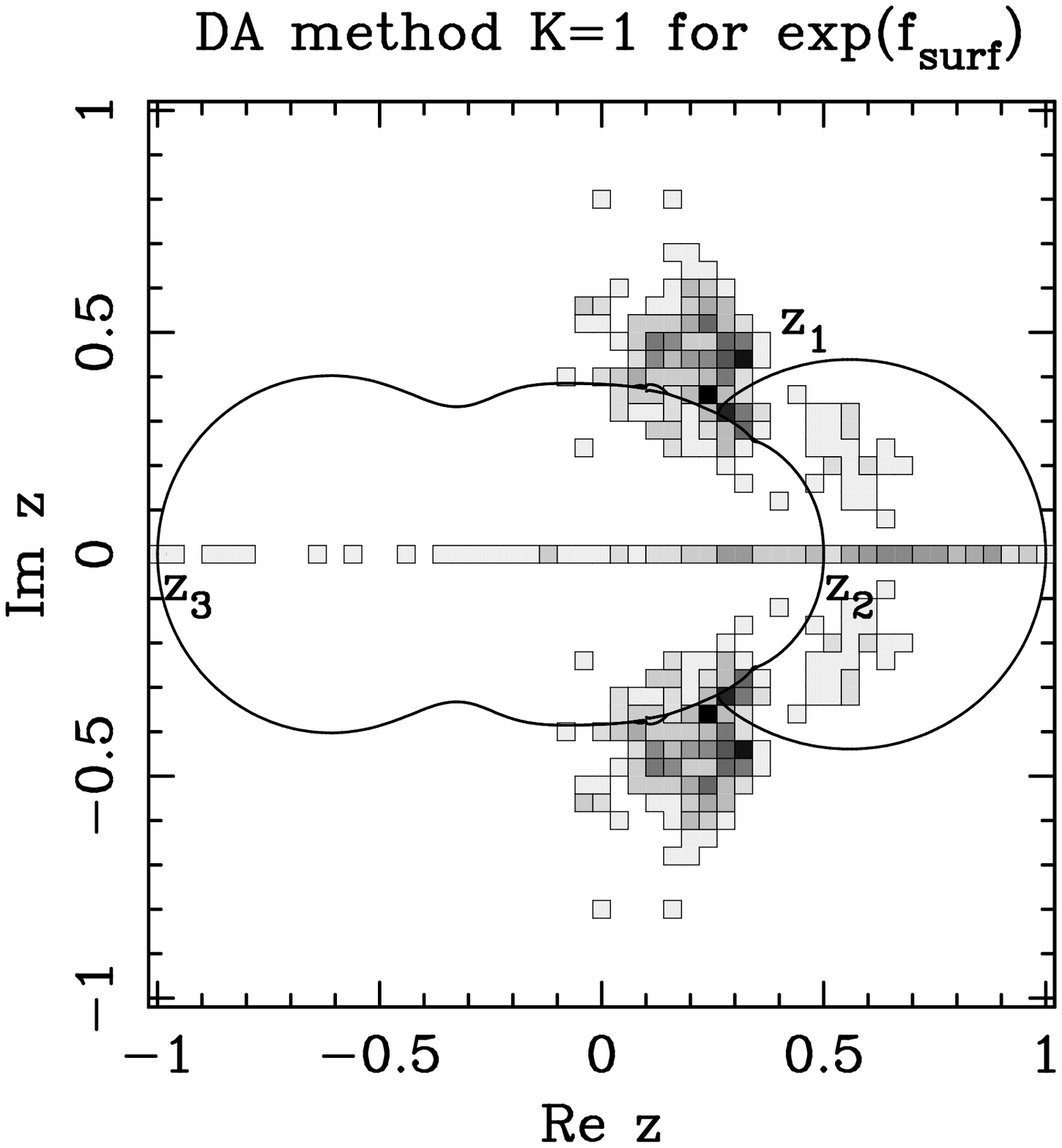} &
  \includegraphics[width=200pt]{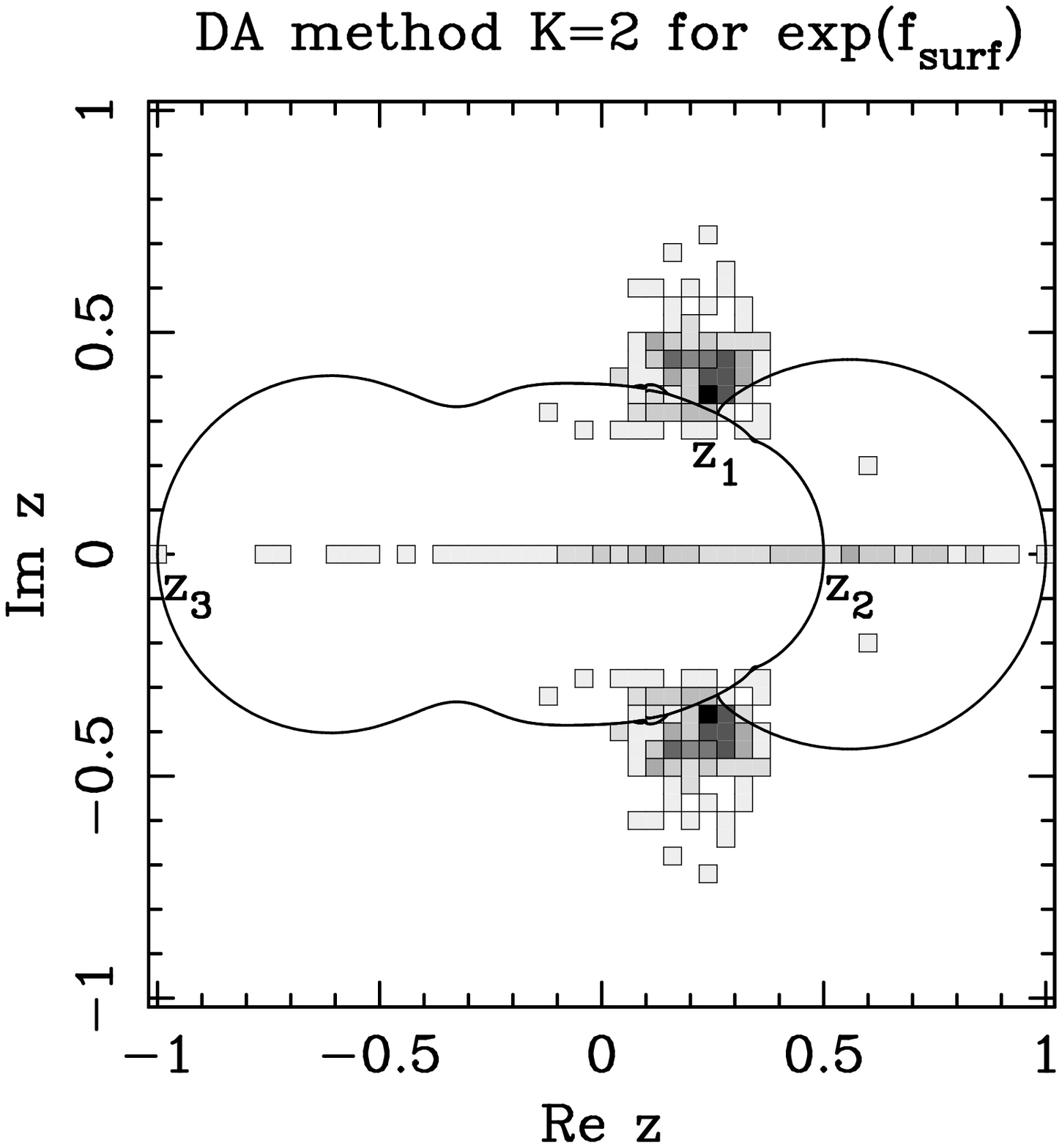} \\
  \phantom{(((a)}(a) & \phantom{(((a)}(b)\\[1mm]
\end{tabular}
\caption{\label{figure_histo_surf}
  Histograms of the non-defective zeros for the surface-free-energy  
  series \protect\reff{series_fsurf} using $K=1$ (a) and $K=2$ (b) differential
  approximants with $N=47$ coefficients.
  Codes are as in Figure~\protect\ref{figure_histo_3F}.
  We show the limiting curve $\scrb_m$
  for the square-lattice strip of width $L=11$ with
  toroidal boundary conditions as a solid black curve.
}
\end{figure}

For the surface free energy, we have $N=47$ coefficients
(assuming the correctness of the conjectural last coefficient).
The histograms for $K=1$ and $K=2$ approximants are displayed in
Figure~\ref{figure_histo_surf}, and they look even broader than those
for the bulk free energy. 
Using second-order ($K=2$) differential approximants we are able to
reliably locate only one singularity, namely
\begin{subeqnarray}
   z_1 & = & 0.25(4) \pm 0.39(4) \,i  \\
   \lambda_1  & = &  -2.0(8)
\end{subeqnarray}
(The estimate for $z_2$ is
\begin{subeqnarray}
   z_2 & = & -1.02(2) \\
   \lambda_2  & = &  -1.9(15)  \;,
\end{subeqnarray}
which we disregard because the error bar on $\lambda_2$ is so large.)
For first-order ($K=1$) approximants, we find two singularities
\begin{subeqnarray}
z_1 &=& 0.25(5) \pm 0.43(9) \,i \\
z_2 &=& 0.7(1)  
\end{subeqnarray}
with exponents $\lambda_1=-1.3(5)$ and $\lambda_2=-1.6(11)$,
respectively.

\begin{figure}
\centering
\begin{tabular}{cc}
  \includegraphics[width=200pt]{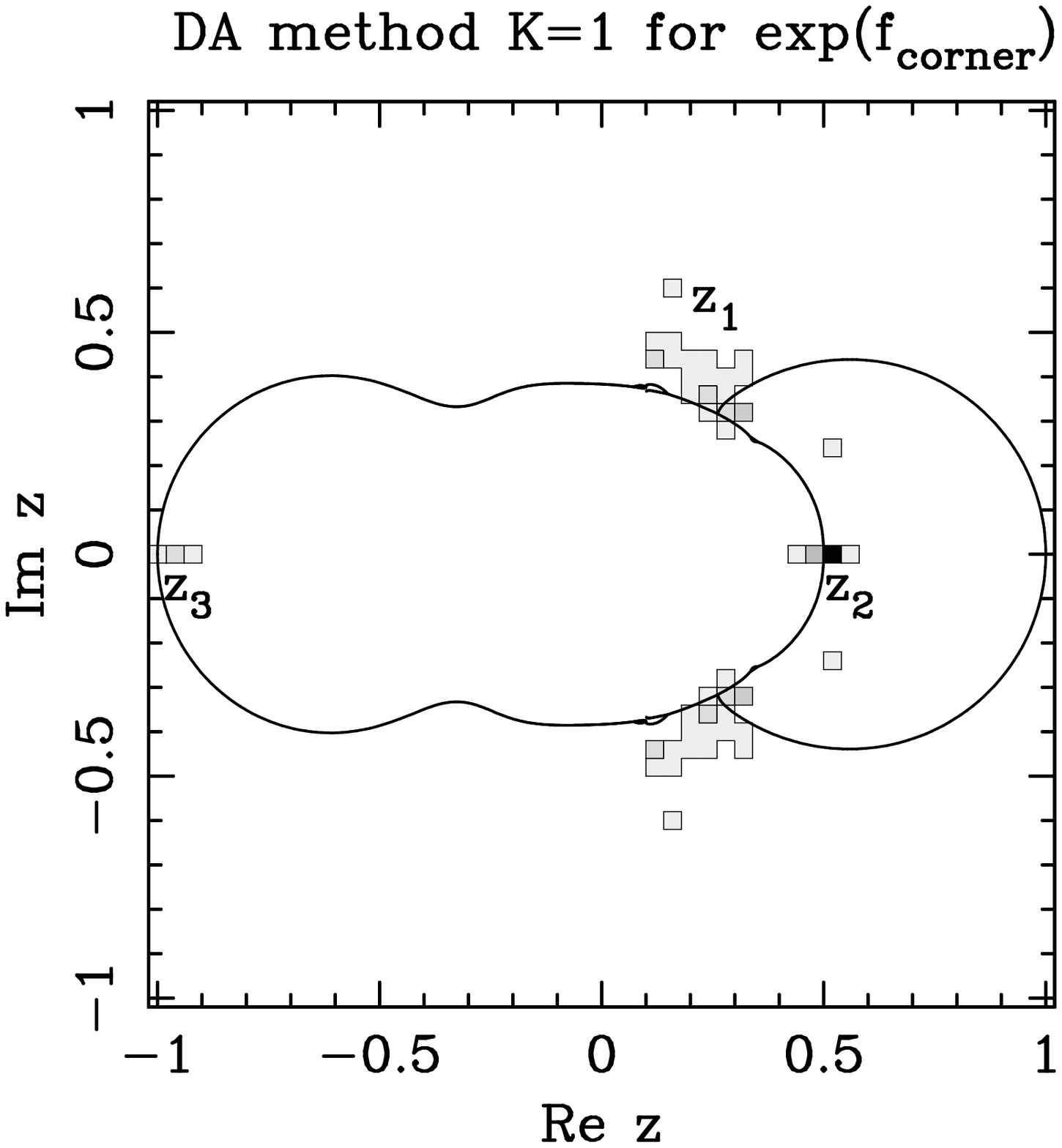} &
  \includegraphics[width=200pt]{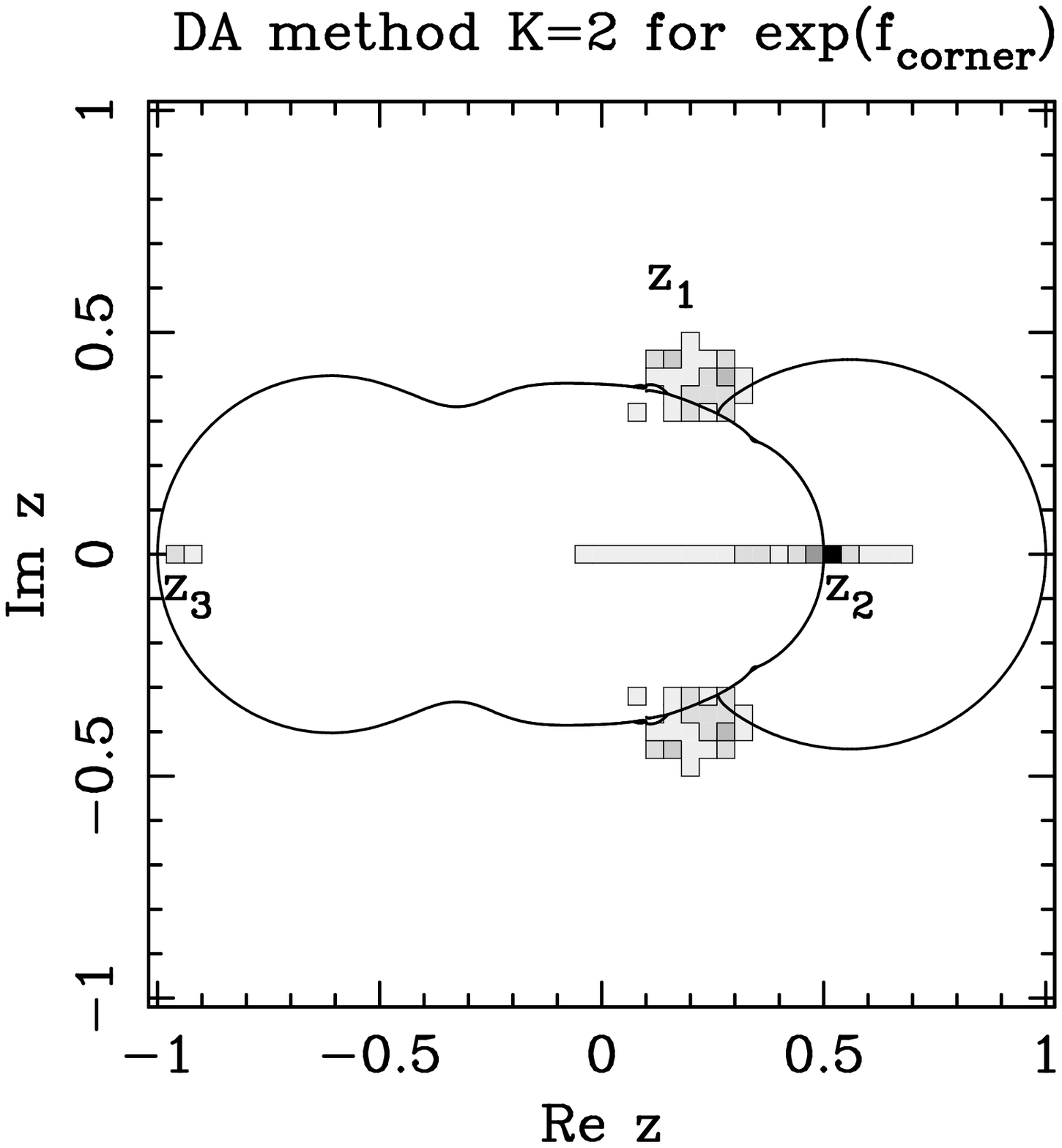} \\
  \phantom{(((a)}(a) & \phantom{(((a)}(b)\\[1mm]
  \multicolumn{2}{c}{\includegraphics[width=200pt]%
    {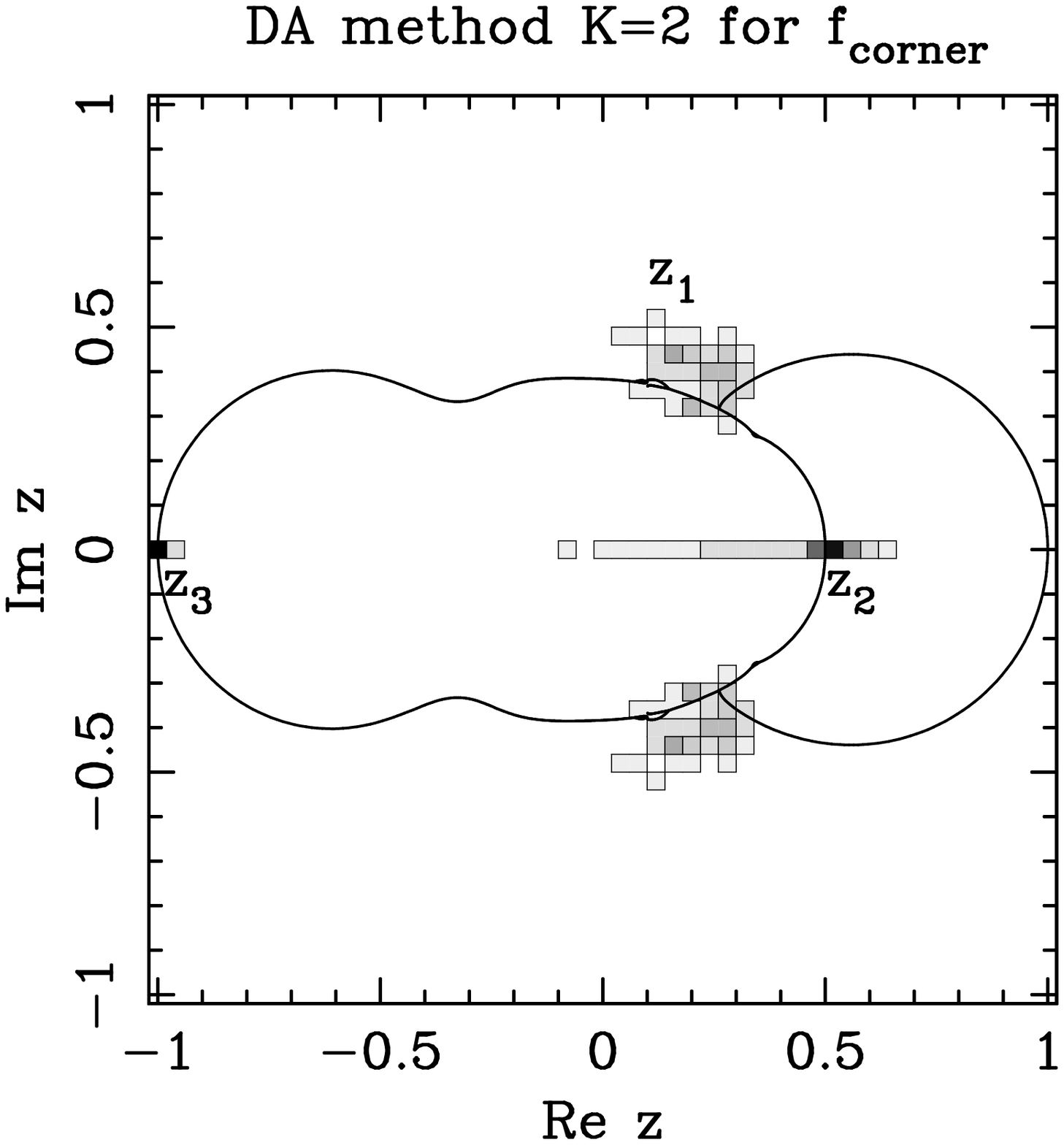}}\\
  \multicolumn{2}{c}{\phantom{(((a)}(c)} \\[1mm]
\end{tabular}
\caption{\label{figure_histo_corner}
  Histograms of the non-defective zeros for the corner-free-energy  
  series \protect\reff{series_exp_fcorner} using $K=1$ (a) and $K=2$ (b) 
  differential approximants with $N=46$ coefficients.
  Panel (c) shows the analogous $K=2$ histogram for the unexponentiated
  series \protect\reff{series_fcorner}.
  Codes are as in Figure~\protect\ref{figure_histo_3F}.
  We show the limiting curve $\scrb_m$
  for the square-lattice strip of width $L=11$ with
  toroidal boundary conditions as a solid black curve.
}
\end{figure}

Finally, for the corner free energy, we have $N=46$ coefficients,
and the corresponding histograms are displayed in
Figure~\ref{figure_histo_corner}(a,b).
In this case, we find a fairly sharp peak around $z_2 \approx 0.5$
(it is especially sharp for $K=1$),
and a complex-conjugate pair of broad peaks around $z\approx 0.2\pm 0.4\,i$
(they are, in fact, slightly less broad than for the bulk and surface
 free energies).
As these peaks are close to those found for the other two free energies,
let us try to locate them carefully. 
Using second-order ($K=2$) differential approximants we locate
two singularities:
\begin{subeqnarray}
z_1 &=& 0.22(4) \pm 0.36(5) \,i \\
z_2 &=& 0.51(1)  
\end{subeqnarray}
with associated critical exponents $\lambda_1=-1.9(9)$ and $\lambda_2=-1.1(3)$. 
These error bars are smaller than those found for the
bulk and surface free energies (this is especially so for $z_2$),
reflecting the sharper peaks in the histogram.
For first-order ($K=1$) approximants, we find the singularities
\begin{subeqnarray}
z_1 &=& 0.23(3) \pm 0.36(2) \,i \\
z_2 &=& 0.509(10)  
\end{subeqnarray}
with exponents $\lambda_1=-1.8(7)$ and $\lambda_2=-1.0(1)$.
As for bulk case, the analysis of series for $f_{\rm corner}(z)$
[rather than its exponential]
leads to a new singularity $z_3 \approx -1$,
with a quite sharp peak:
see Figure~\ref{figure_histo_corner}(c) for the histogram corresponding
to the $K=2$ approximants. Our analysis of this peak yields
\begin{equation}
z_3 \;=\; -1.007(8) \,,
\end{equation}
with exponent $\lambda_3=-0.8(2)$. 
The error bar here is about half of that found in
\reff{z3.K=1.expfbulk}/\reff{z3.K=1.fbulk}
for the bulk and surface free energies,
reflecting the sharper peak in the histogram.

One possible reason for the better behavior of the corner-free-energy series,
at least as concerns the real zeros $z_2$ and $z_3$,
comes from the empirical fact that the series for $f_{\rm corner}(q)$,
written in powers of $1/q$, has all {\em nonnegative}\/ coefficients
(as does, therefore, the series for its exponential).
By contrast, all the other series analyzed here have
coefficients of both signs, with no clear sign pattern.

In order to obtain more realistic error bars, we have repeated the
analysis using the following variations (in addition to the order 
of the differential approximant $K=1,2$): 
\begin{itemize}
  \item Using the free energy $f_i$ instead of its exponential $e^{f_i}$.
  \item Using the variable $1/q$ instead of the variable $z=1/(q-1)$.
  \item ``Truncating'' the series expansions of $e^{f_i(z)}$ 
        by deleting the terms of order $z^0, \ldots, z^9$
        and then dividing by $z^{10}$.
        This truncation ought to have little influence on the estimates
        of the singularities, as the relevant information 
        is encoded in the higher-order terms. 
\end{itemize} 
Alas, the estimates obtained with these procedures
deviate among themselves by about twice 
the error bars provided by each method.
For instance, the position of the singularity $z_1$
for the bulk free energy takes values 
ranging from $0.23(3)\pm 0.39(3)\,i$
[for the truncated series expansion of $e^{f_{\rm bulk}(z)}$ with $K=2$]
to $0.30(4)\pm 0.51(5)\,i$
[for the series expansion of $f_{\rm bulk}(1/q)$ with $K=1$]. 

Taking all this information into account, our final estimates for 
the location of the singularities are 
\begin{subeqnarray}
\text{Bulk} &\Rightarrow& \begin{cases} z_1 \;=\; 0.26(4)\pm 0.42(4)\,i & \\
                                        z_2 \;=\; 0.63(10) \\
                                        z_3 \;=\; -1.01(2)
                          \end{cases} \\[3mm]
\text{Surface} &\Rightarrow& \begin{cases} z_1 \;=\; 0.24(4)\pm 0.39(5)\,i & \\
                                           z_2 \;=\; 0.67(14) 
                          \end{cases} \\[3mm]
\text{Corner} &\Rightarrow& \begin{cases} z_1 \;=\; 0.23(4)\pm 0.37(4)\,i & \\
                                          z_2 \;=\; 0.51(1) \\
                                          z_3 \;=\; -1.01(1)
                          \end{cases} 
 \label{eq.sing_estimates.final}
\end{subeqnarray}
(We refrain from further discussion of the exponents $\lambda_i$,
 because the error bars are so large.)

We see from \reff{eq.sing_estimates.final}
that the estimated locations of the singularities
associated to the three free energies are (as expected)
compatible within errors.
Combining these estimates, we obtain
\begin{subeqnarray}
   z_1 & = & 0.24(4) \pm 0.39(4) \,i
   \qquad
   [q_1 \;=\; 2.14(19) \pm 1.86(19) \,i]
       \\[1mm]
   z_2 & = & 0.51(2)
   \qquad\qquad\quad\qquad\!
   [q_2 \;=\; 2.96(8)]
       \\[1mm]
   z_3 & = & -1.01(1)
   \qquad\qquad\quad\quad\,\! 
   [q_3 \;=\; 0.01(1)]
 \label{eq.sing_estimates.final.combined}
\end{subeqnarray}
The absolute values are therefore $|z_1|=0.46(5)$, $|z_2|=0.51(2)$ 
and $|z_3|=1.01(1)$.

\begin{figure}
\centering
\vspace*{-2cm}
\begin{tabular}{c@{\hspace{1cm}}c}
  \includegraphics[width=200pt]{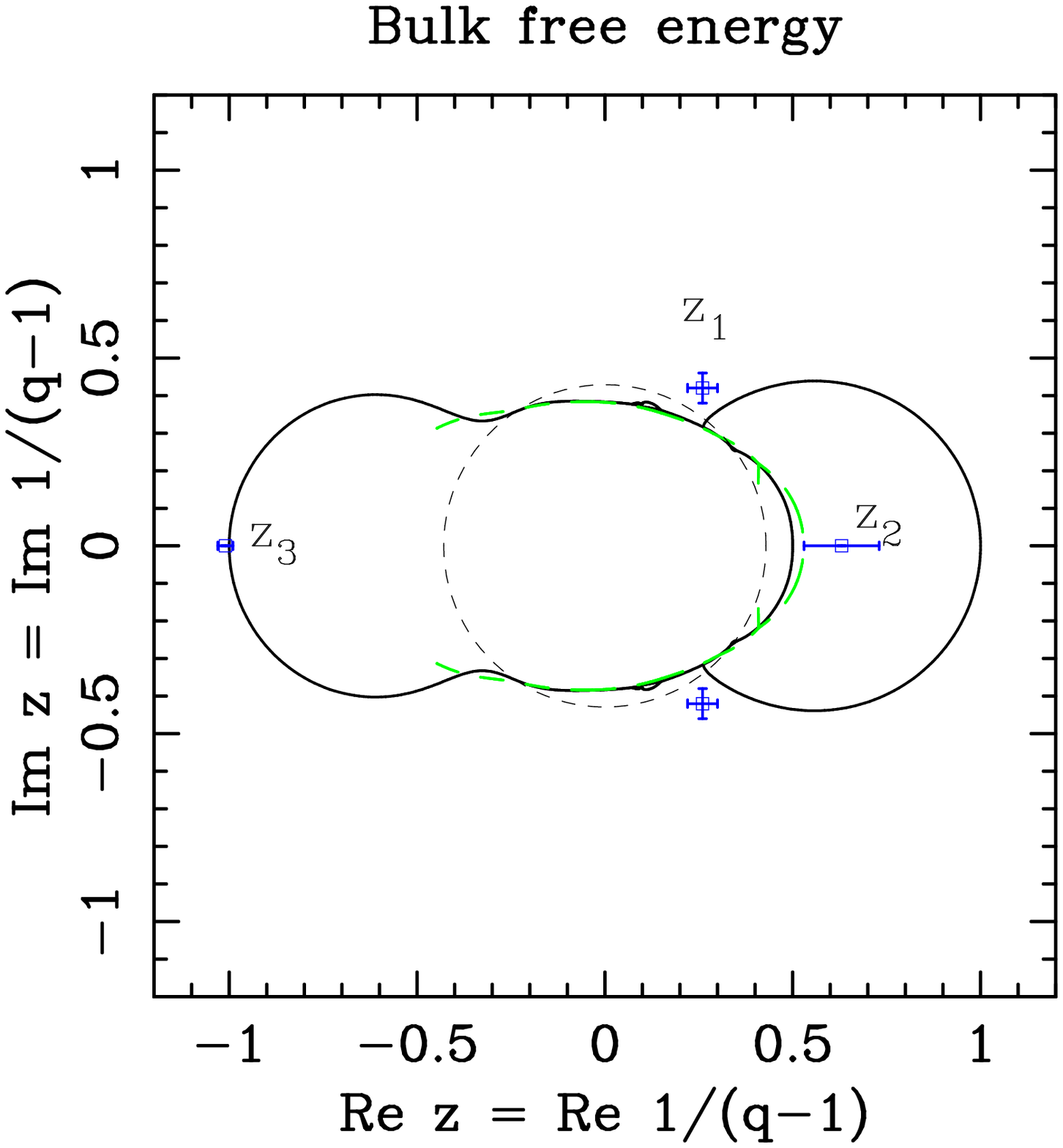} &  
  \includegraphics[width=200pt]{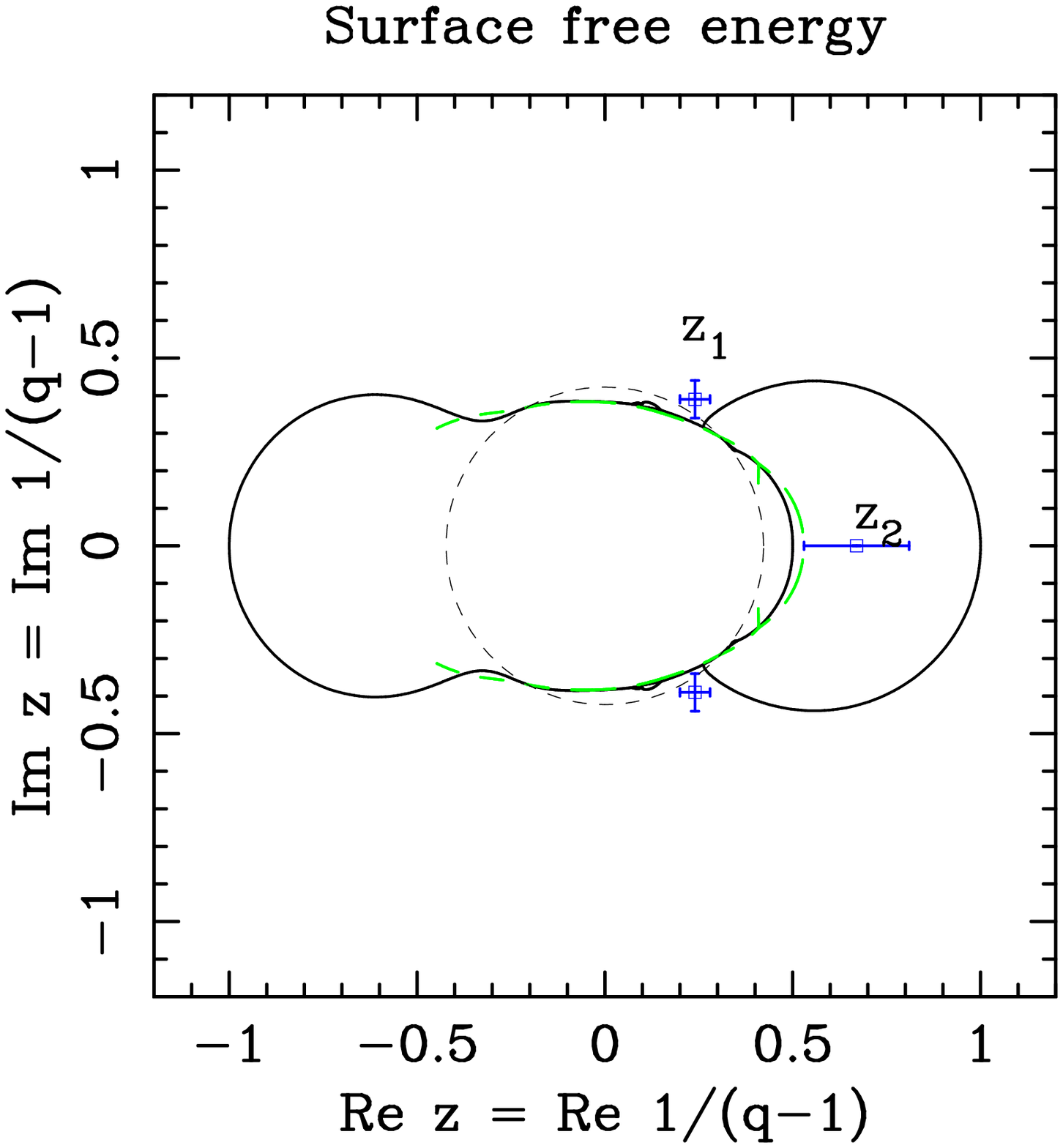} \\
  \phantom{(((a)}(a) & \phantom{(((a)}(b)\\[1mm]
  \multicolumn{2}{c}{\includegraphics[width=200pt]{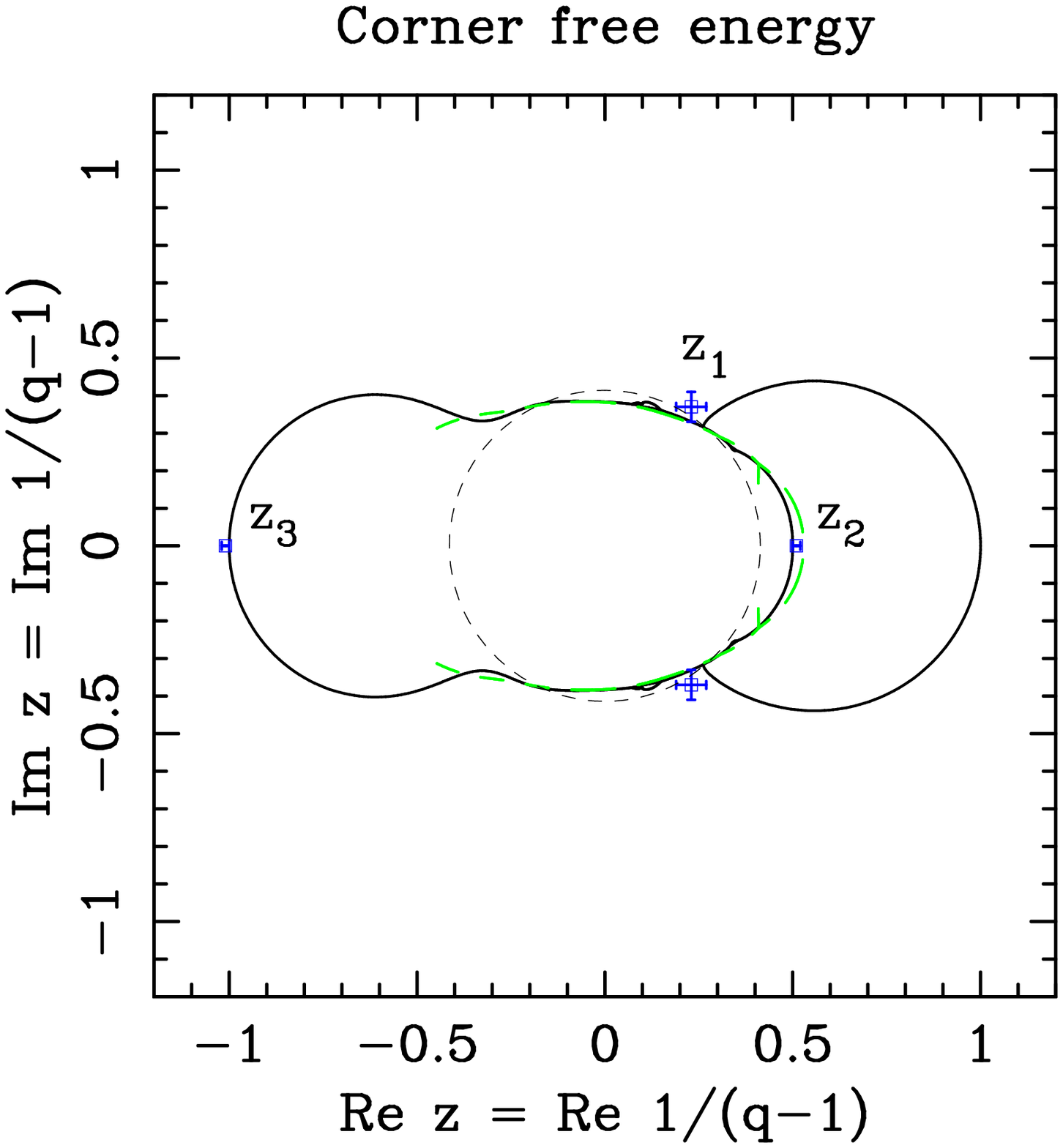}}
  \\ 
  \multicolumn{2}{c}{\phantom{(((a)}(c)} \\[1mm]
\end{tabular}
\caption{\label{figure_singularities}
 Singularities for the bulk (a), surface (b), and corner (c( free energies
 in the complex plane of $z=1/(q-1)$.
 In each panel, we show the limiting curve $\scrb_m$
 for the square-lattice strip of width $L=7$ (resp.\ $L=11$) with
 toroidal (resp.\ cylindrical) boundary conditions as a solid black
 (resp.\ dashed green) curve.
 We also show the location of the singularities found
 in the series analysis (blue $\square$).
 The dashed black line shows the circle with the radius of convergence
 estimated from
 $r_{\rm conv} = \liminf_{n \to\infty} |a_n|^{-1/n}$.
}
\end{figure}

In Figure~\ref{figure_singularities}(a,b,c) we compare
the estimates \reff{eq.sing_estimates.final} for $z_1,z_2,z_3$
with the two best currently available approximations
to the square-lattice limiting curve $\scrb_\infty$
(namely, the curves $\scrb_m$ coming from
 $m=7$ with toroidal boundary conditions \cite{Jacobsen-Salas_toroidal}
 and $m=11$ with cylindrical boundary conditions \cite{transfer2}).
We also show a circle with radius $r_{\rm conv}\approx 0.4290$
[resp.\ $r_{\rm conv}\approx 0.4225$, $r_{\rm conv}\approx 0.4139$]
indicating the radius of convergence obtained by estimating
$r_{\rm conv} = \liminf_{n \to\infty} |a_n|^{-1/n}$.

Let us discuss the three singularities in order.

The singularity of smallest modulus is the complex point $z_1$,
given by (\ref{eq.sing_estimates.final.combined}a).
{}From Figure~\ref{figure_singularities} we see that
the estimated location $z_1$
lies fairly close to the most prominent T point of the limiting curve 
$\mathcal{B}_{7_{\rm tor}}$ (solid black line), namely
\cite[sec.~3.6]{Jacobsen-Salas_toroidal}
\begin{equation}
   z_{\rm T, 7_{\rm tor}}  \;=\; 
      0.26158090412 \pm 0.3162213774 \, i
   \quad
   [q_{\rm T, 7_{\rm tor}} \;=\; 
      2.5531414480 \pm 1.8775702667 \, i]
   \;,
\end{equation}
and one can conjecture that this is not an accident.
On the other hand, this singularity lies rather far from
the T point of the limiting curve $\mathcal{B}_{11_{\rm cyl}}$
(dashed green curve), namely \cite[sec.~3.3]{transfer2}
\begin{equation}
   z_{\rm T, 11_{\rm cyl}}  \;=\; 0.409 \pm 0.218 \, i
   \qquad
   [q_{\rm T, 11_{\rm cyl}} \;=\; 2.902 \pm 1.015 \, i]
   \;.
\end{equation}
Thus, we are unable to draw any firm conclusion
about the physical nature of this singularity.
It does, however, seem likely that this singularity lies
somewhere on the limiting curve $\mathcal{B}_\infty$ ---
possibly (but not necessarily) at a ``special'' point
of $\mathcal{B}_\infty$ such as a T point.

The two singularities on the real axis are given by
(\ref{eq.sing_estimates.final.combined}b,c). 
The estimate for $z_2$ is compatible within errors with $q_2 = 3$,
which is the physical critical point $q_c({\rm sq}) = 3$
of the square-lattice zero-temperature Potts antiferromagnet
\cite{Lenard_67,Baxter_70,Baxter_82b,Nijs_82,Park_89,Burton_Henley_97,%
Salas_98,Ferreira_99}
and is expected to be the uppermost real crossing point
of the limiting curve $\mathcal{B}_\infty$
(see Section~\ref{sec.q0} below).
The estimate for $z_3$ is compatible with $q_3 = 0$,
which is expected to be the lowermost real crossing point
of the limiting curve $\mathcal{B}_\infty$.

In conclusion, the analysis of the free-energy series expansions
has led to the location of four singularities in the complex $q$-plane:
two of them are the real points $q=0$ and $q=3$,
which correspond to the transitions between
the disordered and critical phases,
while the other two are a pair of complex-conjugate points
that presumably lie somewhere on the limiting curve $\mathcal{B}_\infty$,
{\em possibly}\/ at a ``special'' point such as a T point.

Our difficulties in locating the singularities of the free energies 
--- in particular, the wide dispersion of estimates visible in
the histograms
(Figures~\ref{figure_histo_bulk}--\ref{figure_histo_corner}),
which contrasts strongly with the behavior observed in our
test cases $3_{\rm F}$ and $4_{\rm P}$
(Figure~\ref{figure_histo_3F}) ---
suggest that the singularity structure of the infinite-volume free energies
may be quite different from that of the finite-width strips
that we used as test cases.
In particular, it is possible that the entire limiting curve
$\mathcal{B}_\infty$ is a natural boundary for the
infinite-volume free energies --- possibly with a very soft singularity ---
and that the differential approximants are trying, with limited success,
to mimic this behavior.  In this interpretation,
the points $z_1$, $z_2$ and $z_3$ would simply be approximations
to the places on $\mathcal{B}_\infty$ where the free energies
are most strongly singular;  and one would furthermore expect that as the
number $N$ of series coefficients increases, the number of peaks
in the histograms will increase, and the zeros will gradually
condense on the whole of the curve $\mathcal{B}_\infty$
(perhaps with high density on some parts of the curve
and low density on others).
It would be interesting to try further numerical tests of this scenario;
but they will probably require fairly radical extensions of our
large-$q$ series, e.g.\ to order $N \approx 100$.

%
%
\section{Numerical results for widths
           $\bm {m = 9_{\rm F}, 10_{\rm F}, 11_{\rm F}, 12_{\rm F}}$}
\label{sec.numer}

As part of this work, we have computed the transfer matrices
for square-lattice strips with free boundary conditions
and widths $m=9,10,11,12$.
(Widths $m \le 8$ were computed in a previous paper \cite{transfer1}.)
In this section we use these newly-computed transfer matrices
to study the real and complex roots of the chromatic polynomials
$P_{m_{\rm F}\times n_{\rm F}}(q)$ for $m=9,10,11,12$,
focusing on the behavior in the infinite-length limit ($n \to \infty$).
In Section~\ref{sec.numer.curves} we discuss the limiting curve
$\scrb_m$ and the isolated limiting points for each width $m$.
In Section~\ref{sec.q0} we study the behavior of the
real crossing points $q_0(m)$ and attempt to extract
$q_0({\rm sq}) = \lim\limits_{m \to\infty} q_0(m)$.
In Section~\ref{sec.real} we discuss the {\em real}\/ chromatic roots
for finite lattices $m_{\rm F}\times n_{\rm F}$.

%
%
\subsection{Limiting curves and isolated limiting points}
  \label{sec.numer.curves}

\begin{figure}
  \vspace*{-1cm}
  \centering
  \begin{tabular}{cc}
  \includegraphics[width=200pt]{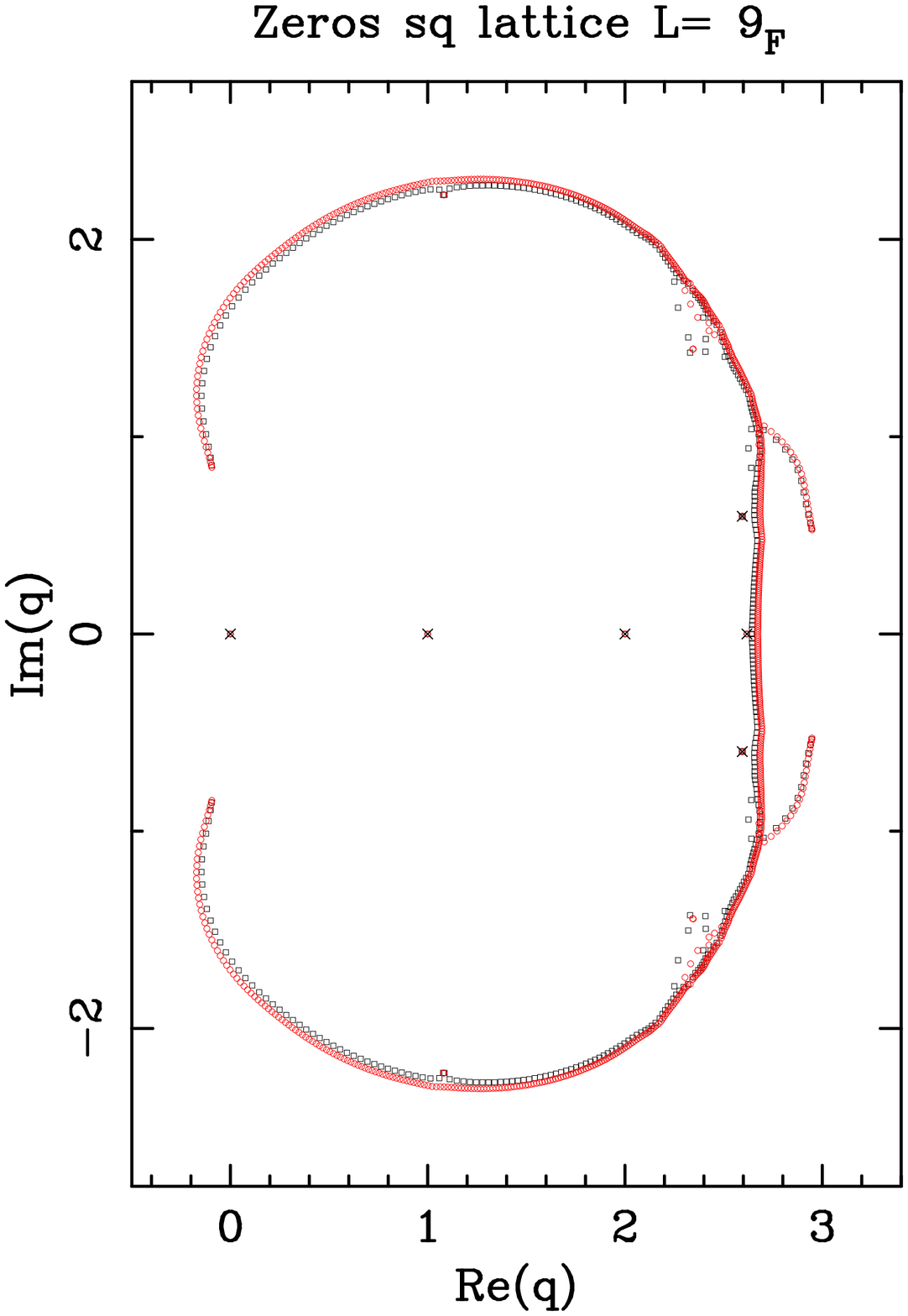} & 
  \includegraphics[width=200pt]{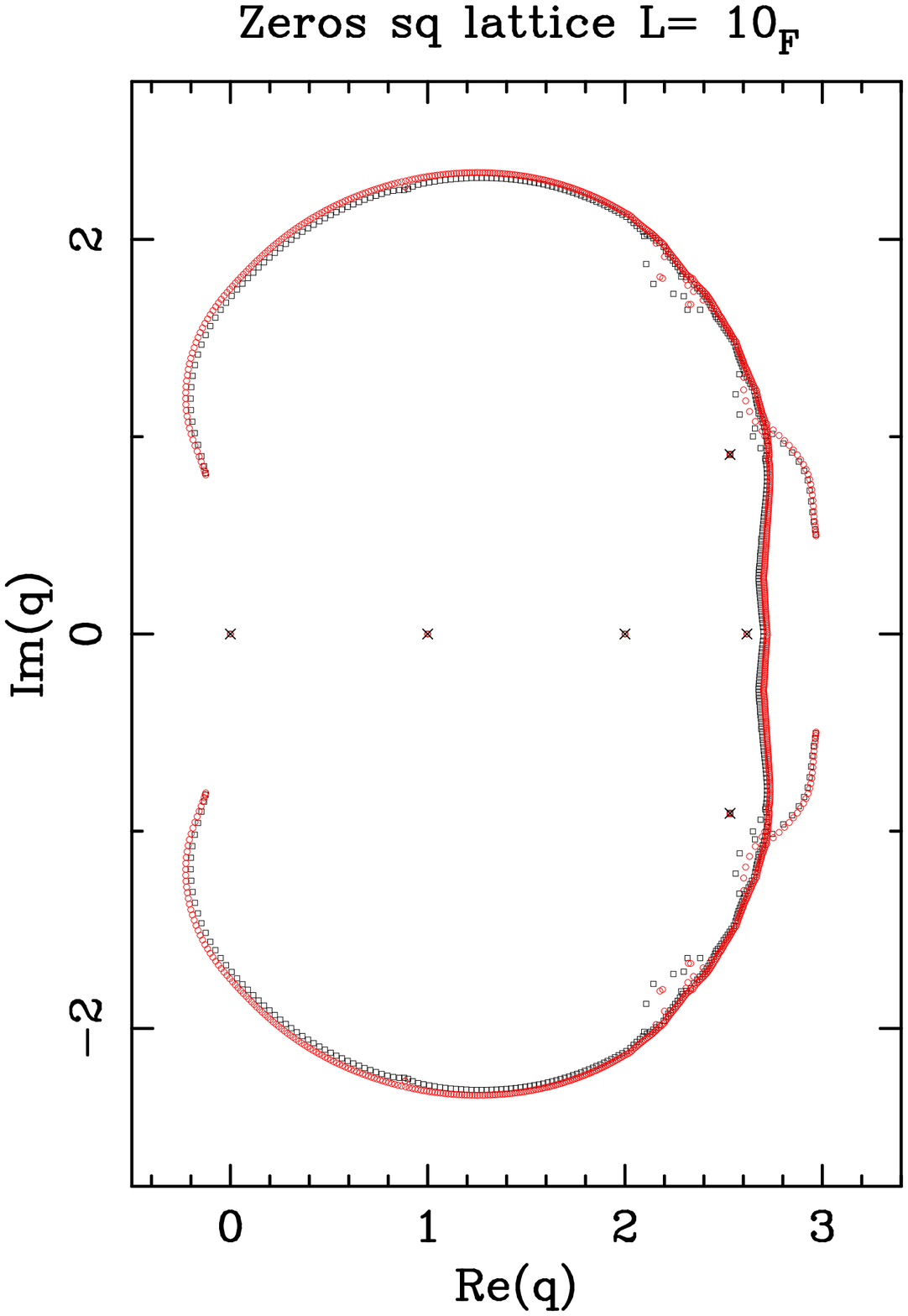} \\
  \qquad (a) & \qquad (b) \\[2mm]
  \includegraphics[width=200pt]{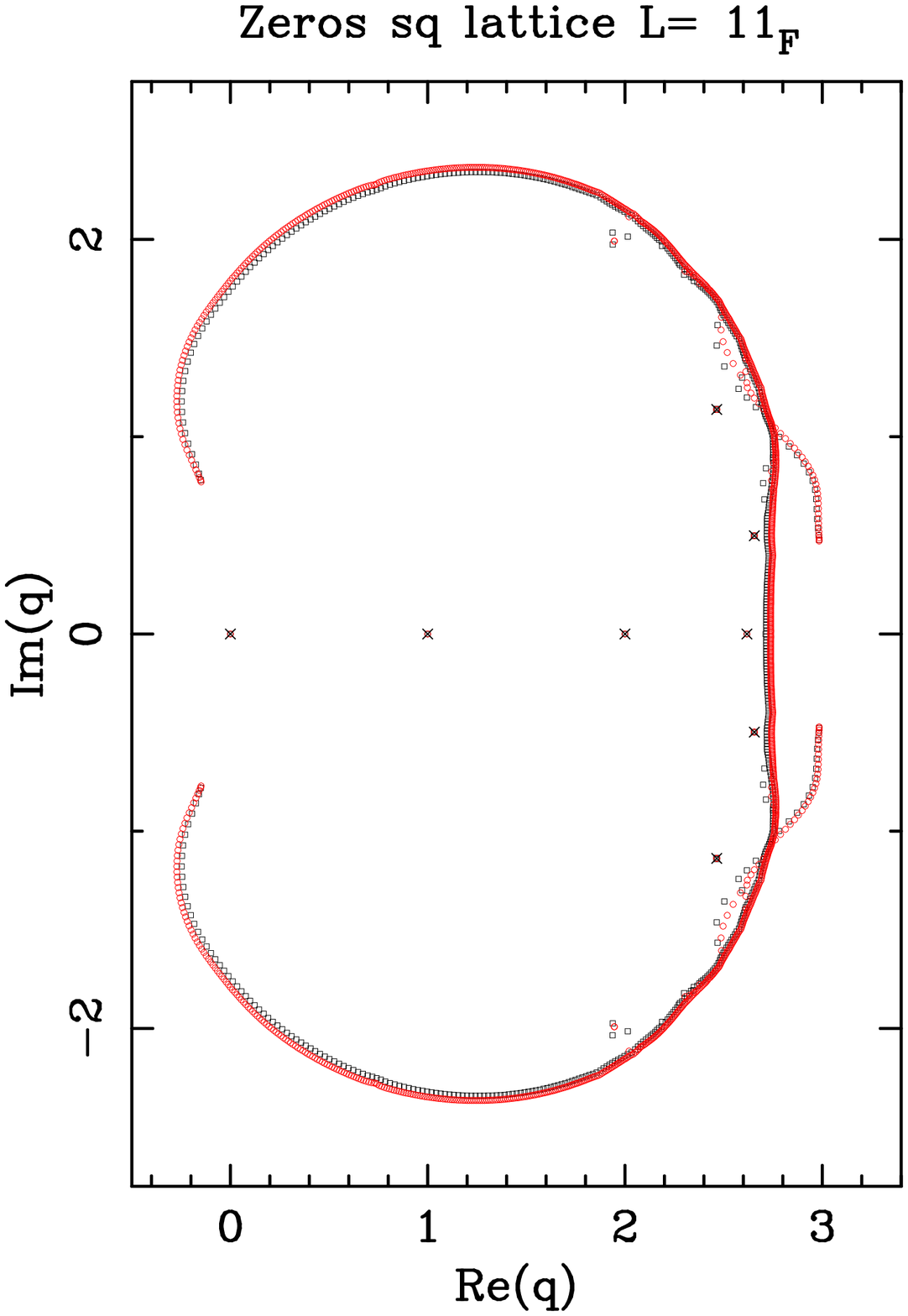} & 
  \includegraphics[width=200pt]{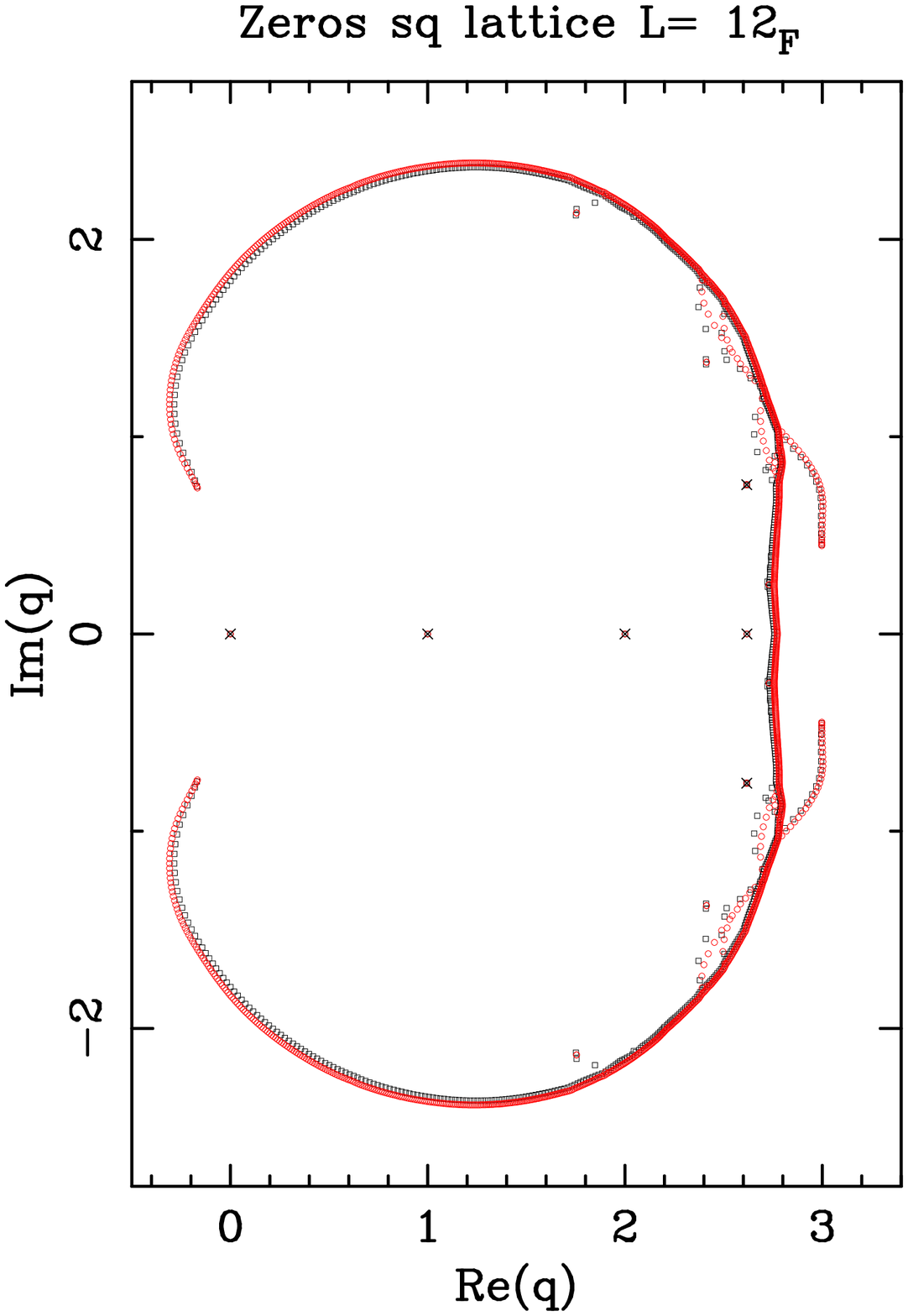} \\
  \qquad (c) & \qquad (d) 
  \end{tabular}
  \caption{
  Zeros of the chromatic polynomial
  for the square lattices of widths $9_{\rm F}$ (a), $10_{\rm F}$ (b), 
  $11_{\rm F}$ (c), and $12_{\rm F}$ (d). 
  For each width $L$, we show the chromatic zeros of the strips 
  $L_{\rm F} \times (5L)_{\rm F}$ ($\Box$ black) and 
  $L_{\rm F} \times (10L)_{\rm F}$ ($\circ$ red). 
  }
\label{Figure_sqF}
\end{figure}

Having computed the transfer matrices for
$m = 9_{\rm F}, 10_{\rm F}, 11_{\rm F}, 12_{\rm F}$,
we can analyze them
as in \cite{transfer1,transfer2,transfer3,transfer4,Jacobsen-Salas_toroidal}
to extract the chromatic roots for finite-length lattices ($n < \infty$)
as well as the limiting curves and isolated limiting points
in the infinite-length limit ($n \to \infty$).
Unfortunately, we have not been able to compute the full limiting curves
$\scrb_m$, as this would have required a major computational effort.
However, we have tried to locate the most important points on each of them,
i.e.\ crossings of the real axis and endpoints near the real axis,
using the direct-search method \cite{transfer1}.
These results are summarized in Table~\ref{table_summary}.\footnote{
  For widths $m \le 11$, we implemented the direct-search method
  in {\sc Mathematica}, using 100-digit internal precision.
  For width $m=12$, we used the double-precision {\sc Fortran} subroutines
  in the {\sc arpack} package \cite{arpack}.
  We also checked our results for $7 \le m \le 11$ using {\sc arpack};
  the results are consistent with those obtained using {\sc Mathematica}
  but are less precise.

  Let us also warn the reader that the value for $m=8$ reported in
  \cite{transfer1} is wrong; the correct value is displayed here
  in Table~\ref{table_summary}.
}
(Furthermore, the chromatic roots for $n=5m$ and $n=10m$
shown in Figures~\ref{Figure_sqF}(a)--(d)
give a fairly good idea of the general shape of the curve $\scrb_m$.)
Finally, for all these strips there are real isolated
limiting points at the Beraha numbers $q=0,1,2, B_5$,
and for $m=9,11,12$ there also appear to be complex isolated limiting points
(we cannot, however, guarantee that we have found all of them).
In Figures~\ref{Figure_sqF}(a)--(d) these points are
marked with a $\times$ sign. 

For $m=9$, the dimension of the transfer matrix is $179$.
In Figure~\ref{Figure_sqF}(a) 
we have plotted the chromatic zeros in the complex $q$-plane for the
strips $9_{\rm F}\times 45_{\rm F}$ and $9_{\rm F}\times 90_{\rm F}$. 
The limiting curve crosses the real $q$-axis at a single point: 
\begin{equation}
q_0(9) \;\approx \; 2.70165995678  \;.
\end{equation}
Figure~\ref{Figure_sqF}(a) also suggests that there is a
complex-conjugate pair of isolated limiting points at 
$q\approx 2.5946 \pm 0.5963\, i$.

For $m=10$, the dimension of the transfer matrix is $435$.
In Figure~\ref{Figure_sqF}(b) 
we have plotted the chromatic zeros for the strips 
$10_{\rm F}\times 50_{\rm F}$ and $10_{\rm F}\times 100_{\rm F}$. 
In this case the limiting curve does not cross the real $q$-axis;
rather, there is a pair of complex-conjugate endpoints
very close to the real $q$-axis, at
\begin{equation}
q_0(10) \;\approx \; 2.7343903604 \pm 0.0003924978\; i   \;.
\end{equation}

For $m=11$, the dimension of the transfer matrix is $1142$. 
In Figure~\ref{Figure_sqF}(c) we have plotted the chromatic zeros for the 
strips $11_{\rm F}\times 55_{\rm F}$ and $11_{\rm F}\times 110_{\rm F}$. 
The limiting curve crosses the real $q$-axis at
\begin{equation}
q_0(11) \;\approx \; 2.7608973951  \;.
\end{equation}
Figure~\ref{Figure_sqF}(c) also suggests that there are
two complex-conjugate pairs of isolated limiting points at
$q\approx 2.6555 \pm 0.4978\, i$ and $q\approx 2.4648 \pm 1.1380\, i$.

Finally, for $m=12$, the dimension of the transfer matrix is $2947$. 
In Figure~\ref{Figure_sqF}(d) we have plotted the chromatic zeros for the
strips $12_{\rm F}\times 60_{\rm F}$ and $12_{\rm F}\times 120_{\rm F}$.
In this case the limiting curve does not seem to cross the real $q$-axis;
rather, there is a pair of complex-conjugate endpoints
very close to the real $q$-axis, at
\begin{equation}
q_0(12) \;\approx \; 2.782817590 \pm 0.00018700\; i   \;.
\end{equation}
Figure~\ref{Figure_sqF}(d) also suggests that there is a
complex-conjugate pair of isolated limiting points at
$q\approx 2.6172 \pm 0.7562\, i$.

In view of the results reported in Table~\ref{table_summary},
we conjecture that the limiting curve $\scrb_m$ crosses the real axis
for all odd $m$ and has a complex-conjugate pair of endpoints
very near the real axis for all even $m \ge 8$.

%
%
\subsection[Value of ${q_0({\rm sq})}$]{Value of $\bm{q_0({\bf sq})}$}
\label{sec.q0}

We can try to use the results reported in Table~\ref{table_summary}
to obtain the value of $q_0({\rm sq}) = \lim\limits_{m \to\infty} q_0(m)$.
Of course, we expect $q_0({\rm sq})= 3$
\cite{Lenard_67,Baxter_70,Baxter_82b,Nijs_82,Park_89,Burton_Henley_97,%
Salas_98,Ferreira_99}, 
but it will be interesting to see with what accuracy this result can be
obtained. 

As just discussed, there are clear parity effects in $q_0(m)$:
the limiting curve $\mathcal{B}_m$ crosses the real $q$-axis for odd $m$
but not for even $m$.
We have therefore split the data into two sets according to the parity of $m$,
and analyzed each set separately.
Please note that the data for $q_0(m)$ are essentially exact:
there is a tiny {\em non-statistical}\/ error of order $10^{-10}$
in their numerical estimates.
In order to keep to the standard notation in finite-size-scaling theory,
we here denote the strip width by $L$ instead of $m$.

For each data set we have considered several Ans\"atze, and for a given
Ansatz with $k$ free parameters, we have taken into account the points with 
$L= L_{\rm min}, L_{\rm min}+2,\ldots,L_{\rm min}+2(k-1)$. So in each fit 
there are no degrees of freedom. From the variation of the estimates as
$L_{\rm min}$ is increase, we can roughly estimate the error bar. 

If we use the power-law Ansatz
\begin{equation}
q_0(L) \;=\; q_0({\rm sq})  + B L^{-\Delta}
\label{def_power_law_q0}
\end{equation}
for the data coming from odd $L$,
we obtain the following estimates using $L_{\rm min}=7$  
\begin{subeqnarray}
q_0({\rm sq})  &=& 2.999(5) \\
\Delta         &=& 1.108(16) 
\label{fit.q0.odd.power}
\end{subeqnarray}
where the error bars are defined to be twice the distance between
these estimates and those obtained with $L_{\rm min}=5$. 

If we play the same game for the quantity $\Re q$ for even $L$, we arrive
at the estimates for $L_{\rm min}=8$ 
\begin{subeqnarray}
q_0({\rm sq})  &=& 3.00(7) \\
\Delta         &=& 1.10(29) 
\label{fit.q0.even.power}
\end{subeqnarray}
where again the error bars are twice the difference from the
estimates for $L_{\rm min}=6$. The convergence in this case is not
very good, so we cannot trust these results too much. 

The results \reff{fit.q0.odd.power}/\reff{fit.q0.even.power} are 
consistent with the behavior
\begin{subeqnarray}
q_0({\rm sq})  &=& 3   \slabel{conj.q0.a} \\
\Delta         &=& 1
\label{conj.q0}
\end{subeqnarray}
Of course, we expect that $q_0({\rm sq})=q_c({\rm sq})=3$;
moreover, the correction exponent $\Delta=1$ is to be expected
for {\em free}\/ b.c.\ because of the surface effect.
For cylindrical boundary conditions \cite[Section~4]{transfer2},
the estimates are less well converged but are consistent with
$q_0({\rm sq}) = 3$ and suggest that $\Delta \gtapprox 1.7$
(e.g.\ perhaps $\Delta=2$).
Finally, very recent work \cite[Section 5.1]{Jacobsen-Salas_toroidal}
gives strong evidence that $q_0({\rm sq}) = 3$ holds also for toroidal 
boundary conditions.\footnote{
   The authors of Ref.~\protect\cite{Jacobsen-Salas_toroidal} 
   found that for toroidal square-lattice strips 
   of {\em odd}\/ widths $3\leq L\leq 11$,
   the value of $q_0(L)$ is {\em exactly}\/ equal to
   the expected limiting value $q_0({\rm sq})=3$,
   while for toroidal strips of {\em even}\/ widths $2\leq L\leq 12$,
   the fits of the numerical data
   to the Ansatz~\protect\reff{def_power_law_q0} gave 
   $q_0(\text{sq})=2.999\pm 0.012$ and $\Delta = 2.04 \pm 0.08$,
   in agreement with the conjectured behavior \protect\reff{conj.q0.a}
   but suggesting that here $\Delta=2$.
}

Better fits can in principle be obtained by imposing the values \reff{conj.q0}. 
In particular, we expect the following Ansatz to describe better the 
data points:
\begin{equation}
q_0(L) \;=\; 3  - B L^{-1} - C L^{-2} \,.  
\label{def_power_law_q0_bis}
\end{equation}
However the results are not very stable: for odd $L$ we find that
for $L_{\rm min}=7$, $B\approx 2.436$, and $C\approx 2.243$, while
for $L_{\rm min}=9$, $B\approx 2.283$, and $C\approx 2.719$.
A similar behavior is found for even $L$: 
for $L_{\rm min}=8$, $B\approx 2.409$, and $C\approx 2.473$, while 
for $L_{\rm min}=10$, $B\approx 2.357$, and $C\approx 2.994$.
We conclude that our numerical data are not accurate enough to be able to
accurately determine corrections to scaling to the behavior of 
$q_0(L)$.

For odd $L$, we can also look at $\Im q_0(L)$. 
In Ref.~\cite[Conjecture 4.2]{transfer2} it was conjectured that for 
square-lattice strips with cylindrical boundary conditions 
$\Im q_0(L)\sim B_5^{-L/2}=\tau^{-L}$ where $\tau=(1+\sqrt{5})/2$ is
the golden ratio. We have fitted our numerical results for $\Im q_0(L)$
with $L=8,10,12$ to check whether this conjecture holds or not also
for free boundary conditions. A 3--parameter fit to a power-law Ansatz
$\Im q_0(L)=A + BL^{-\Delta}$ shows that $A\approx -0.00042$ is indeed
a small number. We also obtain that $B\approx 0.033$ and 
$\Delta\approx 1.606$. As we have only three data points, there are no
degrees of freedom in the fit and the error bars cannot be reliably 
estimated. In conclusion, we expect that $\Im q_0(L)$ decays to zero
exponentially fast. Thus, we use the improved Ansatz:  
$\Im q_0(L)=A \times B^L$. The results for $L_{\rm min}=10$ are
$A = 0.16(13)$ and $B=0.690(75)$. The error bars are computed as
twice the difference between the estimates for $L_{\rm min}=10$ and
$L_{\rm min}=8$. If we compare $B^{-1}=1.44(16)$ to  
$\tau \approx 1.6180339887$, we see that they are barely 
compatible within errors.
As the error bar for $B^{-1}$ is $20$ times larger than the corresponding
error for cylindrical boundary conditions \cite[Section 4]{transfer2}, we
cannot draw any firm conclusion about the validity of the above conjecture
for square-lattice strips with free boundary conditions.   

%
%
\subsection{Real chromatic roots}   \label{sec.real}

It is also of interest to study the real chromatic roots
for lattices of finite length $n$.
For every length $n$ there are, of course, roots at $q=0$ and $q=1$;
and there are also roots converging (exponentially rapidly) as $n \to\infty$
to the isolated limiting point at $q=2$.
Here we shall concentrate on the roots converging to the
isolated limiting point at $q=B_5 = (3 + \sqrt{5})/2$
and, for $m=9$ and 11, to the real crossing point $q_0(m)$.

For width $m=10,12$, the real roots converging to $B_5$
do so monotonically from below;
we refrain from presenting the details.
There are no real roots above $B_5$.

For widths $m=9$ and 11, by contrast, two interesting things happen
(see Tables~\ref{table_zeros_9F} and \ref{table_zeros_11F}).
On the one hand, the real roots converging to $B_5$
do so with parity $(-1)^{n+1}$, i.e.\ alternating from above and below
(starting from $n=39$ and $n=23$, respectively).
On the other hand, for odd lengths $n$, there are also real roots
converging to $q_0(m)$ from below.
We conjecture that these behaviors will persist for all larger
odd widths $m$.

These results for $m=9,11$ provide a (presumably infinite)
family of counterexamples to a conjecture made in \cite{transfer1}
concerning the real chromatic roots of bipartite planar graphs.
For further discussion and for a revised conjecture,
see Appendix~\ref{appendix.upper}.

An important question is the rate of convergence of these real zeros
towards the limiting values $B_5$ and $q_0(m)$.
For $B_5$ we expect an exponentially rapid convergence
(as with all isolated limiting points in the Beraha--Kahane--Weiss theorem);
for $q_0(m)$, by contrast, we expect a $1/n$ convergence.
We have tested these predictions by fitting the given real roots $q_n$
to a power-law Ansatz
\begin{equation}
q \;=\; q_\infty + A n^{-\Delta}   \;,
\label{def_power_law_real_zeros}
\end{equation}
using the data points $n \ge$ a variable threshold $n_{\rm min}$.
For the sequences of zeros converging to $B_5$,
the estimates of the power $\Delta$ appear to increase without bound
as $n_{\rm min}$ is increased (e.g.\ $\Delta \gtapprox 45$),
suggesting that the convergence is indeed exponentially fast.
For the sequences converging to $q_0(m)$, by contrast,
we obtain powers $\Delta\approx 1.1$ for small values of $n_{\rm min}$,
which slowly decrease toward $1$ as we increase $n_{\rm min}$.
In particular, for $n_{\rm min} = 99$
we obtain $\Delta \approx 1.023$ for $m=9$
and  $\Delta\approx 1.037$ for $m=11$.
This is consistent with the $1/n$ behavior expected from the
Beraha--Kahane--Weiss theorem.

%
%
\section{Summary and open problems}  \label{sec.summary}

In this paper we have derived some new structural results
for the transfer matrix
of square-lattice Potts models with free and cylindrical boundary conditions.
In particular, we have obtained explicit closed-form expressions for the
dominant (at large $|q|$) diagonal entry in the transfer matrix,
for arbitrary widths $m$,
as the solution of a special one-dimensional polymer model.
We have also obtained the first 47 (resp.\ 46) terms
in the large-$q$ expansion of the bulk and surface (resp.\ corner) free energies
for the zero-temperature antiferromagnet (= chromatic polynomial).
Finally, we have computed chromatic roots for strips of width $m=9,10,11,12$
with free boundary conditions and located roughly the limiting curves.

Let us end by listing some open problems for future research.
The first group of such problems
would be to understand theoretically (and ideally to prove rigorously)
some of the regularities that were observed empirically in
Sections~\ref{sec.eigen} and \ref{sec.thermo}, for instance:

1) The empirically observed property that the polynomials
$p_s^{\rm F}(k)$ and $p_s^{\rm P}(k)$ arising in
\reff{eq.empirical.1}/\reff{eq.empirical_P.1} ff.\ 
have for even $s\ge 2$ a factor $k-s/2$. 

2) The formulae \reff{eq.fibonacci.F.s=1} and (\ref{eq.empirical_P.2}a,b)
relating the coefficients $a_k^{\rm F}(k+s)$ and $\tilde{a}_k^{\rm P}(k+s)$
for $s=1$ and $s=0,1$, respectively, to Fibonacci numbers
(and possible extensions of these formulae to higher values of $s$).

3) The fact that $a_{k,\ell}^{\rm F}$ and $\tilde{a}_{k,\ell}^{\rm P}$
[defined in \reff{def_akl_bis_F} and \reff{eq.atilde_kl}]
are, for each fixed $\ell \ge 0$, the restriction to integers $k \ge \ell$
of a {\em polynomial}\/ in $k$ of degree $\ell$.

4) The behavior \reff{eq.bkm.surprise}/\reff{eq.ckm.surprise}
relating the leading eigenvalue $\lambda_\star^{\rm F/P}(m)$
to the dominant diagonal entry $t_{\rm F/P}(m)$.

5) The fact that the coefficients $c_k^{\rm F/P}(m)$
in the large-$q$ expansion of $\log \lambda_\star^{\rm F/P}(m)$
[cf.\ \reff{def_ckm}] are
the restriction to integers $m \ge m_{\rm min}^{\rm F/P}(k)$
of a polynomial of degree 1 in $m$.
Finite-size-scaling theory (Section~\ref{sec.thermo.gen})
gives a nonrigorous (but physically intuitive)
explanation of why the $c_k^{\rm F/P}(m)$ are
polynomials of degree 1 in $m$ for {\em large enough $m$}\/.
But we do not really understand why the cutoffs
$m_{\rm min}^{\rm F/P}(k)$ take the values they do;
even less do we understand the improvements found empirically in
Sections~\ref{sec.eigen.F} and \ref{sec.eigen.P}.
And of course we lack a rigorous proof for all of this.

6) The correction term of order $z^{4L_{\rm max}-1}$
found empirically for the surface free energy
[cf.\ the paragraph preceding \reff{series_exp_fsurf}].

\bigskip

Among the more fundamental extensions of this work, we would like to propose
the following:

\medskip

1. {\em Obtain longer large-$q$ series.}\/
It is natural to ask whether our large-$q$ series for the
bulk, surface and corner free energies (Section~\ref{sec.free_energy_series})
can be extended to higher order.
Modest extensions are no doubt possible over the next few years,
with increased computer power;
but the memory and CPU-time requirements are heavy,
even for width $m=12$ or 13.
It seems to us that radical algorithmic improvements
in the implementation of the transfer-matrix method
will be required in order to go to significantly larger widths $m$.
Unfortunately, our results from the series analysis
(Section~\ref{subsec.series_analysis})
suggest that these series are badly behaved,
and that it will be necessary to obtain {\em many}\/ more terms
--- for instance, doubling the length of our series
from $N=47$ to $N \approx 100$ ---
in order to improve significantly our understanding of the
analytic structure.

\medskip

2. {\em Extend this work to the triangular lattice.}\/
Extension of this work to the triangular lattice
is particularly interesting in view of Baxter's \cite{Baxter_86,Baxter_87}
conjectured exact solution for the bulk free energy
of the triangular-lattice chromatic polynomial
(see also \cite[Section~6]{transfer3} for a critical discussion).
It ought to be fairly easy to obtain exact formulae
for the dominant transfer-matrix entry,
analogous to those obtained here in Section~\ref{sec.struc}
for the square lattice.
Likewise, the transfer-matrix calculations
in Sections~\ref{sec.eigen} and \ref{sec.thermo}
can easily be extended to the triangular lattice,
as discussed in \cite{transfer3},
albeit possibly for slightly smaller widths.
The main trouble arises in the use of the finite-lattice method:
using transfer matrices of widths up to $L_{\rm max}$,
we obtain the large-$q$ series through order $4 L_{\rm max} - 2$
on the square lattice (Section~\ref{sec.free_energy_series})
but only order $\approx 2 L_{\rm max}$ on the triangular lattice 
\cite[Section~3.1]{forest}.

But the calculation of the large-$q$ series for the {\em bulk}\/
free energy is expected to be superfluous, as Baxter \cite{Baxter_86,Baxter_87}
has an exact expression
\be
g_1(q) \;=\; -{1\over x}\prod\limits_{j=1}^\infty
            {(1-x^{6j-3})(1-x^{6j-2})^2(1-x^{6j-1}) \over
             (1-x^{6j-5})(1-x^{6j-4})(1-x^{6j})(1-x^{6j+1}) }
 \label{eq.baxter.g1}
\ee
(where $q = 2 - x - x^{-1}$ and $|x| < 1$)
that is claimed to represent the exponential of the bulk free energy
at large enough $|q|$ (namely, outside the limiting curve $\scrb_\infty$).
The finite-lattice calculations would thus serve only to check
\reff{eq.baxter.g1} through order $x^{\approx 20}$
(and also to compute the series for the surface and corner free energies).
Assuming that this check confirms \reff{eq.baxter.g1} ---
as we expect that it will\footnote{
   See also the numerical confirmations in \cite[Section~6.3]{transfer3}.
}
--- we will learn nothing new!
One could equally well carry out a differential-approximant analysis
like that of Section~\ref{subsec.series_analysis} directly on the
series obtained by expanding \reff{eq.baxter.g1},
for which it is easy to obtain 1000 or more terms.

On the other hand, there is little point in carrying out such a
differential-approximant analysis, since we already know the
analytic structure of \reff{eq.baxter.g1}:
namely, it is analytic and nonvanishing in the disc $|x| < 1$,
or equivalently in the complex $q$-plane with a cut
along the real interval $0 \le q \le 4$.
It follows that, if Baxter's formula \reff{eq.baxter.g1} is correct
at large $|q|$, then the limiting curve $\scrb_\infty$
for the triangular-lattice chromatic polynomial
is {\em not}\/ a natural boundary for the bulk free energy
--- contrary to what we suspect to be the case
for the square-lattice chromatic polynomial
(see the discussion at the end of Section~\ref{subsec.series_analysis}).

When all is said and done,
our lack of understanding of the triangular-lattice chromatic polynomial
concerns {\em small}\/ $|q|$
--- namely, the region {\em inside}\/ the limiting curve $\scrb_\infty$ ---
and not large $|q|$
(see \cite[Section~6]{transfer3} for discussion).
This issue is probably best addressed by trying to understand
better the structure of Baxter's \cite{Baxter_86,Baxter_87}
Bethe-Ansatz solution for finite widths $m$,
including the effect of boundary conditions.

\medskip

3. {\em Carry out transfer-matrix calculations for the diced lattice.}\/
The discussion in Appendix~\ref{appendix.upper} and in \cite{MC_diced}
suggests that it would be very interesting
to carry out transfer-matrix calculations analogous to those of
\cite{transfer1,transfer2,transfer3,transfer4,Jacobsen-Salas_toroidal}
and the present paper for the {\em diced lattice}\/
(namely, the dual of the kagom\'e lattice).
We have established in \cite{MC_diced} that $q_c(\text{diced}) > 3$,
but we do not know whether the limiting curves $\scrb_m$
intersect the real axis at some value $q_0(m) > 3$,
or merely have complex-conjugate endpoints that tend to $q_c(\text{diced})$
as $m \to\infty$.

\medskip

4. {\em Understand the analytic nature of complex phase boundaries.}\/
As discussed at the beginning of Section~\ref{sec.series},
it is proven in some cases, and expected in general
for statistical-mechanical models in dimension $d \ge 2$,
that the infinite-volume free energy has a ``soft'' essential singularity
(where it is infinitely differentiable but not analytic)
whenever a first-order-phase-transition point in the physical region
is approached.
But it is unclear whether the {\em complex}\/ phase boundaries
are natural boundaries of the free energies defined in each phase.
(In our case, we are uncertain whether the limiting curve $\scrb_\infty$
is a natural boundary for the bulk free energy of the square-lattice
chromatic polynomial.)
It would be useful to resolve this question in at least one model
(for instance, the analyticity or nonanalyticity at pure-imaginary
magnetic field of the low-temperature Ising ferromagnet),
ideally by a mathematically rigorous proof.

\appendix

%
%
\section{Some combinatorial identities} \label{appendix.combinBis}
\label{appendix.combin}

In this appendix we will prove
Propositions~\ref{prop.eq.empirical.1} and \ref{prop.eq.empirical_P.1}
concerning the quantities $a_k^{\rm F}(k-s)$ and $\tilde{a}_k^{\rm P}(k-s)$,
respectively. We will also prove eq.~\reff{eq.fibonacci.F.s=0} for the 
quantity $a_k^{\rm F}(k)$.  

We shall use Knuth's notation for falling powers \cite{Graham_94}
\begin{equation}
   x^{\underline{n}} \;=\; x(x-1) \cdots (x-n+1)
   \;.
\end{equation}
We also use the standard convention \cite{Graham_94}
for the definition of binomial coefficients:
\begin{equation}
\binom{x}{k}  \;=\;
\begin{cases}
   x^{\underline{k}}/k! & \mbox{\rm for integer $k \ge 0$} \\[1mm]
   0                    & \mbox{\rm for integer $k < 0$} 
\end{cases}
 \label{def.binom}
\end{equation}
where $x$ can be a real or complex number
(or more generally an algebraic indeterminate)
while $k$ is always an integer.
Finally, we adopt the convention that $k! = \Gamma(k+1) = \infty$
for integer $k<0$, so that \reff{def.binom} could be written simply as
$\binom{x}{k} = x^{\underline{k}}/k!$.

%
%
\subsection{Some preliminary lemmas} \label{appendix.A1}

Let us start with a simple lemma that will make the analysis easier: 

\begin{lemma} \label{lemma.appendix.1}
For any integers $s,a\ge 0$, let us define the polynomial
\begin{equation}
 T_{s,a}(\lambda) \;=\; \sum\limits_{r=0}^{s} \lambda^r \,
        \frac{r^{\underline{a}}}{r! \, (s-r)!}  \;.
\label{def_Tsa}
\end{equation}
Then
\begin{equation}
 T_{s,a}(\lambda) \;=\; \frac{ \lambda^a \, (\lambda+1)^{s-a} }{(s-a)!} 
\label{def_Tsa.eq}
\end{equation}
(hence in particular $T_{s,a}(\lambda) = 0$ for $a > s$).
\end{lemma}

\proof
Expanding $r^{\underline{a}}$ in \reff{def_Tsa}
and using the fact that $r^{\underline{a}} = 0$ for $a>r$, we obtain
\begin{equation}
 T_{s,a}(\lambda) \;=\; \sum\limits_{r=a}^{s} \lambda^r \,
        \frac{r(r-1)(r-2)\ldots (r-a+1)}{r! \, (s-r)!}\,.
\label{def_Tsa.2}
\end{equation}
If $a>s$, then the sum is empty, so let us suppose that $a \le s$.
Then we can rearrange the factorials to obtain
\begin{eqnarray}
 T_{s,a}(\lambda) &=& \sum\limits_{r=a}^{s} \lambda^r \,
                    \frac{1}{(r-a)! \, (s-r)!} \nonumber \\
        &=& \frac{1}{(s-a)!} \,
            \sum\limits_{r=a}^{s} \lambda^r \binom{s-a}{r-a} \nonumber \\
        &=& \frac{\lambda^a}{(s-a)!} \,
            \sum\limits_{r'=0}^{s-a} \lambda^{r'} \binom{s-a}{r'} \nonumber \\
        &=& \frac{\lambda^a}{(s-a)!} \, (\lambda+1)^{s-a} \;.
\label{def_Tsa.3}
\end{eqnarray}
\qed

\bigskip

The next lemma will allow us to perform the inner sums
in \reff{eq.akm_final_F} and \reff{eq.atildekm_final_P}:

\begin{lemma} \label{lemma.appendix.2}
For any integers $\ell,a,b \ge 0$, let us define the polynomial
\begin{equation}
 F_{\ell,a,b}(\lambda) \;=\; \sum\limits_{r=0}^{\ell} \lambda^r \,
       \binom{\ell-a-b-1}{r}\, \binom{b}{\ell-r} \,. 
\label{def_Fl}
\end{equation}
Then
\begin{equation}
 F_{\ell,a,b}(\lambda)
  \;=\; \frac{ (-1)^\ell \, b!} {\ell! \, (a+b-\ell)!} \, Q_{\ell,a}(b,\lambda)
  \;,
\label{def_Fl.eq}
\end{equation}
where
\begin{equation}
  Q_{\ell,a}(b,\lambda) \;=\;
  \sum\limits_{m=0}^{\min(a,\ell)}
   \binom{a}{m} \, (a+b-\ell)^{\underline{a-m}} \: \ell^{\underline{m}} \:
      \lambda^m \, (\lambda-1)^{\ell-m}
\label{def_Ql}
\end{equation}
is a polynomial in $b,\lambda$
that is of degree $a$ in the variable $b$
and of degree $\ell$ in $\lambda$,
with leading coefficient
\be
    [b^a] Q_{\ell,a}(b,\lambda) \;=\;  (\lambda-1)^\ell
 \label{eq.leading.Qla}
\ee
and hence $[b^a \lambda^\ell] Q_{\ell,a}(b,\lambda) = 1$.
\end{lemma}

\proof
Let us first consider the case $\ell-a-b-1\ge 0$.
Then ${\ell-a-b-1 \choose r}=0$ whenever $r>\ell-a-b-1$.
On the other hand, ${b \choose \ell-r}=0$ whenever $\ell-r > b$,
i.e.\ $r<\ell -b$.
But these two conditions on $r$ involve all possible cases, as $a\ge 0$.
Therefore, $F_{\ell,a,b}(\lambda) = 0$ for $\ell-a-b-1\ge 0$.

Let us now consider the nontrivial case $\ell-a-b-1 < 0$.
As the upper index in ${\ell-a-b-1 \choose r}$ is now negative,
it is convenient to rewrite \reff{def_Fl} as
\begin{equation}
F_{\ell,a,b}(\lambda) \;=\; \sum_{r=0}^{\ell} (-\lambda)^r \,
       \binom{r-\ell+b+a}{r}\, \binom{b}{\ell-r} \,, 
\label{def.Fl.Bis}
\end{equation}
where now all the indexes are nonnegative.
We can then rewrite the binomial coefficients in terms of factorials as 
follows:
\begin{eqnarray}
F_{\ell,a,b}(\lambda) &=& \sum_{r=0}^{\ell} (-\lambda)^r  \,
   \frac{ (r-\ell+b+a)! }{r!\, (a+b-\ell)!}
\: \frac{ b! }{(\ell-r)! \, (b-\ell+r)! } \nonumber \\
         &=& \frac{ b! }{(a+b-\ell)!} 
         \sum_{r=0}^{\ell} \frac{ (-\lambda)^r }{r! \, (\ell-r)!} 
         \prod_{m=1}^{a} (r+b-\ell+m) \;.
\label{def.Fl.Tris}
\end{eqnarray}
The product in \reff{def.Fl.Tris} can be written as a sum using 
the ``binomial theorem'' for falling powers \cite[Exercise 5.37]{Graham_94}.
Defining $x=b-\ell$, we have
\begin{subeqnarray}
\prod_{m=1}^{a} (r+x+m) &=& (r+x+a)^{\underline{a}} \\[-2mm]
   &=& \sum\limits_{m=0}^a \binom{a}{m} \, r^{\underline{m}} \,  
       (x+a)^{\underline{a-m}}
\end{subeqnarray}
and therefore
\begin{subeqnarray}
 F_{\ell,a,b}(\lambda) &=& \frac{ b! }{(a+b-\ell)!} 
         \sum_{m=0}^{a} \binom{a}{m} \, (a+b-\ell)^{\underline{a-m}} \,
         \sum_{r=0}^{\ell} 
         \frac{ (-\lambda)^r \, r^{\underline{m}} }{r! \, (\ell-r)!} 
            \\ 
         &=& \frac{ b! }{(a+b-\ell)!}  
         \sum_{m=0}^{a} \binom{a}{m} \, (a+b-\ell)^{\underline{a-m}} \;
         \frac{(-1)^\ell \, \lambda^m \, (\lambda-1)^{\ell-m} }{(\ell-m)!}
            \qquad \\
         &=& \frac{ b! }{(a+b-\ell)!} \:
         \frac{ (-1)^\ell }{\ell!} \:
              Q_{\ell,a}(b,\lambda)  \;,
\label{def.Fl.Quatro}
\end{subeqnarray}
where we have used Lemma~\ref{lemma.appendix.1}
and the definition \reff{def_Ql} of $Q_{\ell,a}(b,\lambda)$
[note that the sum in \reff{def_Ql} can be stopped at $m=\ell$ because
 $\ell^{\underline{m}} = 0$ for $m>\ell$].
Clearly, $Q_{\ell,a}(b,\lambda)$
is a polynomial in $b,\lambda$ that is
of degree at most $a$ in the variable $b$
and of degree at most $\ell$ in $\lambda$.
Indeed, this is its exact degree, as the leading coefficient is
$[b^a] Q_{\ell,a}(b,\lambda) = (\lambda-1)^\ell$
[from the $m=0$ term in \reff{def_Ql}]
and hence $[b^a \lambda^\ell] Q_{\ell,a}(b,\lambda) = 1$.
\qed

\bigskip

\noindent
{\bf Remarks.}
1.  When $\lambda=1$, \reff{def_Fl} reduces to the Vandermonde convolution
\begin{equation}
 F_{\ell,a,b}(1) \;=\; \binom{\ell-a-1}{\ell} \;,
\end{equation}
while \reff{def_Ql} reduces to
\be
   Q_{\ell,a}(b,1)
   \;=\;
   \binom{a}{\ell} \, (a+b-\ell)^{\underline{a-\ell}} \: \ell!
   \;=\;
   \frac{a!}{(a-\ell)!} \: \frac{(a+b-\ell)!}{b!}
   \;,
\ee
so that \reff{def_Fl.eq} gives
\be
   F_{\ell,a,b}(1)  \;=\; (-1)^\ell \, \binom{a}{\ell}
                    \;=\; \binom{\ell-a-1}{\ell}
   \;.
\ee

2. In our applications we will always take $\ell=k-p\ge 0$ and $\lambda=3$.
In addition, in Theorem~\ref{lemma.appendix.4} we will take
$(a,b)=(s-1,p+1)$ with $p\ge0$, and $(a,b)=(s,p-1)$ with $p\ge1$; 
finally in Theorem~\ref{lemma.appendix.6}, we will take 
$(a,b)=(s-1,p-1)$ with $p\ge1$. 

%
%
\subsection[Proof of formula for $a_k^{\rm F}(k-s)$]%
{Proof of formula for $\bm{a_k^{\rm F}(k-s)}$} \label{appendix.A2}

The next lemma deals with the outer sums we find in 
\reff{eq.akm_final_F}: 

\begin{lemma} \label{lemma.appendix.3}
Let $\alpha=0$ or $1$.
Then, for any integers $s\ge 1-\alpha$ and $k\ge s+\alpha$, let us define
\begin{equation}
A_{\alpha,s,k}(\lambda) \;=\;  
\sum\limits_{p=\alpha}^k (-1)^p \binom{k-s-p-1}{p-\alpha} \, 
F_{k-p,s+\alpha-1,p+1-2\alpha}(\lambda) \,,  
\label{def_Ak}
\end{equation}
where $F_{\ell,a,b}(\lambda)$ is defined by \reff{def_Fl}. We then have
\begin{equation}
A_{\alpha,s,k}(\lambda) \;=\; \begin{cases} 
  \displaystyle
  \sum\limits_{q=0}^s \frac{ (-1)^q\, (k+1-q)}{q! \, (s-q)!} \,
  Q_{q,s-1}(k+1-q,\lambda) & \text{if $\alpha=0$}   \\[6mm]
  \displaystyle
  \sum\limits_{q=0}^s \frac{ (-1)^{q+1}}{q! \, (s-q)!} \,
  Q_{q,s}(k-1-q,\lambda) & \text{if $\alpha=1$}
\end{cases} 
\label{def_outer.eq}
\end{equation}
where the polynomial $Q_{\ell,a}(b,\lambda)$ is defined by \reff{def_Ql}. 
In particular, $A_{\alpha,s,k}(\lambda)$ is the restriction
to integers $k \ge s+\alpha$ of a polynomial in $k,\lambda$
that is of degree $s$ in $k$ and $\lambda$ separately,
with leading coefficient
\be
   [k^s] \, A_{\alpha,s,k}(\lambda) 
   \;=\;
   \frac{(-1)^\alpha}{s!} \, (2-\lambda)^s
 \label{def_outer.eq_leading}
\ee
and hence
$[k^s \lambda^s] \, A_{\alpha,s,k}(\lambda) = (-1)^{s+\alpha}/s!$. 
The next subleading coefficients are given by 
\begin{subeqnarray}
[k^{s-1}] \, A_{\alpha,s,k}(\lambda) 
   &=& \frac{(2-\lambda)^{s-1}}{2 (s-1)!} \times  
\begin{cases}
 (4-\lambda-2s+\lambda s) & \text{if $\alpha=0$} \\[3mm]
 (2-\lambda)(s+1)         & \text{if $\alpha=1$}
\end{cases} 
\slabel{def_outer.eq_leading2}  \\[5mm]
[k^{s-2}] \, A_{\alpha,s,k}(\lambda) 
   &=&  
\frac{(2-\lambda)^{s-1}}{24\, (s-2)!} \times 
\begin{cases}
-(3\lambda s^2 -6s^2 -7\lambda s +2s +2\lambda -4)& \text{if $\alpha=0$}\\[3mm]
 3\lambda s^2 -6s^2 +5\lambda s-34s +2\lambda -4 & \text{if $\alpha=1$} 
\end{cases} \nonumber \\ 
  & & \slabel{def_outer.eq_leading3} \\[2mm]
[k^{s-3}] \, A_{\alpha,s,k}(\lambda)
   &=&
\frac{(2-\lambda)^{s-2}}{48\, (s-3)!} \times
\begin{cases}
  (-48+80s-10\lambda s -3\lambda^2 s -20s^2 & \\
    \quad +2\lambda s^2 +4\lambda^2 s^2 -4s^3 +4\lambda s^3 -\lambda^2 s^3)  
  & \text{if $\alpha=0$}\\[3mm]
 (2-\lambda)s(s+1)(26-\lambda+2s-\lambda s)& \text{if $\alpha=1$}
\end{cases} \nonumber \\
  & & \slabel{def_outer.eq_leading4} \\[2mm]
[k^{s-4}] \, A_{\alpha,s,k}(\lambda)
   &=&
\frac{(2-\lambda)^{s-2}}{5760\, (s-4)!} \times
\begin{cases}
   (2-\lambda)(-16+8\lambda+924s+18\lambda s & \\ 
    \quad -1310 s^2 -125\lambda s^2 +420s^3 & \\
    \quad +90\lambda s^3 +30s^4 -15 \lambda s^4)
  & \text{if $\alpha=0$}\\[3mm]
  32-32\lambda +8\lambda^2 +1032s +888\lambda s & \\
    \quad + 18\lambda^2 s -5300 s^2 +1220 \lambda s^2 & \\
    \quad -5\lambda^2 s^2 -1560 s^3 + 840 \lambda s^3 & \\
    \quad -30\lambda^2 s^3 -60 s^4 +60 \lambda s^4 
    -15 \lambda^2 s^4 & \text{if $\alpha=1$}
\end{cases} \nonumber \\
  & & \slabel{def_outer.eq_leading5} 
 \label{def_outer.eq_subleading}
\end{subeqnarray} 
\end{lemma}

\proof 
In \reff{def_Ak} let us insert the result \reff{def_Fl.eq}
of Lemma~\ref{lemma.appendix.2} for 
$F_{k-p,s+\alpha-1,p+1-2\alpha}(\lambda)$;\footnote{
  Please note that the hypotheses of Lemma~\ref{lemma.appendix.3}
  ($\alpha \in \{0,1\}$, $s\ge 1-\alpha$ and $k\ge s+\alpha$)
  together with the summation limits $\alpha \le p \le k$
  imply that the three subscripts in 
  $F_{k-p,s+\alpha-1,p+1-2\alpha}$ are nonnegative;
  this justifies the use of Lemma~\ref{lemma.appendix.2}.
}
we get
\begin{eqnarray}
A_{\alpha,s,k}(\lambda) &=&  
  \sum\limits_{p=\alpha}^k \binom{k-s-p-1}{p-\alpha}  \,
  \frac{ (-1)^{k} \, (p+1-2\alpha)!}{(2p+s-k-\alpha)!\, (k-p)!} \nonumber \\
  & & \qquad\qquad  
        \times\, Q_{k-p,s+\alpha-1}(p+1-2\alpha,\lambda)   \;. 
\label{def_outer.eq2}
\end{eqnarray}
First of all, we notice that all terms with $p \le k-s-1$ vanish:
for if $0 \le k-s-p-1 < p-\alpha$, then $\binom{k-s-p-1}{p-\alpha} = 0$,
while if $2p+s-k-\alpha < 0$, then $1/(2p+s-k-\alpha)! = 0$;
and these two cases cover all values of $p$ satisfying $p \le k-s-1$.
%
%
So the only nonzero contributions to the sum \reff{def_outer.eq2} come
from $p\ge k-s\ge \alpha$.  Using the variable $q=k-p$, we can
rewrite \reff{def_outer.eq2} as
\begin{eqnarray} 
A_{\alpha,s,k}(\lambda) &=& 
  \sum\limits_{q=0}^s \binom{q-s-1}{k-q-\alpha} 
  \frac{ (-1)^{k} \, (k+1-q-2\alpha)!}{(k+s-2q-\alpha)!\, q!} \nonumber \\
   & &  \qquad\qquad \times\, Q_{q,s+\alpha-1}(k+1-q-2\alpha,\lambda) 
  \nonumber \\[2mm]
  &=& \sum\limits_{q=0}^s \binom{k+s-2q-\alpha}{k-q-\alpha} 
  \frac{ (-1)^{q+\alpha} \, (k+1-q-2\alpha)!}{(k+s-2q-\alpha)!\, q!} 
  \nonumber \\
   & &  \qquad\qquad \times\,  Q_{q,s+\alpha-1}(k+1-q-2\alpha,\lambda) 
  \nonumber \\[2mm]
  &=& \sum\limits_{q=0}^s 
  \frac{ (-1)^{q+\alpha}}{q!\, (s-q)!} \frac{ (k-q-2\alpha+1)!}{(k-q-\alpha)!}
    \nonumber \\ 
  & &  \qquad\qquad \times\, Q_{q,s+\alpha-1}(k+1-q-2\alpha,\lambda) \;. 
\label{def_outer.eq3}
\end{eqnarray}
Evaluating this for $\alpha=0,1$ yields \reff{def_Ak}.
This expression for $A_{\alpha,s,k}(\lambda)$
is manifestly a polynomial in $k$ and $\lambda$
of degree at most $s$ in each of the variables separately.
Indeed, using \reff{eq.leading.Qla} and performing a binomial sum over $q$,
we obtain \reff{def_outer.eq_leading}.

The computation of the subleading terms \reff{def_outer.eq_subleading} is
more involved. Let us start with the case $\alpha=0$.  
Our goal is to extract the leading powers of $k$ in \reff{def_outer.eq}, 
so our first task is to convert the falling power  
$(k+s-2q)^{\underline{s-1-m}}$ that appears in the definition of 
$Q_{q,s-1}(k+1-q,\lambda)$ [cf.~\reff{def_Ql}] to a regular power 
using the Stirling cycle numbers \cite{Graham_94}.
The key formula is obtained from \reff{def_xfalling} and the binomial
theorem: 
\be
(k+x)^{\underline{m}} \;=\; \sum\limits_{r=0}^m k^{m-r} 
 \sum\limits_{p=0}^r (-1)^p \, \stirlingcycle{m}{m-p}\, \binom{m-p}{r-p} \,
    x^{r-p} \,.  
\label{appendix.expansion}
\ee
Using this formula in the definition of $Q_{q,s-1}(k+1-q,\lambda)$ allows
us to obtain the first five leading terms in the expansion of the summand
in \reff{def_outer.eq}:
\begin{eqnarray}
   & & (k+1-q) \, Q_{q,s-1}(k+1-q,\lambda)
         \nonumber \\[2mm]
   & & \qquad\quad =\;
 (k+1-q)
 \sum\limits_{m=0}^q \binom{s-1}{m} (k+s-2q)^{\underline{s-1-m}} \,
q^{\underline{m}}\, \lambda^m \, (\lambda-1)^{q-m} \nonumber \\ 
   & & \qquad\quad =\;
  \sum\limits_{i=0}^4 k^{s-i} \, a^{(0)}_i \;+\; O(k^{s-5}) \;,
\end{eqnarray}
where the coefficients $a^{(0)}_i$ are given by
\begin{subeqnarray}
a^{(0)}_0 &=& (\lambda-1)^q  \\[4mm]
a^{(0)}_1 &=& (\lambda-1)^{q-1}\, \left[
      - q\, (1-2s+\lambda s) 
      + \frac{(\lambda-1)}{2}\, s(s+1)\right]  \\[4mm] 
a^{(0)}_2 &=& (\lambda-1)^{q-2} \binom{s-1}{1} \left[  
        q^{\underline{2}}\, \frac{(\lambda-2)}{2} \, (2-2s+\lambda s) 
        \right. \nonumber \\[2mm]
          & & \qquad 
      - q \, \frac{(\lambda-1)}{2} \, (-2+3s-2s^2+\lambda s^2) 
              \nonumber \\[2mm]
          & & \qquad \left.  
             +  \frac{(\lambda-1)^2}{24}\, s(s+1)(2+3s) \right]   \\[4mm]
a^{(0)}_3 &=& (\lambda-1)^{q-3} \binom{s-1}{2} \left[
      - q^{\underline{3}}\, \frac{(\lambda-2)^2}{3} (3-2s+\lambda s) \right. 
      \nonumber \\[2mm]
          & & \qquad  
      + q^{\underline{2}}\, \frac{(\lambda-1)}{2}\,
        (16 -6\lambda - 16 s + 10 \lambda s - \lambda^2 s + 4 s^2 
        - 4 \lambda s^2 + \lambda^2 s^2) \nonumber \\[2mm]
          & & \qquad  
      - q \, \frac{(\lambda-1)^2}{12}\,
          (12 - 25 s + 17 s^2 - \lambda s^2 - 6 s^3 + 3\lambda s^3)
           \nonumber \\[2mm]
          & & \qquad \left.
      + \frac{(\lambda-1)^3}{24}\, s^2 (1+s)^2 \right]   \\[4mm]
a^{(0)}_4 &=& (\lambda-1)^{q-4} \binom{s-1}{3} \left[
        q^{\underline{4}}\, \frac{(\lambda-2)^3}{4} \, (4-2s+\lambda s)\right.
        \nonumber \\[2mm]
          & & \qquad 
      - q^{\underline{3}}\, \frac{(\lambda-1)(\lambda-2)}{2} 
          (36 - 12\lambda - 26s + 17\lambda s - 2\lambda^2 s + 4 s^2 - 
           4\lambda s^2 + \lambda^2 s^2)\nonumber \\[2mm]
          & & \qquad  
      + q^{\underline{2}} \, \frac{(\lambda-1)^2}{8}\,
       (-208 +56\lambda s + 236s - 102\lambda s + 4 \lambda^2 s - 88 s^2 + 
         58\lambda s^2  \nonumber \\[2mm]
        & & \qquad \qquad - 7\lambda^2 s^2 + 12 s^3 - 12\lambda s^3 + 
         3\lambda^2 s^3)\nonumber \\[2mm]
          & & \qquad 
      - q\, \frac{(\lambda-1)^3}{8}\, (s-1) 
                (8 - 14 s + 7 s^2 - 2s^3 + \lambda s^3)\nonumber \\[2mm]
         & & \qquad \left.  + \frac{(\lambda-1)^4}{960} 
         s(s+1)(-8 - 10s + 15 s^2 + 15s^3)\right] 
\label{def_a0i}
\end{subeqnarray}
In the derivation of these formulae we have used the following special
values of the Stirling cycle numbers \cite[Chapter~5, Exercise~16]{Comtet} 
(see also \cite{Graham_94,Sloane_on-line}): 
\begin{subeqnarray}
\stirlingcycle{n}{n-1} &=& \binom{n}{2}   \\[3mm] 
\stirlingcycle{n}{n-2} &=& \binom{n}{3} \frac{3n-1}{4}  \\[3mm]
\stirlingcycle{n}{n-3} &=& \binom{n}{4} \binom{n}{2}   \\[3mm] 
\stirlingcycle{n}{n-4} &=& \binom{n}{5} \frac{15n^3-30n^2+5n+2}{48}
\end{subeqnarray}

The second task is to perform the sum over $q$ in \reff{def_outer.eq}. 
It is clear from \reff{def_a0i} that the coefficients $a^{(0)}_i$
are polynomials in $q$; and we have chosen to write them in terms of 
the falling powers $q^{\underline{m}}$
in order to facilitate the use of
Lemma~\ref{lemma.appendix.1} to perform the sum over $q$ in
\reff{def_outer.eq}. Each coefficient $a^{(0)}_i$
in \reff{def_a0i} contributes only to the coefficient
$[k^{s-i}] A_{0,s,k}(\lambda)$:
\begin{equation}
[k^{s-i}]A_{0,s,k}(\lambda)\;=\;  
\sum\limits_{q=0}^s \frac{(-1)^q }{q! \,(s-q)!} a^{(0)}_i \quad
\text{for $i\ge 0$.} 
\end{equation}
Using the definition \reff{def_Tsa} of $T_{s,a}(\lambda)$,
we can rewrite \reff{def_a0i} as
\begin{subeqnarray}
[k^s]A_{0,s,k}(\lambda) &=&  T_{s,0}(1-\lambda)  \\[2mm]  
[k^{s-1}]A_{0,s,k}(\lambda) &=& (\lambda-1)^{-1} \left[  
- (1-2s+\lambda s)  T_{s,1}(1-\lambda)  \nonumber \phantom{\frac12}\right. \\
  & & \qquad \left.  
+ \frac{(\lambda-1)}{2} s(s+1)T_{s,0}(1-\lambda) 
  \right] 
\end{subeqnarray}
along with similar formulae for 
$[k^{s-i}]A_{0,s,k}(\lambda)$ with $i=2,3,4$.
{}From these equations together with Lemma~\ref{lemma.appendix.1}
we can derive after some algebra the results 
\reff{def_outer.eq_leading}/\reff{def_outer.eq_subleading} for $\alpha=0$.  

The computation for $\alpha=1$ is similar. The five leading terms for 
$Q_{q,s}(k-1-q,\lambda)$ can be written as  
\begin{eqnarray}
Q_{q,s}(k-1-q,\lambda) &=& 
 \sum\limits_{m=0}^q \binom{s}{m} (k+s-2q-1)^{\underline{s-m}} \,
q^{\underline{m}}\, \lambda^m \, (\lambda-1)^{q-m} \nonumber \\ 
  &=& \sum\limits_{i=0}^4 k^{s-i} \, a^{(1)}_i \;+\; O(k^{s-5}) \;,
\end{eqnarray}
where the coefficients $a^{(1)}_i$ are given by
\begin{subeqnarray}
a^{(1)}_0 &=& (\lambda-1)^q   \\[4mm]
a^{(1)}_1 &=& (\lambda-1)^{q-1}\, \binom{s}{1} \, 
     \left[ q\, (2-\lambda) + \frac{(\lambda-1)}{2}(s-1)  \right] \\[4mm]
a^{(1)}_2 &=& (\lambda-1)^{q-2} \binom{s}{2} \left[  
               q^{\underline{2}}\, (2-\lambda)^2 
             - q \, (\lambda-1) (6-2\lambda-2s+\lambda s) 
               \phantom{\binom{s}{2}} 
       \right. \nonumber \\
          & & \qquad \left.  
             + \frac{(\lambda-1)^2}{12}\, (s-2)(3s-1)\right]    \\[4mm]
a^{(1)}_3 &=& (\lambda-1)^{q-3} \binom{s}{3} \left[
               q^{\underline{3}}\, (2-\lambda)^3
             - q^{\underline{2}} \, \frac{3(2-\lambda)(\lambda-1)}{2}\,
                (10-3\lambda-2s+\lambda s) \right.  \nonumber \\
          & & \qquad -q \, \frac{(\lambda-1)^2}{4}\, 
              (s-3)(20-4\lambda-6s+3\lambda s) \nonumber \\ 
          & & \qquad \left.
             + \frac{(\lambda-1)^3}{8}\, s(s-1)(s-3)\right]    \\[4mm]
a^{(1)}_4 &=& (\lambda-1)^{q-4} \binom{s}{4} \left[
               q^{\underline{4}}\, (2-\lambda)^4
             - 2q^{\underline{3}} \, (2-\lambda)^2(\lambda-1)\,
                (14-4\lambda-2s+\lambda s) \phantom{\frac{1}{1}}
          \right.  \nonumber \\
          & & \qquad 
             +q^{\underline{2}} \, \frac{(\lambda-1)^2}{2}\,  
              (328-208\lambda+28\lambda^2 -124s+100\lambda s-19\lambda^2 s
              \nonumber \\
          & & \qquad \qquad 
              +12s^2-12\lambda s^2 +3\lambda^2 s^2) \nonumber \\
          & & \qquad 
             -q \, \frac{(\lambda-1)^3}{2}\,  
              (s-4) (-18+2\lambda +12s- 3\lambda s- 2s^2 +\lambda s^2) 
              \nonumber \\
          & & \qquad \left.
             + \frac{(\lambda-1)^4}{240}\, (s-4)(2+5s-30s^2+15s^3)\right] 
\label{def_a1i}
\end{subeqnarray}

We now perform the sum over $q$ in \reff{def_outer.eq}. 
Again, each coefficient $a^{(1)}_i$ 
in \reff{def_a1i} contributes only to the coefficient 
$[k^{s-i}] A_{1,s,k}(\lambda)$: 
\begin{equation}
[k^{s-i}]A_{1,s,k}(\lambda)\;=\;  
- \sum\limits_{q=0}^s \frac{(-1)^q }{q! \,(s-q)!} a^{(1)}_i \quad
\text{for $i\ge 0$.} 
\end{equation}
Using the definition \reff{def_Tsa} of $T_{s,a}(\lambda)$,
we can rewrite \reff{def_a1i} as
\begin{subeqnarray}
[k^s]A_{1,s,k}(\lambda) &=&  -T_{s,0}(1-\lambda)   \\[3mm]  
[k^{s-1}]A_{1,s,k}(\lambda) &=& 
  - (\lambda-1)^{-1} \binom{s}{1} \left[ (2-\lambda)T_{s,1}(1-\lambda)  
    \phantom{\frac12} \right. \nonumber \\[2mm]
  & & \qquad \left.  
    + \frac{(\lambda-1)}{2}(s-1) T_{s,0}(1-\lambda) \right]  
\end{subeqnarray}
along with similar formulae for $[k^{s-i}]A_{1,s,k}(\lambda)$
with $i=2,3,4$.
These equations lead via Lemma~\ref{lemma.appendix.1} 
to \reff{def_outer.eq_leading}/\reff{def_outer.eq_subleading} for $\alpha=1$.  
\qed

\bigskip

We now proceed to prove the main result of this subsection:

\begin{theorem}[= Proposition~\protect\ref{prop.eq.empirical.1}]
 \label{lemma.appendix.4} 
For each integer $s\ge 1$, the quantity
\begin{eqnarray}
   a^{\rm F}_k(k-s)  & \!=\! &
   \sum_{p=0}^{k}  (-1)^p {k-s-1-p\choose p} \sum_{r=0}^{k-p} 3^r
       {k-s-1-2p \choose r} {p+1 \choose k-p-r}
   \nonumber \\
& & \, +
   \sum_{p=1}^{k}  (-1)^p {k-s-1-p\choose p-1} \sum_{r=0}^{k-p} 3^r
       {k-s-2p \choose r} {p-1 \choose k-p-r}
   \qquad\quad
\label{appendix.akm_F}
\end{eqnarray}
is, when restricted to $k \ge s+1$,
given by a polynomial in $k$ of degree $\max(0,s-3)$,
with leading coefficient
\begin{eqnarray}
[k^{s-3}] a^{\rm F}_k(k-s) &=& \frac{(-1)^{s+1}}{(s-3)!}  \quad 
  \text{for $s\ge 3$}
 \label{appendix.akm_F.i=3}
\end{eqnarray}
and first subleading coefficient
\begin{eqnarray}
[k^{s-4}] a^{\rm F}_k(k-s) &=& \frac{(-1)^s s}{2(s-4)!}  \quad 
  \text{for $s\ge 4$ \;.} 
 \label{appendix.akm_F.i=4}
\end{eqnarray}
\end{theorem}

\proof 
Using Lemma~\ref{lemma.appendix.3} we can rewrite 
\reff{appendix.akm_F} as
\begin{equation}
a^{\rm F}_k(k-s) \;=\; A_{0,s,k}(3)+A_{1,s,k}(3) \;,
\end{equation}
where each of the two terms is, for integers $k \ge s+1$, the restriction
of a polynomial in $k$ of degree $s$.
Moreover, it follows from 
\reff{def_outer.eq_leading}/\reff{def_outer.eq_subleading}
that for a general value of $\lambda$ we have
\begin{subeqnarray}
[k^s]     [A_{0,s,k}(\lambda)+A_{1,s,k}(\lambda)] &=& 0 \\[1mm]
[k^{s-1}] [A_{0,s,k}(\lambda)+A_{1,s,k}(\lambda)]
           &=& \frac{(2-\lambda)^{s-1}}{(s-1)!} (3-\lambda)  \\[1mm] 
[k^{s-2}] [A_{0,s,k}(\lambda)+A_{1,s,k}(\lambda)]
           &=& \frac{s\, (2-\lambda)^{s-1}}{2(s-2)!}(3-\lambda) \\[1mm]
[k^{s-3}] [A_{0,s,k}(\lambda)+A_{1,s,k}(\lambda)]
           &=& \frac{(2-\lambda)^{s-2}}{24(s-3)!}
           (3\lambda^2 s^2 -15\lambda s^2 +18s^2 -\lambda^2 s \nonumber \\
           & & \qquad \qquad -19 \lambda s  +66s -24) \\[1mm]
[k^{s-4}] [A_{0,s,k}(\lambda)+A_{1,s,k}(\lambda)]
           &=& -\,\frac{(2-\lambda)^{s-2}s}{48(s-4)!}
           (\lambda^2 s^2 -5\lambda s^2 +6s^2 -\lambda^2 s \nonumber \\
           & & \qquad \qquad -19 \lambda s +66s -24) 
\end{subeqnarray}
Specializing to $\lambda=3$, we see that the first three coefficients vanish,
while the next two coefficients are nonzero
and given by \reff{appendix.akm_F.i=3}/\reff{appendix.akm_F.i=4}.
Therefore, $a^{\rm F}_k(k-s)$ is the restriction to integers $k \ge s+1$
of a polynomial in $k$ of degree exactly $\max(0,s-3)$.
\qed

%
%
\subsection[Proof of formula for $\tilde{a}_k^{\rm P}(k-s)$]%
{Proof of formula for $\tilde{\bm{a}}\bm{{}_k^{\rm P}(k-s)}$}
   \label{appendix.A3}

In this subsection we shall carry out an analogous analysis
for the case of periodic transverse boundary conditions.
It is convenient to prove first a
technical lemma (analogous to Lemma~\ref{lemma.appendix.3}) that deals
with the outer sums we find in \reff{eq.atildekm_final_P}:

\begin{lemma} \label{lemma.appendix.5}
For any integers $s\ge 1$ and $k\ge s$, let us define
\begin{equation}
B_{s,k}(\lambda) \;=\;  
(k-s)\, \sum\limits_{p=1}^k \frac{(-1)^p}{p} \binom{k-s-p-1}{p-1} \, 
F_{k-p,s-1,p}(\lambda) \,,  
\label{def_Bsk}
\end{equation}
where $F_{\ell,a,b}(\lambda)$ is defined by \reff{def_Fl}. We then have
\begin{equation}
B_{s,k}(\lambda) \;=\; 
  (k-s)\, \sum\limits_{q=0}^s \frac{ (-1)^{q+1}}{q! \, (s-q)!} \,
  Q_{q,s-1}(k-q,\lambda) \,, 
\label{def_outerBis.eq}
\end{equation}
where the polynomial $Q_{\ell,a}(b,\lambda)$ is defined by \reff{def_Ql}. 
In particular, $B_{s,k}(\lambda)$ is the restriction
to integers $k \ge s$ of a polynomial in $k,\lambda$
that is of degree $s$ in $k$ and $\lambda$ separately  
with leading coefficient
\be
   [k^s] \, B_{s,k}(\lambda) 
   \;=\;
   -\,\frac{1}{s!} \, (2-\lambda)^s
 \label{def_outerBis.eq_leading}
\ee
and hence
$[k^s \lambda^s] \, B_{s,k}(\lambda) = (-1)^{s+1}/s!$. 
The next subleading coefficient is given by 
\be
[k^{s-1}] \, B_{s,k}(\lambda) 
  \;=\; \frac{s+1}{2(s-1)!} \, (2-\lambda)^s \,. 
 \label{def_outerBis.eq_subleading}
\ee
\end{lemma}

\proof
In \reff{def_Bsk} let us insert the result \reff{def_Fl.eq}
of Lemma~\ref{lemma.appendix.2} for
$F_{k-p,s-1,p}(\lambda)$; we get
\begin{eqnarray}
B_{s,k}(\lambda) &=& (k-s) 
  \sum\limits_{p=1}^k \binom{k-s-p-1}{p-1}  \,
  \frac{ (-1)^{k} \, (p-1)!}{(2p+s-k-1)!\, (k-p)!} \nonumber \\
  & & \qquad\qquad  
        \times\, Q_{k-p,s-1}(p,\lambda)   \;. 
\label{def_outerBis.eq2}
\end{eqnarray}
First of all, we see that for $k=s$, the sum is finite, and therefore
$B_{k,k}(\lambda)=0$. So, we can restrict ourselves to the nontrivial cases
$k\ge s+1$. Secondly, we notice, 
as in the proof of Lemma~\ref{lemma.appendix.3}, that all terms with 
$p \le k-s-1$ vanish. So the only nonzero contributions to the sum 
\reff{def_outerBis.eq2} come from $p\ge k-s\ge 1$.  
Using the variable $q=k-p$, we can rewrite \reff{def_outerBis.eq2} as
\begin{eqnarray} 
B_{s,k}(\lambda) &=& (k-s)  
  \sum\limits_{q=0}^s \binom{q-s-1}{k-q-1} 
  \frac{ (-1)^{k} \, (k-q-1)!}{(k+s-2q-1)!\, q!} \,  Q_{q,s-1}(k-q,\lambda) 
  \nonumber \\[2mm]
  &=& (k-s) \sum\limits_{q=0}^s \binom{k+s-2q-1}{k-q-1} 
  \frac{ (-1)^{q+1} \, (k-q-1)!}{(k+s-2q-1)!\, q!} \,  Q_{q,s-1}(k-q,\lambda) 
  \nonumber \\[2mm]
  &=& (k-s) \sum\limits_{q=0}^s 
  \frac{ (-1)^{q+1}}{q!\, (s-q)!} \, Q_{q,s-1}(k-q,\lambda) \;. 
\label{def_outerBis.eq3}
\end{eqnarray}
This expression for $B_{s,k}(\lambda)$
is manifestly a polynomial in $k$ and $\lambda$
of degree at most $s$ in each of the variables separately.
Indeed, using \reff{eq.leading.Qla} and performing a binomial sum over $q$,
we obtain \reff{def_outerBis.eq_leading}.

The computation of the subleading term \reff{def_outerBis.eq_subleading} is
similar to the one in Lemma~\ref{lemma.appendix.3}.
Using the expansion \reff{appendix.expansion} in the definition of 
$Q_{q,s-1}(k-q,\lambda)$ that appears in \reff{def_outerBis.eq3}, we
obtain the two leading terms: 
\begin{eqnarray}
Q_{q,s-1}(k-q,\lambda) &=& 
 \sum\limits_{m=0}^{s-1} \binom{s-1}{m} (k+s-2q-1)^{\underline{s-1-m}} \,
q^{\underline{m}}\, \lambda^m \, (\lambda-1)^{q-m} \nonumber \\ 
 &=& k^{s-1} a^{(2)}_1 + k^{s-2} a^{(2)}_2 + O\left(k^{s-3}\right) \,,
\end{eqnarray}
where the coefficients $a^{(2)}_k$ are given by
\begin{subeqnarray}
a^{(2)}_1 &=& (\lambda-1)^q  \\
a^{(2)}_2 &=& (\lambda-1)^{q-1}\, (s-1) \left[ \frac{s(\lambda-1)}{2}
                   + q(2-\lambda) \right]  
\label{def_a2i}
\end{subeqnarray}

The second task is to perform the sum over $q$ in \reff{def_outerBis.eq}.
Because of the prefactor $k-s$, each coefficient
$a^{(2)}_i$ in \reff{def_a2i} gives a nonzero contribution to both
$[k^{s-i+1}] B_{s,k}(\lambda)$ and $[k^{s-i}] B_{s,k}(\lambda)$.
After some algebra we find
\begin{subeqnarray}
[k^s]B_{s,k}(\lambda) &=&
\sum\limits_{q=0}^s \frac{(-1)^q }{q! \,(s-q)!} (-a^{(2)}_1)\nonumber \\
  &=& T_{s,0}(1-\lambda)   \\[2mm]
     [k^{s-1}]B_{s,k}(\lambda) &=&
\sum\limits_{q=0}^s \frac{(-1)^q}{q! \,(s-q)!}
     \left[- a^{(2)}_2 + s a^{(2)}_1\right]\nonumber \\
  &=& \frac{s(3-s)}{2} T_{s,0}(1-\lambda) - \frac{(s-1)(2-\lambda)}{\lambda-1}
   T_{s,1}(1-\lambda)   
\end{subeqnarray}
where the $T_{s,a}$ are given in \reff{def_Tsa.eq}. From
these equations we can derive after some algebra the results
\reff{def_outerBis.eq_leading}/\reff{def_outerBis.eq_subleading}.
\qed
 
Now we are ready to prove the main result of this subsection:
\begin{theorem}[= Proposition~\protect\ref{prop.eq.empirical_P.1}]
  \label{lemma.appendix.6} 
For each integer $s\ge 1$, the quantity
\begin{equation}
   \!\!
   \tilde{a}^{\rm P}_k(k-s) \;=\; 
   3^k \binom{k-s}{k} \,+\,
   \sum_{p=1}^k (-1)^p \, \frac{k-s}{p} \, {k-s-p-1 \choose p-1}
   \sum_{r=0}^{k-p}  3^r {k-s-2p \choose r} {p \choose k-p-r}
\label{appendix.akm_P}
\end{equation}
is, when restricted to $k \ge s$,
given by a polynomial in $k$ of degree $s$,
with leading coefficient
\begin{eqnarray}
[k^s] \tilde{a}^{\rm P}_k(k-s) &=& \frac{(-1)^{s+1}}{s!}
\end{eqnarray}
and first subleading coefficient
\begin{eqnarray}
[k^{s-1}] \tilde{a}^{\rm P}_k(k-s)  &=& \frac{(-1)^s (s+1)}{2(s-1)!} \;. 
\end{eqnarray}  
Furthermore, $\tilde{a}^{\rm P}_k(0)=0$,
so that the polynomial representing $\tilde{a}^{\rm P}_k(k-s)$
for $k \ge s$ has a factor $k-s$.
\end{theorem}

\proof
The term $3^k \binom{k-s}{k}$ in \reff{appendix.akm_P}
vanishes for integers $s \ge 1$ and $k \ge s$, so we have
\be
\tilde{a}^{\rm P}_k(k-s) \;=\; B_{s,k}(3) \,.
\ee
Therefore, the main assertions of Theorem~\ref{lemma.appendix.6}
are an immediate consequence of Lemma~\ref{lemma.appendix.5}.
Finally, it is known from \reff{eq.APconst}
that $\tilde{a}^{\rm P}_k(0)=0$ for $k \ge 1$;
and this corresponds to the case $k=s$,
which lies within the regime $k \ge s$
where $\tilde{a}^{\rm P}_k(k-s)$ is given by a polynomial in $k$.
\qed

%
%
\subsection[Proof of formula for $a_k^{\rm F}(k)$]%
{Proof of formula for $\bm{a_k^{\rm F}(k)}$} \label{appendix.A4}

Our last goal is to prove \reff{eq.fibonacci.F.s=0}.
Let us start by proving
two lemmas similar to Lemmas~\ref{lemma.appendix.3} and~\ref{lemma.appendix.5}. 

\begin{lemma} \label{lemma.appendix.7} 
For any integer $k\ge 1$, let $a_{1,k}(\lambda)$ be defined as
\begin{equation}
a_{1,k}(\lambda) \;=\; 
\sum_{p=0}^{k}  (-1)^p {k-1-p\choose p} \sum_{r=0}^{k-p} \lambda^r \,
          {k-2p-1 \choose r} {p+1 \choose k-p-r} \,.
\label{def_a1k}
\end{equation}
Then 
\begin{equation}
a_{1,k}(\lambda) \;=\; 
1 + \sum_{p=0}^{k-1}  (-1)^p {k-1-p\choose p} \lambda^{k-2p-1} \,. 
\label{def_a1k.eq}
\end{equation}
\end{lemma}

\proof
The inner sum in \reff{def_a1k},
\be
S_{k,p}(\lambda) \;=\; \sum_{r=0}^{k-p} \lambda^r \,
          {k-2p-1 \choose r} {p+1 \choose k-p-r} \;,
\label{def_a1k.inner.eq1}
\ee
can be performed analogously to the proof of Lemma~\ref{lemma.appendix.2}:
If $k-2p-1\ge 0$, then ${k-1-2p \choose r}=0$ for all $r>k-2p-1\ge 0$. 
Moreover, ${p+1 \choose k-p-r}=0$ whenever $k-p-r > p+1\ge 1$, i.e.\ 
for all $r<k-2p-1$. Thus, when $k-2p-1\ge 0$
 there is a single nonzero term coming from
$r=k-2p-1$ and whose contribution is $\lambda^{k-2p-1}$.  
On the other hand, if $k-2p-1<0$,
then following similar steps to those of the proof of 
Lemma~\ref{lemma.appendix.2}, we obtain a finite sum. Putting these
two contributions together we get  
\begin{eqnarray}
S_{k,p}(\lambda) &=& \frac{(p+1)!}{(2p-k)!} \sum_{r=0}^{k-p} 
\frac{ (-\lambda)^r}{r!\, (k-p-r)! (2p+r+1-p)} 
 \nonumber \\ 
  & & \quad  +  \lambda^{k-2p-1} I[k\ge 2p+1] \,, 
\label{def_a1k.inner.eq2}
\end{eqnarray}
where $I[A]$ is the indicator function of the event $A$, and in the
first term it is not necessary to add $I[k<2p+1]$ because of the
factor $1/(2p-k)!$.  

The outer sum in \reff{def_a1k} is easy:
\begin{eqnarray}
a_{1,k}(\lambda) &=& 
\sum\limits_{p=0}^k (-1)^p \binom{k-1-p}{p} \,
  \frac{(p+1)!}{(2p-k)!} \sum_{r=0}^{k-p} 
\frac{ (-\lambda)^r}{r!\, (k-p-r)! (2p+r+1-p)} \nonumber \\ 
  & &  \quad  + 
\sum\limits_{p=0}^k (-1)^p \binom{k-1-p}{p} \,  
  \lambda^{k-2p-1} \, I[k\ge 2p+1] \,,  
\end{eqnarray}
as the only nonzero contribution of the second term corresponds to
$p=k$. (The contributions of the terms with $0\le p\le k-1$ vanish because
Euler's reflection formula.) We then obtain,
\be
a_{1,k}(\lambda) \;=\; 1 + \sum\limits_{p=0}^{\lfloor (k-1)/2 \rfloor} 
 (-1)^p \binom{k-1-p}{p} \, \lambda^{k-2p-1} \,. 
\ee
Notice that the upper limit $\lfloor (k-1)/2 \rfloor$ comes from the 
condition $k\ge 2p+1$; but we can replaced it by the simpler term 
$k-1$, as $\binom{k-1-p}{p}=0$ whenever $p>k-1-p$ (i.e., $2p>k-1$). 
We then obtain the claimed formula \reff{def_a1k.eq}. 
\qed

\begin{lemma} \label{lemma.appendix.8}
For any integer $k\ge 1$, let $a_{2,k}(\lambda)$ be defined as
\begin{equation}
a_{2,k}(\lambda) \;=\; 
\sum_{p=0}^{k}  (-1)^p {k-1-p\choose p-1} \sum_{r=0}^{k-p} \lambda^r \,
          {k-2p \choose r} {p-1 \choose k-p-r} \,.
\label{def_a2k}
\end{equation}
Then
\begin{equation}
a_{2,k}(\lambda) \;=\; -1 \,. 
\label{def_a2k.eq}
\end{equation}
\end{lemma}

\proof
By inspection it is clear that $a_{2,k}(\lambda)=A_k(1,0;\lambda)$
[cf.~\reff{def_Ak}]. Then Lemma~\ref{lemma.appendix.3} ensures that  
for any $k\ge 1$
\begin{equation}
A_k(1,0;\lambda) \;=\; -Q_{0,0}(k-1,\lambda) \;=\; -1 \,,
\end{equation}
where we have also used \reff{def_Ql}. \qed

Now the strategy is to show that the quantities 
$C_k(\lambda)=a_{1,k}(\lambda)+a_{2,k}(\lambda)$ satisfy a 
second-order linear recursion, which reduces to that of the 
Fibonacci numbers $F_{2k}$ when $\lambda=3$.  
In particular we prove the following lemma:

\begin{lemma} \label{lemma.appendix.9}
For any integer $k\ge 1$, let $C_{k}(\lambda)$ be defined as
\begin{equation}
C_{k}(\lambda) \;=\; a_{1,k}(\lambda) + a_{2,k}(\lambda)\,,
\label{def_Ck}
\end{equation}
where the quantities $a_{i,k}(\lambda)$ are defined in 
\reff{def_a1k}/\reff{def_a2k}. Then, for any $k\ge 3$, the quantity
\begin{equation}
f_{k}(\lambda) \;=\; C_k(\lambda) - \lambda C_{k-1}(\lambda) +
  C_{k-2}(\lambda)
\label{def_fk}
\end{equation}
vanishes identically (i.e., $f_k(\lambda)=0$ for all $k\ge 3$).
\end{lemma}

\proof
{}From Lemmas~\ref{lemma.appendix.7} and~\ref{lemma.appendix.8} it is 
clear that
\begin{equation}
C_{k}(\lambda) \;=\; \sum\limits_{p=0}^{k-1}(-1)^p 
      \binom{k-1-p}{p} \, \lambda^{k-2p-1} \,. 
\label{def_Ck.eq}
\end{equation}
Then, the expression for $f_k(\lambda)$ for any $k\ge 3$ is given by
\begin{eqnarray}
f_k &=& \sum\limits_{p=0}^{k-1} (-1)^p     \lambda^{k-2p-1} \binom{k-1-p}{p}
       +\sum\limits_{p=0}^{k-2} (-1)^{p+1} \lambda^{k-2p-1} \binom{k-2-p}{p}
      \nonumber \\
    & & \quad 
       +\sum\limits_{p=0}^{k-3} (-1)^{p} \lambda^{k-2p-3} \binom{k-3-p}{p} \,.
\label{def_fk.eq1}
\end{eqnarray}
The last term can be rewritten in a more convenient way if we change
the summation variable to $p+1$:
\be
\sum\limits_{p=0}^{k-3} (-1)^{p} \lambda^{k-2p-3} \binom{k-3-p}{p} \;=\;
\sum\limits_{p=1}^{k-2} (-1)^{p+1} \lambda^{k-2p-1} \binom{k-2-p}{p-1} 
\ee
This term is essentially the same as the second term in the sum 
\reff{def_fk.eq1}, except for the binomial coefficients. Using the Pascal
identity
\be
\binom{k-p-2}{p} + \binom{k-p-2}{p-1} \;=\; \binom{k-p-1}{p}\,,
\ee
we obtain that
\begin{eqnarray}
f_k &=& \sum\limits_{p=0}^{k-1} (-1)^p     \lambda^{k-2p-1} \binom{k-1-p}{p}
       +\sum\limits_{p=0}^{k-2} (-1)^{p+1} \lambda^{k-2p-1} \binom{k-2-p}{p}
     \nonumber \\
    &=& (-1)^{k-1} \binom{0}{k-1} \lambda^{1-k} \nonumber \\
    &=& (-\lambda)^{1-k} \, \delta_{k,1} \;=\; 0 \,,
\label{def_fk.eq2}
\end{eqnarray}
which is zero because we are assuming that $k\ge 3$. This completes the
proof. 
\qed

Now we are ready to prove the main result \reff{eq.fibonacci.F.s=0}:

\begin{theorem} 
    \label{lemma.fibonacci}
For any integer $k\ge 1$ let us define the quantity [cf.~\reff{eq.akm_final_F}]
\begin{eqnarray}
   a^{\rm F}_k(k)  & =&
   \sum_{p=0}^{k}  (-1)^p {k-1-p\choose p} \sum_{r=0}^{k-p} 3^r
       {k-1-2p \choose r} {p+1 \choose k-p-r}
   \nonumber \\
& & \quad +
   \sum_{p=1}^{k}  (-1)^p {k-1-p\choose p-1} \sum_{r=0}^{k-p} 3^r
       {k-2p \choose r} {p-1 \choose k-p-r} \,.  \qquad
\label{appendix.akk_F}
\end{eqnarray}
Then for any $k\ge 1$, $a^{\rm F}_k(k) = F_{2k}$, where $F_{2k}$ is the
$(2k)$--th Fibonacci number \cite[sequences A001906/A088305]{Sloane_on-line}.  
\end{theorem} 

\proof
The quantity $a^{\rm F}_k(k)$ can be written 
as 
\be
a^{\rm F}_k(k) \;=\; C_k(3) \,,
\ee
where $C_k$ is given in Lemma~\ref{lemma.appendix.9}
(see also Lemmas~\ref{lemma.appendix.7} and~\ref{lemma.appendix.8}).
For any $k\ge 3$, the quantities $a^{\rm F}_k(k)$ satisfy the recurrence
\reff{def_fk} for $\lambda=3$, i.e.
\begin{equation}
a^{\rm F}_k(k) - 3 a^{\rm F}_{k-1}(k-1) + a^{\rm F}_{k-2}(k-2) \;=\; 0\,,
\label{def_recurrence.aFkk}
\end{equation}
which is the same recurrence as the Fibonacci numbers $F_{2n}$ 
\cite{Sloane_on-line}.
Furthermore, a direct computation shows that the initial conditions for
the two sequences are the same: 
$a^{\rm F}_1(1)=1 = F_2$ and $a^{\rm F}_2(2)= 3 = F_4$.
Therefore, the sequences $a^{\rm F}_k(k)$ and $F_{2k}$ coincide.  \qed 

%
%
\section{Upper zero-free interval for bipartite planar graphs}
\label{appendix.upper}

Let $G$ be a loopless planar graph.
Then it is not hard to prove that $P_G(q) > 0$
for all {\em integers}\/ $q \ge 5$;\footnote{
   This is the Five-Color Theorem, which goes back to Heawood in 1890.
   For a proof, see e.g.\ \cite[Theorem V.8, pp.~154--155]{Bollobas_98};
   or for an elegant alternate proof of an even stronger result,
   see \cite[Theorem V.12, pp.~161--163]{Bollobas_98}.
}
moreover, one of the most famous theorems of graph theory
--- the Four-Color Theorem
\cite{Appel_77a,Appel_77b,Appel_89,Robertson_97,Thomas_98} ---
asserts that $P_G(q) > 0$ holds in fact for all integers $q \ge 4$.

It is natural to ask whether these results can be extended from
integer $q$ to {\em real}\/ $q$.
The answer is yes, at least in part:
Birkhoff and Lewis \cite{Birkhoff_46} proved in 1946
that if $G$ is a loopless planar graph,
then $P_G(q) > 0$ for all real numbers $q \ge 5$.\footnote{
  See also Woodall \cite[Theorem 1]{Woodall_97} and
  Thomassen \cite[Theorem 3.1 ff.]{Thomassen_97}
  for alternate proofs of a more general result.
  These theorems are, in turn, consequences of a stronger
  and even more general result for matroids
  that was proved but not stated (!)\ twenty years earlier by
  Oxley \cite{Oxley_78};  see Jackson \cite[Theorem~38]{Jackson_03}.
}
Furthermore, they conjectured that $P_G(q) > 0$ also for $4 < q < 5$;
and while no one has yet found a proof,
no one has found a counterexample either,
so it seems plausible (in the light of the Four-Color Theorem)
that the conjecture is true.

Now, some planar graphs can be colored with three or even two colors;
this means that their chromatic polynomials $P_G(q)$ are strictly positive
for {\em integers}\/ $q \ge 3$ or $q \ge 2$, respectively.
Can {\em these}\/ bounds can be extended to real $q$?
That is, if $G$ is a $k$-colorable planar graph,
do we have $P_G(q) > 0$ for all real $q \ge k$?
Woodall \cite[p.~142]{Woodall_97} conjectured that the answer is yes.
For $k=4$, this is the conjecture of Birkhoff and Lewis mentioned above.
For $k=3$, however, Thomassen \cite[pp.~505--506]{Thomassen_97}
has shown that Woodall's conjecture is false:
there exist 3-colorable (and in fact 2-degenerate\footnote{
   A graph $G$ is said to be {\em $k$-degenerate}\/
   if every subgraph $H \subseteq G$ has at least one vertex
   of degree $\le k$.
   It is not difficult to prove \cite[Theorem V.1, p.~148]{Bollobas_98}
   that every $k$-degenerate graph is $(k+1)$-colorable.
})
planar graphs with real chromatic roots greater than 3.
Indeed, by combining Thomassen's construction
\cite[proof of Theorem~3.9]{Thomassen_97}
with Royle's \cite{Royle_TRI} recent construction of planar graphs
with real chromatic roots arbitrarily close to 4,
we see that there exist 3-colorable (and in fact 2-degenerate)
planar graphs with real chromatic roots arbitrarily close to 4.

In Ref.~\cite{transfer1} we showed that Woodall's conjecture is false also 
for $k=2$: there exist 2-colorable (i.e.\ bipartite) planar graphs
with real chromatic roots greater than 2.
For example, the $4_{\rm P} \times 6_{\rm F}$ square lattice
has chromatic roots at $q \approx 2.009978$ and $q \approx 2.168344$.
For the cases $8_{\rm F} \times n_{\rm F}$ and
$8_{\rm P} \times n_{\rm F}$,
we also observed numerically \cite{transfer1}  
that there are real chromatic roots tending to
$B_5 = (3+\sqrt{5})/2 \approx 2.618034$ from below as $n \to \infty$.
This led us to modify Woodall's conjecture as follows
\cite[Conjecture~7.5]{transfer1}:

\begin{conjecture}{\bf \protect\cite{transfer1}}
  \label{conj.1}
Let $G$ be a bipartite planar graph.
Then $P_G(q) > 0$ for real $q \ge B_5 = (3+\sqrt{5})/2$.
\end{conjecture}

However, the numerical results reported in the present paper
(see Tables~\ref{table_zeros_9F} and \ref{table_zeros_11F})
now show that this conjecture is also false.
Indeed, it appears that all strip graphs 
$9_{\rm F}\times n_{\rm F}$ (with {\em odd}\/ $n\geq 39$) and 
$11_{\rm F}\times n_{\rm F}$ (with {\em odd}\/ $n\geq 23$)
have {\em two}\/ real roots greater than $B_5$.  

So, not only is Conjecture~\ref{conj.1} false;
it is actually false for {\em two distinct reasons}\/.
First of all, the real roots converging to the isolated limiting point $B_5$
need not do so only from below.
(Empirically we find that, for $m=9$ and 11, these roots converge to $B_5$
 with parity $(-1)^{n+1}$.)
Secondly, real roots can converge to
the real crossing point $q_0(m) > B_5$ when one exists.\footnote{
   A similar phenomenon was exploited recently by Royle \cite{Royle_TRI}
   to provide examples of plane triangulations with real chromatic roots
   converging to 4 from below.  His graphs are $4_{\rm P} \times n_{\rm F}$
   triangular lattices with carefully chosen endgraphs
   adjoined at top and bottom
   (see \cite{Rocek_98} for a general analysis of such
    strips-with-endgraphs).
   The key fact underlying this construction is that
   $q_0({\rm tri}, 4_{\rm P}) = 4$.
}
(Empirically we find that such a crossing point exists at least for
 {\em odd}\/ widths $m$ and that the real roots converging to it exist for
 sufficiently large {\em odd}\/ lengths $n$.)

We expect both of these behaviors to persist for all larger
{\em odd}\/ widths $m$.  In particular, since we expect that
$q_0(m) \uparrow 3$ as $m \to\infty$ (see Section~\ref{sec.q0}),
we expect that $m_{\rm F} \times n_{\rm F}$ square-lattice strips
with $m,n$ both odd and large can have real chromatic roots
arbitrarily close to 3.
However, in most cases of which we are aware
\cite{transfer1,transfer2,transfer3,transfer4,Jacobsen-Salas_toroidal,%
Royle_TRI}, the convergence of real chromatic roots to $q_0(m)$
occurs only {\em from below}\/.\footnote{
  The only exceptions we know are as follows:
  \begin{itemize}
     \item[(a)] For square-lattice strips of width $m=2$
        with cyclic boundary conditions \protect\cite{Shrock_97},
        and triangular-lattice strips of width $m=2,3$
        with cyclic and toroidal boundary conditions 
        \protect\cite{transfer4,Jacobsen-Salas_toroidal},
        it appears that there exists an infinite sequence of lengths $n$
        with real zeros converging to $q_0(m)$ {\em from above}\/.
     \item[(b)] For triangular-lattice strips of width $m=4,5$
        with cyclic boundary conditions \protect\cite{transfer4}
        and width $m=5,7$ with toroidal boundary conditions
        \protect\cite{Jacobsen-Salas_toroidal},
        and for square-lattice strips of width $m=3$ with cyclic
        boundary conditions \protect\cite{transfer4},
        there are real zeros $q > q_0(m)$ for some small lengths $n$,
        but the ultimate convergence of real zeros to $q_0(m)$
        appears to be only from below.
  \end{itemize}
  There are also some cases in which the maximal real zero is
  exactly equal to $q_0(m)$.
  Finally, in some cases (e.g., triangular-lattice strips of width
  $m=4,5,6,7$ with toroidal boundary conditions 
  \protect\cite{Jacobsen-Salas_toroidal})
  there are sequences of {\em complex}\/ zeros converging to $q_0(m)$
  but apparently {\em no}\/ sequences of real zeros converging to $q_0(m)$. 

  Let us remark that the tables of real chromatic roots in
  Refs.~\protect\cite{transfer1,transfer2,transfer3,transfer4}
  show only the cases of strips $m\times (km)$ with $k=1,\ldots,10$.
  Moreover, in Ref.~\protect\cite{Jacobsen-Salas_toroidal}
  such tables are absent.
  Therefore, we have here used the transfer matrices reported in those
  papers to recompute all the chromatic polynomials for strips
  $m\times n$ with $1\le n \le 10m$
  (and in some cases for even larger lengths)
  and their corresponding chromatic roots.
}
Therefore, the following weakened version of Conjecture~\ref{conj.1}
is plausible:

\begin{conjecture}
  \label{conj.2} 
Let $G$ be a bipartite planar graph. Then $P_G(q) > 0$ for real $q\geq 3$. 
\end{conjecture}

\noindent
(Note that the honeycomb lattice, which is bipartite and planar,
 has $q_c = B_5 < 3$.)

We can also give some plausible ``physics'' considerations
that provide additional support for a slightly weakened version
of Conjecture~\ref{conj.2}.
Some two-dimensional antiferromagnetic models at zero temperature
have the remarkable property that they can be mapped onto a ``height''
(or ``interface'' or ``SOS-type'') model
(see e.g.\ \cite{Salas_98} and the references cited there).
Experience tells us that when such a representation exists,
the corresponding zero-temperature spin model
is most often critical.\footnote{
  Some exceptions are the constrained square-lattice 
  4-state antiferromagnetic Potts model \protect\cite{Burton_Henley_97}
  and the triangular-lattice antiferromagnetic spin-$s$ Ising model for large
  enough $s$ \protect\cite{Zeng_97},
  both of which appear to lie in a non-critical ordered phase
  at zero temperature.
} 
In particular, when the $q$-state zero-temperature Potts antiferromagnet
on a lattice ${\cal L}$ admits a height representation,
one expects that $q = q_c({\cal L})$.\footnote{
   Here $q_c({\cal L})$ is the value (which is conjectured to exist)
   such that for $q > q_c({\cal L})$ the antiferromagnetic Potts model
   has exponential decay of correlations uniformly at all temperatures,
   including zero temperature, while for $q = q_c({\cal L})$
   the model has a zero-temperature critical point.
}
This prediction is confirmed in all heretofore-studied cases:
3-state square-lattice \cite{Nijs_82,Kolafa_84,Burton_Henley_97,Salas_98},
3-state kagom\'e \cite{Huse_92,Kondev_96},
4-state triangular \cite{Henley_unpublished},
and 4-state on the line graph (= covering lattice) of the square lattice
\cite{Kondev_95,Kondev_96}.
Now, the height representation of the
square-lattice zero-temperature 3-state Potts antiferromagnet
(i.e., the square-lattice chromatic polynomial at $q=3$),
as presented in \cite{Salas_98},
generalizes immediately from the square lattice
to an arbitrary finite or infinite plane quadrangulation $G$.
This suggests that $q_c(\scrl) = 3$ for all infinite periodic lattices
that are plane quadrangulations.
It in turn provides some support for the following
weakened version of Conjecture~\ref{conj.2}:

\begin{conjecture}
  \label{conj.3}
Let $G$ be a plane quadrangulation. Then $P_G(q) > 0$ for real $q\geq 3$.
\end{conjecture}

Surprisingly, however, we know of a potential counterexample to 
Conjectures~\ref{conj.2} and \ref{conj.3}!
Consider the diced lattice \cite[Figure~15]{Syozi_72},
which is the dual of the kagom\'e lattice and is a plane quadrangulation.
Even though there exists a height representation for the
3-state Potts antiferromagnet on the diced lattice,
this model is in a non-critical ordered phase at zero temperature.
Indeed, Feldmann {\em et al}.\/ \cite{Feldmann_97} found
(by dualizing the numerical data of Ref.~\cite{Jensen_97}
 for the kagom\'e-lattice Potts model)
that the 3-state diced-lattice Potts model
has a critical point at $v_c = -0.8607\pm 0.0008$,
which lies at {\em nonzero}\/ temperature
in the physical antiferromagnetic region $-1 \le v \le 0$.
(By contrast, the 4-state model was found to have
its antiferromagnetic critical point
in the unphysical region at $v_c = -1.18 \pm 0.02$,
so that the 4-state antiferromagnet is
disordered for all temperatures, including zero temperature.)
We have very recently \cite{MC_diced} confirmed
this scenario by Monte Carlo simulations:
for $q=3$ we find a critical point at
$v_c = -0.860599 \pm 0.000004$,
while for $q=4$ we find a finite correlation length
uniformly in the region $-1 \le v \le 0$.
Furthermore, for $q=3$ we are able to give a mathematically rigorous proof
of the existence of a phase transition at nonzero temperature,
using a (computer-assisted) Peierls argument.
These results indicate that $3 < q_c({\rm diced}) < 4$,
contrary to our prediction that $q_c = 3$
for all (regular) plane quadrangulations.
Indeed, linear interpolation from the predictions of \cite{Feldmann_97}
would suggest $q_c({\rm diced}) \approx 3.44$.

Now, one strongly expects that finite pieces of the diced lattice
with free or cylindrical boundary conditions
will have chromatic roots tending to $q_c({\rm diced})$
in the infinite-volume limit.
What is less clear, however, is whether the roots
converging to $q_c$ will necessarily be {\em real}\/ roots.
If they are,
then a sufficiently large finite piece of the diced lattice,
with boundary conditions arranged so that the graph is a
plane quadrangulation (or at least planar and bipartite),
would provide a counterexample to Conjecture~\ref{conj.3}
(or at least to Conjecture~\ref{conj.2}).
But if they are not, then the fact that $q_c > 3$
is in no way incompatible with Conjectures~\ref{conj.2} and \ref{conj.3}.
Please note also that the roots converging to $q_c$ can be complex
for some choices of endgraphs and real for others;
this possibility was recently exploited by Royle \cite{Royle_TRI}
to provide examples of plane triangulations
(based on $4_{\rm P} \times n_{\rm F}$ triangular-lattice strips
 with suitable endgraphs \cite{Rocek_98})
having real chromatic roots converging from below to $q_c({\rm tri}) = 4$.
It would be very interesting to undertake a transfer-matrix study
of the diced-lattice chromatic polynomial along the lines of
\cite{transfer1,transfer2,transfer3,transfer4,Jacobsen-Salas_toroidal,%
Rocek_98,Royle_TRI},
in order to test whether the roots converging to $q_c$ can be real,
at least for {\em some}\/ choices of planar bipartite endgraphs.

%
%
\section*{Acknowledgments}

We wish to thank Jesper Jacobsen for many helpful conversations,
Chris Henley for correspondence concerning height representations,
Tony Guttmann for help with series analysis,
and Roman Koteck\'y and Charles Pfister for discussions
concerning the nature of complex phase boundaries.
We also thank two anonymous referees for many helpful suggestions
that strongly influenced the revised version of this paper.

J.S.\ is grateful for the kind hospitality of
the Physics Department of New York University
and the Mathematics Department of University College London,
where part of this work was done. 
Likewise, A.D.S.\ is grateful for the kind hospitality of
the M.S.M.I.\  group of the Universidad Carlos III de Madrid.
Both authors also thank the Isaac Newton Institute for Mathematical Sciences,
University of Cambridge, for hospitality during the programme on
Combinatorics and Statistical Mechanics (January--June 2008);
and A.D.S.\ thanks the Institut Henri Poincar\'e
for hospitality during the programme on
Interacting Particle Systems, Statistical Mechanics and Probability Theory
(September--December 2008).

The authors' research was supported in part
by U.S.\ National Science Foundation grants
PHY--0116590 and PHY--0424082 (A.D.S.),
and by Spanish MEC grants MTM2005--08618, MTM2008-03020, and FIS2004-03767 
(J.S.).

%
%

%
%
%
%
%
%
\clearpage
\begin{sidewaystable}
\centering
\begin{tabular}{r|rrrrrrrrrrrrrrrr}
\hline\hline\\[-4mm]
$m$& $b^{\rm F}_0$   & $b^{\rm F}_1$   & $b^{\rm F}_2$   & $b^{\rm F}_3$& 
     $b^{\rm F}_4$   & $b^{\rm F}_5$   & $b^{\rm F}_6$   & $b^{\rm F}_7$& 
     $b^{\rm F}_8$   & $b^{\rm F}_9$   & $b^{\rm F}_{10}$& $b^{\rm F}_{11}$& 
     $b^{\rm F}_{12}$& $b^{\rm F}_{13}$& $b^{\rm F}_{14}$& $b^{\rm F}_{15}$ 
\\[2mm]
\hline
$1$ & $1$ & $1$ & $0$ & $0$ & $0$ & $0$ & $0$ & \multicolumn{1}{|r}{$0$} & $0$ & $0$ & $0$ & $0$ & $0$ & $0$ & $0$ & $0$\\
\cline{9-10}
$2$ & $1$ & $3$ & $3$ & $0$ & $0$ & $0$ & $0$ & $0$ & $0$ & \multicolumn{1}{|r}{$0$} & $0$ & $0$ & $0$ & $0$ & $0$ & $0$\\
\cline{11-12}
$3$ & $1$ & $5$ & $10$ & $8$ & $1$ & $-1$ & $1$ & $-2$ & $6$ & $-16$ & $35$ & \multicolumn{1}{|r}{$-61$} & $69$ & $42$ & $-583$ & $2371$\\
\cline{13-14}
$4$ & $1$ & $7$ & $21$ & $32$ & $23$ & $3$ & $-2$ & $-1$ & $6$ & $-14$ & $28$ & $-54$ & $102$ & \multicolumn{1}{|r}{$-172$} & $145$ & $695$\\
\cline{15-16}
$5$ & $1$ & $9$ & $36$ & $80$ & $102$ & $66$ & $10$ & $-9$ & $6$ & $-6$ & $14$ & $-38$ & $97$ & $-218$ & $361$ & \multicolumn{1}{|r}{$-75$}\\
\cline{17-17}
$6$ & $1$ & $11$ & $55$ & $160$ & $290$ & $322$ & $192$ & $26$ & $-19$ & $2$ & $15$ & $-35$ & $77$ & $-160$ & $241$ & $-5$\\
$7$ & $1$ & $13$ & $78$ & $280$ & $655$ & $1017$ & $1011$ & $556$ & $75$ & $-59$ & $21$ & $-6$ & $32$ & $-84$ & $103$ & $107$\\
$8$ & $1$ & $15$ & $105$ & $448$ & $1281$ & $2541$ & $3486$ & $3153$ & $1617$ & $201$ & $-151$ & $22$ & $64$ & $-73$ & $24$ & $132$\\
$9$ & $1$ & $17$ & $136$ & $672$ & $2268$ & $5460$ & $9492$ & $11741$ & $9785$ & $4697$ & $550$ & $-436$ & $96$ & $103$ & $-97$ & $67$\\
$10$ & $1$ & $19$ & $171$ & $960$ & $3732$ & $10548$ & $22128$ & $34468$ & $39006$ & $30223$ & $13652$ & $1461$ & $-1190$ & $229$ & $316$ & $-221$\\
$11$ & $1$ & $21$ & $210$ & $1320$ & $5805$ & $18819$ & $46149$ & $86346$ & $122436$ & $128142$ & $92975$ & $39640$ & $3874$ & $-3318$ & $650$ & $881$\\
$12$ & $1$ & $23$ & $253$ & $1760$ & $8635$ & $31559$ & $88462$ & $192787$ & $327120$ & $427276$ & $417066$ & $284954$ & $115022$ & $10141$ & $-9225$ & $1865$\\
\hline\hline
\end{tabular}
\vspace{1cm}
\caption{
   Coefficients $b^{\rm F}_k(m)$ of the large-$q$ expansion of the
   dominant eigenvalue $\lambda_{\star}^{\rm F}$
   for free boundary conditions. For each $1\leq m \leq 12$, we include
   all coefficients $b^{\rm F}_k(m)$ up to $k=15$.
   For the whole data set up to $k=40$,
   see the {\sc Mathematica} file {\tt data\_FREE.m}
   included in the on-line version of the paper at arXiv.org.
   Those data points below the stair-case-like line satisfy  
   $m\geq m_{\rm min}^{\rm F}(k)$ [cf.\ \protect\reff{def_m_min_F}].
}
\label{table_coef_b_F}
\end{sidewaystable}

%
%
\clearpage
\begin{sidewaystable}
\centering
\begin{tabular}{r|rrrrrrrrrrrrrrr}
\hline\hline\\[-4mm]
$m$& $ c^{\rm F}_1$     & $2c^{\rm F}_2$     &  $3c^{\rm F}_3$    & 
     $4c^{\rm F}_4$     & $5c^{\rm F}_5$     & $6c^{\rm F}_6$     &  
     $7c^{\rm F}_7$     & $8c^{\rm F}_8$     & $9c^{\rm F}_9$     & 
     $10c^{\rm F}_{10}$ & $11c^{\rm F}_{11}$ & $12c^{\rm F}_{12}$ & 
     $13c^{\rm F}_{13}$ & $14c^{\rm F}_{14}$ & $15c^{\rm F}_{15}$ \\[2mm]
\hline
$1$ & $-1$ & $-1$ & $-1$ & $-1$ & $-1$ &  $-1$ & \multicolumn{1}{|r}{$-1$} & $-1$ & $-1$ & $-1$ & $-1$ & $-1$ & $-1$ & $-1$ & $-1$\\
\cline{8-9}
$2$ & $-3$ & $-3$ & $0$ & $9$ & $27$ & $54$ & $81$ & $81$ & \multicolumn{1}{|r}{$0$} & $-243$ & $-729$ & $-1458$ & $-2187$ & $-2187$ & $0$\\
\cline{10-11}
$3$ & $-5$ & $-5$ & $1$ & $19$ & $55$ & $109$ & $170$ & $243$ & $496$ & $1685$ & \multicolumn{1}{|r}{$5957$} & $17449$ & $39463$ & $57430$ & $-21119$\\
\cline{12-13}
$4$ & $-7$ & $-7$ & $2$ & $29$ & $83$ & $164$ & $259$ & $405$ & $1001$ & $3743$ & $13688$ & $42548$ & \multicolumn{1}{|r}{$111078$} & $240401$ & $393557$\\
\cline{14-15}
$5$ & $-9$ & $-9$ & $3$ & $39$ & $111$ & $219$ & $348$ & $567$ & $1506$ & $5801$ & $21430$ & $67839$ & $184617$ & $437890$ & \multicolumn{1}{|r}{$898368$}\\
\cline{16-16}
$6$ & $-11$ & $-11$ & $4$ & $49$ & $139$ & $274$ & $437$ & $729$ & $2011$ & $7859$ & $29172$ & $93130$ & $258169$ & $635645$ & $1406389$\\
$7$ & $-13$ & $-13$ & $5$ & $59$ & $167$ & $329$ & $526$ & $891$ & $2516$ & $9917$ & $36914$ & $118421$ & $331721$ & $833400$ & $1914425$\\
$8$ & $-15$ & $-15$ & $6$ & $69$ & $195$ & $384$ & $615$ & $1053$ & $3021$ & $11975$ & $44656$ & $143712$ & $405273$ & $1031155$ & $2422461$\\
$9$ & $-17$ & $-17$ & $7$ & $79$ & $223$ & $439$ & $704$ & $1215$ & $3526$ & $14033$ & $52398$ & $169003$ & $478825$ & $1228910$ & $2930497$\\
$10$ & $-19$ & $-19$ & $8$ & $89$ & $251$ & $494$ & $793$ & $1377$ & $4031$ & $16091$ & $60140$ & $194294$ & $552377$ & $1426665$ & $3438533$\\
$11$ & $-21$ & $-21$ & $9$ & $99$ & $279$ & $549$ & $882$ & $1539$ & $4536$ & $18149$ & $67882$ & $219585$ & $625929$ & $1624420$ & $3946569$\\
$12$ & $-23$ & $-23$ & $10$ & $109$ & $307$ & $604$ & $971$ & $1701$ & $5041$ & $20207$ & $75624$ & $244876$ & $699481$ & $1822175$ & $4454605$\\
\hline\hline
\end{tabular}
\vspace{1cm}
\caption{
   Coefficients $k c^{\rm F}_k(m)$ of the large-$q$ expansion of 
   $\log(q^{-m} \lambda_{\star}^{\rm F})$ where $\lambda_{\star}^{\rm F}$ is
   the dominant eigenvalue for free boundary conditions. 
   For each $1\leq m \leq 12$, we include
   all coefficients $c^{\rm F}_k(m)$ up to $k=15$.
   For the whole data set up to $k=40$,
   see the {\sc Mathematica} file {\tt data\_FREE.m}
   included in the on-line version of the paper at arXiv.org.
   Those data points below the stair-case-like line satisfy
   $m\geq m_{\rm min}^{\rm F}(k)$ [cf.\ \protect\reff{def_m_min_F}].
}
\label{table_coef_c_F}
\end{sidewaystable}

%
%
\clearpage
\begin{sidewaystable}
\centering
\begin{tabular}{r|rrrrrrrrrrrrrrrr}
\hline\hline\\[-4mm]
$m$& $\Delta^{\rm F}_6$    & $\Delta^{\rm F}_7$& $\Delta^{\rm F}_8$& 
     $\Delta^{\rm F}_9$    & $\Delta^{\rm F}_{10}$& 
     $\Delta^{\rm F}_{11}$ & $\Delta^{\rm F}_{12}$& $\Delta^{\rm F}_{13}$& 
     $\Delta^{\rm F}_{14}$ & $\Delta^{\rm F}_{15}$ &
     $\Delta^{\rm F}_{16}$ & $\Delta^{\rm F}_{17}$ & $\Delta^{\rm F}_{18}$ & 
     $\Delta^{\rm F}_{19}$ & $\Delta^{\rm F}_{20}$ & $\Delta^{\rm F}_{21}$ 
\\[2mm]
\hline\\[-4mm]
$1$ & $0$ & \multicolumn{1}{|r}{$1$} & $10$ & $57$ & $243$ & $867$ & $2777$ & $8430$ & $\frac{50447}{2}$ & $75586$ & $224939$ & $650699$ & $\frac{3583917}{2}$ & $4637373$ & $\frac{22526941}{2}$ & $\frac{78375472}{3}$\\[2mm]
\cline{3-4}
$2$ & $0$ & $0$ & $0$ & \multicolumn{1}{|r}{$1$} & $13$ & $97$ & $548$ & $2604$ & $10942$ & $41717$ & $146333$ & $476291$ & $\frac{2896101}{2}$ & $4142266$ & $\frac{22531163}{2}$ & $29609731$\\[2mm]
\cline{5-6}
$3$ & $0$ & $0$ & $0$ & $0$ & $0$ & \multicolumn{1}{|r}{$1$} & $16$ & $150$ & $1075$ & $6440$ & $33513$ & $154727$ & $643296$ & $2438443$ & $8522559$ & $27772788$\\[2mm]
\cline{7-8}
$4$ & $0$ & $0$ & $0$ & $0$ & $0$ & $0$ & $0$ & \multicolumn{1}{|r}{$1$} & $19$ & $216$ & $1883$ & $13595$ & $84238$ & $458660$ & $2235421$ & $9900665$\\[2mm]
\cline{9-10}
$5$ & $0$ & $0$ & $0$ & $0$ & $0$ & $0$ & $0$ & $0$ & $0$ & \multicolumn{1}{|r}{$1$} & $22$ & $295$ & $3031$ & $25574$ & $183804$ & $1155646$\\[2mm]
\cline{11-12}
$6$ & $0$ & $0$ & $0$ & $0$ & $0$ & $0$ & $0$ & $0$ & $0$ & $0$ & $0$ & \multicolumn{1}{|r}{$1$} & $25$ & $387$ & $4578$ & $44167$\\[2mm]
\cline{13-14}
$7$ & $0$ & $0$ & $0$ & $0$ & $0$ & $0$ & $0$ & $0$ & $0$ & $0$ & $0$ & $0$ & $0$ & \multicolumn{1}{|r}{$1$} & $28$ & $492$\\[2mm]
\cline{15-16}
$8$ & $0$ & $0$ & $0$ & $0$ & $0$ & $0$ & $0$ & $0$ & $0$ & $0$ & $0$ & $0$ & $0$ & $0$ & $0$ & \multicolumn{1}{|r}{$1$}\\[2mm]
\cline{17-17}
$9$ & $0$ & $0$ & $0$ & $0$ & $0$ & $0$ & $0$ & $0$ & $0$ & $0$ & $0$ & $0$ & $0$ & $0$ & $0$ & $0$\\[2mm]
$10$ & $0$ & $0$ & $0$ & $0$ & $0$ & $0$ & $0$ & $0$ & $0$ & $0$ & $0$ & $0$ & $0$ & $0$ & $0$ & $0$\\[2mm]
$11$ & $0$ & $0$ & $0$ & $0$ & $0$ & $0$ & $0$ & $0$ & $0$ & $0$ & $0$ & $0$ & $0$ & $0$ & $0$ & $0$\\[2mm]
$12$ & $0$ & $0$ & $0$ & $0$ & $0$ & $0$ & $0$ & $0$ & $0$ & $0$ & $0$ & $0$ & $0$ & $0$ & $0$ & $0$\\[2mm]
\hline\hline
\end{tabular}
\vspace{1cm}
\caption{
   Coefficients $\Delta^{\rm F}_k(m)$ 
   for free boundary conditions [cf.\ \protect\reff{def_Delta}]
   for $6\leq k \leq 21$ and $1\leq m\leq 12$. 
   For $0\leq k\leq 6$ and $1\leq m \leq 12$, the coefficients 
   $\Delta_k^{\rm F}(m)$ vanish. 
   For the whole data set up to $k=33$,
   see the {\sc Mathematica} file {\tt data\_FREE.m}
   included in the on-line version of the paper at arXiv.org.
   Those data points below the stair-case-like line satisfy
   $m\geq m_{\rm min}^{\rm F}(k)$ [cf.\ \protect\reff{def_m_min_F}]
   and are therefore zero.
}
\label{table_diff_coef_c_F}
\end{sidewaystable}

%
%
%
%
\clearpage
\thispagestyle{empty}
\begin{sidewaystable}
\hspace*{-1cm}
\small
\begin{tabular}{r|rrrrrrrrrrrrrrrr}
\hline\hline\\[-4mm]
$m$& $b^{\rm P}_0$   & $b^{\rm P}_1$   & $b^{\rm P}_2$   & $b^{\rm P}_3$&
     $b^{\rm P}_4$   & $b^{\rm P}_5$   & $b^{\rm P}_6$   & $b^{\rm P}_7$&
     $b^{\rm P}_8$   & $b^{\rm P}_9$   & $b^{\rm P}_{10}$& $b^{\rm P}_{11}$&
     $b^{\rm P}_{12}$& $b^{\rm P}_{13}$& $b^{\rm P}_{14}$& $b^{\rm P}_{15}$
\\[2mm]
\hline
$1$ & $0$ & $0$ & $0$ & $0$ & $0$ & $0$ & $0$ & $0$ & $0$ & $0$ & $0$ & $0$ & $0$ & $0$ & $0$ & $0$\\
\cline{2-2}
$2$ & $1$ & \multicolumn{1}{|r}{$3$} & $3$ & $0$ & $0$ & $0$ & $0$ & $0$ & $0$ & $0$ & $0$ & $0$ & $0$ & $0$ & $0$ & $0$\\
\cline{3-3}
$3$ & $1$ & $6$ & \multicolumn{1}{|r}{$14$} & $13$ & $0$ & $0$ & $0$ & $0$ & $0$ & $0$ & $0$ & $0$ & $0$ & $0$ & $0$ & $0$\\
\cline{4-4}
$4$ & $1$ & $8$ & $28$ & \multicolumn{1}{|r}{$51$} & $45$ & $2$ & $-8$ & $10$ & $2$ & $2$ & $-192$ & $980$ & $-2942$ & $6164$ & $-7566$ & $-9986$\\
\cline{5-5}
$5$ & $1$ & $10$ & $45$ & $115$ & \multicolumn{1}{|r}{$174$} & $141$ & $20$ & $-45$ & $20$ & $50$ & $15$ & $-400$ & $670$ & $930$ & $-8155$ & $27400$\\
\cline{6-6}
$6$ & $1$ & $12$ & $66$ & $214$ & $441$ & \multicolumn{1}{|r}{$575$} & $428$ & $81$ & $-119$ & $-45$ & $210$ & $35$ & $-396$ & $-122$ & $1075$ & $2106$\\
\cline{7-7}
$7$ & $1$ & $14$ & $91$ & $357$ & $924$ & $1617$ & \multicolumn{1}{|r}{$1868$} & $1275$ & $273$ & $-287$ & $-210$ & $294$ & $532$ & $-679$ & $-539$ & $-609$\\
\cline{8-8}
$8$ & $1$ & $16$ & $120$ & $552$ & $1716$ & $3744$ & $5748$ & \multicolumn{1}{|r}{$5991$} & $3777$ & $812$ & $-636$ & $-634$ & $280$ & $1096$ & $724$ & $-3022$\\
\cline{9-9}
$9$ & $1$ & $18$ & $153$ & $807$ & $2925$ & $7623$ & $14505$ & $19962$ & \multicolumn{1}{|r}{$19034$} & $11140$ & $2313$ & $-1497$ & $-1374$ & $-360$ & $1662$ & $4377$\\
\cline{10-10}
$10$ & $1$ & $20$ & $190$ & $1130$ & $4675$ & $14144$ & $32005$ & $54340$ & $68085$ & \multicolumn{1}{|r}{$59999$} & $32790$ & $6375$ & $-3660$ & $-2760$ & $-1792$ & $215$\\
\cline{11-11}
$11$ & $1$ & $22$ & $231$ & $1529$ & $7106$ & $24453$ & $64009$ & $128777$ & $198330$ & $228866$ & \multicolumn{1}{|r}{$187901$} & $96306$ & $17336$ & $-9537$ & $-5093$ & $-4367$\\
\cline{12-12}
$12$ & $1$ & $24$ & $276$ & $2012$ & $10374$ & $39984$ & $118702$ & $275460$ & $501213$ & $708868$ & $760164$ & \multicolumn{1}{|r}{$585131$} & $282358$ & $46638$ & $-25822$ & $-9177$\\
\cline{13-13}
$13$ & $1$ & $26$ & $325$ & $2587$ & $14651$ & $62491$ & $207285$ & $544232$ & $1139450$ & $1899300$ & $2490423$ & $2499471$ & \multicolumn{1}{|r}{$1813122$} & $826359$ & $124540$ & $-71760$\\
\hline\hline
\end{tabular}
\vspace{1cm}
\caption{
   Coefficients $b^{\rm P}_k(m)$ of the large-$q$ expansion of the
   dominant eigenvalue $\lambda_{\star}^{\rm P}$
   for cylindrical boundary conditions.
   For each $1\leq m \leq 13$, we include
   all coefficients $b^{\rm F}_k(m)$ up to $k=15$.
   For the whole data set up to $k=40$,
   see the {\sc Mathematica} file {\tt data\_CYL.m}
   included in the on-line version of the paper at arXiv.org.
   Those data points below the stair-case-like line satisfy
   $m\geq m_{\rm min}^{\rm P}(k)=k+2$.
}
\label{table_coef_b_P}
\end{sidewaystable}

%
%
\clearpage
\thispagestyle{empty}
\begin{sidewaystable}
\centering
\begin{tabular}{r|rrrrrrrrrrrrrrr}
\hline\hline\\[-4mm]
$m$& $ c^{\rm P}_1$     & $2c^{\rm P}_2$     &  $3c^{\rm P}_3$    &
     $4c^{\rm P}_4$     & $5c^{\rm P}_5$     & $6c^{\rm P}_6$     &
     $7c^{\rm P}_7$     & $8c^{\rm P}_8$     & $9c^{\rm P}_9$     &
     $10c^{\rm P}_{10}$ & $11c^{\rm P}_{11}$ & $12c^{\rm P}_{12}$ &
     $13c^{\rm P}_{13}$ & $14c^{\rm P}_{14}$ & $15c^{\rm P}_{15}$ \\[2mm] 
\hline 
$2$ & $-3$ & $-3$ & $0$ & $9$ & $27$ & $54$ & $81$ & $81$ & $0$ & $-243$ & $-729$ & $-1458$ & $-2187$ & $-2187$ & $0$\\
\cline{2-2}
$3$ & $-6$ & \multicolumn{1}{|r}{$-8$} & $-3$ & $16$ & $34$ & $-59$ & $-622$ & $-2464$ & $-6843$ & $-14648$ & $-24118$ & $-28595$ & $-24342$ & $-59256$ & $-386483$\\
\cline{3-3}
$4$ & $-8$ & $-8$ & \multicolumn{1}{|r}{$7$} & $52$ & $162$ & $493$ & $1595$ & $4764$ & $11383$ & $15722$ & $-31303$ & $-360503$ & $-1801119$ & $-6704573$ & $-19871718$\\
\cline{4-4}
$5$ & $-10$ & $-10$ & $5$ & \multicolumn{1}{|r}{$46$} & $120$ & $155$ & $-325$ & $-2914$ & $-10984$ & $-27430$ & $-37245$ & $62971$ & $649587$ & $2559421$ & $6057120$\\
\cline{5-5}
$6$ & $-12$ & $-12$ & $6$ & $60$ & \multicolumn{1}{|r}{$173$} & $360$ & $765$ & $2644$ & $11922$ & $49003$ & $170840$ & $505212$ & $1251225$ & $2440055$ & $2788766$\\
\cline{6-6}
$7$ & $-14$ & $-14$ & $7$ & $70$ & $196$ & \multicolumn{1}{|r}{$379$} & $581$ & $742$ & $385$ & $-4074$ & $-32123$ & $-167537$ & $-694721$ & $-2340051$ & $-6229223$\\
\cline{7-7}
$8$ & $-16$ & $-16$ & $8$ & $80$ & $224$ & $440$ & \multicolumn{1}{|r}{$719$} & $1352$ & $4652$ & $21864$ & $96762$ & $384848$ & $1421716$ & $4975913$ & $16390643$\\
\cline{8-8}
$9$ & $-18$ & $-18$ & $9$ & $90$ & $252$ & $495$ & $801$ & \multicolumn{1}{|r}{$1450$} & $4473$ & $17622$ & $61032$ & $166635$ & $307107$ & $-34941$ & $-3730356$\\
\cline{9-9}
$10$ & $-20$ & $-20$ & $10$ & $100$ & $280$ & $550$ & $890$ & $1620$ & \multicolumn{1}{|r}{$5059$} & $20670$ & $78685$ & $266050$ & $836335$ & $2622628$ & $8719855$\\
\cline{10-10}
$11$ & $-22$ & $-22$ & $11$ & $110$ & $308$ & $605$ & $979$ & $1782$ & $5555$ & \multicolumn{1}{|r}{$22628$} & $85052$ & $276485$ & $789910$ & $2016223$ & $4478441$\\
\cline{11-11}
$12$ & $-24$ & $-24$ & $12$ & $120$ & $336$ & $660$ & $1068$ & $1944$ & $6060$ & $24696$ & \multicolumn{1}{|r}{$92915$} & $303624$ & $884886$ & $2400080$ & $6337977$\\
\cline{12-12}
$13$ & $-26$ & $-26$ & $13$ & $130$ & $364$ & $715$ & $1157$ & $2106$ & $6565$ & $26754$ & $100646$ & \multicolumn{1}{|r}{$328771$} & $956020$ & $2567903$ & $6567418$\\
\hline\hline
\end{tabular}
\vspace{1cm}
\caption{
   Coefficients $k c^{\rm P}_k(m)$ of the large-$q$ expansion of 
   $\log(q^{-m} \lambda_{\star}^{\rm P})$ where $\lambda_{\star}^{\rm P}$ is
   the dominant eigenvalue for cylindrical boundary conditions. 
   For each $2\leq m \leq 13$, we include
   all coefficients $c^{\rm P}_k(m)$ up to $k=15$.
   For the whole data set up to $k=40$,
   see the {\sc Mathematica} file {\tt data\_CYL.m}
   included in the on-line version of the paper at arXiv.org.
   Those data points below the stair-case-like line satisfy
   $m\geq m_{\rm min}^{\rm P}(k)=k+2$.
}
\label{table_coef_c_P}
\end{sidewaystable}

\clearpage
%
%
\thispagestyle{empty}
\begin{sidewaystable}
\hspace*{-1cm}
\small
\begin{tabular}{r|rrrrrrrrrrrrrrrr}
\hline\hline\\[-4mm]
$m$& $\tilde{b}^{\rm P}_0$   & $\tilde{b}^{\rm P}_1$   
   & $\tilde{b}^{\rm P}_2$   & $\tilde{b}^{\rm P}_3$ 
   & $\tilde{b}^{\rm P}_4$   & $\tilde{b}^{\rm P}_5$   
   & $\tilde{b}^{\rm P}_6$   & $\tilde{b}^{\rm P}_7$ 
   & $\tilde{b}^{\rm P}_8$   & $\tilde{b}^{\rm P}_9$   
   & $\tilde{b}^{\rm P}_{10}$& $\tilde{b}^{\rm P}_{11}$ 
   & $\tilde{b}^{\rm P}_{12}$& $\tilde{b}^{\rm P}_{13}$
   & $\tilde{b}^{\rm P}_{14}$& $\tilde{b}^{\rm P}_{15}$
\\[2mm]
\hline
$0$ & $1$ & \multicolumn{1}{|r}{$0$} & $0$ & $0$ & $0$ & $0$ & $0$ & $0$ & $0$ & $0$ & $0$ & $0$ & $0$ & $0$ & $0$ & $0$\\
\cline{3-3}
$1$ & $1$ & $2$ & \multicolumn{1}{|r}{$0$} & $0$ & $0$ & $0$ & $0$ & $0$ & $0$ & $0$ & $0$ & $0$ & $0$ & $0$ & $0$ & $0$\\
\cline{4-4}
$2$ & $1$ & $4$ & $6$ & \multicolumn{1}{|r}{$0$} & $0$ & $0$ & $0$ & $0$ & $0$ & $0$ & $0$ & $0$ & $0$ & $0$ & $0$ & $0$\\
\cline{5-5}
$3$ & $1$ & $6$ & $15$ & $17$ & \multicolumn{1}{|r}{$0$} & $0$ & $0$ & $0$ & $0$ & $0$ & $0$ & $0$ & $0$ & $0$ & $0$ & $0$\\
\cline{6-6}
$4$ & $1$ & $8$ & $28$ & $52$ & $50$ & \multicolumn{1}{|r}{$2$} & $-8$ & $10$ & $2$ & $2$ & $-192$ & $980$ & $-2942$ & $6164$ & $-7566$ & $-9986$\\
\cline{7-7}
$5$ & $1$ & $10$ & $45$ & $115$ & $175$ & $147$ & \multicolumn{1}{|r}{$20$} & $-45$ & $20$ & $50$ & $15$ & $-400$ & $670$ & $930$ & $-8155$ & $27400$\\
\cline{8-8}
$6$ & $1$ & $12$ & $66$ & $214$ & $441$ & $576$ & $435$ & \multicolumn{1}{|r}{$81$} & $-119$ & $-45$ & $210$ & $35$ & $-396$ & $-122$ & $1075$ & $2106$\\
\cline{9-9}
$7$ & $1$ & $14$ & $91$ & $357$ & $924$ & $1617$ & $1869$ & $1283$ & \multicolumn{1}{|r}{$273$} & $-287$ & $-210$ & $294$ & $532$ & $-679$ & $-539$ & $-609$\\
\cline{10-10}
$8$ & $1$ & $16$ & $120$ & $552$ & $1716$ & $3744$ & $5748$ & $5992$ & $3786$ & \multicolumn{1}{|r}{$812$} & $-636$ & $-634$ & $280$ & $1096$ & $724$ & $-3022$\\
\cline{11-11}
$9$ & $1$ & $18$ & $153$ & $807$ & $2925$ & $7623$ & $14505$ & $19962$ & $19035$ & $11150$ & \multicolumn{1}{|r}{$2313$} & $-1497$ & $-1374$ & $-360$ & $1662$ & $4377$\\
\cline{12-12}
$10$ & $1$ & $20$ & $190$ & $1130$ & $4675$ & $14144$ & $32005$ & $54340$ & $68085$ & $60000$ & $32801$ & \multicolumn{1}{|r}{$6375$} & $-3660$ & $-2760$ & $-1792$ & $215$\\
\cline{13-13}
$11$ & $1$ & $22$ & $231$ & $1529$ & $7106$ & $24453$ & $64009$ & $128777$ & $198330$ & $228866$ & $187902$ & $96318$ & \multicolumn{1}{|r}{$17336$} & $-9537$ & $-5093$ & $-4367$\\
\cline{14-14}
$12$ & $1$ & $24$ & $276$ & $2012$ & $10374$ & $39984$ & $118702$ & $275460$ & $501213$ & $708868$ & $760164$ & $585132$ & $282371$ & \multicolumn{1}{|r}{$46638$} & $-25822$ & $-9177$\\
\cline{15-15}
$13$ & $1$ & $26$ & $325$ & $2587$ & $14651$ & $62491$ & $207285$ & $544232$ & $1139450$ & $1899300$ & $2490423$ & $2499471$ & $1813123$ & $826373$ & \multicolumn{1}{|r}{$124540$} & $-71760$\\
\hline\hline
\end{tabular}
\vspace{1cm}
\caption{
   Coefficients $\tilde{b}^{\rm P}_k(m)$ of the large-$q$ expansion of the
   dominant eigenvalue $\lambda_{\star}^{\rm P}$
   for cylindrical boundary conditions.
   For each $1\leq m \leq 13$, we include
   all coefficients $\tilde{b}^{\rm F}_k(m)$ up to $k=15$.
   For the whole data set up to $k=40$,
   see the {\sc Mathematica} file {\tt data\_CYL.m}
   included in the on-line version of the paper at arXiv.org.
   Those data points below the stair-case-like line satisfy
   $m\geq k$.
}
\label{table_coef_btilde_P}
\end{sidewaystable}

%
%
\clearpage
\thispagestyle{empty}
\begin{sidewaystable}
\hspace*{-1cm}
\begin{tabular}{r|rrrrrrrrrrrrrrr}
\hline\hline\\[-4mm]
$m$& $ \tilde{c}^{\rm P}_1$     & $2\tilde{c}^{\rm P}_2$     &  
     $3\tilde{c}^{\rm P}_3$    &
     $4\tilde{c}^{\rm P}_4$     & $5\tilde{c}^{\rm P}_5$     & 
     $6\tilde{c}^{\rm P}_6$     &
     $7\tilde{c}^{\rm P}_7$     & $8\tilde{c}^{\rm P}_8$     & 
     $9\tilde{c}^{\rm P}_9$     &
     $10\tilde{c}^{\rm P}_{10}$ & $11\tilde{c}^{\rm P}_{11}$ & 
     $12\tilde{c}^{\rm P}_{12}$ &
     $13\tilde{c}^{\rm P}_{13}$ & $14\tilde{c}^{\rm P}_{14}$ & 
     $15\tilde{c}^{\rm P}_{15}$ \\[2mm] 
\hline 
$2$ & $-4$ & $-4$ & \multicolumn{1}{|r}{$8$} & $56$ & $176$ & $368$ & $416$ & $-544$ & $-4672$ & $-15424$ & $-33664$ & $-42112$ & $33536$ & $386816$ & $1346048$\\
\cline{4-4}
$3$ & $-6$ & $-6$ & $3$ & \multicolumn{1}{|r}{$6$} & $-111$ & $-705$ & $-2463$ & $-6090$ & $-11580$ & $-20001$ & $-49836$ & $-195861$ & $-767643$ & $-2515155$ & $-6905922$\\
\cline{5-5}
$4$ & $-8$ & $-8$ & $4$ & $40$ & \multicolumn{1}{|r}{$182$} & $880$ & $3674$ & $12096$ & $30586$ & $51852$ & $-5024$ & $-534920$ & $-3020324$ & $-12012568$ & $-38506486$\\
\cline{6-6}
$5$ & $-10$ & $-10$ & $5$ & $50$ & $140$ & \multicolumn{1}{|r}{$125$} & $-1130$ & $-8030$ & $-32125$ & $-92520$ & $-188000$ & $-169195$ & $584925$ & $3408080$ & $8106710$\\
\cline{7-7}
$6$ & $-12$ & $-12$ & $6$ & $60$ & $168$ & $330$ & \multicolumn{1}{|r}{$807$} & $4148$ & $23001$ & $101778$ & $360304$ & $1038354$ & $2394614$ & $3965607$ & $1844286$\\
\cline{8-8}
$7$ & $-14$ & $-14$ & $7$ & $70$ & $196$ & $385$ & $623$ & \multicolumn{1}{|r}{$686$} & $-2198$ & $-25704$ & $-148624$ & $-642719$ & $-2239797$ & $-6350183$ & $-13989773$\\
\cline{9-9}
$8$ & $-16$ & $-16$ & $8$ & $80$ & $224$ & $440$ & $712$ & $1296$ & \multicolumn{1}{|r}{$4724$} & $26024$ & $135790$ & $619232$ & $2490264$ & $8897768$ & $28169498$\\
\cline{10-10}
$9$ & $-18$ & $-18$ & $9$ & $90$ & $252$ & $495$ & $801$ & $1458$ & $4545$ & \multicolumn{1}{|r}{$17532$} & $54663$ & $100467$ & $-131058$ & $-2239899$ & $-12715191$\\
\cline{11-11}
$10$ & $-20$ & $-20$ & $10$ & $100$ & $280$ & $550$ & $890$ & $1620$ & $5050$ & $20580$ & \multicolumn{1}{|r}{$78795$} & $275410$ & $943000$ & $3394462$ & $12965290$\\
\cline{12-12}
$11$ & $-22$ & $-22$ & $11$ & $110$ & $308$ & $605$ & $979$ & $1782$ & $5555$ & $22638$ & $85162$ & \multicolumn{1}{|r}{$276353$} & $776611$ & $1851289$ & $3184016$\\
\cline{13-13}
$12$ & $-24$ & $-24$ & $12$ & $120$ & $336$ & $660$ & $1068$ & $1944$ & $6060$ & $24696$ & $92904$ & $303492$ & \multicolumn{1}{|r}{$885042$} & $2418448$ & $6584247$\\
\cline{14-14}
$13$ & $-26$ & $-26$ & $13$ & $130$ & $364$ & $715$ & $1157$ & $2106$ & $6565$ & $26754$ & $100646$ & $328783$ & $956176$ & \multicolumn{1}{|r}{$2567721$} & $6542653$\\
\hline\hline
\end{tabular}
\vspace{1cm}
\caption{
   Coefficients $k \tilde{c}^{\rm P}_k(m)$ defined in 
   \protect\reff{def_ctildekm}. 
   For each $2\leq m \leq 13$, we include
   all coefficients $c^{\rm P}_k(m)$ up to $k=15$.
   For the whole data set up to $k=40$,
   see the {\sc Mathematica} file {\tt data\_CYL.m}
   included in the on-line version of the paper at arXiv.org.
   Those data points below the stair-case-like line satisfy
   $m\geq \widetilde{m}_{\rm min}^{\rm P}(k)=\max(k,2)$.
}
\label{table_coef_ctilde_P}
\end{sidewaystable}

%
%
\clearpage
\thispagestyle{empty}
\begin{sidewaystable}
\centering
\begin{tabular}{r|rrrrrrrrrrrrrrrr}
\hline\hline\\[-4mm]
$m$& $\Delta^{\rm P}_2$& $\Delta^{\rm P}_3$& 
     $\Delta^{\rm P}_4$    & $\Delta^{\rm P}_5$& 
     $\Delta^{\rm P}_6$    & $\Delta^{\rm P}_7$& $\Delta^{\rm P}_8$& 
     $\Delta^{\rm P}_9$    & $\Delta^{\rm P}_{10}$ &
     $\Delta^{\rm P}_{11}$ & $\Delta^{\rm P}_{12}$ & $\Delta^{\rm P}_{13}$ & 
     $\Delta^{\rm P}_{14}$ & $\Delta^{\rm P}_{15}$ & $\Delta^{\rm P}_{16}$  
\\[2mm]
\hline\\[-4mm]
$2$ & $0$ & \multicolumn{1}{|r}{$2$} & $9$ & $24$ & $43$ & $34$ & $-\frac{217}{2}$ & $-\frac{1894}{3}$ & $-1954$ & $-4468$ & $-\frac{15449}{2}$ & $-8736$ & $-621$ & $\frac{109992}{5}$ & $\frac{133695}{4}$\\[2mm]
\cline{3-3}
$3$ & $0$ & $0$ & \multicolumn{1}{|r}{$-6$} & $-39$ & $-145$ & $-390$ & $-822$ & $-1455$ & $-\frac{5235}{2}$ & $-6642$ & $-\frac{45289}{2}$ & $-76023$ & $-222030$ & $-562002$ & $-1284432$\\[2mm]
\cline{4-4}
$4$ & $0$ & $0$ & $0$ & \multicolumn{1}{|r}{$14$} & $110$ & $474$ & $1431$ & $3174$ & $4362$ & $-3272$ & $-53007$ & $-254964$ & $-914542$ & $-\frac{8107726}{3}$ & $-\frac{12999385}{2}$\\[2mm]
\cline{5-5}
$5$ & $0$ & $0$ & $0$ & $0$ & \multicolumn{1}{|r}{$-25$} & $-225$ & $-1105$ & $-3850$ & $-10281$ & $-20610$ & $-\frac{49275}{2}$ & $16705$ & $\frac{345615}{2}$ & $371102$ & $-\frac{916875}{2}$\\[2mm]
\cline{6-6}
$6$ & $0$ & $0$ & $0$ & $0$ & $0$ & \multicolumn{1}{|r}{$39$} & $397$ & $2219$ & $8943$ & $28532$ & $73884$ & $150254$ & $\frac{397011}{2}$ & $-80262$ & $-\frac{3442249}{2}$\\[2mm]
\cline{7-7}
$7$ & $0$ & $0$ & $0$ & $0$ & $0$ & $0$ & \multicolumn{1}{|r}{$-56$} & $-637$ & $-4011$ & $-18438$ & $-68313$ & $-211897$ & $-552462$ & $-1169735$ & $-1737302$\\[2mm]
\cline{8-8}
$8$ & $0$ & $0$ & $0$ & $0$ & $0$ & $0$ & $0$ & \multicolumn{1}{|r}{$76$} & $956$ & $6714$ & $34742$ & $146296$ & $522552$ & $1607014$ & $4237541$\\[2mm]
\cline{9-9}
$9$ & $0$ & $0$ & $0$ & $0$ & $0$ & $0$ & $0$ & $0$ & \multicolumn{1}{|r}{$-99$} & $-1365$ & $-10596$ & $-61002$ & $-287121$ & $-1152501$ & $-4024251$\\[2mm]
\cline{10-10}
$10$ & $0$ & $0$ & $0$ & $0$ & $0$ & $0$ & $0$ & $0$ & $0$ & \multicolumn{1}{|r}{$125$} & $1875$ & $15960$ & $101208$ & $525662$ & $2333070$\\[2mm]
\cline{11-11}
$11$ & $0$ & $0$ & $0$ & $0$ & $0$ & $0$ & $0$ & $0$ & $0$ & $0$ & \multicolumn{1}{|r}{$-154$} & $-2497$ & $-23144$ & $-160292$ & $-909502$\\[2mm]
\cline{12-12}
$12$ & $0$ & $0$ & $0$ & $0$ & $0$ & $0$ & $0$ & $0$ & $0$ & $0$ & $0$ & \multicolumn{1}{|r}{$186$} & $3242$ & $32521$ & $244227$\\[2mm]
\cline{13-13}
$13$ & $0$ & $0$ & $0$ & $0$ & $0$ & $0$ & $0$ & $0$ & $0$ & $0$ & $0$ & $0$ & \multicolumn{1}{|r}{$-221$} & $-4121$ & $-44499$\\[2mm]
\hline\hline
\end{tabular}
\vspace{1cm}
\caption{
   Modified coefficients $\Delta^{\rm P}_k(m)$ for 
   cylindrical boundary conditions 
   for $2\leq k \leq 16$ and $2\leq m\leq 13$.
   For $k=1$ and $2\leq m \leq 13$, the coefficients 
   $\Delta^{\rm P}_k(m)$ vanish. 
   Those data points below the stair-case-like line satisfy
   $m\geq \widetilde{m}_{\rm min}^{\rm P}(k)=\max(k,2)$
   and are therefore zero.
}
\label{table_diff_coef_ctilde_P}
\end{sidewaystable}

%
%
\def\pp{\phantom{$-$}}
\clearpage
\begin{table}
\centering
\scriptsize
\begin{tabular}{|l|l|l|l|}
\hline\hline
\multicolumn{1}{|c|}{$k$} & 
\multicolumn{1}{|c|}{$\alpha_k$} &  
\multicolumn{1}{|c|}{$\beta_k$} &
\multicolumn{1}{|c|}{$\gamma_k$}\\
\hline 
0  &  \pp1               & \pp1               & \pp1                \\
1  &  \pp0               & \pp1               & \pp0                \\
2  &  \pp0               & \pp0               & \pp0                \\
3  &  \pp1               & $-$1                 & \pp1                \\
4  &  \pp0               & $-$1                 & \pp0                \\
5  &  \pp0               & \pp0               & \pp0                \\
6  &  \pp0               & \pp1               & \pp0                \\
7  &  \pp1               & $-$1                 & \pp4                \\
8  &  \pp3               & $-$8                 & \pp12               \\
9  &  \pp4               & $-$16                & \pp20               \\
10  &  \pp3              &  $-$16               &  \pp28              \\
11  &  \pp3              &  $-$12               &  \pp67              \\
12  &  \pp11             &  $-$41               &  \pp208             \\
13  &  \pp24             &  $-$138              &  \pp484             \\
14  &  \pp8              &  $-$210              &  \pp753             \\
15  &  $-$91               &  \pp47             &  \pp750             \\
16  &  $-$261              &  \pp849            &  \pp679             \\
17  &  $-$290              &  \pp1471           &  \pp2320            \\
18  & \pp254             &  $-$493              &  \pp10020           \\
19  &  \pp1671           &  $-$8052             &  \pp30548           \\
20  &  \pp3127           &  $-$19901            &  \pp68832           \\
21  &  \pp786            &  $-$19966            &  \pp108744          \\
22  &  $-$13939            &  \pp37556          &  \pp65229           \\
23  &  $-$49052            &  \pp223807         &  $-$236055            \\
24  &  $-$80276            &  \pp508523         &  $-$739289            \\
25  &  \pp21450          &  \pp321314         &  \pp101404          \\
26  &  \pp515846         &  $-$2052462          &  \pp7201383         \\
27  &  \pp1411017        &  $-$8417723          &  \pp26255714        \\
28  &  \pp1160761        &  $-$13374892         &  \pp43505098        \\
29  &  $-$4793764          &  \pp10841423       &  $-$17552274          \\
30  &  $-$20340586         &  \pp112595914      &  $-$291420026         \\
31  &  $-$29699360         &  \pp260687001      &  $-$674637832         \\
32  &  \pp33165914       &  \pp70989018       &  \pp27442           \\
33  &  \pp256169433      &  $-$1341964856       &  \pp4426763291      \\
34  &  \pp495347942      &  $-$4108283969       &  \pp12910062402     \\
35  &  $-$127736296        &  $-$3304416038       &  \pp9737827437      \\
36  &  $-$3068121066       &  \pp14960606999    &  $-$49131891078       \\
37  &  $-$7092358808       &  \pp58237169596    &  $-$184470253912      \\
38  &  $-$1024264966       &  \pp65268280922    &  $-$183956055539      \\
39  &  \pp35697720501    &  $-$162368154719     &  \pp621518352427    \\
40  &  \pp91243390558    &  $-$767619757924     &  \pp2660084937207   \\
41  &  \pp25789733672    &  $-$975329692910     &  \pp3075466954690   \\
42  &  $-$420665229170     &  \pp1872486336165  &  $-$7500763944932     \\
43  &  $-$1089052872105    &  \pp9701425034093  &  $-$34822638005931    \\
44  &  $-$238516756366     &  \pp12262136381593 &  $-$38841312202313    \\
45  &  \pp5101697398582  &  $-$24192583755347   &  \pp104412348649015 \\
46  &  \pp12146149238921 &  $-$118764516172484  &  \pp448320847685638 \\
47  &  $-$598931311074     &  $-$130312353695974  &                     \\
\hline\hline
\end{tabular}
\caption{\label{table_series}
Large-$q$ series expansions for the
bulk, surface and corner free energies
of the square-lattice zero-temperature Potts antiferromagnet,
in terms of the variable $z=1/(q-1)$.
The coefficients are defined by
$e^{f_{\rm bulk}}= [(q-1)^2/q] \sum_{k=0}^\infty \alpha_k z^k$,
$e^{f_{\rm surf}}=\sum_{k=0}^\infty \beta_k z^k$ 
and $e^{f_{\rm corner}}=\sum_{k=0}^\infty \gamma_k z^k$. 
The coefficients $\alpha_k$ for $k \le 36$
were obtained in Ref.~\cite{Bakaev_94};
all the other coefficients are new.
The coefficient $\beta_{47}$ is conjectural.
}
\end{table}

%
%
\clearpage
\begin{sidewaystable}
\centering
\begin{tabular}{|r||c|c|c|}
\hline\hline
$N$ & $z_1$ & $z_2$ & $z_3$ \\
\hline
30 & $-0.27(1) \pm 0.773(7)\,i$ &  & $0.502(3) \pm 0.292(5)\,i$\\ 
    & $0.6(2)$ &  & $0.2(2)$\\ 
\hline
40 & $-0.2811(5) \pm 0.7757(7)\,i$ & $0.45(2) \pm 0.53(1)\,i$ & $0.501(2) \pm 0.289(2)\,i$\\ 
    & $0.47(3)$ & $0.4(5)$ & $0.4(1)$\\ 
\hline
50 & $-0.2811(2) \pm 0.7752(1)\,i$ & $0.452(6) \pm 0.538(5)\,i$ & $0.49999(2) \pm 0.28868(2)\,i$\\ 
    & $0.50(1)$ & $0.3(2)$ & $0.501(5)$\\ 
\hline
60 & $-0.28112(3) \pm 0.77520(2)\,i$ & $0.4480(9) \pm 0.538(1)\,i$ & $0.499997(5) \pm 0.288676(6)\,i$\\ 
    & $0.500(3)$ & $0.48(7)$ & $0.501(2)$\\ 
\hline
70 & $-0.281118(3) \pm 0.775201(4)\,i$ & $0.44778(4) \pm 0.53827(8)\,i$ & $0.499999(2) \pm 0.288675(1)\,i$\\ 
    & $0.4999(6)$ & $0.500(4)$ & $0.5001(5)$\\ 
\hline
80 & $-0.281118(2) \pm 0.775201(2)\,i$ & $0.447783(5) \pm 0.538251(8)\,i$ & $0.500000(1) \pm 0.2886751(1)\,i$\\ 
    & $0.5000(3)$ & $0.5000(7)$ & $0.5001(4)$\\ 
\hline
90 & $-0.281118(2) \pm 0.775201(2)\,i$ & $0.447784(1) \pm 0.538250(2)\,i$ & $0.4999997(9) \pm 0.2886752(8)\,i$\\ 
    & $0.5000(3)$ & $0.4999(3)$ & $0.5001(3)$\\ 
\hline
100 & $-0.2811174(5) \pm 0.7752009(6)\,i$ & $0.447784(1) \pm 0.538249(2)\,i$ & $0.4999998(8) \pm 0.2886752(8)\,i$\\ 
    & $0.5000(1)$ & $0.5000(2)$ & $0.5000(2)$\\ 
\hline
Exact& $-0.2811172691 \pm 0.7752009092\,i$ &$0.44778393657 \pm 0.5382490441\,i$ &$0.5\pm 0.2886751346\,i$ \\ 
     & $0.5$ &$0.5$ &$0.5$ \\ 
\hline
$|z_i|$ & $0.8245989138$ &$0.7001589015$ &$0.5773502692$ \\ 
\hline\hline
\end{tabular}
\caption{\label{table_sq_3F_series1}
Estimates of the position and exponent of the ``correct'' singularities for the
square-lattice strip of width $L=3$ and free boundary conditions.
The column $N$ indicates the maximum order of the series in $z=1/(q-1)$
 that was used in the analysis.
The row marked `Exact' shows the known exact results.
The last row (marked $|z_i|$) shows
the absolute value of the corresponding exact singularity.
Blank entries mean that we were unable to obtain the corresponding estimates.
}
\end{sidewaystable}

\clearpage
%
%
\begin{table}
\centering
\begin{tabular}{|r||c|c|c|c|}
\hline\hline
$N$ & $z_4$ & $z_5$ & $z_6$ & $z_7$\\
\hline
30 & $0.36(1) \pm 0.66(2)\,i$ & $-1.000(5)$ &  & \\ 
    & $0.4(5)$ & $-0.1(2)$ &  & \\ 
\hline
40 & $0.368(7) \pm 0.654(8)\,i$ & $-1.0000(1)$ &  & \\ 
    & $0.0(2)$ & $-0.001(6)$ &  & \\ 
\hline
50 & $0.3695(6) \pm 0.6524(1)\,i$ & $-1.0000(1)$ &  & \\ 
    & $-0.03(5)$ & $0.000(6)$ &  & \\ 
\hline
60 & $0.3695(2) \pm 0.6516(4)\,i$ & $-1.00000(6)$ &  & \\ 
    & $-0.01(3)$ & $0.000(4)$ &  & \\ 
\hline
70 & $0.3695(1) \pm 0.6514(2)\,i$ & $-1.00000(2)$ &  & $0.487(5) \pm 0.427(4)\,i$\\ 
    & $0.00(1)$ & $0.000(2)$ &  & $2.03(9)$\\ 
\hline
80 & $0.36947(2) \pm 0.65136(2)\,i$ & $-1.00000(2)$ & $0.59(5)$ & $0.488(2) \pm 0.428(1)\,i$\\ 
    & $-0.001(3)$ & $0.000(2)$ & $0.0(6)$ & $2.01(4)$\\ 
\hline
90 & $0.369460(3) \pm 0.651363(2)\,i$ & $-1.00000(2)$ & $0.595(1)$ & $0.4878(5) \pm 0.4277(2)\,i$\\ 
    & $0.0001(6)$ & $0.000(2)$ & $0.001(1)$ & $2.000(2)$\\ 
\hline
100 & $0.369460(3) \pm 0.651363(3)\,i$ & $-1.000000(9)$ & $0.5945(4)$ & $0.4877(4) \pm 0.4277(2)\,i$\\ 
    & $0.0000(6)$ & $0.0000(9)$ & $0.000(8)$ & $2.000(2)$\\ 
\hline
\hline
\end{tabular}
\caption{\label{table_sq_3F_series2}
Estimates of the position and exponent of the spurious singularities for the
square-lattice strip of width $L=3$ and free boundary conditions.
The column $N$ indicates the maximum order of the series in $z=1/(q-1)$
that was used in the analysis.
Blank entries mean that we were unable to obtain the corresponding estimates.
}
\end{table}

%
%
\clearpage
\begin{sidewaystable}
\centering
\scriptsize
\begin{tabular}{|r||c|c|c|c|c|}
\hline\hline
$N$ & $z_1$ & $z_2$ & $z_3$ & $z_4$ & $z_5$ \\
\hline
30 & $0.331(2) \pm 0.238(2)\,i$ & $-0.068(1) \pm 0.4795(8)\,i$ & $0.29(2) \pm 0.49(3)\,i$ &  & \\ 
    & $0.5(2)$ & $0.53(8)$ & $-0.3(3)$ &  & \\ 
\hline
40 & $0.3315(4) \pm 0.2370(4)\,i$ & $-0.0682(2) \pm 0.4798(2)\,i$ & $0.30(2) \pm 0.48(2)\,i$ &  & \\ 
    & $0.53(8)$ & $0.51(3)$ & $-0.4(6)$ &  & \\ 
\hline
50 & $0.33162(6) \pm 0.23712(5)\,i$ & $-0.068171(3) \pm 0.479859(4)\,i$ & $0.30(1) \pm 0.48(3)\,i$ &  & \\ 
    & $0.50(2)$ & $0.5005(9)$ & $-0.2(6)$ &  & \\ 
\hline
60 & $0.331635(2) \pm 0.237117(2)\,i$ & $-0.0681711(4) \pm 0.4798607(9)\,i$ & $0.30(2) \pm 0.46(3)\,i$ &  & \\ 
    & $0.4995(9)$ & $0.5001(3)$ & $0.3(13)$ &  & \\ 
\hline
70 & $0.3316353(8) \pm 0.237116(1)\,i$ & $-0.0681712(2) \pm 0.4798609(5)\,i$ & $0.29(2) \pm 0.46(3)\,i$ &  & \\ 
    & $0.4998(4)$ & $0.5000(1)$ & $0.4(12)$ &  & \\ 
\hline
80 & $0.3316354(7) \pm 0.2371156(1)\,i$ & $-0.0681712(2) \pm 0.4798609(2)\,i$ & $0.29(1) \pm 0.45(2)\,i$ & $0.76(3)$ & \\ 
    & $0.4998(4)$ & $0.50000(6)$ & $0.6(10)$ & $-0.9(6)$ & \\ 
\hline
90 & $0.3316354(6) \pm 0.2371155(9)\,i$ & $-0.0681712(2) \pm 0.4798609(2)\,i$ & $0.28144(4) \pm 0.45208(5)\,i$ & $0.75(1)$ & \\ 
    & $0.4999(3)$ & $0.50001(7)$ & $0.503(6)$ & $-0.7(4)$ & \\ 
\hline
100 & $0.3316354(7) \pm 0.237116(1)\,i$ & $-0.0681712(1) \pm 0.4798609(3)\,i$ & $0.281427(4) \pm 0.452108(3)\,i$ & $0.754(9)$ & \\ 
    & $0.4999(4)$ & $0.50000(9)$ & $0.5000(4)$ & $-0.7(3)$ & \\ 
\hline
Exact& $0.3316354418\pm 0.2371152471\,i$ &$-0.0681712693 \pm 0.4798609413\,i$ &$0.2814289723 \pm 0.4521062477\,i$ &$0.7398155434$ &$0.7978491474$ \\ 
     & $0.5$ &$0.5$ &$0.5$ &$0.5$ &$0.5$ \\ 
\hline
$|z_i|$ & $0.4076833412$ &$0.4846791154$ &$0.5325432618$ &$0.7398155434$ &$0.7978491474$ \\ 
\hline\hline
\end{tabular}
\caption{\label{table_sq_4P_series1}
Estimates of the position and exponent of the ``correct'' singularities for the
square-lattice strip of width $L=4$ and periodic boundary conditions.
The column $N$ indicates the maximum order of the series in $z=1/(q-1)$
 that was used in the analysis.
The row marked `Exact' shows the known exact results.
The last row (marked $|z_i|$) shows
the absolute value of the corresponding exact singularity.
}
\end{sidewaystable}

\clearpage
%
%
\begin{table}
\centering
\begin{tabular}{|r||c|c|}
\hline\hline
$N$ & $z_6$ & $z_7$\\
\hline
30 & $-0.999(4)$ & \\ 
    & $0.0(1)$ & \\ 
\hline
40 & $-1.000(1)$ & \\ 
    & $0.00(3)$ & \\ 
\hline
50 & $-1.00000(4)$ & $0.30(2) \pm 0.47(2)\,i$\\ 
    & $0.000(2)$ & $0.3(12)$\\ 
\hline
60 & $-1.00000(1)$ & $0.29(1) \pm 0.46(2)\,i$\\ 
    & $0.0002(9)$ & $0.5(10)$\\ 
\hline
70 & $-0.999999(9)$ & $0.29(2) \pm 0.46(3)\,i$\\ 
    & $0.0001(5)$ & $0.3(10)$\\ 
\hline
80 & $-1.00000(1)$ & $0.286(2) \pm 0.517(2)\,i$\\ 
    & $0.0001(7)$ & $0.0(1)$\\ 
\hline
90 & $-1.000000(2)$ & $0.2854(4) \pm 0.5180(4)\,i$\\ 
    & $0.0000(2)$ & $0.00(3)$\\ 
\hline
100 & $-1.000000(4)$ & $0.28528(4) \pm 0.51814(6)\,i$\\ 
    & $0.0000(4)$ & $0.000(4)$\\ 
\hline
\hline
\end{tabular}
\caption{\label{table_sq_4P_series2}
Estimates of the position and exponent of the spurious singularities for the
square-lattice strip of width $L=4$ and periodic boundary conditions.
The column $N$ indicates the maximum order of the series in $z=1/(q-1)$
 that was used in the analysis.
Blank entries mean that we were unable to obtain the corresponding estimates.
}
\end{table}

%
%
%
%
\clearpage
\begin{table}
\centering
\begin{tabular}{rllc}
\hline\hline
$m_{\rm F}$  & \multicolumn{1}{c}{$\Re q_0$} & 
               \multicolumn{1}{c}{$|\Im q_0|$} & Type \\
\hline  
$3_{\rm F}$  &$2$            & $0$            & S\\ 
$4_{\rm F}$  &$2.2283590792$ & $0$            & +\\ 
$5_{\rm F}$  &$2.4284379020$ & $0$            & S\\ 
$6_{\rm F}$  &$2.5286467909$ & $0$            & +\\
$7_{\rm F}$  &$2.6062482130$ & $0$            & S\\ 
$8_{\rm F}$  &$2.6602596967$ & $0.0007413717$ & D\\ 
$9_{\rm F}$  &$2.7016599568$ & $0$            & S\\ 
$10_{\rm F}$ &$2.7343903604$ & $0.0003924978$ & D\\ 
$11_{\rm F}$ &$2.7608973951$ & $0$            & S\\ 
$12_{\rm F}$ &$2.782817590$  & $0.00018700$   & D\\ 
\hline\hline
\end{tabular}

\vspace{1cm}
\caption{
   Values of $q_0(m)$ for square-lattice strips with free boundary conditions.
   Type ``S'' means that the limiting curve $\scrb_m$ crosses the real axis
   at a single point, ``D'' means that the limiting curve does not cross
   the real $q$-axis but has two complex-conjugate endpoints nearby,  
   and ``+'' means that the limiting curve contains a segment
   on the real $q$-axis (we define $q_0$ to be the lower end of that segment).
   For points of type ''D'', we also show the imaginary part.
   The data for $m\leq 7$ are taken from \protect\cite{transfer1}.
   Notice that the value for $m=8$ reported in \protect\cite{transfer1}
   is wrong; the correct value is displayed here.
}
\label{table_summary}
\end{table}

%
%
\clearpage
\begin{table}
\centering
\begin{tabular}{rll}
\hline\hline
\multicolumn{1}{c}{$n$} & \multicolumn{1}{c}{4th Zero - $B_5$} & 
                          \multicolumn{1}{c}{5th Zero} \\
\hline 
 30  & $-8.9459389900\times 10^{-3}$ &  \\
 32  & $-5.6487223660\times 10^{-3}$ &  \\
 34  & $-3.1962919070\times 10^{-3}$ &  \\
 36  & $-1.5712164060\times 10^{-3}$ &  \\
 38  & $-6.6617400470\times 10^{-4}$ &  \\
 39  & \phantom{$-$}$5.8575529850\times 10^{-4}$ & $2.6298556468$ \\ 
 40  & $-2.5161135990\times 10^{-4}$ &  \\
 41  & \phantom{$-$}$1.7270275260\times 10^{-4}$ & $2.6345567384$ \\ 
 42  & $-8.8890257250\times 10^{-5}$ &  \\
 43  & \phantom{$-$}$5.5346483490\times 10^{-5}$ & $2.6384105545$ \\ 
 44  & $-3.0434570840\times 10^{-5}$ &  \\
 45  & \phantom{$-$}$1.8199961300\times 10^{-5}$ & $2.6417380816$ \\ 
 46  & $-1.0281765140\times 10^{-5}$ &  \\
 47  & \phantom{$-$}$6.0420071920\times 10^{-6}$ & $2.6446771074$ \\ 
 48  & $-3.4539716110\times 10^{-6}$ &  \\
 49  & \phantom{$-$}$2.0135551740\times 10^{-6}$ & $2.6473078802$ \\ 
 50  & $-1.1574596530\times 10^{-6}$ &  \\
 51  & \phantom{$-$}$6.7216127470\times 10^{-7}$ & $2.6496845069$ \\ 
 52  & $-3.8744221380\times 10^{-7}$ &  \\
 53  & \phantom{$-$}$2.2455458550\times 10^{-7}$ & $2.6518466022$ \\ 
 54  & $-1.2962041990\times 10^{-7}$ &  \\
 55  & \phantom{$-$}$7.5047689080\times 10^{-8}$ & $2.6538246782$ \\ 
 56  & $-4.3353170180\times 10^{-8}$ &  \\
 57  & \phantom{$-$}$2.5086443380\times 10^{-8}$ & $2.6556430336$ \\ 
 58  & $-1.4497910190\times 10^{-8}$ &  \\
 59  & \phantom{$-$}$8.3866240660\times 10^{-9}$ & $2.6573214747$ \\ 
 60  & $-4.8479232330\times 10^{-9}$ &  \\
 61  & \phantom{$-$}$2.8038883660\times 10^{-9}$ & $2.6588764247$ \\ 
 62  & $-1.6210151380\times 10^{-9}$ &  \\
 63  & \phantom{$-$}$9.3745081850\times 10^{-10}$ & $2.6603216811$ \\ 
 64  & $-5.4201053100\times 10^{-10}$ &  \\
 65  & \phantom{$-$}$3.1343270220\times 10^{-10}$ & $2.6616689582$ \\ 
 66  & $-1.8122675760\times 10^{-10}$ &  \\
 67  & \phantom{$-$}$1.0479599710\times 10^{-10}$ & $2.6629282884$ \\ 
 68  & $-6.0594535690\times 10^{-11}$ &  \\
 69  & \phantom{$-$}$3.5038676450\times 10^{-11}$ & $2.6641083276$ \\ 
 70  & $-2.0260148680\times 10^{-11}$ &  \\
\hline
$\infty$ & $\phantom{-}0$                      & $2.7016599568$ \\
\hline\hline
\end{tabular}
\vspace{1cm}
\caption{
   Fourth and fifth real chromatic zeros for a strip 
   $9_{\rm F} \times n_{\rm F}$ with free boundary conditions. 
   As the fourth zero converges rapidly to $B_5$, we show their 
   difference. 
   The last row (labelled with $n=\infty$) shows the
   infinite-length limit; in particular, the value of $q_0(9)$.
}
\label{table_zeros_9F}
\end{table}

%
%
\clearpage
\begin{table}
\centering
\begin{tabular}{rll}
\hline\hline
\multicolumn{1}{c}{$n$} & \multicolumn{1}{c}{4th Zero - $B_5$} & 
                          \multicolumn{1}{c}{5th Zero} \\
\hline 
 14  & $-3.6516138320\times 10^{-2}$ &  \\
 16  & $-1.9375298640\times 10^{-2}$ &  \\
 18  & $-8.1115634110\times 10^{-3}$ &  \\
 20  & $-2.2141892360\times 10^{-3}$ &  \\
 22  & $-3.6952234840\times 10^{-4}$ &  \\
 23  & \phantom{$-$}$2.7283542170\times 10^{-4}$ & $2.6360127941$ \\ 
 24  & $-4.7440730960\times 10^{-5}$ &  \\
 25  & \phantom{$-$}$2.3461500250\times 10^{-5}$ & $2.6467690175$ \\ 
 26  & $-5.5660625770\times 10^{-6}$ &  \\
 27  & \phantom{$-$}$2.2973640450\times 10^{-6}$ & $2.6554417441$ \\ 
 28  & $-6.2842068380\times 10^{-7}$ &  \\
 29  & \phantom{$-$}$2.3500000310\times 10^{-7}$ & $2.6628475620$ \\ 
 30  & $-6.9454825890\times 10^{-8}$ &  \\
 31  & \phantom{$-$}$2.4549166270\times 10^{-8}$ & $2.6693333011$ \\ 
 32  & $-7.5798800910\times 10^{-9}$ &  \\
 33  & \phantom{$-$}$2.5935418370\times 10^{-9}$ & $2.6750988717$ \\ 
 34  & $-8.2103756740\times 10^{-10}$ &  \\
 35  & \phantom{$-$}$2.7571779370\times 10^{-10}$ & $2.6802756665$ \\ 
 36  & $-8.8542077070\times 10^{-11}$ &  \\
 37  & \phantom{$-$}$2.9415010300\times 10^{-11}$ & $2.6849567990$ \\ 
 38  & $-9.5240493620\times 10^{-12}$ &  \\
 39  & \phantom{$-$}$3.1444559750\times 10^{-12}$ & $2.6892116307$ \\ 
 40  & $-1.0229343110\times 10^{-12}$ &  \\
 41  & \phantom{$-$}$3.3652761350\times 10^{-13}$ & $2.6930937453$ \\ 
 42  & $-1.0977435000\times 10^{-13}$ &  \\
 43  & \phantom{$-$}$3.6039742950\times 10^{-14}$ & $2.6966457592$ \\ 
 44  & $-1.1774405610\times 10^{-14}$ &  \\
 45  & \phantom{$-$}$3.8610612020\times 10^{-15}$ & $2.6999024444$ \\ 
 46  & $-1.2625636940\times 10^{-15}$ &  \\
 47  & \phantom{$-$}$4.1373846520\times 10^{-16}$ & $2.7028928687$ \\ 
 48  & $-1.3536187010\times 10^{-16}$ &  \\
 49  & \phantom{$-$}$4.4340363840\times 10^{-17}$ & $2.7056419158$ \\ 
 50  & $-1.4511035420\times 10^{-17}$ &  \\
 51  & \phantom{$-$}$4.7522986580\times 10^{-18}$ & $2.7081713766$ \\ 
 52  & $-1.5555245880\times 10^{-18}$ &  \\
 53  & \phantom{$-$}$5.0936147060\times 10^{-19}$ & $2.7105007206$ \\ 
 54  & $-1.6674076830\times 10^{-19}$ &  \\
\hline
$\infty$ & $\phantom{-}0$   & $2.7608973951$ \\
\hline\hline
\end{tabular}
\vspace{1cm}
\caption{
   Fourth and fifth real chromatic zeros for a strip 
   $11_{\rm F} \times n_{\rm F}$ with free boundary conditions. 
   As the third zero converges rapidly to $B_5$, we show their 
   difference. The last row (labelled with $n=\infty$) shows the
   infinite-length limit; in particular, the value of $q_0(11)$. 
}
\label{table_zeros_11F}
\end{table}

%
%
%
%
\clearpage

%
%
\clearpage

%
%
\clearpage

%
%
\clearpage

%
%
\clearpage

%
%
\clearpage

%
%
\clearpage

%
%
\clearpage


\end{document}